%% file: WertepnyPhDThesis.tex
\pdfoutput=1 % added in 
\documentclass[11pt, phd]{osudissert96} 

\usepackage{amsmath, amssymb, amsthm,mathtools}
\usepackage{graphicx} 
\usepackage{float}
\usepackage{rotating} % for rotating figures/tables

\usepackage{multirow} % for multirow figures
\usepackage{caption}
\usepackage{subcaption}
\usepackage{bm,bbm} % for bold math---useful
\usepackage{booktabs} % for better looking tables
\usepackage{slashed} % Dirac slash notation
\usepackage{dsfont}
\usepackage{bbm}
\usepackage{bbold}
\usepackage[silent,margin]{fixme}
\usepackage{bookmark} % bookmarks in pdf file
\hypersetup{colorlinks=true,linkcolor=blue} 

\usepackage{cite} % for backreferenced citations - supercedes \usepackage[numbers, sort&compress]{natbib} 
\usepackage[xindy,acronym, section=chapter]{glossaries}
\glsdisablehyper
\makeglossaries

\graphicspath{{figures/}}

\input{macros.tex}
\input{lengths.tex}

\newcommand{\TmpCiteM}[1]{}

\renewcommand{\eqref}[1]{\textup{\ref{#1}}}
\title{
	\texorpdfstring{
		Two-Particle Correlations in Heavy-Light Ion Collisions
	}{
		Two-Particle Correlations in Heavy-Light Ion Collisions
	}
}
\author{Douglas Evan Wertepny}
\advisorname{Professor Yuri Kovchegov}
\degree{Doctor of Philosophy} % Default value
\member{Professor Ulrich Heinz}
\member{Professor Michael Lisa}
\member{Professor Ciriyam Jayaprakash}
\authordegrees{B.S.}
\graduationyear{2016}
\unit{Physics} % defaults to ``Graduate Program in Physics''

\begin{document}

\maketitle
\disscopyright
% \frontmatter
\clearpage
% \listoffixmes
% \clearpage

\include{Wertepny_Abs}
\dedication{To my loving family. I couldn't have done this without your support.}
\include{Wertepny_Ack}
\include{Wertepny_Vita}

\tableofcontents 

% list of figures (comment out if you don't have any figures)
\clearpage
\listoffigures 

% list of tables (comment out if you don't have any tables)
\clearpage
\listoftables 

% \mainmatter
\startdoublespace
% \startsinglespace

\include{Chapter_Introduction}
\include{Chapter_Setup}
\include{Chapter_pA_review}
\include{Chapter_Calculation}
\include{Chapter_Properties} 
\include{Chapter_Conclusions}

% \backmatter
% We use BIBTeX for the bibliography---you don't have to
% \bibliographystyle{natureURL} % use your favorite BIBTeX style
\bibliographystyle{unsrt} % use your favorite BIBTeX style
%\bibliographystyle{opcit} % use your favorite BIBTeX style
%\renewcommand\bibname{Master list of references} % rename bibliography 
% "You may decide how this section should be titled. The terms References or Bibliography are the most commonly chosen titles."
%\nocite{*} % To display all refs, even uncited refs (useful when editting)
\bibliography{references}
% can use multiple .bib files, but no white space allowed for multiple included references, and ".bib" cannot be included

% If for some reason you are anti-BIBTeX, then you would use the% following instead of the above:
%\begin{thebibliography}{99}
% ...
%\end{thebibliography}

% Note: GS 2010 requires bibliography/references _before_ the appendix
% if you believe their guidelines; however, conversations with GS
% staff suggests _they don't care_. Go figure. So do what you like.

\appendix
\input{Appendix_Wilson}

\end{document}

%% file: macros.tex
\newcommand{\as}{\alpha_s}

\newcommand{\bea}{\begin{eqnarray}}
\newcommand{\eea}{\end{eqnarray}}
\newcommand{\beq}{\begin{equation}}
\newcommand{\eeq}{\end{equation}}
\newcommand{\ben}{\begin{equation*}}
\newcommand{\een}{\end{equation*}}
\newcommand{\be}{\begin{equation}}
\newcommand{\ee}{\end{equation}}

\newcommand{\fig}[1]{\ref{#1}}

\renewcommand{\imath}{\ensuremath{i}}

%% file: lengths.tex
\newlength{\movieplotwidth}
\newlength{\phaseplotwidth}
\newlength{\universalityplotwidth}
\newlength{\wideplotwidth}

%% file: Wertepny_Abs.tex
% !TEX root = WertepnyPhDThesis.tex

% Need to focus on what we did which is new
% Change it so not exactly same as papers

\begin{abstract}

We studied the initial, high-energy scatterings in heavy ion collisions
using the saturation/Color Glass Condensate framework, with a focus on two-particle long-range rapidity correlations.
These two-particle correlations are modeled as two-gluon correlations, assuming that the kinetic information of the gluons survives the evolution of the system and presents itself in the final state particles.
We calculate the two-gluon production cross section using the saturation framework in the heavy-light ion regime, where we consider all-order saturation effects in the heavy nucleus while considering only two-orders in saturation effects in the light ion.

  The two-gluon production cross section
  produces four types of correlations: (i)
  geometric correlations, (ii) Hanbury Brown and Twiss (HBT) like
  correlations accompanied by a back-to-back maximum,
  (iii) near-side correlations, and (iv)
  away-side azimuthal correlations. All of these correlations are long-range in
  rapidity. The geometric correlations (i) are due to the fact that
  nucleons are correlated by simply being confined within the same
  nucleus and may lead to long-range rapidity correlations for the
  produced particles without strong azimuthal angle dependence.
  Long-range rapidity correlations (iii) and
  (iv) have exactly the same amplitudes along with azimuthal and
  rapidity shapes: one centered around $\Delta \phi =0$ and the
  other one centered around $\Delta \phi =\pi$ (here $\Delta \phi$ is
  the azimuthal angle between the two produced gluons).

  The geometry dependence of the correlation function
  (not to be confused with the geometric correlations) leads
  to stronger azimuthal near- and away-side correlations in the
  tip-on-tip U+U collisions than in the side-on-side U+U collisions,
  an exactly opposite behavior from the correlations generated by the
  elliptic flow of the quark-gluon plasma: a study of azimuthal
  correlations in the U+U collisions may thus help to disentangle the
  two sources of correlations.
  Finally we rewrite our result for the two-gluon production cross-section in a
  $k_T$-factorized form resulting in a factorized expression
  involving a convolution of one- and two-gluon Wigner distributions
  over both the transverse momenta and impact parameters.
  This differs from the normal $k_T$-factorized forms used in the literature and may have implications for the nature of $k_T$-factorization beyond the two-gluon production case.

\end{abstract}

%% file: Wertepny_Ack.tex
% !TEX root = WertepnyPhDThesis.tex
\begin{acknowledgments}

I would like to thank Professor Yuri Kovchegov for all of the hard work and time he has invested into my physics career.
Starting from when I first took your Quantum Field Theory class I have always respected you as a physicist.
I was thrilled when you took me on as one of your students and have since learned much from you.
Without your guidance I do not know where I would be at this time.
I would also like to thank all of the other professors at OSU whom I have learned much from.

I would like to thank all of the support staff in the physics department.
You all work behind the scenes to make the entire department function.
I would like to give a special thanks to Kris Dunlap for all of the help you have given me over the years.
From switching research groups to dealing with other incidents you have been vital in helping me get through it all and for that I thank you.

I would also like to thank my family for all the support they have given me over the years.
From the time I was born you have always been supportive of me and have pushed me to work hard.
You have always been there to help in good times and bad and I love you all for it.
Thanks for being there for me, I could not have done this without you.

This material is based upon work supported by the U.S. Department of Energy, Office of Science, Office
of Nuclear Physics under Award Number DE-SC0004286.
\end{acknowledgments}

%% file: Wertepny_Vita.tex
% !TEX root = WertepnyPhDThesis.tex
\begin{vita}
% \small
\dateitem{May 2009}{B.S. in Physics and Music, The University of Dayton, Dayton, Ohio}
\begin{publist}
\raggedright
\pubitem{
  \noindent
  \textit{
    Regularization of the Light-Cone Gauge Gluon Propagator Singularities Using Sub-Gauge Condition
  }\\
  G.~A.~Chirilli,~Y.~V.~Kovchegov,~and~D.~E.~Wertepny,~JHEP.~\textbf{1512}, 138, (2015)
}
\pubitem{
  \noindent
  \textit{
    Classical Gluon Production Amplitude for Nucleus-Nucleus Collisions: First Saturation Correction in the Projectile
  }\\
  G.~A.~Chirilli,~Y.~V.~Kovchegov,~and~D.~E.~Wertepny,~JHEP.~\textbf{1503},~015~(2015)
}
\pubitem{
  \noindent
  \textit{
    Two-Gluon Correlations in Heavy-Light Ion Collisions: Energy and Geometry Dependence, IR Divergences, and kT-Factorization
  }\\
  Y.~V.~Kovchegov~and~D.~E.~Wertepny,~Nucl.~Phys.~A~\textbf{925},~254-295~(2014)
}
\pubitem{
  \noindent
  \textit{
    Fast Neutron Production from Lithium Converters and Laser Driven Protons
  }\\
  M.~Storm,~S.~Jiang,~D.~E.~Wertepny,~et.~al.,~Phys.~Plasmas~\textbf{20},~053106~(2013)
}
\pubitem{
  \noindent
  \textit{
    Long-Range Rapidity Correlations in Heavy-Light Ion Collisions
  }\\
  Y.~V.~Kovchegov~and~D.~E.~Wertepny,~Nucl.~Phys.~A~\textbf{906},~50-83~(2013)
}
\pubitem{
  \noindent
  \textit{
    Pressure broadening and shift of the cesium D1 transition by the noble gases and N2, H2, HD, D2, CH4, C2H6, CF4 and 3He
  }\\
  G.~A.~Pitz,~D.~E.~Wertepny,~and~G.~P.~Perram,~Nucl.~Phys.~A~\textbf{80},~062718~(2009)
}
\end{publist}
\begin{fieldsstudy}
\majorfield{Physics}
\end{fieldsstudy}

\end{vita}

%% file: Chapter_Introduction.tex
% !TEX root = WertepnyPhDThesis.tex
\cleardoublepage
\chapter{Introduction}
\label{ch:Introduction}

This dissertation is based on and borrows heavily from the work done by the author in \cite{Kovchegov:2012nd,Kovchegov:2013ewa}.

%%%%%%%%%%%%%%%%%%%%%%%%%%%%%%%%%%%%%%%%%%%%%%%%%%%%%%%%
\input{Section_IntroQCD}
\input{Section_AA_Collision}
\input{Section_Saturation}
\input{Section_IntroCorrelations}
%%%%%%%%%%%%%%%%%%%%%%%%%%%%%%%%%%%%%%%%%%%%%%%%%%%%%%%%
\section{Goals and organization of thesis}

The aim of this work is to begin to analytically include saturation
effects into the two-gluon correlation function in nucleus--nucleus
collisions. Indeed full analytic inclusion of saturation effects
originating in both nuclei would be a very hard problem: even the
single gluon production in the quasi-classical MV limit of $AA$
collisions can be dealt with only numerically at present
\cite{Krasnitz:1999wc,Krasnitz:2003jw,Lappi:2003bi,Blaizot:2010kh}. To
make the problem more tractable we assume that one of the colliding
nuclei is much larger than the other one, such that saturation effects
are important only in interactions with the larger nucleus.
This is known as the heavy-light ion regime.

Before going into the calculation itself, there are a few topics that need to be addressed.
In Ch.~\ref{ch:Setup} we will cover some of these topics in detail.
In Sec.~\ref{sec:QCD} we review more in depth the QCD Lagrangian and discuss some aspects of gauge fixing.
Sec.~\ref{sec:aAregime} is reserved to describe and justify the heavy-light ion regime used in this work and to show exactly how this affects the two-gluon correlation function.

In Ch.~\ref{ch:pAreview} we review the known result for single-gluon production in $pA$ collisions within the saturation framework in the classical MV limit.
Doing this in detail serves as a pedagogical way of seeing how exactly this framework is used in practice and its overall implications.
At the same time many of the results derived here can be easily extended to more complicated situations, allowing us to derive the two-gluon correlations more efficiently.
The major topics covered are shock wave generation in the target nucleus (Sec.~\ref{sec:target}), with the analysis of the resulting shock wave presented in (Sec.~\ref{sec:Dipole}), the $pA$ single-gluon production cross section (Sec.~\ref{sec:single}) and its $k_T$-factorized form (Sec.~\ref{sec:pAfact}).

In Ch.~\ref{ch:Calculation}, using the same formalism described in the previous chapter (Ch.~\ref{ch:pAreview}), we calculate the two-gluon production cross section in the heavy-light ion regime.
There are three types of contributions.
The first type is known as the ``square" terms (diagrams), Sec.~\ref{subsec:ggsquare}, while the remaining two are called the ``crossed" terms (diagrams), Sec.~\ref{subsec:ggcrossed}.
With a close analysis of the resulting correlation function in Sec.~\ref{sec:lrrc}, we see three distinct types of long-range rapidity correlations:
one which is peaked in azimuthal angle on the near-side ($\Delta \phi = 0$),
an equal-and-opposite correlation on the away-side ($\Delta \phi = \pi$),
and a correlation that seems to be similar to a Hanbury Brown and Twiss (HBT) correlation which are peaked when the transverse momenta of the particles are the same and when they are equal and opposite ($\bm k_1 = \bm k_2$ and $\bm k_1 = - \bm k_2$).
All of the two-gluon correlations produced in this heavy-light ion regime are symmetric in azimuthal angle ($\Delta \phi$).

In Ch.~\ref{ch:Properties} we cover two more implications of the two-gluon production cross section and correlation function derived in Ch.~\ref{ch:Calculation}.
Sec.~\ref{sec:Geo} shows that in collisions of asymmetric ions (examples being uranium and gold) the effects of saturation give rise to correlations that behave differently than those originating from hydrodynamical flow.
We focus on uranium--uranium (U+U) collisions and show that there is an enhancement of two-particle correlations in tip-on-tip collisions when compared with side-on-side collisions, which is opposite of that predicted by hydrodynamics.
In Sec.~\ref{sec:fact} we derive the $k_T$-factorized form of the two-gluon production cross section in the heavy-light ion limit.
The end result is that, unlike the $pA$ case, the cross section cannot be written in terms of the usual unintegrated gluon distributions.
Instead we must introduce gluon distributions which are in fact Wigner distributions, depending on both transverse momentum and transverse position.
Finally in Ch.~\ref{ch:Conclusions} we conclude with a brief overview of the major results presented herein.

%% file: Section_IntroQCD.tex
% !TEX root = WertepnyPhDThesis.tex

\section{Introduction to QCD}
\label{sec:IntroQCD}

This section closely follows the pedagogical outline found in \cite{Yagi:2005yb}.

Quantum Chromodynamics (QCD) is the quantum theory that governs the behavior of the strong force
\cite{Yang:1954vj,GellMann:1964nj,Fritzsch:1973pi,Gross:1973id,Politzer:1973fx,Weinberg:1973un}.
Before getting into the details of QCD it is helpful to examine the theory of Quantum Electrodynamics (QED), the quantum mechanical extension of classical electrodynamics (for more on classical electrodynamics see \cite{jackson_classical_1999}).
Despite the fact that the physics of QCD is very different from the physics of QED there are many similarities between the two theories.

QED is an abelian U$(1)$ gauge theory and consists of two types of fundamental particles, photons and charged leptons.
The charged leptons are massive spin-$\frac{1}{2}$ particles (fermions). 
There are three types of charged leptons: electrons ($e^-$), muons ($\mu^-$) and tau leptons ($\tau^-$), each of which has a corresponding antiparticle ($e^+$, $\mu^+$, $\tau^+$) that has the same mass and spin but opposite charge.
Each of these are part of a different family and the mass of each particle is larger than the preceding one, $\tau$ being the heaviest ($1.78 \, \frac{GeV}{c^2}$ \cite{Agashe:2014kda}) and $e$ being the lightest ($0.511 \, \frac{MeV}{c^2}$ \cite{Agashe:2014kda}).
These particles each have electric charge $-e$ (the antiparticles have charge $+e$) and couple to electromagnetic fields which, when quantized, are photons ($\gamma$).

Photons are spin-1 massless particles (bosons) and are vector fields notated by $A_\mu$.
These are the gauge fields of QED and also known as the carriers of the electromagnetic force.
It is important to note that photons do not carry charge which means that photons cannot directly interact with each other, they can only interact through an intermediating particle, charged leptons in this case.
The coupling constant associated with QED is $\alpha_{EM} = \frac{e^2}{4 \pi \hbar c}$ and is fairly weak, on the order of $\mathcal{O}(10^{-2})$.
It should be noted that $\alpha_{EM}$, due to the quantum mechanical nature of the theory, changes as a function of energy or momentum scale.
This property is known as the running of the coupling.
In QED the running isn't very strong: $\alpha_{EM}$ increases with energy and, in experimentally relevant situations, it always remains small.

QCD follows a similar pattern with a few important distinctions.
It is a non-abelian SU$(3)$ gauge theory that contains of two types of fundamental particles, quarks and gluons.
There are 6 different types of quarks (known as flavors), each of which are massive spin-$\frac{1}{2}$ particles (fermions).
There are 3 families each consisting of two different quarks (each of which have their own anti-particles); up and down ($u$, $d$), charm and strange ($c$, $s$), and top and bottom ($t$, $b$).
Each family is heavier than the preceding family with $t$ being the heaviest ($170 \, \frac{GeV}{c^2}$ from direct measurements \cite{Agashe:2014kda}) and $u$ being the lightest ($2.3 \, \frac{MeV}{c^2}$ \cite{Agashe:2014kda}).
Quarks carry color charge, which has charge constant $g$, and couple to the color fields known as gluons.
What makes color charge different from electric charge is that there are three different types of color charge (also known simply as colors) red, blue and green ($r$, $g$, $b$) each of which has a corresponding anti-charge ($\bar r$, $\bar g$, $\bar b$).
This is in contrast to QED which has only a single type, the electric charge.
Quarks carry electric charge as well but this is unimportant for the current discussion, due to the fact that here we are focusing on the differences between QED and QCD.

Gluons are spin-1 massless particles (bosons) and makeup the gauge field of the theory.
They are vector fields notated by $A_\mu^a$, where $a$ refers to the color charge of the gluon.
Gluons, unlike quarks, have 8 possible color charges.
These can be thought of as a linear superposition of the color and anti-color charges quarks carry (for example a gluon can have the following combination: $\frac{r \bar b + b \bar r}{\sqrt 2}$).
One can also think of the different color charges the quarks and gluons carry as the fundamental and the adjoint representations of SU$(3)$ color group respectively.
Since gluons carry charge they can directly interact with one another.
This makes the physical implications of QCD drastically different than QED.
One important consequence of this is the strong coupling constant, defined as $\as = \frac{g^2}{4 \pi \hbar c}$,  runs significantly as a function of energy and momentum (and therefore distance scales) and decreases with energy.
Depending on the energy scale the coupling constant can be on the order of anywhere from $\mathcal{O}(1)$ to $\mathcal{O}(10^{-1})$.
A summary of the similarities and differences between QED and QCD is shown in Table~\ref{table:QEDvQCD}.

\begin{table}[h!]
\centering
\begin{tabular}{||c c c ||} 
 \hline
 & QED & QCD  \\ [0.5ex] 
 \hline\hline
 Gauge Group 			& Abelian		& Non-Abelian \\ [0.3ex] 
 		& U$(1)$ 		& SU$(3)$ \\ [0.3ex] 
 Charged Fermions 		& Charged Leptons 		& Quarks\\ [0.3ex]
 		& $e$, $\mu$, $\tau$ 		& $u$, $d$, $c$, $s$, $t$, $b$ \\ [0.3ex]
 Gauge Bosons 			& Photons, $\gamma$ 		& Gluons, $g$\\ [0.3ex]
 		& $A_\mu$ 	& $A^a_\mu$ \\ [0.3ex]
 Type of Charge 			& Electric 		& Color \\ [0.3ex]
		& $e$ 			& $r$, $b$, $g$ \\ [0.3ex]
 Coulping Constant		& $\alpha_{EM} = \frac{e^2}{4 \pi \hbar c}$ 		& $\as = \frac{g^2}{4 \pi \hbar c}$\\ [0.4ex]
 		& $\mathcal{O}(10^{-2})$		& $\mathcal{O}(1)$ to $\mathcal{O}(10^{-1})$ \\ [1ex] 
 \hline
\end{tabular}
\caption{Comparing various properties of QED and QCD. Adapted from \cite{Yagi:2005yb}.}
\label{table:QEDvQCD}
\end{table}

Since the coupling constant in QED is always small at all experimentally probed energies, $\alpha_{EM} \sim \mathcal{O}(10^{-2})$, it is relatively simple to work with quantitatively.
The coupling constant, $\alpha_{EM}$, is small enough that it is possible to do calculations by perturbatively expanding around a free-field theory where photons and charged leptons decouple.
The potential between two stationary oppositely charged leptons (take an electron and positron) to lowest order in $\alpha_{EM}$ is the well known attractive Coulomb potential, $\propto - \frac{1}{r}$, where $r$ is the distance between the two particles.
With higher order corrections in $\alpha_{EM}$ (known as quantum corrections), one eventually starts to see a screening behavior due to electron and positron pairs (heavier charged lepton pairs also contribute) popping in and out of the vacuum at small distances.
These quantum corrections do not strongly affect the behavior at large distances.
Due to this $\frac{1}{r}$ behavior at large distances, if one pulls the particles far enough apart the potential approaches zero and allows for the existence of free particles.

In QCD the situation is a bit more complicated.
The coupling constant $\as$ varies considerably as a function of the distance between the particles.
It turns out at small distances (large energies) the strong coupling constant is small, $\as \sim \mathcal{O}(10^{-1})$, which allows for a perturbative approach to calculations, similar to QED.
In this regime the potential between two oppositely charged quarks, say for example $r$ and $\bar r$, follows the same behavior as in QED, $\propto - \frac{1}{r}$ \cite{Yagi:2005yb}.
This means at large energies (short distances) the quarks are bound together only by the Coulomb-like potential, and, therefore, can be treated as weakly interacting particles.
This phenomenon is known as asymptotic freedom.

\begin{figure}[H]
\centering
  \includegraphics[width=10cm]{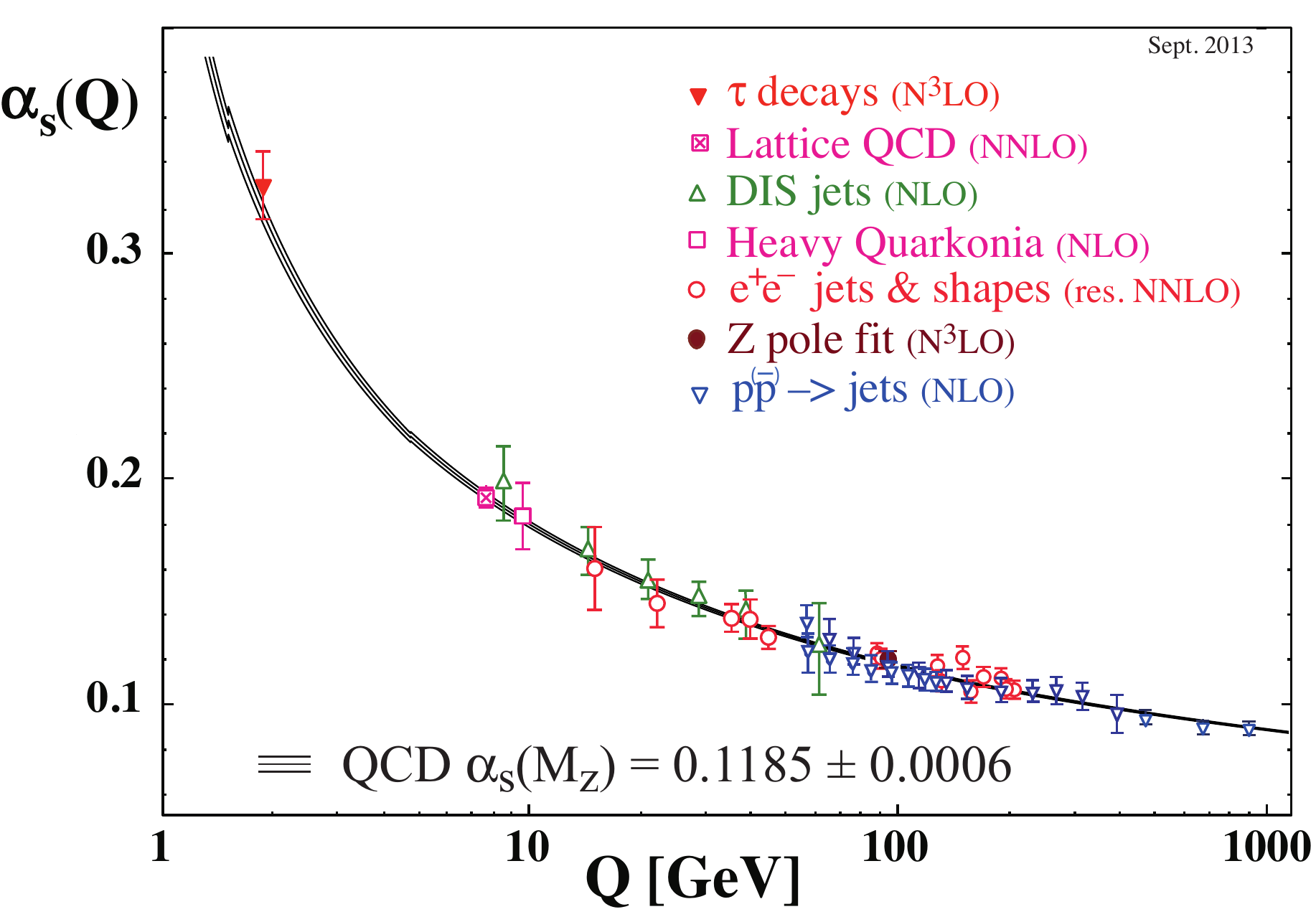}
  \caption{Plot of the running of the strong coupling constant $\as$. Taken from \cite{Agashe:2014kda}.}
\label{running} 
\end{figure}

At larger distances (lower energies) the strong coupling constant becomes large, $\as \sim \mathcal{O}(1)$.
Perturbative calculations are impossible since $\as$ is no longer a valid expansion parameter.
In this regime in order to get quantitative results, one has to use other techniques, the most common being Lattice calculations.
Numerical calculations have shown that at large distances the potential between two quarks increases linearly with distance, $V(r) = \kappa r $ where $\kappa \sim 1 \, \frac{\mbox{GeV}}{\mbox{fm}}$ and is known as the string tension \cite{Agashe:2014kda}.
This property, known as confinement, prevents the existence of free quarks due to the fact that when two quarks are pulled far enough apart the potential energy becomes so great that it creates a new pair of quarks.
The running of the strong coupling constant gives rise to an intrinsic energy scale associated with QCD, known as $\Lambda_{QCD} \approx 250 \, MeV$ \cite{Weinberg:1996kr}, which can roughly be thought of as the energy scale at which process switch from the perturbative to the non-perturbative regime.
A plot of the running of the strong coupling constant is shown in Fig.~\ref{running}.

%%%%%%%%%%%%%%%%%%%%%%%%%%%%%%%%%%%%%%%%%%%%%%%%%%%%%%%%%%%%%%%%%%%%%%%%%%%%
%%%%%%%%%%%%%%%%%%%%%%%%%%%%%%%%%%%%%%%%%%%%%%%%%%%%%%%%%%%%%%%%%%%%%%%%%%%

The lack of free quarks in nature means that we only see bound quarks.
These bound states come in two main types, mesons consisting of a quark and an anti-quark (e.g. mesons) and baryons consisting of three quarks or three anti-quarks (e.g. protons).
The different bound states have different masses which, due to the non-perturbative nature of QCD in this regime, are hard to calculate from first principles.
In fact, calculating the mass spectrum of this zoo of particles from first principles is one of the outstanding problems of QCD.

There are many other aspects of QCD which are currently unknown despite the fact that the Lagrangian governing the theory has been known for decades (for many examples see \cite{Brambilla:2014jmp}).
One such uncertain regime is the limit of high
temperature.
What makes this regime interesting is,
that not only are the quarks and gluons unbound, there are a large number of them in a compact volume.
The large density of color charges combined with the small coupling creates a situation where there is color screening, meaning that particles far away from each other no longer experience a force between them.
This is known as a quark-gluon plasma (QGP) and it's properties have important implications for the nature of QCD.

One of the places QGP is thought to exist are heavy ion collisions.
In heavy-ion collisions, two ions hit each other at such high energies that
a high temperature QCD medium is created which allows for a QGP to exist at its core.
By analyzing these interactions we can learn more about how QCD behaves in a medium at finite temperature.
Another interesting regime that heavy ion collisions probe is the high energy regime.
At the very beginning of the interaction we have a high-energy scattering problem between the constituents of the ions, adding another fascinating dynamic to the phenomenon.

%% file: Section_AA_Collision.tex
% !TEX root = WertepnyPhDThesis.tex

\section{Heavy ion collisions}
\label{sec:AA}

\begin{figure}[H]
\centering
  \includegraphics[width= 1 \textwidth]{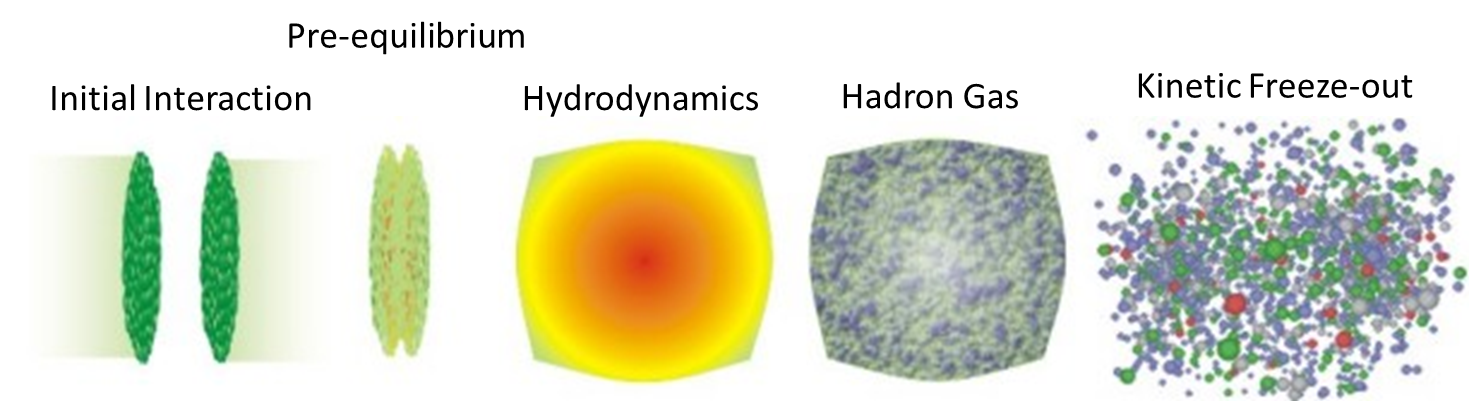}
  \caption{Evolution of the collision of two heavy ions, explained in detail below. Adapted from \cite{Bassevoimage}.}
\label{timeline} 
\end{figure}

One of the main complications of using heavy-ion collisions to study the QGP is the incredibly complex nature of these systems.
There are a wide variety of processes that occur in a single event, each of which probes a different regime of QCD and involves different physics, requiring different computational techniques to understand.
Here we divide up the system into 5 distinct phases; initial interactions, pre-equilibrium, hydrodynamics, hadron gas and kinetic freeze-out.
A general overview is presented below and shown diagrammatically in Fig.~\ref{timeline}.

%%%%%%%%%%%%%%%%%%%%%%%%%%%%%%%%%%%%%%%%%%%%%%%%%%%%%%%%%%%%%%%%%%%%%%%%%%%%
%%%%%%%%%%%%%%%%%%%%%%%%%%%%%%%%%%%%%%%%%%%%%%%%%%%%%%%%%%%%%%%%%%%%%%%%%%%%

\noindent
\textbf{I}: Initial interactions

We have two heavy ions, each of which has an initial distribution of quarks and gluons.
These two ions collide with each other at high energies, they are heavily Lorentz contracted, and their intrinsic dynamics is time dilated.
Lorentz contraction results in the ions appearing as two flat pancakes.
Time dialation slows down the internal fluctuations of the quarks and gluons inside, causing the distribution to appear frozen in place.
The end result is that we have a high-energy scattering problem between two fixed distributions of quarks and gluons which can be described within the saturation/Color Glass Condensate (CGC) formalism, outlined in Sec.~\ref{sec:CGC}.
In this formalism, the gluon fields end up driving the evolution of the system.

\noindent
\textbf{II}: Pre-equilibrium

After the initial interaction, the system continues to evolve as what is known as a ``Glasma," a distribution of gluons in the CGC formalism.
During this phase the system begins to approach a local equilibrium distribution.
This happens through a process known as thermalization which is currently a subject of great interest among theorists as there is currently no universally accepted explanation for how it occurs.
At some point, these fields are close enough to thermal equilibrium that they can be described by dissipative fluid dynamics.

\noindent
\textbf{III}: Hydrodynamics

The onset of the applicability of fluid dynamics,
due to the fact that it requires the system to be in approximate thermal equilibrium,
combined with the large density of quarks and gluons at a high temperature means the system is now a QGP.
The system, at this point, is modeled with various hydrodynamical codes, a review of which is found in \cite{Gale:2013da}.
This hydrodynamical evolution of the system causes the fluid to expand and cool down.

\noindent
\textbf{IV}: Hadron gas and \textbf{V}: Kinetic freeze-out

Eventually, the system will cool down enough so that individual quarks and gluons are no longer free and form bound states (hadrons).
This process is known as hadronization.
From there on these hadrons continue to interact with each other in what is known as a hadron gas until kinetic freeze-out, the point where the hadrons no longer interact, and thus free stream into the detector.

While the hydrodynamic flow (and thermalization to some extent) actually involves the QGP, knowing the physics of all of these different processes is vital in order to understand the data.
Without an accurate description of the physics of all of these regions, it is impossible to accurately reconstruct the collisions from the data.
Understanding the initial interactions is vital for understanding how heavy-ion collisions work, especially since it is from these initial conditions that the rest of the system evolves.
To model these initial interactions we use, as previously mentioned, the saturation framework.

%% file: Section_Saturation.tex
% !TEX root = WertepnyPhDThesis.tex

\section{Saturation physics}
\label{sec:CGC}

The brief review of saturation physics presented here is based on the review presented in \cite{Accardi:2012qut}.
For more detailed reviews see \cite{Jalilian-Marian:2005jf,Weigert:2005us,Iancu:2003xm,Gelis:2010nm,KovchegovLevin}.

As we know, the way QCD behaves varies depending on the distance scale.
We normally think of a proton as purely consisting of its three valence quarks ($uud$).
However, it turns out that this picture is too simple.
As on probes smaller distance scales, the particles making up the nucleon become increasing more complicated.
One starts to see that the proton is composed of, in addition to valance quarks, quark and anti-quark pairs, known as sea quarks, as well as gluons, all of which are commonly referred to as partons.
This phenomenon is important when considering high energy interactions in QCD, which probe such short distance scales.
Here we focus on one particular process, deep-inelastic scattering (DIS).

%%%%%%%%%%%%%%%%%%%%%%%%%%%%%%%%%%%%%%%%%%%%%%%%%%%%%%%%%%%%%%%%%%%%%%%%%%%%
\begin{figure}[H]
\centering
  \includegraphics[width=8cm]{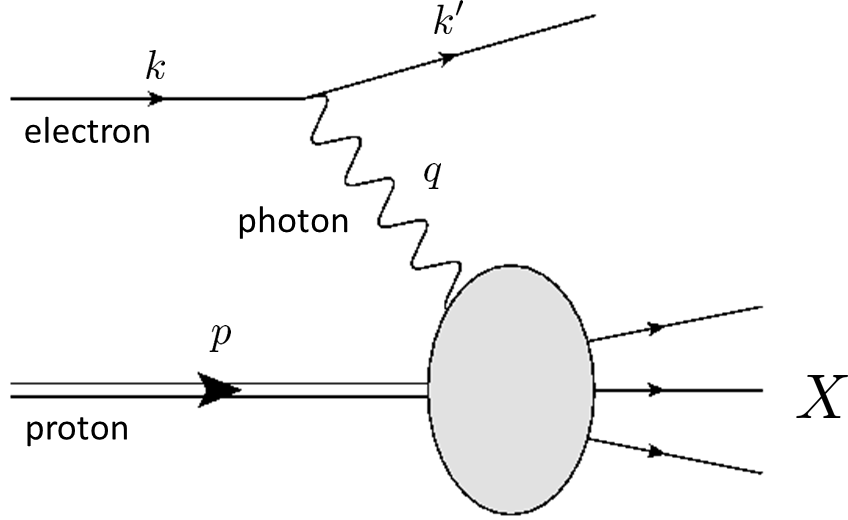}
  \caption{Diagramatic representation of the DIS process, $e + p \rightarrow e + X$. The incoming electron emits a photon (momentum $q$) which interacts with the incoming proton (momentum $p$). This process breaks up the proton producing many particles which are represented by $X$.}
\label{DIS} 
\end{figure}
%%%%%%%%%%%%%%%%%%%%%%%%%%%%%%%%%%%%%%%%%%%%%%%%%%%%%%%%%%%%%%%%%%%%%%%%%%%

In DIS, an electron with momentum $k$ scatters off a proton with momentum $p$.
The scattering process is mediated by a single virtual photon with momentum $q$, which is emitted by the electron and interacts with the proton.
The interaction with the proton breaks it up producing a multitude of particles which we denote as $X$, the exact composition of which is not important for the physics at hand.
A diagram of this process is shown in Fig.~\ref{DIS}.

Depending on the kinematics of the collision, the way the photon observes the target proton changes.
The composition of the proton is described by a parton distribution functions, which describes the number density of a given parton inside the proton.
It turns out, these can be described with only a few variables associated with the kinematics of the collisions.
First off we have the virtuality of the photon $Q^2 = -q^2$.
Secondly we have what is known as Bjorken-x, $x_{B} = \frac{Q^2}{2 p \cdot q}$ (this is often denoted in the literature as just $x$), which, as we can see, is related to $Q^2$ and the center of mass energy squared of the collision between the photon and the proton ($ s = ( p+q)^2 \approx 2 p \cdot q -Q^2$ if we neglect the mass of the proton).

The parton distribution function, given by $f(x, Q^2)$, gives the number density of partons at a particular $x_B$ and $Q^2$.
The parton distribution can be found experimentally.

%%%%%%%%%%%%%%%%%%%%%%%%%%%%%%%%%%%%%%%%%%%%%%%%%%%%%%%%%%%%%%%%%%%%%%%%%%%%
\begin{figure}[H]
\centering
  \includegraphics[width=8cm]{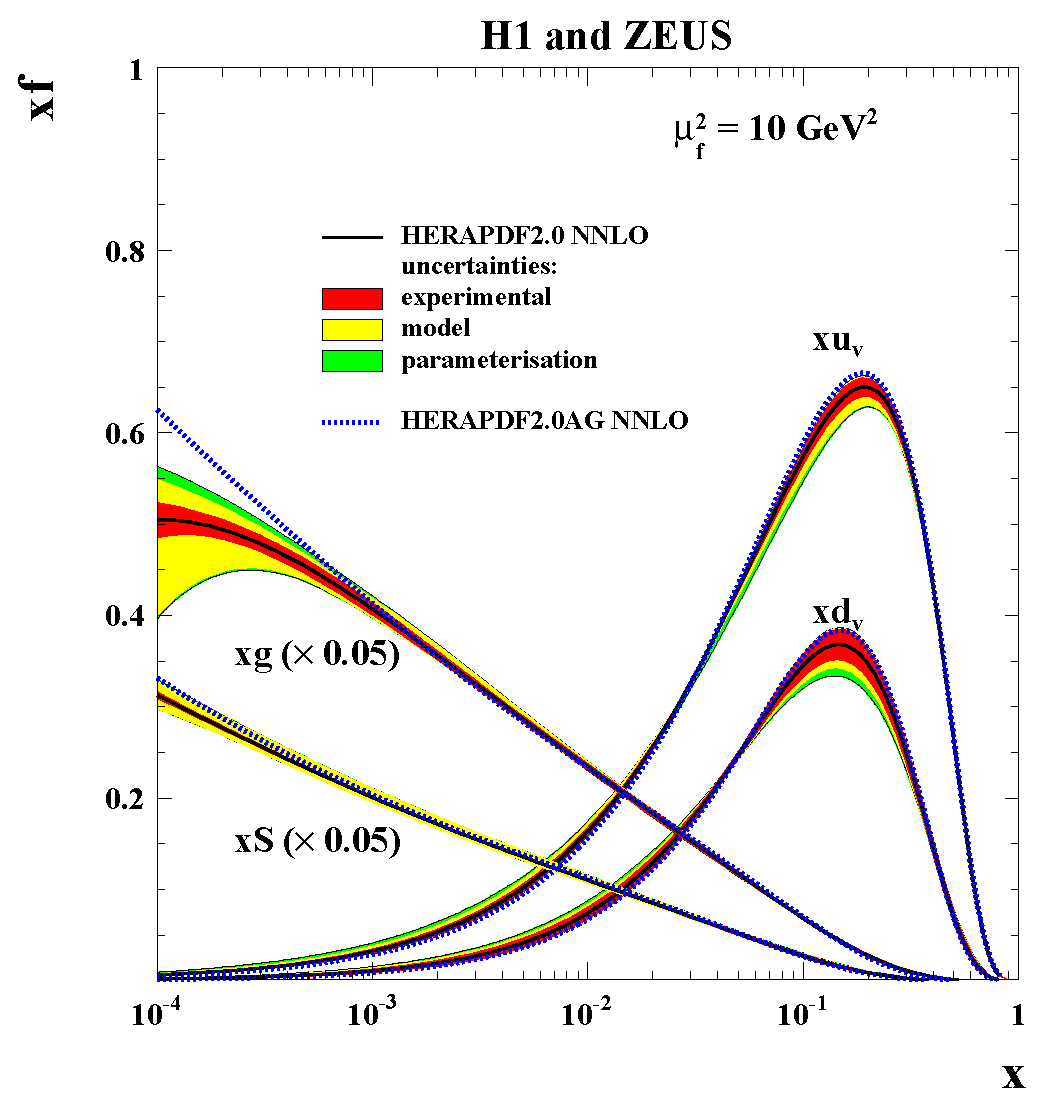}
  \caption{HERA results showing the parton distributions, multiplied by $x_B$), at fixed $Q^2=10 \, GeV^2$ ($\mu^2_f = Q^2$) as a function of $x_B$. The gluon distribution dominates at small-x. Taken from \cite{Abramowicz:2015mha}.}
\label{partonhera} 
\end{figure}
%%%%%%%%%%%%%%%%%%%%%%%%%%%%%%%%%%%%%%%%%%%%%%%%%%%%%%%%%%%%%%%%%%%%%%%%%%%

An experimental determination of the parton distributions for a proton using data from the HERA experiment is shown in Fig.~\ref{partonhera}.
Here $Q^2$ is fixed at $10 \, \mbox{GeV}^2$ while $x_B$ (notated by $x$ in the figure) varies.
Each type of parton distribution is notated differently: $g(x, Q^2)$ for gluons, $S(x, Q^2)$ for the sea quarks, $u_v(x, Q^2)$ for the valence up quarks and $d_v(x, Q^2)$ for the valence down quarks.
In the plot all of them are multiplied by $x_B$ while the gluon and sea quark distribution are further multiplied by $0.05$ so that they fit on the same plot.

As one can see in the small-x regime, $x_B \ll 1$, the gluon probability density is large and thus interactions with gluons dominate scattering processes.
This dense distribution of gluons in the small-x regime is known as the color-glass condensate (CGC).
To describe the change in the gluon distribution at a given $Q^2$ and initial value $x_B$ one uses various evolution equations, either evolving in $x_B$ while keeping $Q^2$ fixed or vice-versa.

The most fundamental evolution equation for evolution in $x_{B}$ at fixed $Q^2$ is the Balisky-Fadin-Kuraev-Lipatov (BFKL) equation \cite{Kuraev:1977fs,Balitsky:1978ic}.
This equation has the physical interpretation that as $x_B$ decreases in value the probability that a gluon in the parton distribution splits into two gluons increases.
It is a linear equation because this splitting process only requires one gluon.
This equation results in the gluon density having power-law dependence on $x_B$.

This is not the whole story, when the density of gluons becomes large enough another effect becomes important, gluon recombination, when two gluons merge into a single gluon.
In order to take this into account we need to use equations that consider this non-linear effect.
The equations that do this are the Balitsky-Kovchegov (BK) \cite{Balitsky:1996ub,Balitsky:1998ya,Kovchegov:1999yj,Kovchegov:1999ua} and Jalilian-Marian-Iancu-McLerran-Weigert-Leonidov-Kovner (JIMWLK) \cite{Jalilian-Marian:1997dw,Jalilian-Marian:1997gr,Iancu:2001ad,Iancu:2000hn} evolution equations.
The small-x evolution of the gluon density is governed by these two effects, gluon splitting which increases the density and gluon recombination with reduces the density.
These effects are shown in Fig~\ref{evolution}.
The point at which gluon recombination becomes important gives rise to what is known as the saturation scale, $Q_s$.

%%%%%%%%%%%%%%%%%%%%%%%%%%%%%%%%%%%%%%%%%%%%%%%%%%%%%%%%%%%%%%%%%%%%%%%%%%%%
\begin{figure}[H]
\centering
  \includegraphics[width=10cm]{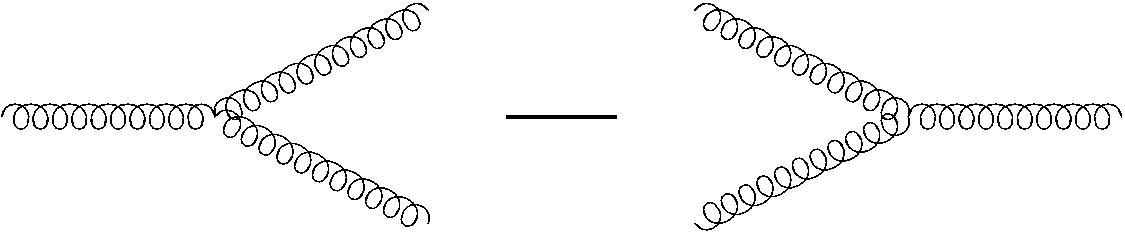} 
  \caption{The evolution of the gluon density is governed by the interplay of two effects, gluon splitting (shown on the left) and gluon recombination (shown on the right).
The gluon splitting increases the density while the recombination reduces the density.}
\label{evolution} 
\end{figure}
%%%%%%%%%%%%%%%%%%%%%%%%%%%%%%%%%%%%%%%%%%%%%%%%%%%%%%%%%%%%%%%%%%%%%%%%%%%

Saturation is not just limited to protons but it happens with heavy ions as well although the situation is slightly different.
Heavy ions, being bound states of protons and neutrons, have a large number of nucleons in them.
The large number of nucleons in the heavy ion gives rise to an enhancement to the number of gluons in the heavy ion.
Due to the high occupation number of the gluons the physics is classical.
Modeling heavy ions in this way, with classical gluon fields, is known as the McLerran-Venugopalan (MV) model \cite{McLerran:1994vd,McLerran:1993ka,McLerran:1993ni}.

To leading order in $A$, the atomic number of nucleus, the nucleons are independent of each other.
In other words the distributions of nucleons in this limit is purely a number density function, $\rho ( \vec r) $, where the probability of finding a nucleon in a given location is independent of the others.

%%%%%%%%%%%%%%%%%%%%%%%%%%%%%%%%%%%%%%%%%%%%%%%%%%%%%%%%%%%%%%%%%%%%%%%%%%%%
\begin{figure}[H]
\centering
  \includegraphics[width=12cm]{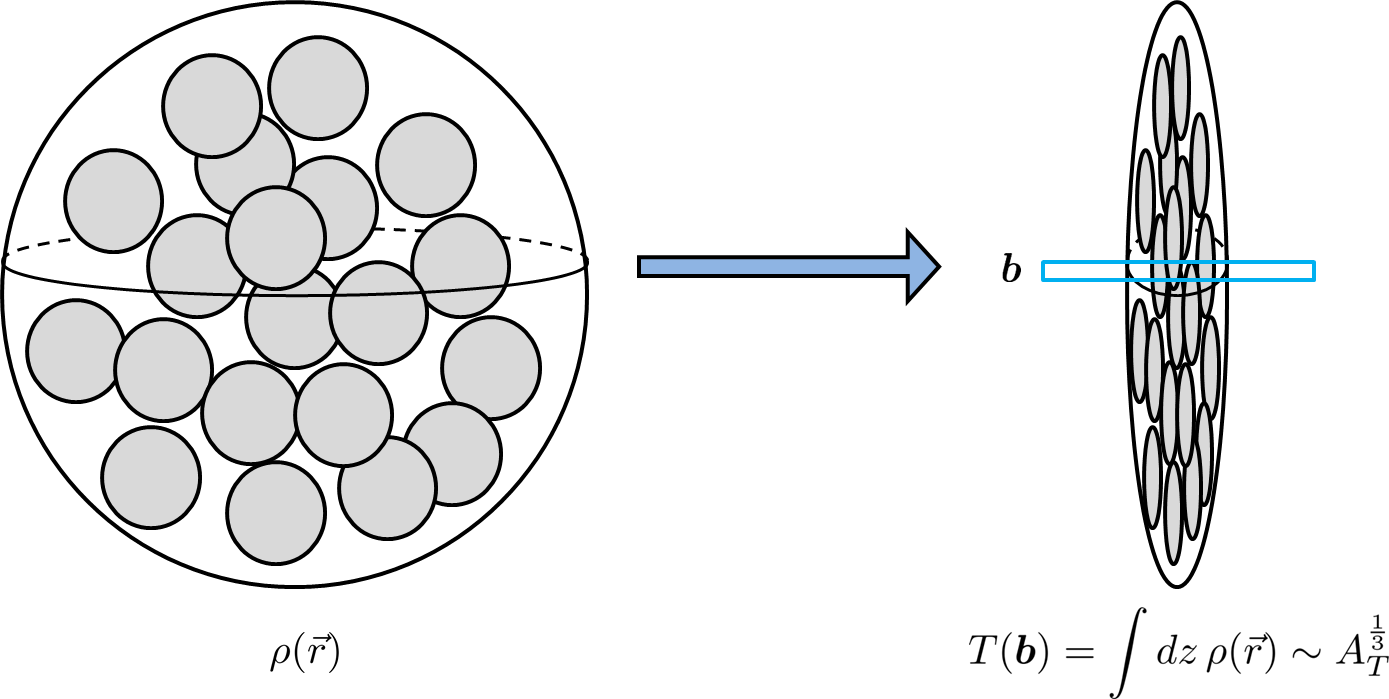} 
  \caption{Due to the Lorenz contraction of the heavy ion any particle passing through it sees the entire nucleus. Everytime the particle interacts it sees the nuclear profile function which scales as $T(\bm b ) \sim A^\frac{1}{3}$. This gives rise an $A^\frac{1}{3}$ for every interaction with the heavy ion.}
\label{evolution} 
\end{figure}
%%%%%%%%%%%%%%%%%%%%%%%%%%%%%%%%%%%%%%%%%%%%%%%%%%%%%%%%%%%%%%%%%%%%%%%%%%%

Heavy ion collisions are initially high energy scattering events, meaning that the ions are heavily Lorentz contracted.
When a particle passes through a heavy ion it roughly sees a flat pancake, so when an incoming particle interacts with nucleons in the heavy ion it probes the entire nuclear profile function (the nuclear profile function is defined as $T(\bm b) = \int dz \, \rho (\vec r)$).
Since the radius of a nucleus scales as $ R \sim A^\frac{1}{3}$ the nuclear profile function scales as $T(\bm b ) \sim A^\frac{1}{3}$
and leads to the $\sim A^\frac{1}{3}$ enhancement of the gluon density in the heavy ion.
The saturation scale also gets enhanced, $Q^2_s \sim A^\frac{1}{3}$, such that saturation effects are stronger in heavy ions than in a proton.
For instance for a gold ion at the relativistic heavy ion collider (RHIC) at $\sqrt{s_{NN}} = 200 \, \mbox{GeV}$ the saturation scale is $Q_s^2 \approx 2 \, \mbox{GeV}^2$ while for a proton at the same energy it is $Q_s^2 \approx .3 \, \mbox{GeV}^2$.
This results in a large occupation number of gluons, justifying the classical description of the MV model.
Thus we can model the initial high energy scatterings of heavy ion collisions using the saturation framework.

%% file: Section_IntroCorrelations.tex
% !TEX root = WertepnyPhDThesis.tex

\section{Two-particle correlations}

In this work we will be focusing on describing the initial conditions of heavy-ion collisions within the saturation framework.
In particular, we will be looking at the effects of the initial conditions on two-particle correlations.
Before doing this, it is helpful to look at the general kinematics of the experimental set-up to get an idea of exactly what the experimental data are telling us.

\subsection{Experimental set-up}
%%%%%%%%%%%%%%%%%%%%%%%%%%%%%%%%%%%%%%%%%%%%%%%%%%%%%%%%%%%%%%%%%%%%%%%%%%%%
\begin{figure}[H]
\centering
  \includegraphics[width= 0.8 \textwidth]{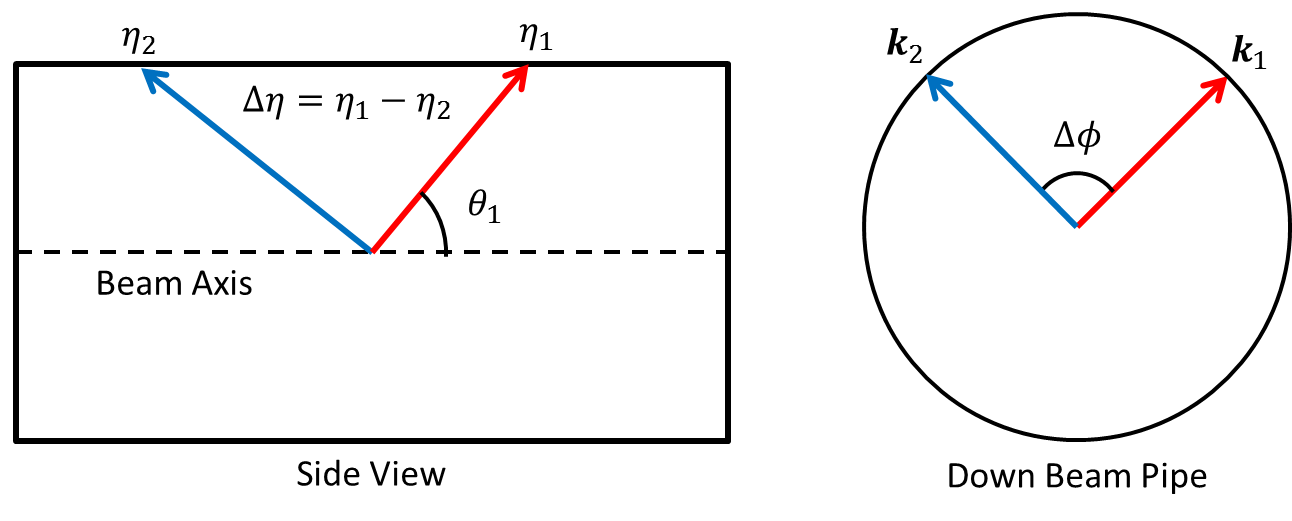}
  \caption{Left: The side view of the detector illustrates the two ion beams traveling in opposite directions along the beam axis.
  At the center they collide and produce particles.
  The kinetic trajectories of two particles are shown in red (particle 1) and blue (particle 2).
  By measuring the angle that the trajectory of particle 1 makes with the beam axis, $\theta_1$, one determines the pseudo rapidity of the particle.
  By subtracting the pseudo rapidity of the two particles one arrives at $\Delta \eta$, one of the variables of interest.
  Right: Looking down the beam pipe at the transverse plane, the two particles with transverse momentums $\bm k_1$ and $\bm k_2$ have an azimuthal angle between them, $\Delta \phi$.}
\label{kinematics}
\end{figure}
%%%%%%%%%%%%%%%%%%%%%%%%%%%%%%%%%%%%%%%%%%%%%%%%%%%%%%%%%%%%%%%%%%%%%%%%%%%%

Fig.~\ref{kinematics} shows the basic set up: we have the two ions traveling along the beam axis, $\hat z$, also called the longitudinal direction which we notate as $x_3$, in opposite directions. Perpendicular to the beam axis we have the transverse coordinates. It is convenient from an experimental point of view to introduce a new variable, pseudo rapidity, represented by $\eta$. For a particle with 3-momentum $\vec k$ it is defined as,
\begin{equation}
\notag
  \eta = \frac{1}{2} \ln \left( \frac{|\vec k|+ k_3}{|\vec k|- k_3} \right).
\end{equation}
It can be related to the angle between the particle's trajectory and the beam pipe axis, $\theta$ by
\begin{equation}
\notag
  \eta_1 = - \ln \left( \tan \frac{\theta_1}{2} \right).
\end{equation}
The experiments can easily find the pseudo rapidity of the particle.
Often the results are reported in terms of pseudo rapidity.

While experimental data may be reported in terms of pseudo rapidity the more useful quantity from a theoretical stand point is known as rapidity. Rapidity, represented by $y$, is defined as 
\begin{equation}
\notag
  y = \frac{1}{2}  \ln \left( \frac{E+k_3}{E-k_3} \right).
\end{equation}
In the limit we are dealing with the particles have a much higher energy than their rest mass.
So we can neglect the mass term when calculating the energy of the problem, $E = \sqrt{m^2 + |\vec k|^2} \approx |\vec k|$. The pseudo rapidity in the limit becomes equal to the rapidity,
\begin{equation}
\notag
  \eta \approx \frac{1}{2} \ln \left( \frac{E+ k_3}{E- k_3} \right) = y.
\end{equation}
Thus when doing theoretical calculations we will consider the rapidity, $y$, and not the pseudo rapidity, $\eta$.

\subsection{Experimental results and relation to the saturation framework}

We are focusing on long-range rapidity correlations between pairs of hadrons that are produced at
small azimuthal angles with respect to each other.
These correlations were discovered in heavy ion ($AA$)
\cite{Adams:2005ph,Adare:2008cqb,Alver:2009id,Abelev:2009af},
proton--proton ($pp$) \cite{Khachatryan:2010gv}, and proton--nucleus
($pA$) collisions \cite{CMS:2012qk}. Due to the particular shape of
the corresponding correlation function, with a narrow correlation in
the azimuthal angle $\Delta \phi$ and a wide correlation in
pseudo-rapidity separation $\Delta \eta$, these correlations are often
referred to as the ``ridge''.

\begin{figure}[H]
\centering
  \includegraphics[width= 0.45 \textwidth]{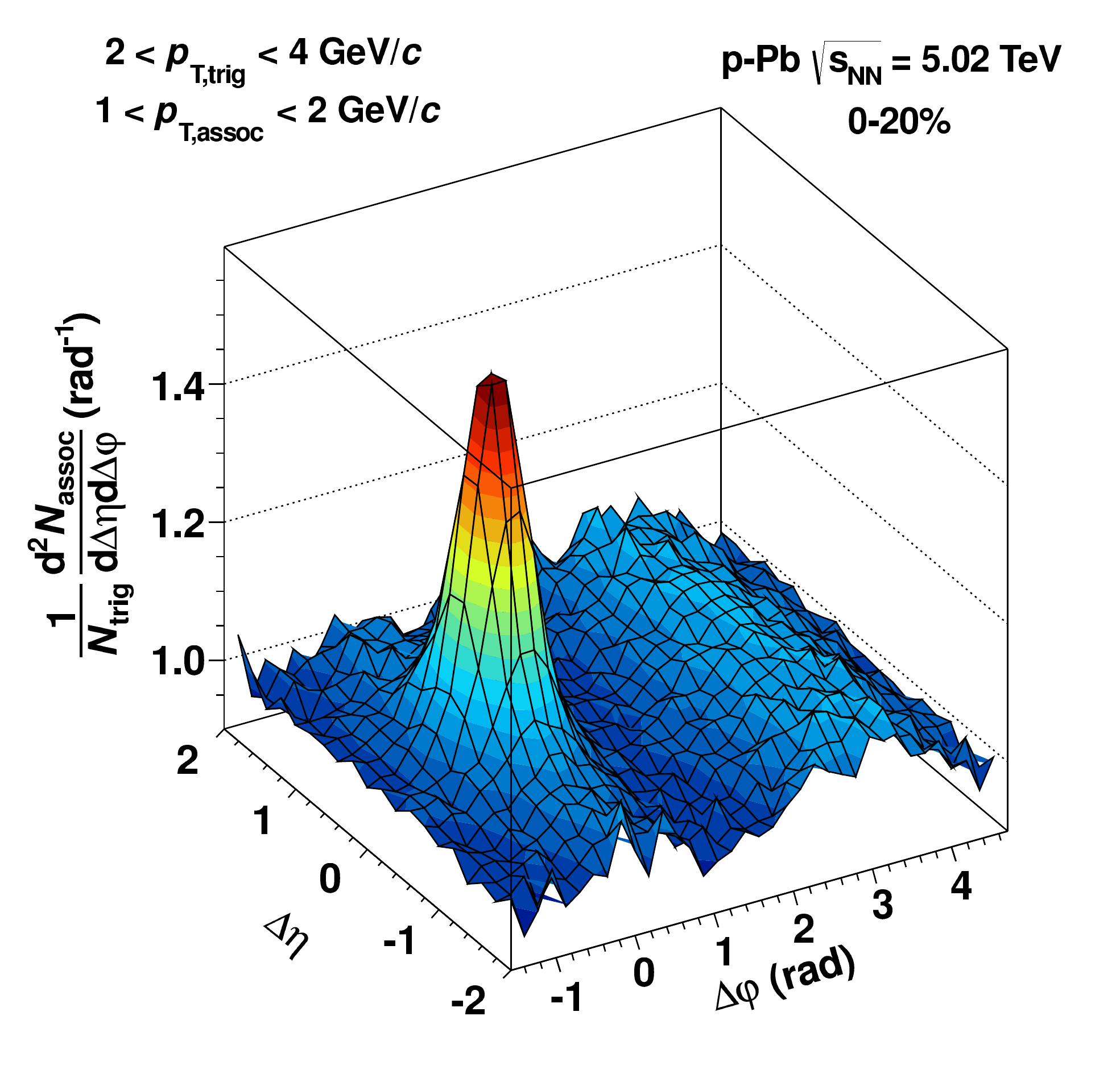}
  \includegraphics[width= 0.45 \textwidth]{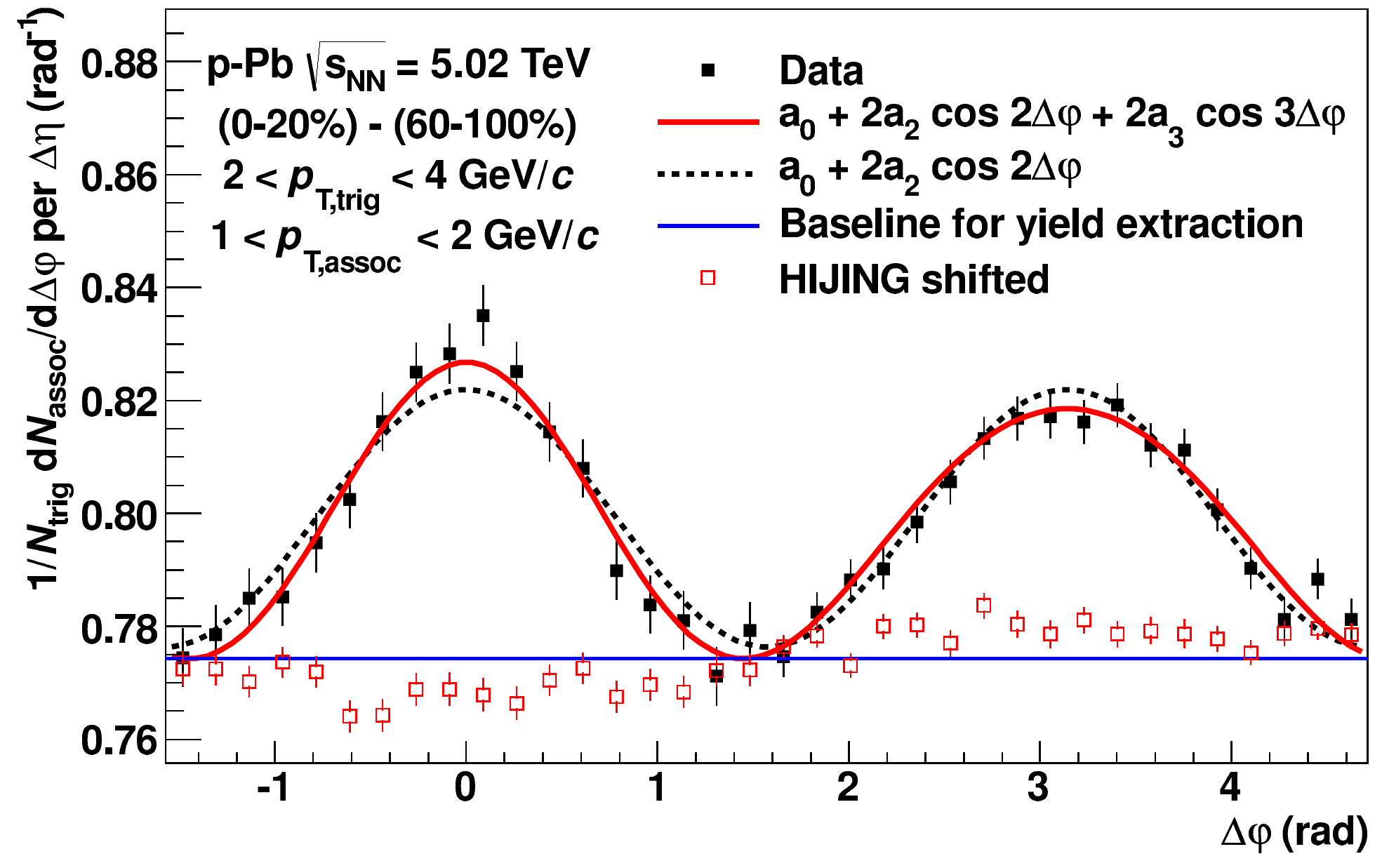}
  \caption{Left: Data for two particle correlations in p--Pb collisions at $\sqrt{s_{NN}} = 5.02 \, TeV$.
  There are two long range in rapidity ``ridges" at at both $\Delta \phi = 0$ and $\Delta \phi = \pi$.
  The peak at $\Delta \phi =0$ and $\Delta \eta = 0$ is due to jet fragmentation.
  Right: Data of the ridge averaged over $0.8 <| \Delta \eta | <1.8$ on the near-side which removes the central peak and $| \Delta \eta | <1.8$ on the away-side.
  Fits containing a $cos (2 \Delta \phi)$ term (dashed black line) and both a $cos (2 \Delta \phi)$ and $cos (3 \Delta \phi)$ (solid red line) term are superimposed.
  Both graphs are from the ALICE collaboration and were taken from \cite{Abelev:2012aa}.}
\label{data} 
\end{figure}

Fig.~\ref{data} shows more recent experimental results from the ALICE collaboration for p--Pb collisions, \cite{Abelev:2012aa}.
We can see that there are two ``ridges" in the results, one on the near-side ($\Delta \phi = 0$) and one on the away-side ($\Delta \phi = \pi$).
These ``ridges" are mostly symmetric except for a small deficit on the away-side, leading to a correlation dominated by even harmonics.

%%%%%%%%%%%%%%%%%%%%%%%%%%%%%%%%%%%%%%%%%%%%%%%%%%%%%%%%%%%%%%%%%%%%%%%%%%%%
%%%%%%%%%%%%%%%%%%%%%%%%%%%%%%%%%%%%%%%%%%%%%%%%%%%%%%%%%%%%%%%%%%%%%%%%%%%%

\begin{figure}[H]
\centering
  \includegraphics[width= 0.6 \textwidth]{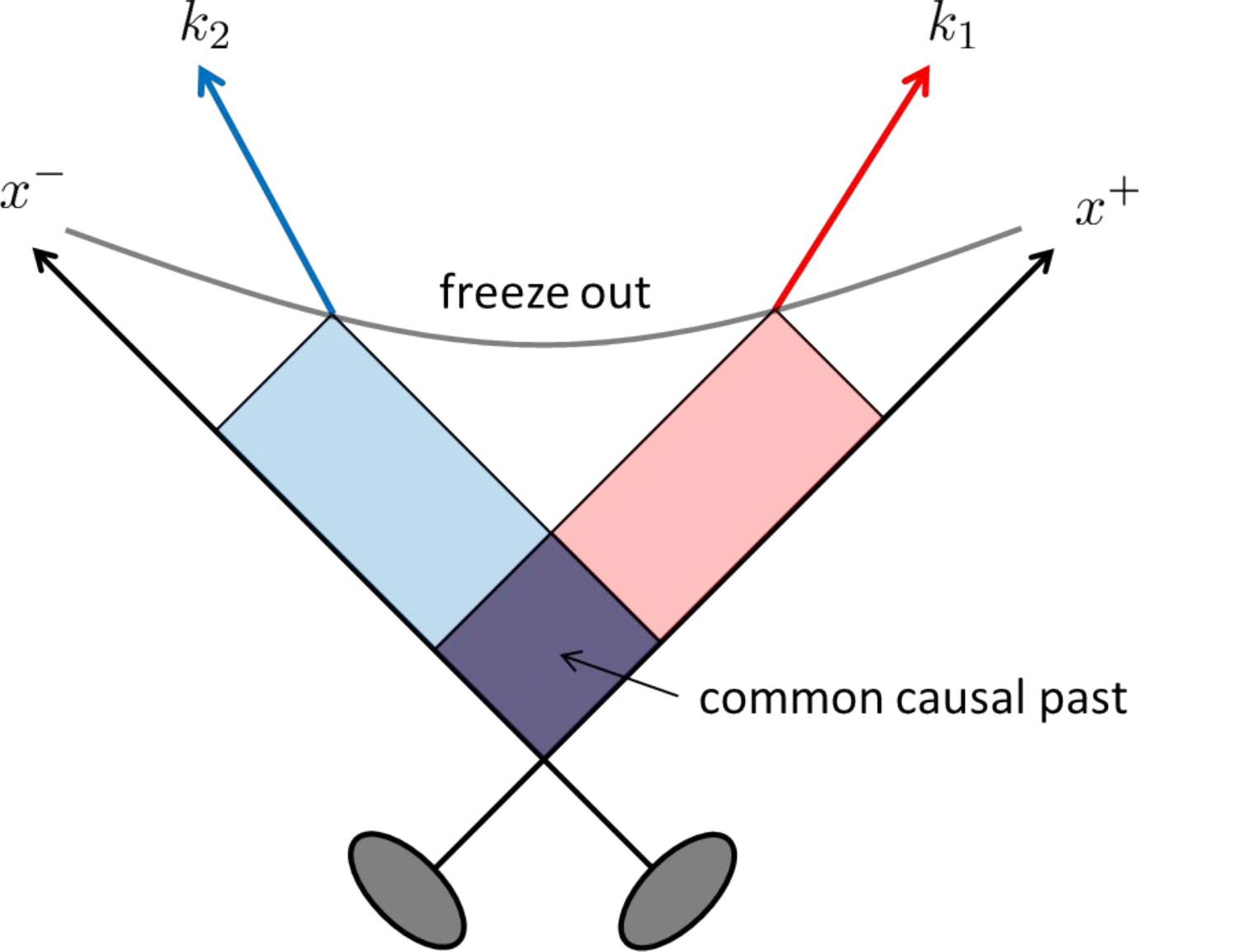}
  \caption{A diagramatic representation of the causality arguement. The two particles with momenta $k_1$ (red) and $k_2$ (blue) have a large difference in rapidity and thus point in different directions inside the light cone. The two particles must originate at the same common causal part, represented by the shaded purple region.}
\label{lightcone} 
\end{figure}

Since these correlated involve particles that are far apart in rapidity they must originate from the early-time dynamics.
The logic goes as follows: But since they are far apart in rapidity they are pointing in different directions inside the light cone, one close to $x^+$ axis the other close to the $x^-$ axis.
Due to causality, any correlation between these two particles must originate in the intersection of their past light-cones, known as their common causal past.
This would place the origin towards the beginning of the interaction, in the initial state \cite{Dumitru:2008wn,Gavin:2008ev}.
This is shown in figure \ref{lightcone}.
Despite knowing that these correlations originate from the initial interactions, the detailed dynamical origin of these correlations is not completely clear.

%%%%%%%%%%%%%%%%%%%%%%%%%%%%%%%%%%%%%%%%%%%%%%%%%%%%%%%%%%%%%%%%%%%%%%%%%%%%
%%%%%%%%%%%%%%%%%%%%%%%%%%%%%%%%%%%%%%%%%%%%%%%%%%%%%%%%%%%%%%%%%%%%%%%%%%%%

It has been proposed in the literature
\cite{Armesto:2006bv,Armesto:2007ia,Dumitru:2008wn,Gavin:2008ev,Dusling:2009ni,Dumitru:2010iy,Dumitru:2010mv,Kovner:2010xk,Gelis:2008sz}
that these ``ridge'' correlations may arise in the classical gluon field
dynamics of the parton saturation physics/CGC.
Indeed classical gluon fields, which in the MV model dominate
gluon production in heavy ion collisions, do lead to a
rapidity-independent distribution of the produced gluons
\cite{Kovner:1995ja,Kovchegov:1997ke,Krasnitz:2003jw,Blaizot:2010kh}
over rapidity intervals of up to $\Delta y \lesssim 1/\as$, which is
the upper limit of their validity (with $\as$ the strong coupling
constant). Correlations between such classical fields, introduced in
the process of averaging over their color sources, do have a long
range in rapidity
\cite{Kovchegov:1999ep,Dumitru:2008wn,Gavin:2008ev}. Moreover, it was
observed in
\cite{Dumitru:2008wn,Dusling:2009ni,Dumitru:2010mv,Dumitru:2010iy}
that the diagrams giving rise to such rapidity correlations also lead
to a narrow correlation in the azimuthal direction, in qualitative
agreement with the shape of the ``ridge'' correlation.

A potential complication with the CGC explanation of the
``ridge'' is the fact that rapidity-dependent corrections of classical
gluon fields do become important once rapidity is on the order of $\Delta y
\sim 1/\as$. These corrections come in through the non-linear BK and JIMWLK
evolution equations. Rapidity-dependent corrections are very important
for describing the hadron multiplicity distribution in rapidity,
$dN/dy$, within the CGC framework \cite{ALbacete:2010ad}. As $dN/dy$ does
depend on rapidity rather strongly in RHIC heavy ion data, and also
strongly (albeit less so) in the LHC data, it is very hard to describe this observable
without the rapidity-dependent nonlinear evolution. It is possible
that similar rapidity-dependent corrections may significantly affect
(and potentially destroy) the long-range structure of the rapidity
correlations present in classical gluon fields of the MV model. Note that
progress on this issue has been made in \cite{Dusling:2009ni},
indicating that inclusion of small-$x$ evolution still leaves the
``ridge'' reasonably flat in rapidity until $\Delta y \sim 1/\as$.

To elucidate the above questions and concerns, and to improve the
precision of the CGC predictions for the ``ridge''-like correlations,
it is important to be able to calculate the two-particle correlation
in the CGC framework beyond the lowest order. While some works do
consider the role of small-$x$ evolution and multiple rescatterings in
the correlation function
\cite{Dusling:2009ni,Dumitru:2010mv,Kovner:2011pe,Levin:2011fb,Kovner:2012jm},
most of the phenomenological approaches
\cite{Dusling:2009ni,Dumitru:2010iy,Dusling:2012iga,Dusling:2012wy,Dusling:2012cg}
simply include the saturation effects into the lowest-order
calculation by evolving the unintegrated gluon distributions with the
running-coupling BK (rcBK) nonlinear evolution
\cite{Gardi:2006rp,Kovchegov:2006vj,Balitsky:2006wa}.
This is not correct due to the fact that these methods do not consider all orders of saturation effects in the ions and it isn't exactly clear that the unintegrated gluon distributions can be used in this case (in fact we show that these distribution do not work, Sec.~\ref{sec:fact}).

One has to keep in mind that in heavy ion collisions the early-stage
azimuthal correlation may be washed out by the final state
interactions leading to thermalization of the produced medium and its
hydrodynamic evolution. (Note though, that the rapidity correlation is
not likely to be strongly affected by such late-time dynamics.) It was
argued, however, that the effect of the elliptic flow in the
hydrodynamic evolution of the quark-gluon plasma (QGP) would be to
(re-)introduce the azimuthal correlations \cite{Gavin:2008ev}.

%% file: Chapter_Setup.tex
% !TEX root = WertepnyPhDThesis.tex
\cleardoublepage
\chapter{Quantitative details of QCD, the heavy--light ion regime and it's effect on the correlation function}
\label{ch:Setup}

As a brief aside, we list the conventions used for the remainder of this dissertation.
From now on, this work uses natural units, where $\hbar=1$ and $c=1$ unless specified otherwise.
Einstein notation, where the sum over repeated indices is implied, is used for both vector and color indices.
Feynman slash notation is used, meaning $\slashed{k}=\gamma^\mu k_\mu$ where $k_\mu$ is some 4-vector and $\gamma^\mu$ are the Dirac matrices.
We work with light-cone coordinates which are $x^\mu = (x^+, x^-, \bm x)$ where $x^+ = \frac{x^0 + x^3}{\sqrt 2}$, $x^- = \frac{x^0 - x^3}{\sqrt 2}$, $\bm x = (x^1, x^2)$ is the transverse vector and we define  $x_\perp = |\bm x|$ (some works in the literature do not have the factor of $\frac{1}{\sqrt 2}$ in the definition of $x^+$ and $x^-$, this is a choice of convention). Our convention results in, for the dot product of two 4 vectors, $x \cdot y = x^+ y^- + x^- y^+ -\bm x \cdot \bm y$.

Before we begin with the rest of the calculation we need to go over a few more topics.
First, we need to cover some quantitative details about QCD which we skipped over in Sec.~\ref{sec:IntroQCD}.
Second, we go over, in detail, what we mean by the heavy-light ion regime and how this affects our approach to the two-particle correlation function calculation.
Last, we look at the quantitative definition of the two-particle correlation function and already, just by using the heavy-light ion regime, notice that we have non-trivial correlations that depend purely on the geometry of the colliding ions.

%%%%%%%%%%%%%%%%%%%%%%%%%%%%%%%%
\input{Section_QCD}
\input{Section_aA_regime}
\input{Section_Def_of_Correlations}
%%%%%%%%%%%%%%%%%%%%%%%%%%%%%%%%

\section{Summary}

In this section we covered a few topics that were needed before we go into detailed derivations of gluon production in heavy-light ion collisions.
First we reviewed some basics of QCD in order to understand some of the mathematics behind it.
Secondly we described in detail what the heavy-light ion regime implies in the calculation at hand.
This regime determined the types of cross sections that dominate the two-gluon correlation functions.
Even without knowing the explicit form of these cross sections we noticed a type of correlations which we called ``geometric" correlations.
These correlations depend only on the initial geometry of the collision and are completely independent of the dynamics of the interaction itself.

%% file: Section_QCD.tex
% !TEX root = WertepnyPhDThesis.tex
\section{Details of QCD}
\label{sec:QCD}

As discussed previously in Sec.~\ref{sec:IntroQCD}, QCD is a non-Abelian SU$(N_c)$ gauge theory.
Here $N_c$ is the number of fundamental color charges and, in the case of QCD, is 3.
It is more convenient to keep track of the number of colors in terms of $N_c$ as opposed to setting this to 3 since later on in this work we will be using 't Hooft's large-$N_c$ limit.

The Lagrangian which governs QCD is \cite{Yang:1954vj}
\begin{equation}
\label{Lqcd}
\mathcal{L}_{QCD} = \sum_{\text{flavors} f} \bar{q}_f  (i \slashed{D}_\mu - m_f ) q_f - \frac{1}{4} F^a_{\mu \nu} F^{a \mu \nu}
\end{equation}
where
\begin{align}
D_\mu & = \partial_\mu -  i t^a A^a_\mu \\
F^a_{\mu \nu} & = \partial_\mu A^a_\nu - \partial_\nu A^a_\mu + g f^{a b c} A^b_\mu A^c_\nu
\end{align}
are known as the covariant derivative and the gluon field strength tensor respectively.

Here $q_f$ and $\bar{q}_f$ are the quark and anti-quark spinor fields, respectively, with flavor $f$ and mass $m_f$.
$A^a_\mu$ is the gauge vector field (in the case of QCD this is the gluon) with color index $a$, which runs from 1 to $N_c^2-1$. $t^a$ are the fundamental generators of the SU($N_c$) group.
These consist of $N_c^2 -1$ different $N_c$ by $N_c$ matrices.
The normalization used for these generators is such that
\begin{equation}
\notag
[t^a,t^b]=i f^{abc} t^c \qquad tr[t^a t^b]=\frac{\delta^{ab}}{2}.
\end{equation}
where $f^{abc}$ are the structure constants. There are also the adjoint generators defined as $(T^a)^{bc} = - i \, f^{abc}$ which are often used in the literature.

One of the primary features of a gauge theory is that it is invariant under a local gauge transformation.
This means that the Lagrangian and the observables of the theory are invariant under the transformation.
For SU($N_c$) theories the quark and gluon fields transform as
\begin{align}
q(x) \rightarrow & \, S(x) q(x)
\notag \\
\bar{q}(x) \rightarrow & \, \bar{q}(x) S^{-1}(x)
\notag \\
A_\mu(x) \rightarrow & \, S(x) A_\mu(x) S^{-1}(x) - \frac{i}{g} [\partial_{\mu} S(x)] S^{-1}(x)
\end{align}
where
\begin{align}
S(x) = e^{i \alpha^a (x) t^a}
\end{align}
Here $\alpha^a (x)$ is a real function with color index $a$.

As mentioned earlier, we are working in the limit where $\as \ll 1$.
This allows us to use perturbation theory to do the calculations, where we expand around the free-theory.
This is done by quantizing the theory using the usual path-integral formulation of quantum field theory.
For the gauge fields (gluons in the case of QCD) we use the usual De Witt--Faddeev--Popov method \cite{Faddeev:1967fc}.
To do this we break the gauge invariance of the system by setting the gauge condition, determining the form of the gluon propagator and the ghost fields.
In this work we are using the light-cone gauge where $A^+=0$.
Normally this method introduces ghosts into the calculations but this is not so in the light-cone gauge as they do not contribute to physical observables \cite{Chirilli:2015fza}.
From here the Feynman rules are derived, which are used for the perturbative calculations.
For more details on this topic see \cite{Sterman:1994ce,Peskin:1995ev,Weinberg:1996kr}.

The Feynman rules we use are the usual rules presented in \cite{Peskin:1995ev} except, since those were derived in the Feynman gauge,
we use a different gluon propagator and do not consider ghosts.
In the light-cone gauge, for a gluon with momentum $k$ and indices $\mu, \, \nu$, the gluon propagator is
\begin{equation}
\frac{-i}{k^2+i \epsilon}\left( g_{\mu \nu} - \frac{\eta_\mu k_\nu +\eta_\nu k_\mu}{k^+} \right),
\end{equation}
where $\eta_\mu$ is defined such that $\eta \cdot k = k^+$ for any 4-vector.
The second term in the bracket has a singularity when $k^+=0$ and is known as the light-cone pole.
In general this pole has to be regulated which introduces new subtleties into the calculation (see \cite{Chirilli:2015fza}).
However it turns out that in the calculations presented here we never have to worry about this pole so these subtleties are neglected.
For an on-shell gluon with momentum $k$ and polarization $\lambda$ we have
\begin{equation}
\epsilon^\mu_\lambda = \left(0,\frac{\bm \epsilon_\lambda \cdot \bm k}{k^+}, \bm \epsilon_\lambda \right).
\end{equation}
where
\begin{equation}
\sum_{\lambda} \bm \epsilon^*_{\lambda, i} \bm \epsilon_{\lambda, j} = \delta_{ij}.
\end{equation}

With this we have enough information to start to understand how the rest of the calculation works.

%% file: Section_aA_regime.tex
% !TEX root = WertepnyPhDThesis.tex
\section{Heavy-light ion regime}
\label{sec:aAregime}

In the saturation framework, Sec.~\ref{sec:CGC}, we view each ion as a large bag full of nucleons, where each of the nucleons is classical color sources which produces the gluon fields.
In order to calculate the full heavy-heavy collision case one would need to consider all of the nucleons of both the target and projectile simultaneously. 
This is an incredibly hard problem and we currently do not have the ability to do this.
However, one can consider the heavy-light ion case, where we consider all of the nucleons in the target but only a finite number of nucleons in the projectile.

Due to the high-energy nature of these collisions we can consider the dynamics of the source nucleons separately from the dynamics of the gluon production itself.
We can calculate the cross sections associated with the processes of interest, single- and two-gluon production, by convoluting the square of the wave functions associated with the nucleon distribution with the cross sections of the gluon production due to the collisions of these individual nucleons.

Denote the 1- and
2-nucleon wave functions of the projectile nucleus $A_1$ by $ \Psi_I (
\bm b )$ and $\Psi_{II} ( {\bm b}_1 , {\bm b}_2 )$: they are
normalized such that
\begin{align}\label{wf_norm}
\int d^2 b \, |\Psi_I ( \bm b )|^2 = A_1, \ \ \ \int d^2 b_1 \, d^2
  b_2 \, |\Psi_{II} ( {\bm b}_1 , {\bm b}_2 )|^2 = A_1 \, (A_1 -1)
  \approx A_1^2.
\end{align}

%%%%%%%%%%%%%%%%%%%%%%%%%%%%%%%%%%%%%%%%%%%%%%%%%%%%%%%%%%%%%%%%%%%%%%%%%%%%
\begin{figure}[H]
\centering
  \includegraphics[width=5cm]{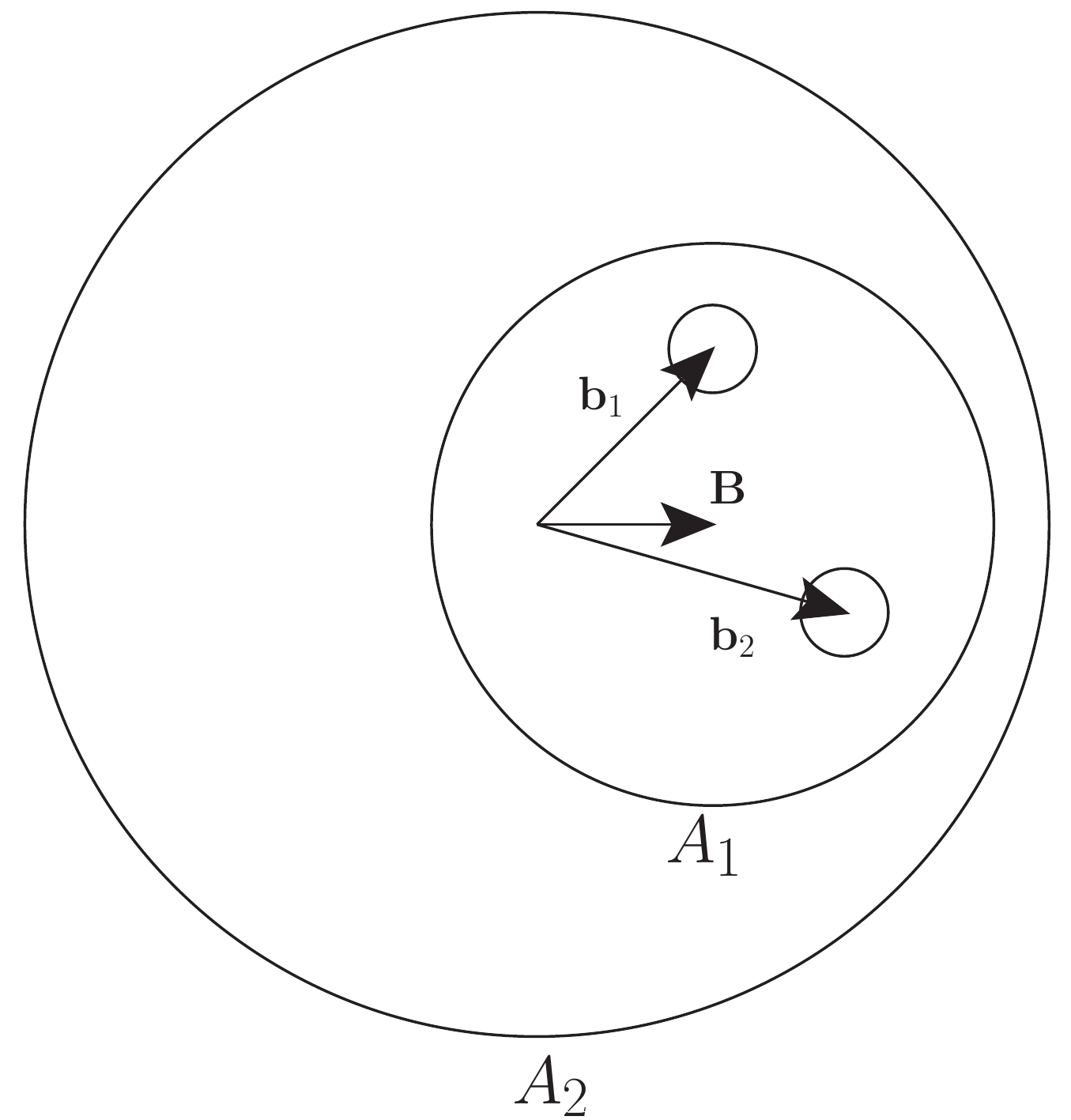}
  \caption{Transverse plane geometry of the two-particle production in
    the collision of a smaller projectile nucleus ($A_1$) with a
    larger target nucleus ($A_2$). The two smaller circles represent
    two nucleons in the nucleus $A_1$ (see text for details).}
\label{nucl_geom} 
\end{figure}
%%%%%%%%%%%%%%%%%%%%%%%%%%%%%%%%%%%%%%%%%%%%%%%%%%%%%%%%%%%%%%%%%%%%%%%%%%%

With the help of these wave functions, contributions to the single-gluon production cross section from the one- and two-nucleon(s) in the projectile case can be written as
\begin{subequations}
\label{gA1}
\begin{align}
\label{pA1}
\frac{d \sigma_{I}}{d^2 k d y}
  = & \int d^2 B \, d^2 b \, | \Psi_I (\bm B - \bm b) |^2 \,
  \left\langle \frac{d \sigma^{p A_2}}{d^2 k d y d^2 b} \right\rangle 
\\
  \frac{d \sigma_{II}}{d^2 k d y}
  = & \int d^2 B \, d^2 b_1 \, d^2 b_2 \, |\Psi_{II} ( \bm B - \bm b_1 , \bm B - \bm b_2 ) |^2 
  \left\langle \frac{d \sigma^{p p + A_2}}{d^2 k d y \, d^2 b_1 d^2 b_2} \right\rangle
\label{ppA1}
\end{align}
\end{subequations}
and contributions to the two-gluon production cross section from the one- and two-nucleon(s) in the projectile case can be written as
\begin{subequations}
\label{NNA2}
\begin{align}
\frac{d \sigma_I}{d^2 k_1 dy_1 \, d^2 k_2 dy_2}
  = & \int d^2 B \, d^2 b \, | \Psi_I (
  \bm B - \bm b) |^2 \,
  \left\langle \frac{d \sigma^{p A_2}}{d^2 k_1  dy_1
  \, d^2 k_2 dy_2 \, d^2 b} \right\rangle
\label{pA2} \\
\frac{d \sigma_{II}}{d^2 k_1 dy_1 \, d^2 k_2 dy_2}
  = & \int d^2 B \, d^2
  b_1 \, d^2 b_2 \, |\Psi_{II} ( \bm B - \bm b_1 , \bm B - \bm b_2 ) |^2 
  \notag \\
& \times \, \left\langle \frac{d \sigma^{p p+A_2}}{d^2  k_1 dy_1 d^2 b_1
  \, d^2 k_2  dy_2 d^2 b_2} \right\rangle \label{ppA2}
\end{align}
\end{subequations}
where $\bm B$ is the impact parameter between the two nuclei and $\bm b$, $\bm b_1$, $\bm b_2$ the transverse positions of the nucleons in the projectile nucleus, all measured with respect to the center of the second (target) nucleus, as shown in Fig.~\fig{nucl_geom} for two nucleons in the projectile. (Transverse vector $\bm b$ labels the position of the incoming nucleon in the single nucleon case, while $\bm b_1$ and $\bm b_2$ label positions of projectile nucleons for the two nucleon case as shown in Fig.~\ref{nucl_geom}.)

%%%%%%%%%%%%%%%%%%%%%%%%%%%%%%%%%%%%%%%%%%%%%%%%%%%%%%%%%%%%%%%%%%%%%%%%%%%%
\begin{figure}[H]
\centering
  \includegraphics[width=14cm]{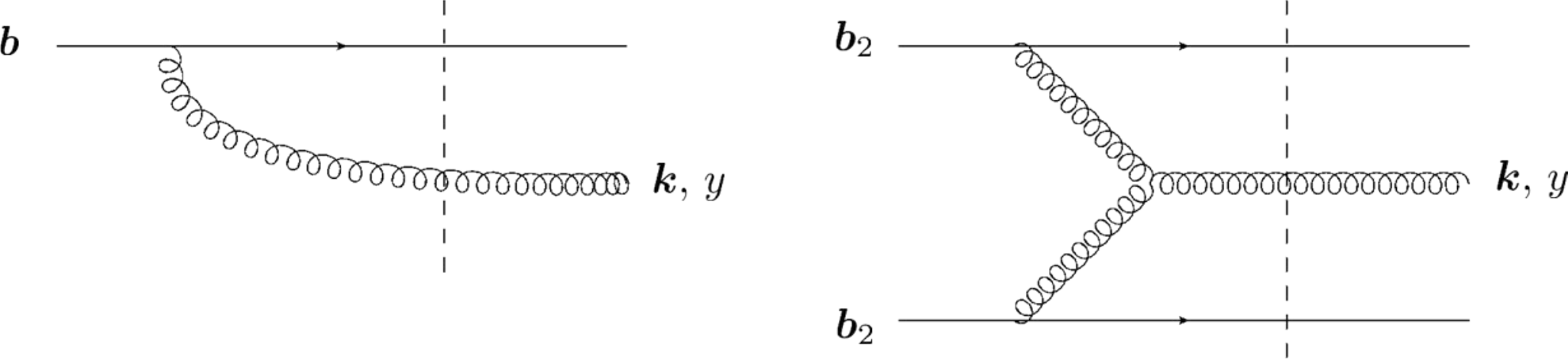}
  \caption{A sample Feynman diagram for each of the single-gluon production processes considered.
  The diagrams show a single gluon being produced by a collision of one- or two-nucleons with the
  target nucleus (left and right diagram respectively). The dashed line represents the interaction with the
  target (also known as the shock wave) which takes into account all of the multiple rescatterings with the target nucleons.}
\label{gexamples} 
\end{figure}
%%%%%%%%%%%%%%%%%%%%%%%%%%%%%%%%%%%%%%%%%%%%%%%%%%%%%%%%%%%%%%%%%%%%%%%%%%%

In Eq.~\ref{gA1}, 
\begin{align}
\nonumber
  \frac{d \sigma^{pA_2}}{d^2 k dy d^2 b}
  \quad \mbox{and} \quad
  \frac{d \sigma^{pp+A_2}}{d^2 k d y \, d^2 b_1 d^2 b_2}
\end{align}
are the cross sections for single-gluon production (with fixed transverse
momentum $\bm k$ and rapidity $y$) in:
the collision of a single nucleon ($p$ with fixed transverse position $\bm b$) with the target nucleus,
and the collision of two nucleons ($pp$ with fixed transverse positions
$\bm b_1$ and $\bm b_2$) with the target nucleus, respectively.
A diagrammatic example of each process is shown in Fig.~\ref{gexamples}.

%%%%%%%%%%%%%%%%%%%%%%%%%%%%%%%%%%%%%%%%%%%%%%%%%%%%%%%%%%%%%%%%%%%%%%%%%%%%
\begin{figure}[H]
\centering
  \includegraphics[width=14cm]{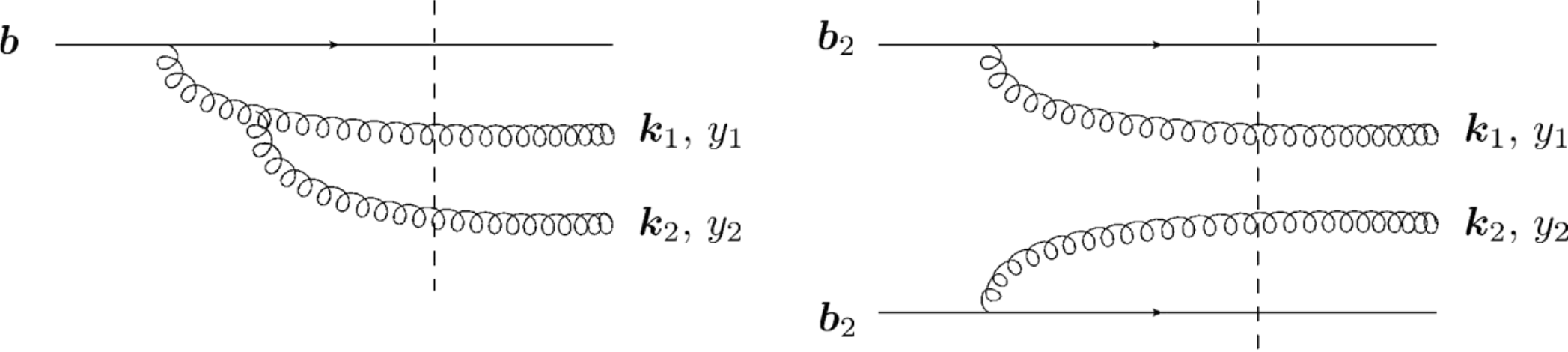}
  \caption{A sample Feynman diagram for each of the two-gluon production processes is considered.
  The diagrams show two gluon being produced by a collision of one- or two-nucleons with the
  target nucleus (left and right diagram respectively). The dashed line represents the interaction with the
  target (also known as the shock wave).}
\label{ggexamples} 
\end{figure}
%%%%%%%%%%%%%%%%%%%%%%%%%%%%%%%%%%%%%%%%%%%%%%%%%%%%%%%%%%%%%%%%%%%%%%%%%%%

In Eq.~\ref{NNA2}, 
\begin{align}  
\nonumber
  \frac{d \sigma^{pA_2}}{d^2 k_1  dy_1 \, d^2 k_2 dy_2 \, d^2 b}
  \quad \mbox{and} \quad
  \frac{d \sigma^{pp+A_2}}{d^2 k_1 dy_1 d^2 b_1 \, d^2 k_2 dy_2 d^2 b_2}
\end{align}
are the cross sections for two-gluon production (with fixed transverse
momenta $\bm k_1$, $\bm k_2$ and rapidities $y_1$, $y_2$) in:
the collision of a single nucleon ($p$ with fixed transverse position $\bm b$) with the target nucleus,
and in the collision of two nucleons ($pp$ with fixed transverse positions
$\bm b_1$ and $\bm b_2$), with the target nucleus, respectively. A diagrammatic example of each process is shown in Fig. \ref{ggexamples}, where the interaction with the target nucleon is represented by a dashed line.
It should be noted that sometimes we use a red box to notate the interaction to emphasize the finite thickness of the target; details of this are presented in Ch.~\ref{ch:pAreview}.
The angle brackets $\langle \ldots \rangle$ in \eqref{NNA2} and \eqref{gA1} denote averaging in
the target nucleus wave function along with summation over all the
nucleons in the target nucleus
\cite{McLerran:1994vd,McLerran:1993ka,McLerran:1993ni,Kovchegov:1996ty,Jalilian-Marian:1997dw,Jalilian-Marian:1997gr,Iancu:2001ad,Iancu:2000hn}.

\begin{figure}[H]
\centering
  \includegraphics[width=5cm]{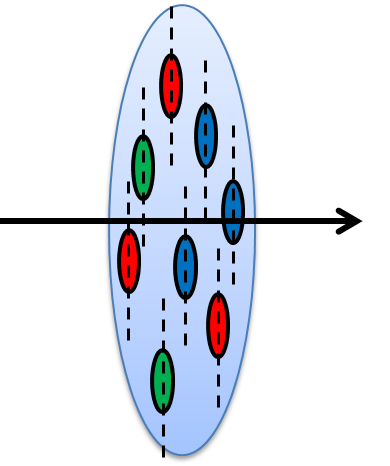}
  \caption{A quark or a gluon (black arrow) passes through the target (grey oval) and interacts with the many fields (dashed lines) originating from the color sources (red, green and blue ovals).
  The kinematics of the interaction are such that it does not change the trajectory of the particle much allowing us to treat the interaction with the target as a shock wave.
  This is described in more detail in Sec.~\ref{sec:target}.}
\label{multi} 
\end{figure}

The idea behind the shock wave representation of the interactions with the target nucleus is as follows:
Since the interaction is at high energies the target is Lorentz contracted and can be thought of approximately as a flat pancake (grouped around $x^+=0$).
The target can be thought of as a source of color fields.
The incoming quark or gluon interacts with the many color fields, referred to as the multiple rescatterings.
Due to the computation details of the calculation, explained in Sec.~\ref{sec:target}, multiple re-scatterings do not change trajectory of the particle much, allowing us to treat the interaction with the target as a shock wave.
Fig.~\ref{multi} shows a pictorial representation of this.

%%%%%%%%%%%%%%%%%%%%%%%%%%%%%%%%%%%%%%%%%%%%%%%%%%%%%%%%%%%%%%%%%%%%%%%%%%%%
%%%%%%%%%%%%%%%%%%%%%%%%%%%%%%%%%%%%%%%%%%%%%%%%%%%%%%%%%%%%%%%%%%%%%%%%%%%

Here we model the nucleus as a bag of independent nucleons, which is a
correct description at the leading order in the atomic number $A$
\cite{Mueller:1989st,McLerran:1994vd,McLerran:1993ka,McLerran:1993ni,Kovchegov:1996ty,Kovchegov:1997pc}. 
In such case the single-nucleon light-cone wave function squared is
simply equal to the nuclear profile function in the projectile
nucleus,\footnote{We do not show the spin and isospin indices
  explicitly in the wave functions: in our notation the wave function
  squared is implicitly averaged over all nucleon polarizations, since
  both colliding nuclei are unpolarized.}
\begin{align}
\label{nmf}
| \Psi_I (\bm b) |^2 = T_1 ({\bm b}).
\end{align}
The nuclear profile function for a nucleus with the nucleon number density $\rho ({\bm b}, z)$ is defined by the integral over the longitudinal coordinate $z$,
\begin{align}\label{nmf_def}
T ({\bm b}) = \int\limits_{-\infty}^\infty dz \, \rho ({\bm b}, z).
\end{align}
This means $T(\bm b ) \sim A^\frac{1}{3}$. For a spherical nucleus of radius $R$ and constant density
$\rho$, $T({\bm b} ) = 2 \, \rho \, \sqrt{R^2 - b^2}$.

For a sufficiently large projectile nucleus, $A_1 \gg 1$, one can
assume that the two-nucleon wave function can be
factorized,
\begin{align}
  \Psi_{II} ( {\bm b}_1 , {\bm b}_2 ) = 
  \Psi_I ({\bm b}_1) \, \Psi_I ({\bm b}_2),
\end{align}
such that, with the help of \eqref{nmf} we can write
\begin{align}
\label{factorization2}
  |\Psi_{II} ( {\bm b}_1 , {\bm b}_2 )|^2 = 
  T_1 ({\bm b}_1) \, T_1 ({\bm b}_2).
\end{align}

The nuclear profile function $T ( \bm b)$ is closely related what is known as the saturation scale $Q^2_s$ (this quantity ends up having important physical significance as will be seen later).
Every nucleon involved in the interaction introduces a power of $Q^2_s$ into the final result (shown in Ch.~\ref{ch:pAreview}).
This is why we refer to considering more nucleons in the process as saturation corrections.
At the classical level, the saturation scale for a nucleon at transverse position $\bm b$ originating from nucleus $i$ is $Q^2_{s, i}(\bm b)= 4 \pi \as^2 T_i (\bm b )$ (it should be noted that when the saturation scale is in the classical limit it is often notated as $Q^2_{s0}$, but in our case, since we are dealing purely at the classical level and have two different saturation scales to worry about, the notation presented here seemed appropriate).
The $\as^2$ in the saturation scale comes from the fact that every interaction with a nucleon involves at least two gluons.

Perturbative dynamics that we employ happens over short transverse distances and, at the leading order, involves a single quark inside nucleon while being insensitive to the non-perturbatively long distances of the order of $\Lambda_{QCD}^{-1}$.
We assume that the quark involved in the interaction is a valence quark, following the original MV model \cite{McLerran:1994vd,McLerran:1993ka,McLerran:1993ni}.
This assumption can easily generalized to a more realistic case if we use the sum of sea and valence quark distributions instead of a simple valance quark distribution implicitly assumed in our calculations.

Since $Q^2_{s, i} \sim \as^2 \, A^{\frac{1}{3}}_i$, every nucleon introduces a power of $\as^2 \, A^{\frac{1}{3}}_i$ ($i=1$ or $2$ for the projectile and target respectively).
Using the heavy-light ion regime ($1 \ll A_1 \ll A_2$, $\as^2 A^\frac{1}{3}_1 \ll 1$ and $\as^2 A^\frac{1}{3}_2 \sim 1$) we can find which of the processes shown in Figs.~\ref{gexamples} and \ref{ggexamples} dominate. Since target nucleon corrections are order of $\mathcal{O}(1)$ we do not include these explicitly while examining the power counting.

Here we are examining the overall power counting of the single-gluon production processes, Eqs.~\eqref{gA1}. Since both of these processes are producing a single classical gluon we gain a factor of $\frac{1}{\as}$ for both.
For the single- and two-nucleon processes with one and two projectile saturation scale corrections in the projectile bringing in $\as^2 A^\frac{1}{3}_1$ and $(\as^2 A^\frac{1}{3}_1)^2$ respectively, the resulting power counting is,
\begin{equation}
\label{gA1scaling}
  \frac{d \sigma_{I}}{d^2 k \, d y} \sim \frac{1}{\as} \left(\as^2 A^{\frac{1}{3}}_1 \right)
\quad \mbox{and} \quad
  \frac{d \sigma_{II}}{d^2 k \, d y} \sim \frac{1}{\as} \left(\as^2 A^{\frac{1}{3}}_1 \right)^2.
\end{equation}
Since the two-nucleon process is suppressed by the saturation scale, $\as^2 A^{\frac{1}{3}}_1 \ll 1$, the single-nucleon process dominates single-gluon production in heavy-light ion collisions.
The single-gluon production cross section is, to leading order, described by \eqref{pA1} where we drop the subscript $I$.

The power counting for the two-gluon production processes, Eqs~\eqref{NNA2}, follows a similar procedure. For the single-nucleon process we have one order in projectile saturation scale, $\as^2 A^{\frac{1}{3}}_1$, but instead of gaining a factor of $\frac{1}{\as}$ for each produced gluon, due to the extra gluon emission from the nucleon, the correction is order $1$. For the two-nucleon process there are two orders of projectile saturation scale, $\left(\as^2 A^{\frac{1}{3}}_1 \right)^2$, and each gluon gives a factor of $\frac{1}{\as}$. This gives rise to the following power counting,
\begin{equation}
\label{gA2scaling}
  \frac{d \sigma_{I}}{d^2 k_1 dy_1 \, d^2 k_2 dy_2}
  \sim \left(\as^2 A^{\frac{1}{3}}_1 \right)
\quad \mbox{and} \quad
  \frac{d \sigma_{II}}{d^2 k_1 dy_1 \, d^2 k_2 dy_2}
  \sim \frac{1}{\as^2} \left(\as^2 A^{\frac{1}{3}}_1 \right)^2 
  = \as^2 \left( A^{\frac{1}{3}}_1 \right)^2
\end{equation}
From the above expression we see that the two-nucleon process has an enhancement of $A^{\frac{1}{3}}_1$ when compared with the single-nucleon process. The two-gluon production cross section is thus dominated by \eqref{ppA2} where we drop the subscript $II$.

Taking the dominant terms for both the single- and two-gluon production cross section and replacing the wave functions squared with the nuclear profile function according to \eqref{nmf} and \eqref{factorization2} we arrive at the corresponding expression for the single- and two-gluon production cross sections respectively.
\eqref{pA2} giving
\begin{subequations}
\label{aAcross}
\begin{align}
\label{p2A1}
  \frac{d \sigma}{d^2 k dy}  = &
  \int d^2 B \, d^2 b \, T_1 ( \bm B - \bm b)
  \left\langle \frac{d \sigma^{p A_2}}{d^2 k
  dy d^2 b } \right\rangle,
\\
\label{p2A}
  \frac{d \sigma}{d^2 k_1 dy_1 \, d^2 k_2 dy_2} = &
  \int d^2 B \, d^2 b_1 \, d^2 b_2 \, T_1 ( \bm B - \bm b_1) \, T_1
  (\bm B - \bm b_2 ) \,
\notag \\
  & \times \left\langle \frac{d \sigma^{pp+A_2}}{d^2 k_1
  dy_1 d^2 b_1 \, d^2 k_2 dy_2 d^2 b_2}
  \right\rangle,
\end{align}
\end{subequations}
where only the nuclear profile function $T_1$ depends on the impact parameter $\bm B$.
It is useful to note that we can perform the integral over $\bm B$ in \eqref{p2A1} and obtain
\begin{align}
\label{pA}
  \frac{d \sigma}{d^2 k \, d y} = A_1 \, \int d^2 b \, \left\langle
  \frac{d \sigma^{p A_2}}{d^2 k d y d^2 b} \right\rangle.
\end{align}
While for pedagogical purposes we use the form in \eqref{p2A1} for much of the single-gluon production cross section derivation in Ch.~\ref{ch:pAreview}, it is more convenient to use \eqref{pA} when calculating the correlation function as will be seen in the next section (Sec.~\ref{sec:Correlation}).

Armed with a knowledge of the general power counting associated with the heavy-light ion regime, thus the dominant contributions, we proceed to examine how these results affect the two-gluon correlation function of interest.

%% file: Section_Def_of_Correlations.tex
% !TEX root = WertepnyPhDThesis.tex
\section{Definition of the correlation function}
\label{sec:Correlation}

\subsection{Definition of the correlator}

Following a standard approach used in experimental analyses of
particle correlations \cite{Eggert:1974ek,Khachatryan:2010gv} the
correlation function can be defined as
\begin{align}\label{corr_def}
C ( {\bm k}_1, {y}_1, {\bm k}_2, {y}_2 ) = {\cal N} \, \frac{\frac{d
   N_{12}}{d^2 k_1 dy_1 \, d^2 k_2 dy_2}}{\frac{d N}{d^2 k_1 d y_1}
  \, \frac{d N}{d^2 k_2 d y_2}} - 1
\end{align}
where
\begin{align}\label{1part_distr}
\frac{d N}{d^2 k_1 d y_1} = \frac{1}{\sigma_{inel}} \, \frac{d
  \sigma}{d^2 k_1 d y_1}
\end{align}
and 
\begin{align}\label{2part_distr}
  \frac{d N_{12}}{d^2 k_1 d y_1 \, d^2 k_2 dy_2} =
  \frac{1}{\sigma_{inel}} \, \frac{d \sigma}{d^2 k_1 d y_1 \, d^2 k_2
  dy_2}
\end{align}
are the single- and double-particle multiplicity distributions with
$\sigma_{inel}$ the net inelastic nucleus--nucleus scattering cross
section.
The normalization factor $\cal N$ in \eqref{corr_def} is fixed by requiring
that the number of particle pairs measured in the same (``real'')
event $N_{12}$ is equal to the number of (``mixed'') pairs with
particles coming from different events $(N)^2$. Here we are interested
in $\Delta \eta$-$\Delta \phi$ correlations.
For such a correlation, with the magnitudes of the transverse momenta $k_{1 \,
  T}$ and $k_{2 \, T}$ constrained to some chosen data bins, the
normalization factor is fixed by
\begin{align}
  \label{eq:norm}
  {\cal N} \int d \phi_1 \, dy_1 \, d \phi_2 \, dy_2 \, \frac{d
   N_{12}}{d^2 k_1 dy_1 \, d^2 k_2 dy_2} = \int d \phi_1 \, d y_1 \,
  \frac{d N}{d^2 k_1 d y_1} \, \int d \phi_2 \, d y_2 \, \frac{d
   N}{d^2 k_2 d y_2}.
\end{align}
(We assume for simplicity that the number of produced particles is
very large, $N \gg 1$, such that $N-1 \approx N$ and $\sigma_{inel}$
is the same in both Eqs. \eqref{1part_distr} and \eqref{2part_distr}
since the production cross section of producing exactly one particle
is negligible.)

Combining Eqs. \eqref{corr_def}, \eqref{1part_distr},
\eqref{2part_distr}, and \eqref{eq:norm} we rewrite the correlation
function in terms of cross sections as
\begin{align}\label{corr_def2}
C ( {\bm k}_1, {y}_1, {\bm k}_2, {y}_2 ) = \frac{\left[ \int d
  \phi_1 \, d y_1 \, \frac{d \sigma}{d^2 k_1 d y_1} \, \int d
  \phi_2 \, d y_2 \, \frac{d \sigma}{d^2 k_2 d y_2} \right] }{
  \left[ \int d \phi_1 \, dy_1 \, d \phi_2 \, dy_2 \, \frac{d
  \sigma}{d^2 k_1 dy_1 \, d^2 k_2 dy_2} \right]} \ \frac{\frac{d
  \sigma}{d^2 k_1 dy_1 \, d^2 k_2 dy_2} }{\frac{d \sigma}{d^2 k_1
  d y_1} \, \frac{d \sigma}{d^2 k_2 d y_2}} - 1.
\end{align}
Using the forms of the cross sections derived earlier, \eqref{pA} and \eqref{p2A}, yields
\begin{align}\label{corr_main}
& C ( {\bm k}_1, {y}_1, {\bm k}_2, {y}_2 ) = \notag \\
& \frac{\left[\int d^2 b_1
  \, d \phi_1 \, d y_1 \, \langle \frac{d \sigma^{p A_2}}{d^2 k_1
  \, dy_1 \, d^2 b_1} \rangle \right] \, \left[\int d^2 b_2 \, d
  \phi_2 \, d y_2 \, \langle \frac{d \sigma^{p A_2}}{d^2 k_2 \,
  dy_2 \, d^2 b_2} \rangle \right]}{\left[ \int d^2 B \, d^2 b_1
  \, d^2 b_2 \, d \phi_1 \, d y_1 \, d \phi_2 \, d y_2 \, T_1 (
  \bm B - \bm b_1) \, T_1 (\bm B - \bm b_2 ) \, \left\langle
  \frac{d \sigma^{pp +A_2}}{d^2 k_1 \, dy_1 \, d^2 b_1 \,
  d^2 k_2 \, dy_2 \, d^2 b_2}\right\rangle \right]}
\notag \\
& \times \, \frac{\int d^2 B \, d^2 b_1 \, d^2 b_2 \, T_1 ( \bm B -
  \bm b_1) \, T_1 (\bm B - \bm b_2 ) \, \left\langle \frac{d
  \sigma^{p p + A_2}}{d^2 k_1 dy_1 d^2 b_1 \, 
  d^2 k_2 dy_2 d^2 b_2} \right\rangle}{ \int d^2 b_1 \,
  \left\langle \frac{d \sigma^{p A_2}}{d^2 k_1 \, d y_1 \, d^2 b_1}
  \right\rangle \, \int d^2 b_2 \, \left\langle \frac{d \sigma^{p
  A_2}}{d^2 k_2 \, d y_2 \, d^2 b_2} \right\rangle} -1.
\end{align}
To complete the calculation one needs the single- and double-gluon
production cross sections which, when used in \eqref{corr_main}, would
give us the correlation function. Before we proceed to construct them,
let us study a simple example elucidating the nature of one of the
correlation types contained in correlator \eqref{corr_main}, which we call ``geometric'' correlations.

%%%%%%%%%%%%%%%%%%%%%%%%%%%%%%%%%%%%%%%%%%%%%%%%%%%%%%%%%%%%%%%%%%%%%%%

\subsection{Geometric correlations}
\label{sub:geocor}

Let us consider the simplest possible example of particle (gluon)
production mechanism where the interaction of the two nucleons in the
first nucleus with the second nucleus in \eqref{p2A} factorizes,
\begin{align}
\label{int_fact}
\left\langle \frac{d \sigma^{p p + A_2}}{d^2 k_1 dy_1 d^2 b_1 \, 
  d^2 k_2 dy_2 d^2 b_2} \right\rangle
  \propto \left\langle \frac{d \sigma^{p A_2}}{d^2 k_1 dy_1 d^2 b_1}
  \right\rangle \, \left\langle \frac{d \sigma^{p A_2}}{d^2 k_2 dy_2
  d^2 b_2} \right\rangle.
\end{align}
This contribution comes from the disconnected Feynman diagrams and is
usually identified as the uncorrelated part of the two-gluon
production cross section. However, it is clear that substituting
\eqref{int_fact} into \eqref{corr_main} does not reduce the correlation
function to zero: instead one gets
\begin{align}\label{corr_fact}
& C ( {\bm k}_1, {y}_1, {\bm k}_2, {y}_2 ) = \notag \\
& \frac{\left[\int d^2 b_1
  \, d \phi_1 \, d y_1 \, \langle \frac{d \sigma^{p A_2}}{d^2 k_1
  \, dy_1 \, d^2 b_1} \rangle \right] \, \left[\int d^2 b_2 \, d
  \phi_2 \, d y_2 \, \langle \frac{d \sigma^{p A_2}}{d^2 k_2 \,
  dy_2 \, d^2 b_2} \rangle \right]}{\left[ \int d^2 B \, d^2 b_1
  \, d^2 b_2 \, d \phi_1 \, d y_1 \, d \phi_2 \, d y_2 \, T_1 (
  \bm B - \bm b_1) \, T_1 (\bm B - \bm b_2 ) \, \left\langle
  \frac{d \sigma^{p A_2}}{d^2 k_1 dy_1 d^2 b_1} \right\rangle \,
  \left\langle \frac{d \sigma^{p A_2}}{d^2 k_2 dy_2 d^2 b_2}
  \right\rangle \right]} \notag \\
& \times \, \frac{\int d^2 B \,
  d^2 b_1 \, d^2 b_2 \, T_1 ( \bm B - \bm b_1) \, T_1 (\bm B - \bm
  b_2 ) \, \left\langle \frac{d \sigma^{p A_2}}{d^2 k_1 dy_1 d^2
  b_1} \right\rangle \, \left\langle \frac{d \sigma^{p A_2}}{d^2
  k_2 dy_2 d^2 b_2} \right\rangle}{ \int d^2 b_1 \, \left\langle
  \frac{d \sigma^{p A_2}}{d^2 k_1 \, d y_1 \, d^2 b_1}
  \right\rangle \, \int d^2 b_2 \, \left\langle \frac{d \sigma^{p
   A_2}}{d^2 k_2 \, d y_2 \, d^2 b_2} \right\rangle} -1,
\end{align}
which, in general, could be non-zero. Notice that the overall proportionality constant from \eqref{int_fact} canceled.

Certainly if the $\bm b$-dependence factorizes from the rapidity and
azimuthal dependence in the single-gluon production cross section in the $pA$ case
\begin{align}\label{gl_prod}
\left\langle \frac{d \sigma^{pA_2}}{d^2 k \, d y \, d^2 b}
  \right\rangle,
\end{align}
then the correlation function \eqref{corr_fact} is zero: however, such
factorization is not always the case. For gluon production in the
saturation framework, the cross section is a complicated function of
$k_T/Q_s ({\bm b}, y)$, which means it is not in a factorized form and
thus the correlator \eqref{corr_fact} is not zero. Note that in the MV
model (which does not contain the small-$x$ evolution), gluon
production is rapidity-independent, and, if one neglects the
dependence of the gluon production cross section on the angle between
$\bm k$ and $\bm b$, the correlator \eqref{corr_fact} becomes
zero. (Dependence of gluon production cross section on the collision
geometry in the MV approximation is not very strong, peaking at
non-perturbatively low momenta \cite{Teaney:2002kn}.)

For the general case in \eqref{corr_fact} we observe a possible
non-trivial correlation in the two-gluon production described by {\sl
  disconnected} Feynman diagrams.  If the gluon production cross
section \eqref{gl_prod} is a slowly varying (but not constant)
function of rapidity, as is the case in the saturation/CGC framework
near mid-rapidity, this correlation would be long-range in
rapidity. The origin of this correlation is somewhat peculiar: even
though the two-nucleon wave function in \eqref{factorization2} is
factorized and, hence, represents uncorrelated nucleons, these two
nucleons are correlated by the simple fact of being parts of the same
bound state, the projectile nucleus. In other words, the probability
of finding two nucleons at the impact parameters $\bm b_1$ and $\bm
b_2$ is proportional to
\begin{align}
\label{eq:prob}
  \sim \, \int d^2 B \, T_1 ( \bm B - \bm b_1) \, T_1 (\bm B - \bm b_2)
\end{align}
and is not a product of two independent probabilities after all impact
parameters $\bm B$ of the incoming nucleus are integrated over: this
is a correlation. Note also that the presence of real wave-function
correlations, that is, non-factorizable corrections to the
right-hand-side of \eqref{factorization2}, would also lead to some
nontrivial two-particle correlations in \eqref{corr_fact}.

If we define the correlation function at the fixed nuclear impact
parameter $\bm B$ by not integrating over $\bm B$ in \eqref{pA1} and \eqref{ppA2}
and using the result in \eqref{corr_def2}, we get
\begin{align}
\label{eq:CfixedB}
& C ( {\bm k}_1, {y}_1, {\bm k}_2, {y}_2 ; {\bm B} ) = \notag \\
& \frac{\left[\int d^2 b_1 \, d \phi_1 \, dy_1 \, |\Psi_I ({\bm B} -
  {\bm b}_1)|^2 \, \langle \frac{d \sigma^{p A_2}}{d^2 k_1 \, dy_1
  \, d^2 b_1} \rangle \right] \, \left[\int d^2 b_2 \, d \phi_2
  \, dy_2 \, |\Psi_I ({\bm B} - {\bm b}_2)|^2 \, \langle \frac{d
  \sigma^{p A_2}}{d^2 k_2 \, dy_2 \, d^2 b_2} \rangle
  \right]}{\left[ \int d^2 b_1 \, d^2 b_2 \, d \phi_1 \, dy_1 \, d
  \phi_2 \, dy_2 \, |\Psi_{II} ( \bm B - \bm b_1, \bm B - \bm b_2
  )|^2 \, \langle \frac{d \sigma^{p p+ A_2}}{d^2 k_1 \, dy_1 \, d^2
  b_1 \, d^2 k_2 \, dy_2 \, d^2 b_2} \rangle \right]} \notag \\
& \times \, \frac{\int d^2 b_1 \, d^2 b_2 \, |\Psi_{II} ( \bm B - \bm
  b_1, \bm B - \bm b_2 )|^2 \, \left\langle \frac{d \sigma^{p p +
  A_2}}{d^2 k_1 dy_1 d^2 b_1 \, d^2
  k_2 dy_2 d^2 b_2} \right\rangle}{ \int d^2 b_1 \, |\Psi_I
  ({\bm B} - {\bm b}_1)|^2 \, \left\langle \frac{d \sigma^{p
  A_2}}{d^2 k_1 \, d y_1 \, d^2 b_1} \right\rangle \, \int d^2
  b_2 \, |\Psi_I ({\bm B} - {\bm b}_2)|^2 \, \left\langle \frac{d
  \sigma^{p A_2}}{d^2 k_2 \, d y_2 \, d^2 b_2} \right\rangle}
  -1,
\end{align}
One can see that this fixed-impact parameter correlation function in
\eqref{eq:CfixedB} is zero, $C ({\bm B}) =0$, for the factorized wave
function from \eqref{factorization2} {\sl and} for disconnected-diagram
interactions from \eqref{int_fact}.  Thus di-gluon correlations due to
our ``geometric'' correlation mechanism seem to also disappear when
the impact parameter is fixed exactly. However, such precise
determination of the impact parameter is typically not done in an experimental
analysis, one fixes the collision centrality $|{\bm
  B}|$ in a certain interval, but usually does not fix the direction of
$\bm B$. The integration over any range of the impact parameter $|{\bm
  B}|$ (or the integration over the angles of $\bm B$ keeping $|{\bm
  B}|$ constant) is likely to introduce these geometric correlations,
as follows from \eqref{corr_main}. Note that the presence of non-trivial
wave function correlations, i.e., correlations beyond the
factorization approximation in \eqref{factorization2}, may also lead to
correlations which may survive in \eqref{eq:CfixedB} even for a fixed
impact parameter $\bm B$ and uncorrelated interactions
\eqref{int_fact}.

Despite its simplicity, the non-vanishing correlation in
\eqref{corr_fact} is one of the main results of this work. In the
saturation/CGC framework it may lead to long-range rapidity
correlations similar to the observed ``ridge'' correlation. Indeed
azimuthal correlations are missing in \eqref{corr_fact}: such
correlations may be formed in heavy ion collisions due to radial flow,
as was argued in \cite{Gavin:2008ev}. Indeed a lot more work is needed
to compare this result with experiment.

%% file: Chapter_pA_review.tex
% !TEX root = WertepnyPhDThesis.tex
\cleardoublepage
\chapter{Review of single-gluon production}
\label{ch:pAreview}

In this chapter we review the calculation for the single inclusive gluon production cross section taking into account a single nucleon in the target.
We use this as a pedagogical way to explain, in detail, how saturation formalism is used in practice.
This serves a number of purposes.
Not only do we need the analytic form of the cross section itself, but the various equations derived here are useful when looking at more complicated situations.
By going through the derivation carefully we can highlight the various approximations and techniques used throughout.
These approximations are often stated in the literature but not clearly justified so it is helpful to go through these in a rigorous manner, not only to justify them but to show exactly how these approximations can break down in more complicated situations.
Various techniques used throughout the derivation can also be extended into other situations, so we will demonstrate them here for later use.

%%%%%%%%%%%%%%%%%%%%%%%%%%%%%%%%%%%%%%%
\input{Section_shock}
\input{Section_pA_cross}
\input{Section_gluon_dipole}
\input{Section_pA_kt}
%%%%%%%%%%%%%%%%%%%%%%%%%%%%%%%%%%%%%%%

\section{Summary}

In this section we reviewed single-gluon production in the heavy-light ion regime.
This served as a pedagogical overview of how the saturation formalism can be applied to gluon production in heavy-light ion collisions.
The quarks and gluons originating from the light ion were found to interact with the the target ion as if it were an instantaneous shock wave.
From here we were able to derive the cross section associated with single-gluon production and then were able to write this cross section in a factorized form.
With the derivations presented here in mind we can now extend the formalism used here to the more complicated case of two-gluon production.

%% file: Section_shock.tex
% !TEX root = WertepnyPhDThesis.tex
% !TEX encoding = UTF-8 Unicode
% !TEX spellcheck = en_US

\section{Target nucleus as a shock wave}
\label{sec:target}

Due to the power counting associated with the target nucleus, $\as^2 A^\frac{1}{3}_2 \sim 1$, we consider all of the nucleons in the target when calculating the cross section.
This seems like a daunting task at first but using the eikonal approximation and the light-cone gauge, $A^+=0$, this becomes feasible.
Under these conditions, each individual nucleon can be viewed as emitting gluon fields which are not only localized at a specific light-cone time $x^+$ but to eikonal accuracy do not change the light-cone energy $k^+$ of either quarks or gluons originating from the projectile.
To examine how this comes about we calculate the gluon field emitted by a single nucleon from the target.

We model the nucleon as a massless eikonal quark, which implies the following:
First, it is massless and has a large $P^-$ momentum ($P^+$ for projectile quarks) which dominates over all other momentum scales in the problem except the momenta of the other color source quarks (either projectile or target) in the problem. We also only keep the leading-order terms in $P^-$.
Second, we view the quark as a classical source which feels no back reaction.
In other words, it has infinite $P^-$ momentum such that when it emits a gluon with momentum $k^-$ it still has momentum $P^-$ even if we integrated over all values of $k^-$, which means $P^- - k^- \approx P^-$.
This may seem like a poor approximation, however it is the same approximation as the one used when solving Maxwell's equations for a given electric current.

%%%%%%%%%%%%%%%%%%%%%%%%%%%%%%%%%%%%%%%%%%%%%%%%%%%%%%%%%%%%%%%%%%%%%%%%%%%%
\begin{figure}[H]
\centering
  \includegraphics[width= 0.5 \textwidth]{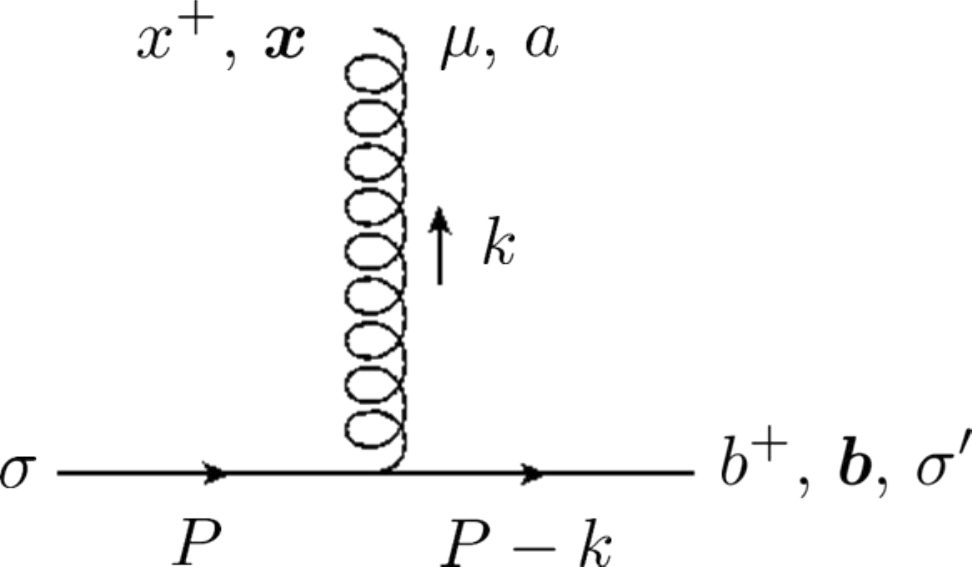}
  \caption{The Feynman diagram associated with the classical gluon field emitted by a target nucleon,
  $A^{a}_{MV, \mu}$. The nucleon is represented by a massless, eikonal valence quark traveling in the $x^-$ direction.
  Due to the fact that the outgoing and incoming quark lines are on mass-shell the outgoing quark has $b^-=\infty$.}
\label{AminusMV} 
\end{figure}
%%%%%%%%%%%%%%%%%%%%%%%%%%%%%%%%%%%%%%%%%%%%%%%%%%%%%%%%%%%%%%%%%%%%%%%%%%%%

To find the produced classical gluon field, notated $A^{a}_{MV, \mu}$, we evaluate the Feynman diagram shown in Fig.~\ref{AminusMV}.
The massless eikonal quark travels in the $x^-$ direction with initial momentum $P=(0,P^-,\bm 0)$ and emits a single off-shell gluon with momentum $k$.
The outgoing quark has momentum $(-k^+, P^- - k^-, -\bm k)$ and is on-shell, resulting in a delta function that sets the momentum squared to zero, $\delta ( (P-k)^2 )$.
The spins of the incoming and outgoing quarks, $\sigma$ and $\sigma'$ respectively, are averaged over.
The outgoing gluon has vector index $\mu$ and color index $a$.
Here we are not keeping track of the quark color indices explicitly.

It ends up being more convenient to have the final result in coordinate space so we include the Fourier transform over $k^-$ and $\bm k$, resulting in an expression which depends on the coordinates $(x^+ -b^+)$ and $(\bm x - \bm b)$. From the  Feynman diagram we arrive at the following expression:
\begin{align}
\label{eq:AMVmu1}
A^{a}_{MV, \mu} = \frac{1}{2} \, \sum_{\sigma, \sigma'} \int & \frac{d^4 k}{(2 \pi)^4}
  e^{-i k^+ x^- - i k^- (x^+ - b^+) + i \bm k \cdot ( \bm x - \bm b )} \,
 \bar u_\sigma' (P - k) i \, g \, t^a \, \gamma^\nu \, u_\sigma (P)
\notag \\
  & \times \frac{-i}{k^2 + i \epsilon} \left( g_{\mu \nu}
  - \frac{1}{k^+}\left( \eta_\mu k_\nu + \eta_\nu k_\mu \right) \right)
  2 \pi \, \delta((P - k)^2).
\end{align}
To further evaluate this equation a few steps are required.

First, we need to evaluate the spinors associated with the quark line which is done by splitting it up into parts.
By using the Dirac equation we can evaluate the contraction of $\bar u_\sigma' (P - k) \, \gamma^\nu \, u_\sigma (P)$ with $(\eta_\mu k_\nu + \eta_\nu k_\mu)$. 
Looking at the $k_\nu$ term we see that by adding and subtracting $P_\nu$,
\begin{equation}
\bar u_\sigma' (P - k) \slashed{k} u_\sigma (P) = \bar u_\sigma' (P - k)
  \left( \slashed{P} - \left( \slashed{P}-\slashed{k}\right) \right) u_\sigma (P) = 0,
\end{equation}
this contribution is exactly zero. Examining the $\eta_\nu$ term and noticing that $\gamma^+ \propto \slashed{P}$ we see this contribution vanishes as well,
\begin{equation}
\eta_\nu \gamma^\nu u_\sigma (P) = \gamma^+ u_\sigma (P) 
  \propto \slashed{P} u_\sigma (P)  =0.
\end{equation}
Analyzing the contraction of $\bar u_\sigma' (P - k) \, \gamma^\nu \, u_\sigma (P)$ with the remaining $g_{\mu \nu}$ term using the spinor identities in \cite{Lepage:1980fj,KovchegovLevin} (where we switch from the $\gamma^+$ basis to the $\gamma^-$ basis), we arrive at
\begin{equation}
g_{\mu \nu} \, \bar u_\sigma' (P - k) \gamma^\nu u_\sigma (P) 
  \approx g_{\mu -} \, \bar u_\sigma' (P - k) \gamma^- u_\sigma (P) 
  = \eta_\mu \, 2 P^- \delta_{\sigma' \sigma}
\end{equation}
where we have used the eikonal limit to neglect  terms subleading in $P^-$.
The delta function can be similarly analyzed and, once again taking the eikonal limit by neglecting $\frac{k^2_\perp}{2 P^-}$, we arrive at 
\begin{equation}
  \delta((P - k)^2) \approx \frac{1}{2 P^-} \delta (k^+)
\end{equation}
Using these results equation \eqref{eq:AMVmu1} becomes
\begin{align}
\label{eq:AMVmu2}
A^{a}_{MV, \mu} = -g \, t^a  \, \eta_\mu \, \frac{1}{2} \, \sum_{\sigma, \sigma'}
  \, \delta_{\sigma' \sigma} \int & \frac{d^4 k}{(2 \pi)^3}
  e^{-i k^+ x^- - i k^- (x^+ - b^+) + i \bm k \cdot ( \bm x - \bm b )}
  \frac{-1}{k^2 + i \epsilon} \, \delta (k^+).
\end{align}
From here, summing and averaging over the spins and integrating over $k^+$ gives
\begin{align}
\label{eq:AMVmu2half}
A^{a}_{MV, \mu} = -g \, t^a  \, \eta_\mu \,
  \int & \frac{d k^-}{2 \pi}\frac{d^2 k}{(2 \pi)^2}
  e^{ - i k^- (x^+ - b^+) + i \bm k \cdot ( \bm x - \bm b )}
  \frac{1}{k^2_\perp}.
\end{align}
At this point we need to evaluate the remaining Fourier transforms.
The $k^-$ integral gives rise to a delta function (with the usual normalization factor of $2 \pi$) while the transverse momentum integral gives rise to a logarithmic term (see Appendix A in \cite{KovchegovLevin}) resulting in the final expression 
\begin{equation}
\label{eq:AMVmu3}
A^{a}_{MV, \mu} = - \eta_\mu \, \frac{g }{2 \pi} t^a \, \delta (x^+ - b^+) \, 
  \ln \frac{1}{|\bm x - \bm b | \Lambda},
\end{equation}
where $\Lambda$ is some IR cutoff which is on the order of $\Lambda_{QCD}$.

This result has a few interesting properties.
First, the vector field only consists of a single component $\eta^\mu$. Second, the field is instantaneous: It is located at a single light-cone time, $b^+$.
It also has no $k^+$ momentum, meaning it does not change the plus momentum of any particle it interacts with. All of these properties end up being useful when trying to sum over an arbitrary number of scatterings.
At this point it is helpful to define the following field:
\begin{equation}
\label{eq:AminusMVi}
A^{a, -}_{MV, i} (x^+, \bm x) = - \frac{g}{2 \pi} t^a_i \, \delta (x^+ - b^+_i) \, 
  \ln \frac{1}{|\bm x - \bm b_i| \Lambda}.
\end{equation}
This is the classical gluon field produced by nucleon $i$, which lies at coordinates $b^+_i$ and $\bm b_i$. As we will see, this notation is useful when considering many scatterings.

\begin{figure}[H]
\centering
  \includegraphics[width= 0.7 \textwidth]{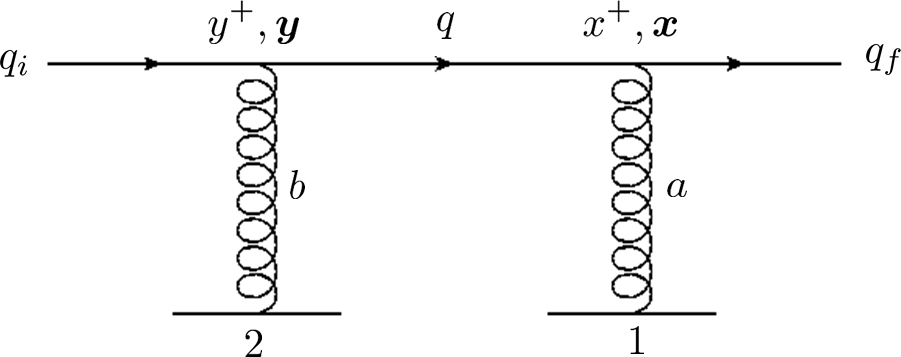}
  \caption{Feynman diagram representing a projectile quark interacting with the classical gluon fields
  emitted from nucleons $1$ and $2$. The external lines of the projectile quark are dropped in the
  calculation.}
\label{fig:2scattering} 
\end{figure}

With an expression for the classical gluon field emitted by a single nucleon in the target in hand, we can now see what happens when a quark interacts with many of these fields.
To start, consider the case of a quark interacting with two different nucleons, one with coordinates $b^+_1$ and $\bm b_1$ where the exchanged gluon has color $a$, the other with coordinates $b^+_2$ and $\bm b_2$ where the exchanged gluon has color $b$, shown in Fig.~\ref{fig:2scattering}.
The quark interacts with the field of nucleon $1$ at position $x^+, \bm x$ and with the field of nucleon $2$ at position $y^+, \bm y$.
The interaction with these fields does not change the plus momentum of the quark, hence $q_i^+= q^+= q_f^+$. The knowledge of these positions implies that the remaining momenta are not conserved at the vertices.

%%%%%%%%%%%%%%%%%%%%%%%%%%%%%%%%%%%%%%%%%%%%%%%%%%%%%%%%%%%%%%%%%%%%%%%%%%%%
%%%%%%%%%%%%%%%%%%%%%%%%%%%%%%%%%%%%%%%%%%%%%%%%%%%%%%%%%%%%%%%%%%%%%%%%%%%%

Since all but the plus momenta are not conserved at the vertices we can analyze the Feynman diagram in Fig.~\ref{fig:2scattering} while ignoring the external quark propagators. We only keep track of the vertices, the quark propagator with momentum $q$ and the fields due to the two nucleons. The expression for this truncated diagram is 
\begin{align}
V_2 = \int & \frac{d q^-}{2 \pi}\frac{d^2 q}{(2 \pi)^2}
  e^{i q^- (x^+ - y^+) - i \bm q \cdot ( \bm x - \bm y )}
  i \, g \, t^a \, \gamma^+ \frac{ i \, \slashed{q}}{q^2 + i \epsilon}
  i \, g \, t^b \, \gamma^+
\notag \\
  & \times A^{a, -}_{MV, 1} (x^+, \bm x)
  A^{b, -}_{MV, 2} (y^+, \bm y),
\end{align}
which we temporarily call $V_2$ for notational purposes. Focusing on only a part of the expression and using $\gamma^+ \slashed{q} \gamma^+ = 2 q^+$, we evaluate the $q^-$ and $\bm q$ integrals:
\begin{align}
\int & \frac{d q^-}{2 \pi}\frac{d^2 q}{(2 \pi)^2}
  e^{- i q^- (x^+ - y^+) + i \bm q \cdot ( \bm x - \bm y )}
  \gamma^+ \frac{ i \, \slashed{q}}{q^2 + i \epsilon} \gamma^+
\notag \\
  & = \int \frac{d q^-}{2 \pi}\frac{d^2 q}{(2 \pi)^2}
  e^{- i q^- (x^+ - y^+) + i \bm q \cdot ( \bm x - \bm y )}
  \gamma^+ \gamma^- \gamma^+ 
  \frac{ i q^+ }{q^2 + i \epsilon}
\notag \\
  & = \gamma^+ \int \frac{d q^-}{2 \pi}\frac{d^2 q}{(2 \pi)^2}
  e^{- i q^- (x^+ - y^+) + i \bm q \cdot ( \bm x - \bm y )}
  \frac{ i }{q^- - \frac{\bm q^2}{2 q^+} + i \epsilon}
\notag \\
  & = \gamma^+ \Theta( x^+ - y^+) \int \frac{d^2 q}{(2 \pi)^2}
  e^{- i \frac{q^2_\perp}{2 q^+} (x^+ - y^+) + i \bm q \cdot ( \bm x - \bm y )}.
\end{align}
At this point, we use the approximation that the separation between the nucleons in the shock wave in the plus direction, $x^+ - y^+$, is small and the plus momentum of the quark is large, $q^+$, such that $\frac{q^2_\perp}{2 q^+} (x^+ - y^+) \approx 0$.
Since it requires a large $q^+$ momentum it is also considered an eikonal approximation.
It should be noted that this approximation gives the condition that $q^2_\perp \ll \frac{2 q^+}{x^+ - y^+}$, but since $\frac{2 q^+}{x^+ - y^+}$ is large we make the approximation that we can integrate $\bm q$ over all values.
Thus, the $\bm q$ integral results in a 2-dimensional delta function of the transverse coordinates.
The entire expression can then be written as
\begin{equation}
\label{eq:2scattering}
V_2 = i \, g \, t^a A^{a, -}_{MV, 1} (x^+, \bm x) 
  i \, g \, t^b A^{b, -}_{MV, 2} (y^+, \bm x)
  \Theta( x^+ - y^+) \, \delta^2 ( \bm x - \bm y ).
\end{equation}

This means that a quark scattering off the classical fields produced by the nucleons results in a time ordered scattering (in the $x^+$ direction) where the transverse position of the quark remains fixed along with its $q^+$ momentum.
This can be generalized for any number of nucleons and results in a path-ordered exponential, a Wilson line, that runs from the point of last scattering, $x^+$, to the point of first scattering, $y^+$ (this is a well known result \cite{Balitsky:1996ub,Weigert:2005us}).
The result for both quarks, $V_{\bm x} [x^+, y^+]$ (known as the fundamental representation), and gluons (known as the adjoint representation), $U^{a b}_{\bm x} [x^+, y^+]$, are shown below.
\begin{align}
\label{eq:Vline}
V_{\bm z} [x^+, y^+]& = \mbox{P} \exp \left\{ i \, g \,
  \int\limits_{y^+}^{x^+} \, d z^+ \, A^{-} (z^+, {\bm z}) \right\}
\\
\label{eq:Uline}
U_{\bm z} [x^+, y^+]& = \mbox{P} \exp \left\{ i \, g \,
  \int\limits_{y^+}^{x^+} \, d z^+ \,  {\cal A}^{ -} (z^+, {\bm z}) \right\},
\end{align}
Here, $A^{-}(x^+, {\bm x})$ and ${\cal A}^{-}(x^+, {\bm x})$ are the total color fields produced by the target in the fundamental and adjoint representations, respectively. In the classical limit, the limit derived above and notated by a $MV$ subscript, these are just the sum of all the fields produced by the nucleons,
\begin{align}
A^-_{MV} (x^+, {\bm x}) & = \sum^N_{i=1} t^a A^{a,-}_{MV,i} (x^+, {\bm x})
\\
{\cal A}^-_{MV}(x^+, {\bm x}) & = \sum^N_{i=1} T^a A^{a,-}_{MV,i} (x^+, {\bm x}),
\end{align}
where $A^{a,-}_{MV,i} (x^+, {\bm x})$ was defined in \eqref{eq:AminusMVi}.

%%%%%%%%%%%%%%%%%%%%%%%%%%%%%%%%%%%%%%%%%%%%%%%%%%%%%%%%%%%%%%%%%%%%%%%%%%%%
\begin{figure}[H]
\centering
  \includegraphics[width= 0.7 \textwidth]{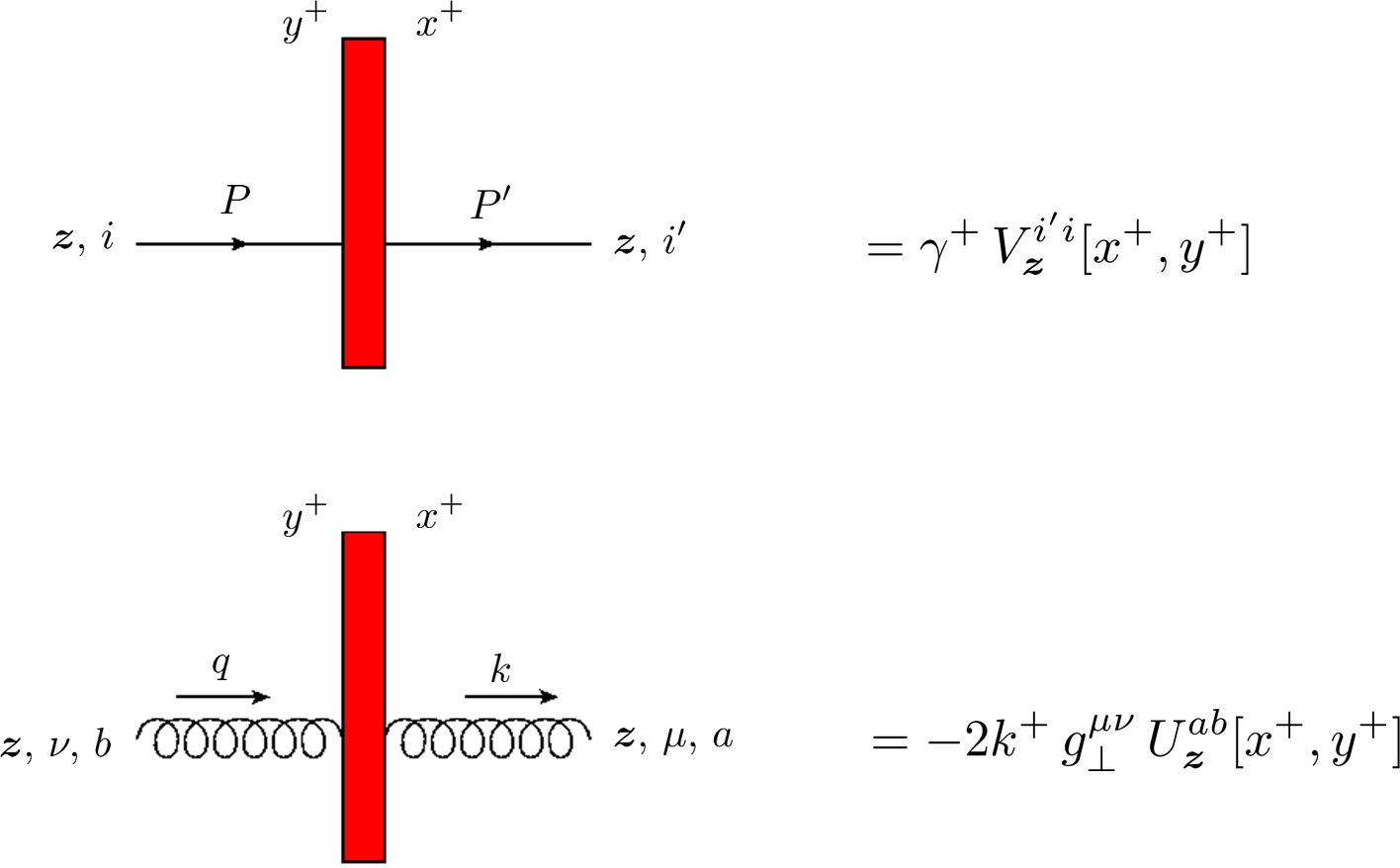}
  \caption{Feynman rules associated with quarks and gluons passing through a shockwave induced by the target nucleus.
  The shockwave is denoted by a red box.}
\label{shockrules} 
\end{figure}
%%%%%%%%%%%%%%%%%%%%%%%%%%%%%%%%%%%%%%%%%%%%%%%%%%%%%%%%%%%%%%%%%%%%%%%%%%%%
The Feynman rules associated with these shockwave interactions are shown in Fig.~\ref{shockrules}.
The shockwaves do not change the transverse position of the particles nor do they change the plus momentum of the particles ($P'^+ = P^+$ and $k^+ = q^+$).
The other momenta of the particles, the minus and transverse, are not conserved. The color indices of the Wilson lines represent the initial and final colors introduced by the scatterings, $(t^a t^b \cdots t^c)^{i' i}$ for quarks and $(T^c T^d \cdots T^e)^{a b}$ for gluons.
We notated the plus position of the final, $x^+$, and initial, $y^+$, scatterings.
At the end of the day, with the approximations used, the Wilson lines are independent of these terms but, at this point in our discussion, we do not know if the finite size of the shock wave could have any repercussions later on in the calculation.

Now with a clear understanding of how the quarks and gluons, originating from the projectile, interact with the gluon fields produced by the target we now approach the pA single gluon production case.

%% file: Section_pA_cross.tex
% !TEX root = WertepnyPhDThesis.tex

\section{Single-gluon production cross section}
\label{sec:single}

\begin{figure}[H]
\centering
  \includegraphics[width= 0.9 \textwidth]{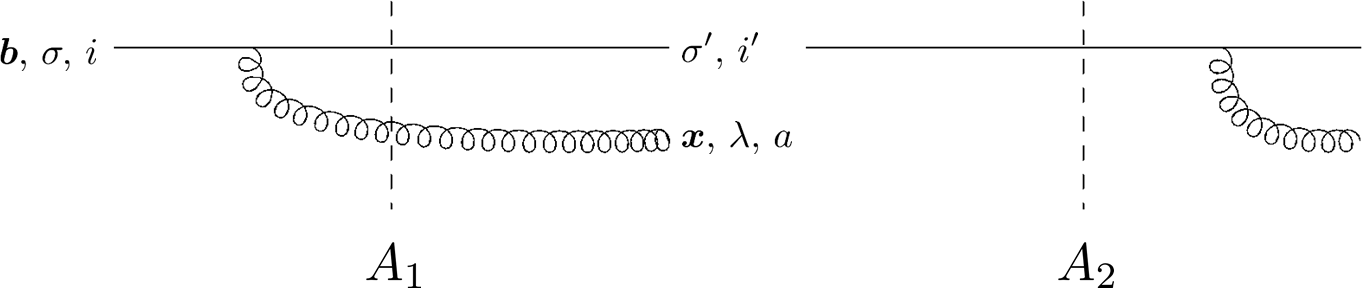}
  \caption{Single-gluon production amplitudes.
  The dashed line represents the shock wave due to the interaction with the target.}
\label{singlediagrams} 
\end{figure}

To calculate the single-gluon production cross section for the $pA$ case we first need to calculate the amplitude of the process, which has two contributing diagrams.
One diagram has the projectile nucleon (modeled as a valence quark) emitting the produced gluon before the interaction such that both the valance quark and the gluon interact with the target (modeled as a shock wave). The other has the gluon emitted after the interaction, only the quark ends up interacting with the target. Both are shown in Fig.~\ref{singlediagrams}.

%%%%%%%%%%%%%%%%%%%%%%%%%%%%%%%%%%%%%%%%%%%%%%%%%%%%%%%%%%%%%%%%%%%%%%%%%%%%
%%%%%%%%%%%%%%%%%%%%%%%%%%%%%%%%%%%%%%%%%%%%%%%%%%%%%%%%%%%%%%%%%%%%%%%%%%%%

\begin{figure}[H]
\centering
  \includegraphics[width= 0.7 \textwidth]{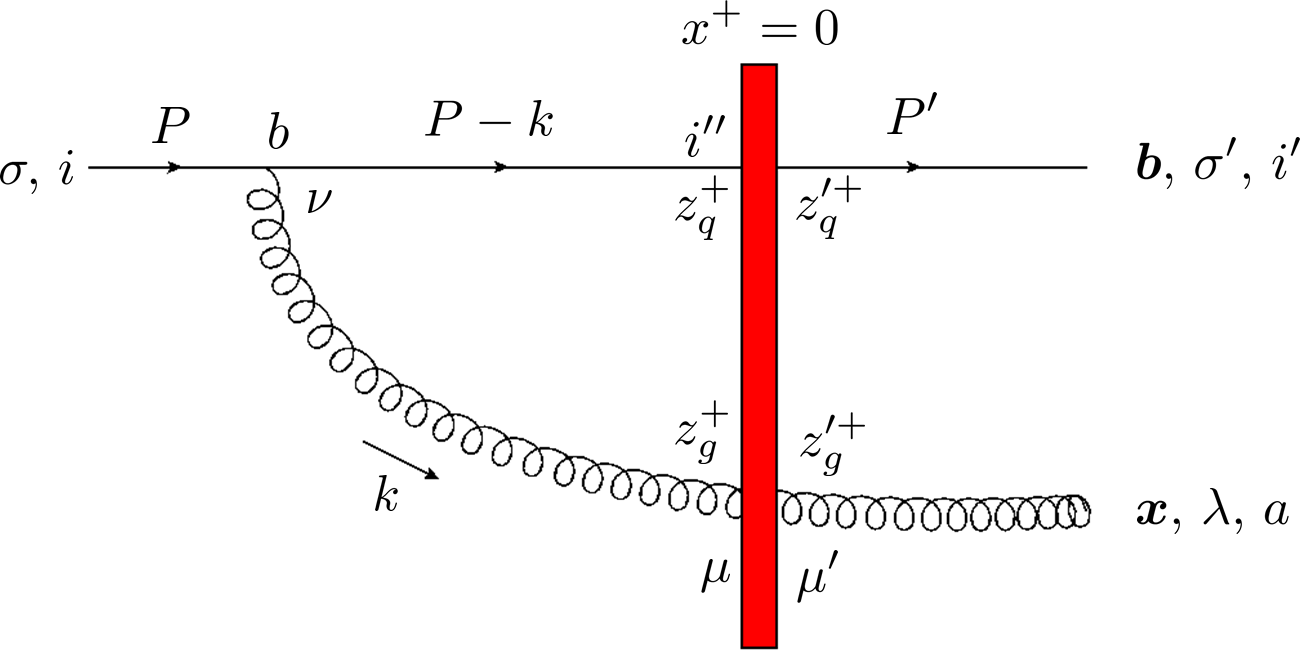}
  \caption{An eikonal quark originating from the projectile emits a gluon before passing through the shock wave.
  The shock wave is a red box to emphasize that the shock wave has some finite thickness.}
\label{pAdiagram} 
\end{figure}

As a pedagogical exercise, which gives more insight into later calculations, we explicitly calculate diagram $A_1$, shown in Fig.~\ref{singlediagrams} and in more detail in Fig.~\ref{pAdiagram}.
We use the usual Feynman rules for QCD in the light-cone gauge while adding in the additional rules associated with the shock wave interactions derived in the previous section, shown in Fig.~\ref{shockrules}.
%%%%%%%%%%%%%%%%%%%%%%%%%%%%%%%%%%%%%%%%%%%%%%%%%%%%%%%%%%%%%%%%%%%%%%%%%%%%
%%%%%%%%%%%%%%%%%%%%%%%%%%%%%%%%%%%%%%%%%%%%%%%%%%%%%%%%%%%%%%%%%%%%%%%%%%%%

The eikonal quark originating from the projectile travels in the $x^+$ direction with momentum $P = (P^+, 0, \bm0)$, spin $\sigma$, color $i$ and at transverse position $\bm b$.
It emits an off shell gluon, with momentum k, and both the quark and emitted gluon pass through the shock wave, which changes neither their transverse coordinates nor their plus momenta, $P^+$ and $k^+$ respectively, due to the eikonal approximation.
The final state quark has spin $\sigma'$, color $i'$ and the same plus momentum $P^+$.
Similar to the calculation done in the previous section, the outgoing quark being on shell gives rise to a factor of $\frac{1}{2 P^+}$.
The final state gluon has transverse coordinate $\bm x$, color $a$, polarization vector $\lambda$ and a plus momentum of $k^+$.
Also as before we want to know the amplitude in terms of the transverse coordinates $\bm b$ and $\bm x$. This requires us to Fourier transform $k^-$ and $\bm k$. Including all of these considerations for diagram $A_1$ we arrive at the expression
\begin{align}
\label{amp_begin}
A_1 = \int & \frac{d k^-}{2 \pi}
  \frac{d^2 k}{(2 \pi)^2} e^{-i  k^- ( z^+_g - z^+_q ) + i \bm k \cdot ( \bm x - \bm b )}
  \frac{1}{2 P^+} \bar u_{\sigma'} \left( P' \right) \gamma^+ V^{i' i''}_{\bm b}[z'^+_q , z^+_q]
  \frac{i \left( \slashed{P} - \slashed{k} \right)}{(P-k)^2 + i \epsilon} 
\notag \\
  \times & i g \, \gamma^\nu (t^b)^{i'' i} 
  \, u_{\sigma} \left( P \right) 
  \frac{-i}{k^2 + i \epsilon}
  \left( g_{\mu \nu} - \frac{\eta_\mu k_\nu +\eta_\nu k_\mu}{k^+} \right)
  \left(- 2 k^+g^{\mu' \mu}_\perp U^{a b}_{\bm x} [z'^+_g , z^+_g] \right)
  \epsilon^*_{\lambda, \mu'}.
\end{align}

First we analyze the quark line associated with the projectile quark, evaluating both the color structure and the spinors in the eikonal limit.
We use the same spinor identities used in the calculation of the classical gluon field presented in the previous section (except this time in the same $\gamma^+$ basis presented in the sources \cite{Lepage:1980fj,KovchegovLevin}) to evaluate the quark line spinors:
\begin{align}
\frac{1}{2 P^+} & \bar u_{\sigma'} \left( P' \right) \gamma^+ V^{i' i''}_{\bm b}[z'^+_q , z^+_q]
  \frac{i \left( \slashed{P} - \slashed{k} \right)}{(P-k)^2 + i \epsilon} i \gamma^\nu (t^b)^{i'' i}
  \, u_{\sigma} \left( P \right)
\notag \\
  & \approx \frac{1}{2 P^+} \bar u_{\sigma'} \left( P' \right) \gamma^+ V^{i' i''}_{\bm b}
  [z'^+_q , z^+_q] \frac{i P^+ \gamma^- }{- P^+ k^- + i \epsilon} i \gamma^\nu (t^b)^{i'' i}
  \, u_{\sigma} \left( P \right)
\notag \\
  & = \frac{1}{2 P^+} \frac{1}{ k^- - i \epsilon} \left( V_{\bm b} [z'^+_q , z^+_q] \, t^b \right)^{i' i}
  \bar u_{\sigma'} \left( P' \right) \gamma^+ \gamma^- \gamma^\nu \, u_{\sigma} \left( P \right)
\notag \\
  & \approx \frac{1}{ k^- - i \epsilon} \left( V_{\bm b} [z'^+_q , z^+_q] t^b \right)^{i' i}
  \delta_{\sigma' \sigma} \, g^{- \nu}.
\end{align}
Plugging this result into \ref{amp_begin} we arrive at
\begin{align}
\label{amp_A1k}
A_1 = 2 g \, \delta_{\sigma' \sigma}
  \int & \frac{d k^-}{2 \pi}
  \frac{d^2 k}{(2 \pi)^2} e^{- i k^- (z^+_g - z^+_q ) + i \bm k \cdot ( \bm x - \bm b )}
  \frac{i}{ k^- - i \epsilon} \frac{\bm \epsilon^*_{\lambda}
  \cdot \bm k }{2 k^+ k^- - k^2_\perp + i \epsilon}
\notag \\
  & \times U^{a b}_{\bm x} [z'^+_g , z^+_g] \, \left( V_{\bm b} [z'^+_q , z^+_q]
  \, t^b \right)^{i' i}.
\end{align}

At this point we first perform the Fourier transform over $k^-$,
\begin{align}
  \int \frac{d k^-}{2 \pi} & e^{-i k^- (z^+_g - z^+_q ) }
  \frac{i}{ k^- - i \epsilon}
  \frac{1}{2 k^+ k^- - k^2_\perp + i \epsilon}
\notag \\
  = \, & \Theta (z^+_g - z^+_q ) \frac{1}{k^2_\perp}
  e^{ - i \frac{k^2_\perp}{2 k^+} (z^+_g - z^+_q )} + \Theta (z^+_q - z^+_g ) \frac{1}{k^2_\perp}
\notag \\
  \approx \, & \frac{1}{k^2_\perp}.
\end{align}
Here we used the approximation that, since the width of the shock wave created by the target is small, $\frac{k^2_\perp}{2 k^+} (z^+_g - z^+_q ) \approx 0$.
It should be noted that this is the exact same approximation used in deriving \eqref{eq:2scattering} in the previous section.
Since the exponential term is suppressed, we often approximate the Fourier transform as just an integral over $k^-$ by setting the coordinates associated with the shock to be the same and at zero, $z^+_g = z'^+_g = z^+_q = z'^+_q =0$, since the target nucleus is centered around $x^+=0$. This causes us to arrive at
\begin{equation}
\label{amp_A?}
A_1 = 2 g \, \delta_{\sigma' \sigma}
  \int \frac{d^2 k}{(2 \pi)^2} e^{i \bm k \cdot ( \bm x - \bm b )}
  \frac{\bm \epsilon^*_{\lambda} \cdot \bm k}{ k^2_\perp } 
  U_{\bm x}^{a b} \left(V_{\bm b} t^b \right)^{i' i}.
\end{equation}
where you will notice we have neglected the $x^+$ dependence of the end points of the Wilson lines.
This is due to the fact that, with the approximations we are using, we can set the end points of the Wilson lines to to $x^+ = \pm \infty$ without loss of generality. This will be further be expanded upon in the next section.

Finally, performing the Fourier transform over $\bm k$, we arrive at
\begin{equation}
\label{amp_A1}
A_1 = \frac{i g}{\pi} \delta_{\sigma' \sigma} 
  \frac{\bm \epsilon^*_{\lambda} \cdot \left( \bm x - \bm b \right)}{ | \bm x - \bm b |^2 }
  U_{\bm x}^{a b} \left( V_{\bm b} t^b \right)^{i' i} .
\end{equation}
This same process can be repeated for diagram $A_2$ giving
\begin{equation}
\label{amp_A2}
A_2 = - \frac{i g}{\pi} \delta_{\sigma' \sigma} 
  \frac{\bm \epsilon^*_{\lambda} \cdot \left( \bm x - \bm b \right)}{ | \bm x - \bm b |^2 }
  \left( t^a V_{\bm b} \right)^{i' i}.
\end{equation}

The total amplitude for the single inclusive gluon production cross section is the sum of these two diagrams, $A_1+A_2$.
By using the Wilson line identity $\left( t^a V_{\bm b} \right)^{i' i} = \left( V_{\bm b} t^b \right)^{i' i}  U^{a b}_{\bm b}$,
the amplitude can be written in the compact form of
\begin{equation}
\label{amp_A}
A^{a,i' i}_{\sigma' \sigma, \lambda} (\bm x, \bm b) = \frac{i g}{\pi} \delta_{\sigma' \sigma} 
  \frac{\bm \epsilon^*_{\lambda} \cdot \left( \bm x - \bm b \right)}{ | \bm x - \bm b |^2 }
  \left[ U_{\bm x} - U_{\bm b} \right]^{a b} \left( V_{\bm b} t^b \right)^{i' i}.
\end{equation}
It turns out this same amplitude also appears in the 2-gluon production case so \ref{amp_A} will be used later, hence why we include all of the indices in the definition.

%%%%%%%%%%%%%%%%%%%%%%%%%%%%%%%%%%%%%%%%%%%%%%%%%%%%%%%%%%%%%%%%%%%%%%%%%%%%
\begin{figure}[H]
\centering
  \includegraphics[width=0.8 \textwidth]{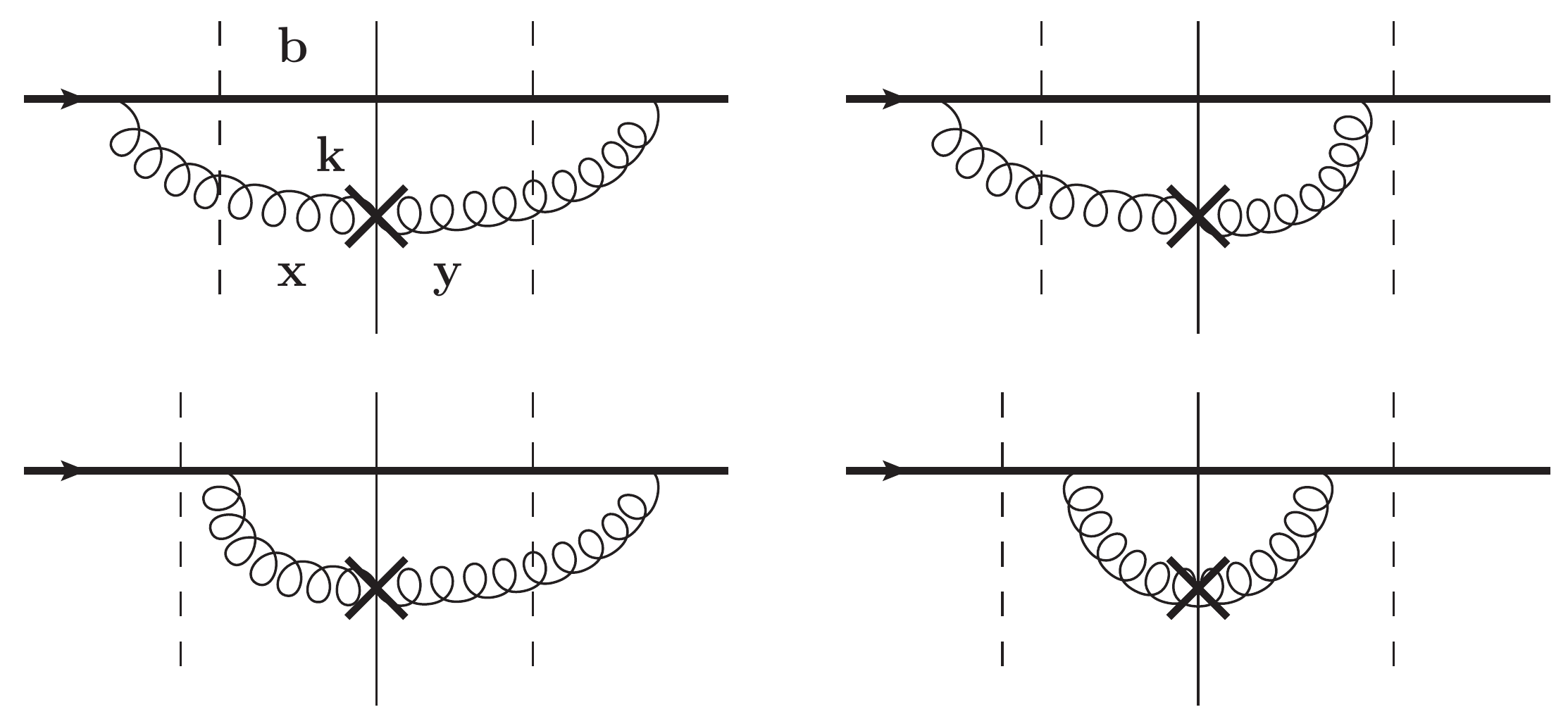}
  \caption{Diagrams contributing to the square of the scattering
    amplitude for the single gluon production in $pA$ collisions. The
    cross denotes the measured produced gluon. }
\label{1Gincl} 
\end{figure}
%%%%%%%%%%%%%%%%%%%%%%%%%%%%%%%%%%%%%%%%%%%%%%%%%%%%%%%%%%%%%%%%%%%%%%%%%%%

To find the cross section, we square the total amplitude while having the gluon's transverse position different on each side of the cut, $\bm x$ and $\bm y$. This is shown in Fig.~\ref{1Gincl}.
To get the cross section in momentum space requires a Fourier transform over each transverse position.
We sum and average over the colors and spins of the quark while only summing over the polarization and color of the produced gluon.
Averaging over all possible charge configuration of the target, represented by $\langle \cdots \rangle$, results in the following expression:
\begin{align}
\label{crosssectionpA1}
\left\langle \frac{d \sigma^{pA_2}}{d^2 k \, dy \, d^2 b}
  \right\rangle = \frac{1}{2 \, (2 \pi)^3} \int & d^2 x \, d^2 y \; e^{- i
  \, {\bm k} \cdot ({\bm x}-{\bm y})}
\notag \\
  & \times \frac{1}{2 N_c} \sum_{\sigma' , \sigma} \sum_\lambda \left\langle
  \left( A^{a,i' i}_{\sigma' \sigma, \lambda} \left( \bm y, \bm b \right)\right)^\dagger
  A^{a,i' i}_{\sigma' \sigma, \lambda} \left( \bm x, \bm b \right) \right\rangle.  
\end{align}

Evaluating the integrand
\begin{align}
\frac{1}{2 N_c} & \sum_{\sigma' , \sigma} \sum_\lambda
  \left( A^{a,i' i}_{\sigma' \sigma, \lambda} \left( \bm y, \bm b \right)\right)^\dagger
  A^{a,i' i}_{\sigma' \sigma, \lambda} \left( \bm x, \bm b \right)
\notag \\
  = & \, \frac{g^2}{2 N_c \pi^2} \sum_{\sigma', \sigma} \delta_{\sigma' \sigma} \delta_{\sigma' \sigma}
  \sum_{\lambda}
  \frac{\bm \epsilon^*_{\lambda} \cdot \left( \bm x - \bm b \right)}{ | \bm x - \bm b |^2 }
  \frac{\bm \epsilon_{\lambda} \cdot \left( \bm y - \bm b \right)}{ | \bm y - \bm b |^2 }
\notag \\
  & \times tr[t^c V^\dagger_b V_b t^b]
  \left[ \left( U^\dagger_{\bm y} - U^\dagger_{\bm b} \right) 
  \left( U_{\bm x} - U_{\bm b} \right) \right]^{c b}
\notag \\ \label{A2pA}
  = & \, \frac{g^2}{2 N_c \pi^2}
  \frac{\left( \bm x - \bm b \right)\cdot \left( \bm y - \bm b \right)}
  { | \bm x - \bm b |^2 | \bm y - \bm b |^2 }
  Tr \left[ \left( U_{\bm x} - U_{\bm b} \right)
  \left( U^\dagger_{\bm y} - U^\dagger_{\bm b} \right)\right]
\end{align}
and plugging this into \ref{crosssectionpA1} we arrive at the known result \cite{Kovchegov:1998bi}
\begin{align}
\label{crosssection12}
\left\langle \frac{d \sigma^{pA_2}}{d^2 k \, dy \, d^2 b}
  \right\rangle & = \frac{\as \, C_F}{4 \, \pi^4} \int d^2 x \, d^2 y \;
  e^{- i \, {\bm k} \cdot ({\bm x}-{\bm y})} \, \frac{ {\bm x} - {\bm
  b}}{ |{\bm x} - {\bm b} |^2 } \cdot \frac{ {\bm y} - {\bm b}}{
  |{\bm y} - {\bm b} |^2 } \notag \\
& \times \; \left[ S_G ( {\bm x} , {\bm y}) -
  S_G ( {\bm x} , {\bm b}) - S_G ( {\bm b} , {\bm y}) + 1
\right].
\end{align}
where $C_F = \frac{N_c^2 -1}{2 N_c}$ is the fundamental-representation
Casimir operator and $S_G ( {\bm x} , {\bm y}) $ is the adjoint (gluon) dipole $S$-matrix
\begin{equation}
\label{gdipole}
S_G ( {\bm x} , {\bm y}) = \frac{1}{N_c^2-1} \;
  \left\langle Tr[ U_{{\bm x}} U_{{\bm y}}^\dagger ] \right\rangle.
\end{equation}
It turns out, although not specified here, that one can include rapidity dependence of the emitted gluon into this expression by changing the form of $S_G$.
In general it is written as $S_G ( {\bm x} , {\bm y}, y)$ which corresponds to the classical case presented here when $y=0$. 
To include rapidity dependence one uses the BK/JIMWLK evolution equations to evolve the dipole \cite{Kovchegov:2001sc,Blaizot:2004wu}.
These effects, while important, are beyond the classical level of approximation we are dealing with here.

To get the single inclusive gluon production cross section for the heavy-light ion case, all that is left is inserting \ref{crosssection12} into \ref{pA}.
\begin{align}
\label{crosssection12aA}
\frac{d \sigma}{d^2 k \, dy}
  & = \frac{\as \, C_F}{4 \, \pi^4} A_1 \int d^2 b \, d^2 x \, d^2 y \;
  e^{- i \, {\bm k} \cdot ({\bm x}-{\bm y})} \, \frac{ {\bm x} - {\bm
  b}}{ |{\bm x} - {\bm b} |^2 } \cdot \frac{ {\bm y} - {\bm b}}{
  |{\bm y} - {\bm b} |^2 } \notag \\
& \times \; \left[ S_G ( {\bm x} , {\bm y}, y ) -
  S_G ( {\bm x} , {\bm b}, y) - S_G ( {\bm b} , {\bm y}, y) + 1 \right].
\end{align}
While the expression above is simple, it is meaningless unless the form of the adjoint (gluon) dipole $S$-matrix, $S_G ( {\bm b} , {\bm y})$, is known.

%% file: Section_gluon_dipole.tex
% !TEX root = WertepnyPhDThesis.tex
\section{Gluon dipole amplitude}
\label{sec:Dipole}

Here we derive the well-known result for the gluon dipole S-Matrix, $S_G (\bm x, \bm y)$, in the quasi-classical MV/Glauber--Mueller (GM) approximation.
This may seem tedious and unnecessary since it has been known for years \cite{Mueller:1989st} but it gives valuable insight to the nature of the eikonal scatterings associated with the target and allows us to generalize to more complicated objects such as the double-dipole operator $\frac{1}{(N_c^2 -1)^2} \left< Tr \left[ U_{\bm x} U_{\bm x} \right] Tr \left[ U_{\bm x} U_{\bm x} \right] \right>$, which is needed for two-gluon production. More about this object is presented in Chapter \ref{ch:Calculation} and Appendix \ref{Appendix:Wilson}.

The exact object that is associated with the single gluon production cross section in the $pA$ case \eqref{crosssection12} is
\begin{align}
\frac{1}{N_c^2-1}
\left< U_{\bm x} [z'^+, z^+] 
  U^\dagger_{\bm y} [w'^+, w^+] \right>,
\end{align}
where the angular brackets represent the averaging over all possible charge configurations of the target nucleus and the coordinates $z'^+, \, z^+, \, w'^+, \, w^+$ are associated with locations of the initial and final scatterings.
Within the approximations used, the cross section is independent of the $x^+$ position of the initial and final scatterings.
This includes both the Fourier transform over the $k^-$ momentum and the Wilson lines themselves.
For these reasons we drop the $x^+$ dependence from the definition of $S_G (\bm x, \bm y)$.

%%%%%%%%%%%%%%%%%%%%%%%%%%%%%%%%%%%%%%%%%%%%%%%%%%%%%%%%%%%%%%%%%%%%%%%%%%%%
\begin{figure}[H]
\centering
  \includegraphics[width= 0.8 \textwidth]{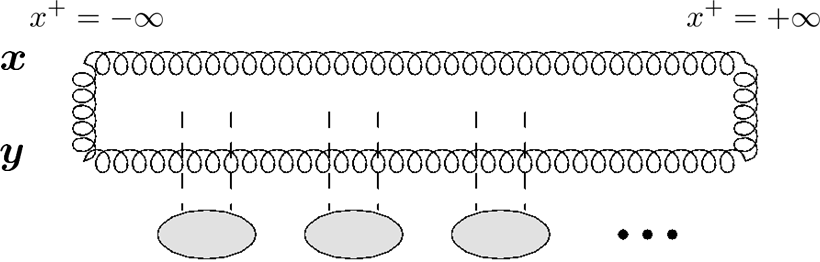}
  \caption{Graphical representation of the gluon dipole S-Matrix, $S_G (\bm x, \bm y)$.
  The gray ovals represent different target nucleons and the dotted lines are the exchanged gluons.
  These gluon can either interact with the top or bottom Wilson line.}
\label{Sdipole} 
\end{figure}
%%%%%%%%%%%%%%%%%%%%%%%%%%%%%%%%%%%%%%%%%%%%%%%%%%%%%%%%%%%%%%%%%%%%%%%%%%%%

The way we interpret this object is presented in figure \ref{Sdipole}.
Since the Wilson lines in $S_G$ only contain interactions with the fields originating from the target we can consider the Wilson lines as traveling from $x^+ = - \infty$ to $x^+ = + \infty$.
All of these target fields come with a vector component of $\eta_\mu$, meaning they do not couple with Wilson lines traveling in the transverse direction.
This allows us to insert two transverse gauge links at $x^+$ infinities into the object, which makes it gauge invariant.
As the dipole passes through the target fields, to leading-order in saturation effects, it can only interact with at least two gluons from a target nucleon because when one averages over the colors of the nucleon source one has a trace of the fundamental generators associated with the quark line, $tr [ t^a t^b ]$.
If only one gluon was exchanged this would give a trace of a single generator which is zero.
Also, interactions with more than two gluons from a single nucleon are suppressed by $\as$ as each gluon interaction gives a power of $\as$.
We also must consider all possible charge configurations and scatterings, meaning that we must consider the fact that only one or many nucleons may interact and that we must average over all possible transverse positions of said nucleons.

%%%%%%%%%%%%%%%%%%%%%%%%%%%%%%%%%%%%%%%%%%%%%%%%%%%%%%%%%%%%%%%%%%%%%%%%%%%%
\begin{figure}[H]
\centering
  \includegraphics[width= 0.6 \textwidth]{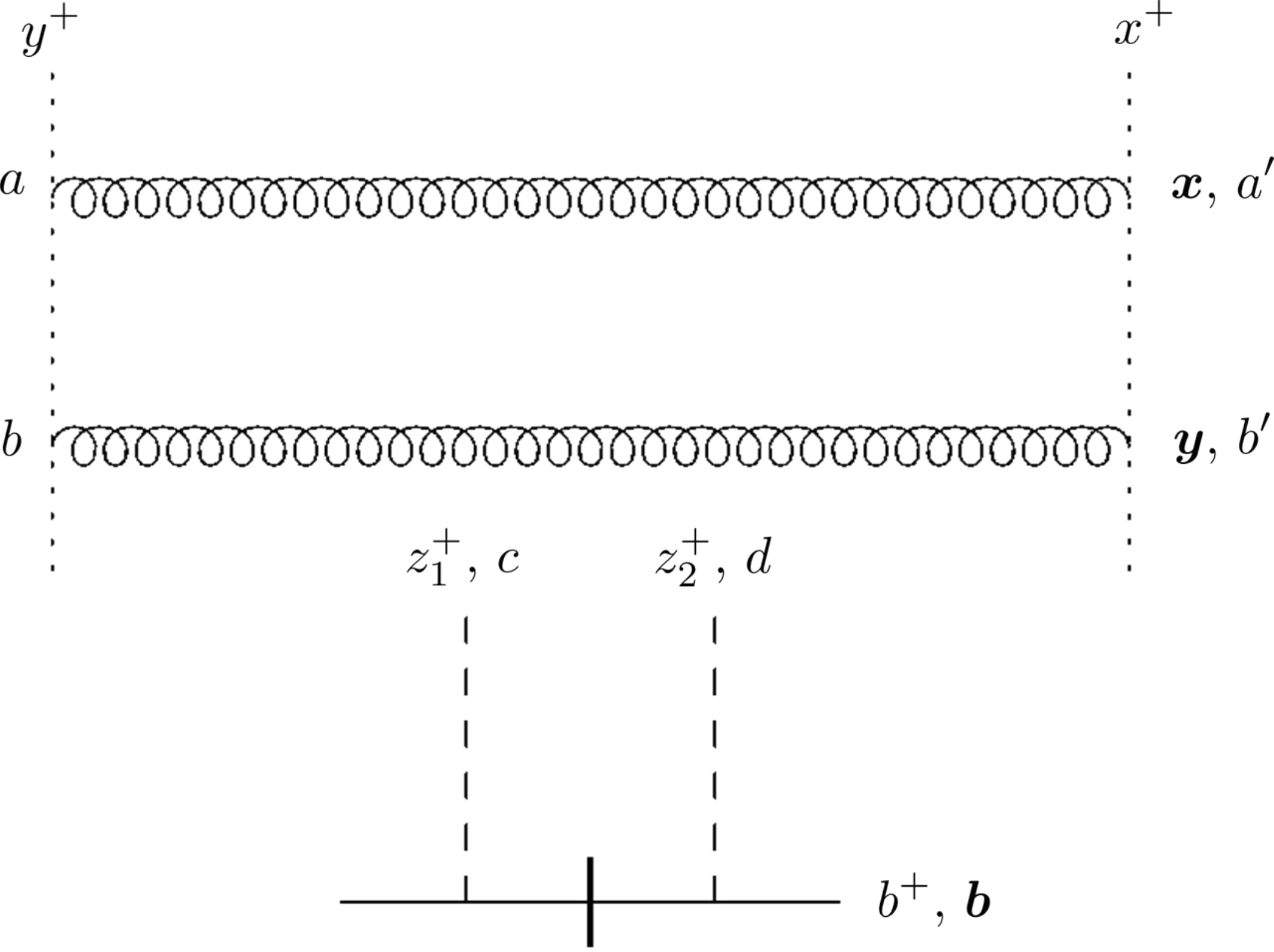}
  \caption{A finite section of the Wilson line object. The nucleon emits two gluon fields which interact with the two Wilson lines in all possible ways.}
\label{dipoletensor} 
\end{figure}
%%%%%%%%%%%%%%%%%%%%%%%%%%%%%%%%%%%%%%%%%%%%%%%%%%%%%%%%%%%%%%%%%%%%%%%%%%%%

To see the effect the multiple rescatterings have on the gluon dipole let us consider how the interaction with a single nucleon works. We consider a finite section of the Wilson line object.
There are two Wilson lines that both run from $x^+$ to $y^+$ and they lie at two different transverse positions, $\bm x$ and $\bm y$.
We keep track of the initial and final colors of the Wilson lines.
These two Wilson lines interact with a single nucleon that lies at position $b^+$ such that $x^+ > b^+ > y^+$.
We consider all possible gluon interactions between the target nucleon and the Wilson lines and average over the color and transverse position of said nucleon. This results in the expression
\begin{align}
\left< U^{a' a}_{\bm x} \right. & [x^+, y^+] \left.
  U^{b' b}_{\bm y} [x^+, y^+] \right>_1 - \delta^{a' a} \delta^{b' b}
\notag \\
  = & \left< g^2 \int^{x^+}_{y^+} d z^+ [T^c]^{a' e} A^{c, -}_{MV, 1} (z^+, \bm x)
  \int^{z^+}_{y^+} d z'^+ [T^d]^{e a} A^{d, -}_{MV, 1} (z'^+, \bm x) \delta^{b' b} \right.
\notag \\
  & + g^2 \delta^{a' a} \int^{x^+}_{y^+} d z^+ [T^c]^{b' e} A^{c, -}_{MV, 1} (z^+, \bm y)
  \int^{z^+}_{y^+} d z'^+ [T^d]^{e b} A^{d, -}_{MV, 1} (z'^+, \bm y)
\notag \\
  & \left. + g^2 \int^{x^+}_{y^+} d z^+ [T^c]^{a' a} A^{c, -}_{MV, 1} (z^+, \bm x)
  \int^{x^+}_{y^+} d z'^+ [T^d]^{b' b} A^{d, -}_{MV, 1} (z'^+, \bm y) \right>_1
\end{align}
where the angular brackets $\left< \cdots \right>_1$ represents averaging over all possible configurations of the nucleon, which we call nucleon 1.
We have subtracted out the case where the dipole does not interact.

Keeping in mind that each field $A^{a,-}_{MV,1} ( z^+, \bm x)$, \eqref{eq:AminusMVi}, comes with a delta function, $\delta (z^+ - b^+)$, which sets the location of the field equal to $b^+_1$, we can replace $\int^{x^+}_{y^+} d z^+ \int^{z^+}_{y^+} d z'^+ \rightarrow \frac{1}{2} \int^{x^+}_{y^+} d z^+ d z'^+$ resulting in
\begin{align}
\label{UUpart}
\left< U^{a' a}_{\bm x} \right. & [x^+, y^+] \left.
  U^{b' b}_{\bm y} [x^+, y^+] \right>_1 - \delta^{a' a} \delta^{b' b}
\notag \\
  = & - \frac{g^2}{2} \, [T^c T^d]^{a' a} \, \delta^{b' b} \int^{x^+}_{y^+} d z^+ d z'^+
  \left<  A^{c, -}_{MV, 1} (z^+, \bm x) A^{d, -}_{MV, 1} (z'^+, \bm x) \right>_1
\notag \\
  & - \frac{g^2}{2} \, \delta^{a' a} \, [T^c T^d]^{b' b} \int^{x^+}_{y^+} d z^+ d z'^+
  \left<  A^{c, -}_{MV, 1} (z^+, \bm y) A^{d, -}_{MV, 1} (z'^+, \bm y) \right>_1
\notag \\
  & - g^2 [T^c]^{a' a} [T^d]^{b' b} \int^{x^+}_{y^+} d z^+ d z'^+
  \left<  A^{c, -}_{MV, 1} (z^+, \bm x) A^{d, -}_{MV, 1} (z'^+, \bm y) \right>_1 .
\end{align}
At this point we need to average over the possible charge configurations of the target nucleon.
This requires averaging over the color of the nucleon and convolving the fields with the nuclear profile function associated with the target nucleus, $T_2 (\bm b)$.
Using the known equation for the classical gluon field $A^{a,-}_{MV,1} ( z^+, \bm x)$, \eqref{eq:AminusMVi}, we have (doing a few simplifications in the process)
\begin{align}
\int^{x^+}_{y^+} & d z^+ d z'^+ \left<  A^{c, -}_{MV, 1} (x^+, \bm x)
  A^{d, -}_{MV, 1} (y^+, \bm y) \right>_1
\notag \\
  &= \frac{1}{N_c} \int^{x^+}_{y^+} d z^+ d z'^+ \int d^2 b \; T_2 (\bm b)
  \; tr \left[ A^{c, -}_{MV, 1} (x^+, \bm x) A^{d, -}_{MV, 1} (y^+, \bm y) \right] 
\notag \\
  & = \frac{g^2}{N_c (2 \pi)^2} tr[ t^c t^d ] 
  \, \int d^2 b \, T_2 \left( \bm b \right) 
  \ln \frac{1}{|\bm x - \bm b| \Lambda} \ln \frac{1}{|\bm y - \bm b| \Lambda}.
\end{align}

When doing the $\bm b$ convolution it is common to use the approximation that the profile changes slowly with respect to the transverse width of the dipole, $|\bm x - \bm y|$.
With this we can use the approximation that $T_2 (\bm b) \approx T_2 \left(\frac{\bm x +\bm y}{2}\right)$, which allows us to pull the nuclear profile function out of the integral.
We also use the Fourier transform of $\ln \frac{1}{|\bm x - \bm b| \Lambda}$ to evaluate the rest of the expression
\begin{align}
\int^{x^+}_{y^+} & d z^+ d z'^+ \left<  A^{c, -}_{MV, 1} (x^+, \bm x)
  A^{d, -}_{MV, 1} (y^+, \bm y) \right>_1
\notag \\
  & = \frac{g^2}{2 N_c} \delta^{cd} \, T_2 \left( \frac{\bm x +\bm y}{2} \right) 
  \int d^2 b \, \int \frac{d^2 k}{(2 \pi)^2} \frac{d^2 q}{(2 \pi)^2}
  e^{-i \bm k \cdot (\bm x -\bm b) - i \bm q \cdot (\bm y -\bm b)} \frac{1}{\bm k^2 \bm q^2}
\notag \\
  & = \frac{g^2}{2 N_c} \delta^{cd} \, T_2 \left( \frac{\bm x +\bm y}{2} \right)
  \int \frac{d^2 k}{(2 \pi)^2} e^{-i \bm k \cdot (\bm x -\bm y)} \frac{1}{|\bm k^2|^2}
\end{align}
Plugging this result in to \eqref{UUpart} we arrive at
\begin{align}
\label{1nucleonint}
\left< U^{a' a}_{\bm x} \right. & \left. [x^+, y^+]
  U^{b' b}_{\bm y} [x^+, y^+] \right>_1  - \delta^{a' a} \delta^{b' b}
  = - \frac{g^4}{4 \pi } \, T_2 \left( \frac{\bm x +\bm y}{2} \right) 
  \int \frac{d^2 k}{2 \pi} \frac{1}{|\bm k^2|^2}
\notag \\
  & \times \left( \frac{1}{2 N_c} \left( [T^c T^c]^{a' a} \, \delta^{b' b} +
  \delta^{a' a} [T^c T^c]^{b' b} \right) + \frac{1}{N_c} [T^c]^{a' a} [T^c]^{b' b}
  e^{-i \bm k \cdot (\bm x -\bm y)} \right)
\notag \\
  & = - \, Q^2_{s,2} \left( \frac{\bm x +\bm y}{2} \right) 
  \int \frac{d^2 k}{2 \pi} \frac{1}{|\bm k^2|^2}
  \left( \delta^{a' a} \, \delta^{b' b} + \frac{1}{N_c} [T^c]^{a' a} [T^c]^{b' b}
  e^{-i \bm k \cdot (\bm x -\bm y)} \right),
\end{align}
where we used $[T^c T^c]^{ab}=N_c \, \delta^{ab}$.
We also used the definition of the saturation scale at the classical level, $Q^2_{s,2} (\bm b)= 4 \pi \, \as^2 T_2 (\bm b)$, to replace the nuclear profile function with the saturation scale, since $T_2 (\bm b) \sim A^\frac{1}{3}_2$.
This allows use to directly see that for every nucleon that interacts with the target we pick up an order of $\as^2 A^\frac{1}{3}_2$ as promised.

The above equation tells us what we gain every time the Wilson lines interact with a single nucleon.
While important, it is not the final result we need; we need to consider what happens after many such scatterings. 
To do this we need to first notice that in the initial and final states, the Wilson lines, for $S_G (\bm x, \bm y)$, are in a color singlet.
This means that the term we pick up for each scattering is given by \ref{1nucleonint} when hit with $\delta^{ab}$. Evaluating this we have
\begin{align}
\delta^{ab} \left< U^{a' a}_{\bm x} \right. & \left. [x^+, y^+]
  U^{b' b}_{\bm y} [x^+, y^+] \right>_1 - \delta^{a' b'}
\notag \\
  & = - \delta^{a'b'}\, Q^2_{s,2} \left( \frac{\bm x +\bm y}{2} \right) 
  \int \frac{d^2 k}{2 \pi} \frac{1}{|\bm k^2|^2}
  \left( 1 - e^{-i \bm k \cdot (\bm x -\bm y)} \right).
\end{align}
Using the relation
\begin{equation}
\int \frac{d^2 k}{2 \pi} \frac{1}{|\bm k^2|^2} \left( 1 - e^{-i \bm k \cdot \bm r} \right) 
  = \frac{1}{4} \bm r^2 \, \ln \frac{1}{|\bm r| \Lambda}
\end{equation}
we can write this result as
\begin{align}
\delta^{ab} \left< U^{a' a}_{\bm x} [x^+, y^+]
  U^{b' b}_{\bm y} [x^+, y^+] \right>_1 - \delta^{a' b'}
  = & - \frac{\delta^{a'b'}}{4} |\bm x -\bm y|^2 \,
  Q^2_{s,2} \left( \frac{\bm x +\bm y}{2} \right) \, \ln \frac{1}{|\bm x -\bm y| \Lambda}
\notag \\
= & - \delta^{a'b'} \, \Gamma_G (\bm x, \bm y) 
\end{align}
which defines $\Gamma_G (\bm x, \bm y)$.

We can see that every time we hit a nucleon we gain a term which is a function of the transverse positions of the dipole, $\bm x$ and $\bm y$, but keep the same color structure.
So when we finally reach the end of the dipole (after scattering off the number of nucleons that are in the target) we end up with the color factor $\delta^{a'b'}\delta^{a'b'}=N^2_c -1$.

In order to find $S_G$ we need to average over all possible charge configurations.
If the dipole interacts with no charges we have a contribution of a factor of 1, while if dipole interacts with 1 charge we gain a factor of $- \Gamma_G (\bm x, \bm y)$.

Now suppose the dipole interacts with two nucleons, labeled 1 and 2.
The dipole could interact with nucleon 1 first then 2 or the other way around.
This causes us to pick up a factor of $\Gamma^2_G (\bm x, \bm y)$ since there are two interactions and a factor of $\frac{1}{2}$ to eliminate double-counting due to the fact that the nucleons are identical.
For three nucleons we have $- \frac{1}{3!} \Gamma^3_G (\bm x, \bm y)$ due to the three scatterings and the 6 possible orderings.
This pattern continues on
\begin{align}
1 -\Gamma_G (\bm x, \bm y) +\frac{1}{2} \Gamma^2_G (\bm x, \bm y)
  -\frac{1}{3 !} \Gamma^3_G (\bm x, \bm y)+\cdots = e^{-\Gamma_G (\bm x, \bm y) }
\end{align}
and results, when one considers all possibilities, in an exponential function.
So for the Gluon Dipole, we have \cite{Mueller:1989st}
\begin{equation}
S_G (\bm x, \bm y) = \exp \left[ - \frac{1}{4} |\bm x -\bm y|^2 \,
  Q^2_{s,2} \left( \frac{\bm x +\bm y}{2} \right) \, \ln \frac{1}{|\bm x -\bm y| \Lambda}\right].
\end{equation}

%%%%%%%%%%%%%%%%%%%%%%%%%%%%%%%%%%%%%%%%%%%%%%%%%%%%%%%%%%%%%%%%%%%%%%%%%%%%
\begin{figure}[H]
\centering
  \includegraphics[width= 0.8 \textwidth]{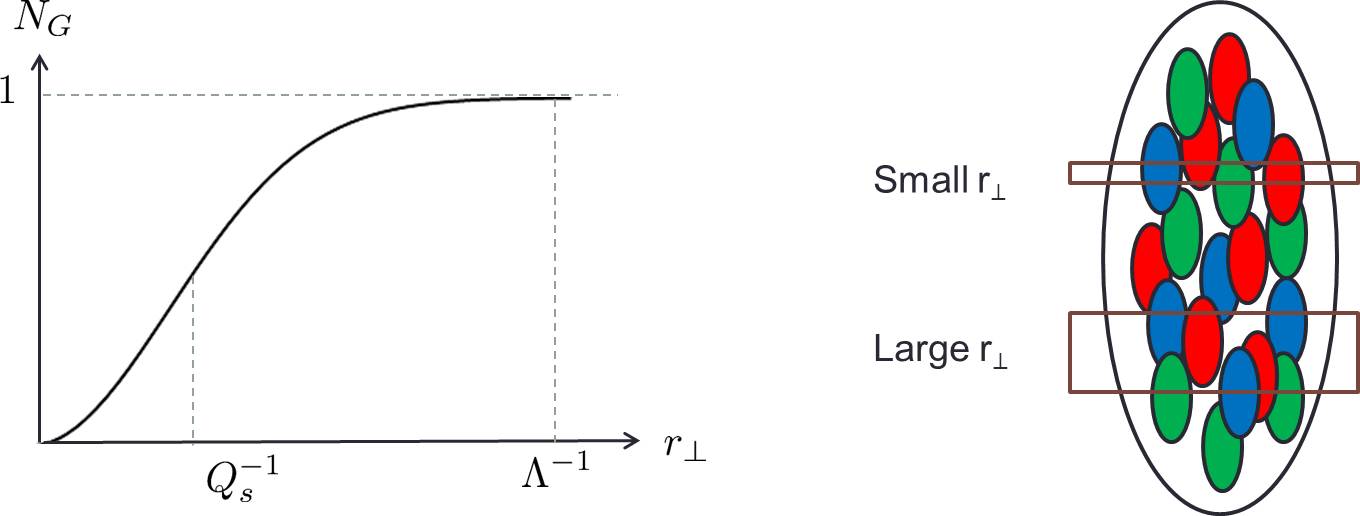}
  \caption{Left: Plot of the forward scattering amplitude, $N_G ({\bm x}, {\bm y})$, of the gluon dipole where
  $r_\perp = |\bm x - \bm y|$ is the transverse size of the dipole. To plot this we neglected
  dependence of the function on the nuclear profile function and the log dependence. Right: A pictorial representation
  of the physical picture. The smaller the dipole the less charge it sees so the nucleus is more transparent.
  The larger the dipole is the more charge it sees and the more opaque the target becomes. This continues until
  the dipole sees so much charge that the charges start to screen one another.}
\label{dipoleplot} 
\end{figure}
%%%%%%%%%%%%%%%%%%%%%%%%%%%%%%%%%%%%%%%%%%%%%%%%%%%%%%%%%%%%%%%%%%%%%%%%%%%%

This result has a clear interpretation once we write this in terms of the forward scattering amplitude for the gluon dipole,
\begin{align} 
\label{ngdipole} 
N_{G} ({\bm x}, {\bm y}) = 1 - S_G (\bm x, \bm y).
\end{align}
This tells us the probability of the gluon dipole passing through the target nucleus. The majority of the dependence comes from the size of the dipole which we write as $r_\perp = | \bm x - \bm y|$.
When the dipole size is much less than the saturation scale, $r_\perp \ll Q_{s,2}^{-1} $, the dipole interacts with a small amount of color charge so the nucleus is fairly transparent.
This causes the amplitude to scale as $\sim r_\perp^2 $.
However, when the dipole size is large when compared with the saturation scale, $r_\perp \gg Q_{s,2}^{-1} $, the dipole interacts with a large amount of color charge: the nucleus become opaque.
This only extends to a certain point, because the charges end up screening each other, causing the target to saturate, reaching the black disc limit.
A pictorial representation of this is seen in figure \ref{dipoleplot}.

%% file: Section_pA_kt.tex
% !TEX root = WertepnyPhDThesis.tex
\section{$k_T$-factorization for single-gluon production}
\label{sec:pAfact}

It is a well-known result of saturation physics that the single-gluon
production cross section in the proton-nucleus ($pA$) collisions
calculated either in the quasi-classical or leading-$\ln 1/x$
evolution approximations can be cast in the form consistent with
$k_T$-factorization \cite{Braun:2000bh,Kovchegov:2001sc,Kharzeev:2003wz}.

Here we are going to derive this $k_T$-factorized expression for the heavy-light collision case in the quasi-classical approximation following the derivations presented in \cite{Jalilian-Marian:2005jf,KovchegovLevin}. Staring by using \eqref{crosssection12} in \eqref{p2A1} we arrive at
\begin{align}
\label{ktstep1}
\frac{d \sigma}{d^2 k \, dy}
  & = \frac{\as \, C_F}{4 \, \pi^4} \int d^2 B \, d^2 b \, d^2 x \, d^2 y \; T_1 (\bm B) \;
  e^{- i \, {\bm k} \cdot ({\bm x}-{\bm y})} \, \frac{ {\bm x} - {\bm
  b}}{ |{\bm x} - {\bm b} |^2 } \cdot \frac{ {\bm y} - {\bm b}}{|{\bm y} - {\bm b} |^2}
\notag \\
  & \times \; \left[ N_G ( {\bm x} , {\bm b}, y) + N_G ( {\bm b} , {\bm y}, y)
  - N_G ( {\bm x} , {\bm y}, y ) \right].
\end{align}
where we replaced the gluon dipole $S$-matrix with the gluon dipole forward scattering amplitude \eqref{ngdipole} and shifted the impact parameter $\bm B$ such that the nuclear profile function no longer depends on $\bm B$.

Notice that each of the gluon dipole forward scattering amplitudes in \eqref{ktstep1} only depending on two of the transverse coordinates $\bm x$, $\bm y$, and $\bm b$. This leaves one coordinate that we are free to integrate out of the expression for each individual term. Doing this and renaming the remaining coordinates such that $\bm y$ is no longer in the expression we arrive at
\begin{align}
\frac{d \sigma}{d^2 k \, dy}
  = & \frac{\as \, C_F}{2 \, \pi^3} \int d^2 B \; T_1 (\bm B)
  \left[ -\int d^2 x \, d^2 y \; e^{- i \, {\bm k} \cdot ({\bm x}-{\bm y})} 
  \ln \frac{1}{|{\bm x} - {\bm y} | \Lambda} N_G (\bm x , \bm y, y ) \right.
\notag \\
  & + i \int d^2 x \, d^2 b \; e^{- i \, {\bm k} \cdot ({\bm x}-{\bm b})}
  \frac{\bm k}{ \bm k^2 } \cdot \frac{\bm x - \bm b}{|\bm x - \bm b |^2}
  N_G (\bm x , \bm b, y ) 
\notag \\
  & - \left.  i \int d^2 y \, d^2 b \; e^{- i \, {\bm k} \cdot ({\bm b}-{\bm y})}
  \frac{\bm y - \bm b}{|\bm y - \bm b |^2} \cdot \frac{\bm k}{ \bm k^2 }
  N_G (\bm b , \bm y, y ) \right]
\notag \\
  = & \frac{\as \, C_F}{2 \, \pi^3} \int d^2 B \, d^2 x \, d^2 b \; T_1 (\bm B) \;
  e^{- i \, {\bm k} \cdot ({\bm x}-{\bm b})} \left(2 i \, \frac{\bm k}{ \bm k^2 }
  \cdot \frac{ {\bm x} - {\bm b}}{|{\bm x} - {\bm b} |^2 } -
  \ln \frac{1}{|{\bm x} - {\bm b} | \Lambda} \right)
\notag \\
  & \times N_G ( {\bm x} , {\bm b}, y )
\notag \\
  = & \frac{\as \, C_F}{2 \, \pi^3} \int d^2 B \, d^2 r \, d^2 b \; T_1 (\bm B) \;
  e^{- i \, \bm k \cdot \bm r} \left(2 i \, \frac{\bm k}{ \bm k^2 }
  \cdot \frac{\bm r}{\bm r^2 } -
  \ln \frac{1}{|\bm r| \Lambda} \right)
\notag \\
  & \times N_G (\bm b+\bm r , \bm b, y ).
\end{align}

Next, since the whole expression is zero when $\bm r = 0$,  we use
\begin{equation}
\nabla^2_{\bm r} \left( e^{- i \, \bm k \cdot \bm r}
  \ln \frac{1}{| \bm r | \Lambda} \right) =
  e^{- i \, \bm k \cdot \bm r} \left(2 i \, \frac{\bm k}{ \bm k^2 }
  \cdot \frac{\bm r}{\bm r^2 } -
  \ln \frac{1}{|\bm r| \Lambda} \right)
\end{equation}
to write the equation as
\begin{align}
\frac{d \sigma}{d^2 k \, dy}
  = \frac{\as \, C_F}{2 \, \pi^3} \frac{1}{\bm k^2} \int & d^2 B \, d^2 r \, d^2 b
  \; T_1 (\bm B) \, N_G (\bm b+\bm r , \bm b, y ) \,
  \nabla^2_{\bm r} \left( e^{- i \, \bm k \cdot \bm r}
  \ln \frac{1}{| \bm r | \Lambda} \right).
\notag
\end{align}
Using integration by parts, we have $\nabla^2_{\bm r}$ go from acting on the term in parentheses to acting on the gluon dipole forward scattering amplitude. Then using the following substitution
\begin{equation}
\notag
\ln \frac{1}{|\bm r| \Lambda} =
  \nabla^2_{\bm r} \left( \frac{1}{4}\bm r^2 \ln \frac{1}{|\bm r| \Lambda} \right)
\end{equation}
the equation becomes
\begin{align}
\label{ktstep2}
\frac{d \sigma}{d^2 k \, dy}
  = \frac{\as \, C_F}{(2 \, \pi)^3} \frac{1}{\bm k^2} \int & d^2 B \, d^2 x \, d^2 y \;
  e^{- i \, {\bm k} \cdot \bm r} \,
  \left( \nabla^2_{\bm r} N_G (\bm b+\bm r , \bm b, y )  \right)
\notag \\
  & \times \left( \nabla^2_{\bm r} \left( \frac{1}{4}\bm r^2 \ln \frac{1}{|\bm r| \Lambda}
  T_1 (\bm B) \right) \right).
\end{align}

Now we introduce $n_G ({\bm b} + {\bm r}, {\bm b}, y)$, which is the gluon dipole
scattering amplitude on the projectile evaluated without saturation
effects (no multiple rescatterings, only linear BFKL evolution). The
two gluons in the dipole are located at transverse positions ${\bm b}
+ {\bm r}$ and ${\bm b}$, and the rapidity interval for the scattering
is $y$. In the quasi-classical limit one has
\begin{align} 
\label{eq:dipole_amp} 
n_G ({\bm b} + {\bm r}, {\bm b}, y=0) = \pi \, \alpha_s^2 \, r_\perp^2 \ln
  \left( \frac{1}{|{\bm r}| \Lambda} \right) T_1 ({\bm b}).
\end{align}

The single-gluon production cross section can be written in terms of gluon dipole scattering amplitudes.
Plugging this into equation \eqref{ktstep2} we arrive at
\begin{align}
\frac{d \sigma}{d^2 k \, dy}
  = \frac{C_F}{\as \pi (2 \, \pi)^3} \frac{1}{\bm k^2} \int & d^2 B \, d^2 r \, d^2 b \;
  e^{- i \, {\bm k} \cdot \bm r} \,
  \left( \nabla^2_{\bm r} \, N_G ( {\bm b} +{\bm r} , {\bm b}, y ) \right)
\notag \\
  & \times \left( \nabla^2_{\bm r} \, n_G ({\bm B} + {\bm r}, {\bm B}, y=0) \right).
\end{align}
In order to get this into the form we want we need to introduce two new dummy variables, transverse coordinate $\bm r'$ and transverse momentum $\bm q$. Doing this by using the formula
\begin{align}
f(\bm r) g(\bm r)
  = & \int d^2 r' \; \delta (\bm r - \bm r') f(\bm r) g(\bm r')
\notag \\
  = & \int \frac{d^2 q}{(2 \pi)^2} \int d^2 r' \;
  e^{- i \, {\bm q} \cdot (\bm r' - \bm r)} f(\bm r) g(\bm r') 
\notag
\end{align}
we have
\begin{align}
\frac{d \sigma}{d^2 k \, dy}
  = \frac{C_F}{\as \pi (2 \, \pi)^3} \frac{1}{\bm k^2}
  \int & \frac{d^2 q}{(2 \pi)^2} \int d^2 B \, d^2 r' \;
  e^{- i \bm r' \cdot \bm q} \; \nabla^2_{\bm r'} \, n_G ({\bm B} + {\bm r'}, {\bm B}, y=0)
\notag \\
  & \times \int d^2 b \, d^2 r \; e^{- i \, \bm r \cdot ( \bm k - \bm q)} \,
  \nabla^2_{\bm r} \, N_G ( {\bm b} +{\bm r} , {\bm b}, y ).
\end{align}
It turns out that this is a convolution of two different unintegrated gluon distributions, one for the light ion and one for the heavy ion.

The unintegrated gluon distribution for the light ion is
\begin{equation} 
\label{eq:dipole_wave_int} 
\left\langle \phi_{A_1} ({\bm q}, y) \right\rangle_{A_1} =
  \frac{C_F}{\alpha_s ( 2 \pi)^3} \int d^2 b \; d^2 r \; 
  e^{-i {\bm q} \cdot {\bm r}} \; \nabla_{{\bm r}}^2 \; n_G ({\bm b} + {\bm r}, {\bm b}, y)
\end{equation}
and the unintegrated gluon distribution for the heavy ion is defined as
\begin{align} 
\label{eq:trace_wave} 
\left\langle \phi_{A_2} ({\bm q}, y) \right\rangle_{A_2} =
  \frac{C_F}{\alpha_s ( 2 \pi)^3} \int d^2 b \; d^2 r \; e^{-i {\bm q} \cdot
  {\bm r}} \; \nabla_{{\bm r}}^2 \; N_{G} ({\bm b} + {\bm r}, {\bm b}, y).
\end{align}
The angle brackets $\langle \ldots
\rangle_{A_1}$ and $\langle \ldots \rangle_{A_2}$ denote averaging in
the projectile and target wave functions respectively.

The cross section for the production of a single gluon in a $pA$
collision, calculated in the quasi-classical and/or leading-$\ln 1/x$
approximations, can be written as a convolution of these two different
unintegrated gluon distributions \cite{Kovchegov:2001sc},
\begin{align} 
\label{eq:singlegluon} 
\frac{d \sigma_g}{d^2 k \, d y} = \frac{2 \alpha_s}{C_F} \frac{1}{{\bm
  k}^2} \int d^2 q \; \left\langle \phi_{A_1} ({\bm q}, Y-y)
  \right\rangle_{A_1} \, \left\langle \phi_{A_2} ({\bm k}-{\bm q}, y)
  \right\rangle_{A_2}
\end{align}
where we replaced the proton by the light ion $A_1$. This substitution implies that no
saturation effects are included in the light ion wave function, which
makes it equivalent to a proton for the purpose of the single-gluon
production calculation. The angle brackets $\langle \ldots
\rangle_{A_1}$ and $\langle \ldots \rangle_{A_2}$ denote averaging in
the projectile and target wave functions respectively.

%% file: Chapter_Calculation.tex
% !TEX root = WertepnyPhDThesis.tex
\cleardoublepage
\chapter{Calculation of the two-gluon correlation function}
\label{ch:Calculation}

In this chapter, we derive the two-gluon correlation function to leading-order in saturation in the heavy-light ion regime.
Here we derive the two-gluon production cross section (the single-gluon production cross section is given in \eqref{crosssection12aA}) and use this result to analyze the azimuthal angular dependence of the correlation function.
This chapter borrows heavily from the work in \cite{Kovchegov:2012nd}.

\input{Section_Cross_section}
\input{Section_Angular_results}

\section{Summary}

In this chapter we calculated the two-gluon production cross section in the heavy-light ion regime.
We found that there were two distinct classes of diagrams contributing to the cross section, ``separated" and ``crossed" diagrams.
Analyzing the correlations associated with these processes we found that they only give even harmonics due to the explicit symmetries of the cross section, the $\bm k_1 \leftrightarrow \bm k_2$ and $\bm k_2 \leftrightarrow -\bm k_2$ symmetries.
These correlations also have an enhancement when the two gluons have the same transverse momenta, $\bm k_1 = \bm k_2$, and when they have opposite transverse momenta, $\bm k_1 = - \bm k_2$.
This means that we have equal ridges on the near and away sides corresponding to peaks at $\Delta  \phi = 0$ and $\Delta \phi = \pi$ where $\Delta  \phi$ is the  azimuthal angle between the two gluons.
It should be noted that the theoretical results presented here, the work done by the author and collaborator in \cite{Kovchegov:2012nd}, predicting an away-side ridge in gluon production were posted on the arXiv before the away-side ridge was discovered by ALICE collaboration in p-Pb collisions at LHC \cite{Abelev:2012aa}. 

%% file: Section_Cross_section.tex
% !TEX root = WertepnyPhDThesis.tex
\section{Two-gluon production cross section}
\label{sec:longrange}

In this section we are going to calculate the two-gluon production
cross section for the heavy--light ion collisions in the
saturation/CGC framework. As described in Sec.~\ref{sec:aAregime} we assume that $A_2 \gg
A_1 \gg 1$, such that the saturation effects are resummed to all
orders only in the target nucleus with atomic number $A_2$. While the
saturation effects are not very important in the projectile nucleus
with the atomic number $A_1$, the fact that $A_1 \gg 1$ implies that
the two gluons are predominantly produced in collisions of different
nucleons in the projectile nucleus with the target nucleus.

Here we follow the same procedure used for the single-gluon production case Sec.~\ref{sec:single} calculation.

\subsection{Amplitudes of the two-gluon production process}

The first step in calculating the two-gluon production cross section is calculating the amplitudes involved.
There are two distinct types of amplitudes involved in this process.
The most obvious one is when the two gluons are emitted from different nucleons originating from the projectile,.

%%%%%%%%%%%%%%%%%%%%%%%%%%%%%%%%%%%%%%%%%%%%%%%%%%%%%%%%%%%%%%%%%%%%%%%%%%%%
\begin{figure}[H]
\centering
  \includegraphics[width=0.8 \textwidth]{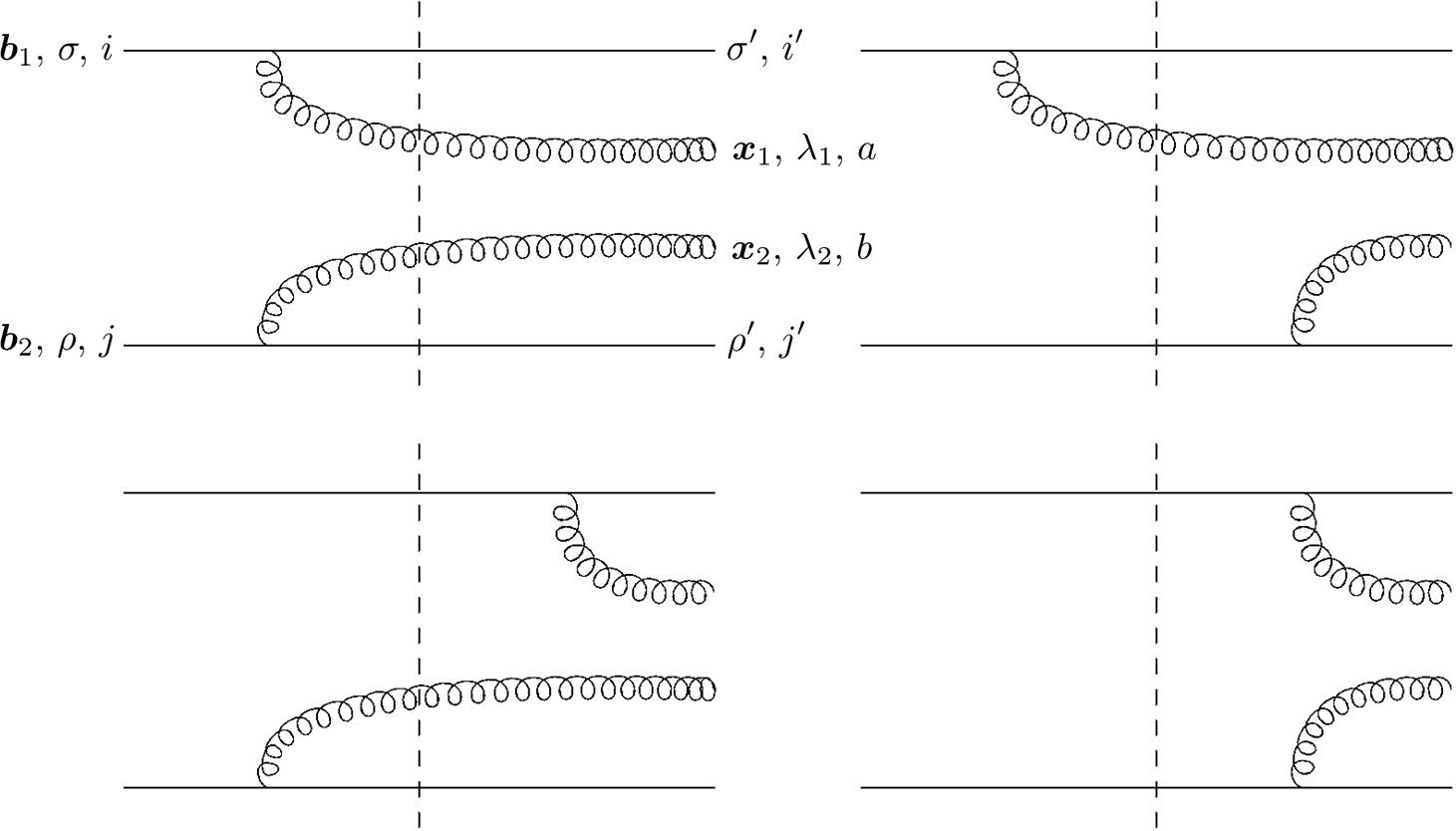}
  \caption{B diagrams. These are the diagrams associated with the gluon emission process where two gluons are emitted from different nucleons in the amplitude. The shock wave is represented by a dashed line.}
\label{ggampB} 
\end{figure}
%%%%%%%%%%%%%%%%%%%%%%%%%%%%%%%%%%%%%%%%%%%%%%%%%%%%%%%%%%%%%%%%%%%%%%%%%%%

The contributing diagrams for the case where the two observed gluons are emitted from different nucleons is shown in figure \ref{ggampB}.
As one might expect, these diagrams are very similar to the diagrams seen in the $pA$ case derived in the previous section and in fact we really just have the amplitude of the $pA$ case twice!
In other words we can write this amplitude, which we call $B$, using the $pA$ result, \eqref{amp_A}, as
\begin{equation}
\label{2nucamp}
B = A^{a,i' i}_{\sigma' \sigma, \lambda_1} (\bm x_1, \bm b_1) 
  A^{b,j' j}_{\rho' \rho, \lambda_2} (\bm x_2, \bm b_2).
\end{equation}
Where the color indices, quark spins, gluon polarizations and transverse coordinates are defined in figure \ref{ggampB}.

%%%%%%%%%%%%%%%%%%%%%%%%%%%%%%%%%%%%%%%%%%%%%%%%%%%%%%%%%%%%%%%%%%%%%%%%%%%%
\begin{figure}[H]
\centering
  \includegraphics[width=0.9 \textwidth]{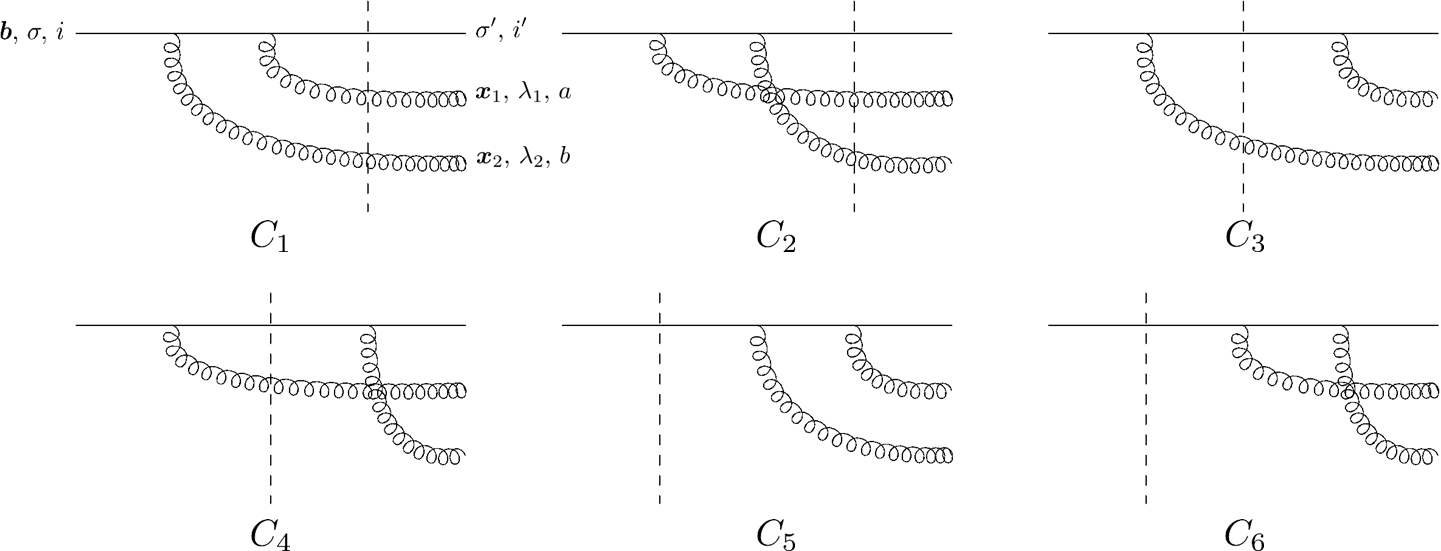}
  \caption{C diagrams. One of the nucleons in the projectile emits both of the observed gluons. Once again the shock wave is represented by a dashed line.}
\label{ggampC} 
\end{figure}
%%%%%%%%%%%%%%%%%%%%%%%%%%%%%%%%%%%%%%%%%%%%%%%%%%%%%%%%%%%%%%%%%%%%%%%%%%%

The other class of diagrams is where one nucleon emits the two observed gluons.
This, at first, seems to be suppressed by an order of saturation scale, however the two gluons end up being connected to the other projectile nucleon once we square the amplitude, keeping it to leading-order.
Unlike the $B$ amplitude case we cannot simply use the $pA$ result to derive this result, we are required to calculate these diagrams explicitly.
As an example, we calculate diagrams $C_1$ and $C_2$.

%%%%%%%%%%%%%%%%%%%%%%%%%%%%%%%%%%%%%%%%%%%%%%%%%%%%%%%%%%%%%%%%%%%%%%%%%%%%
\begin{figure}[H]
\centering
  \includegraphics[width=1 \textwidth]{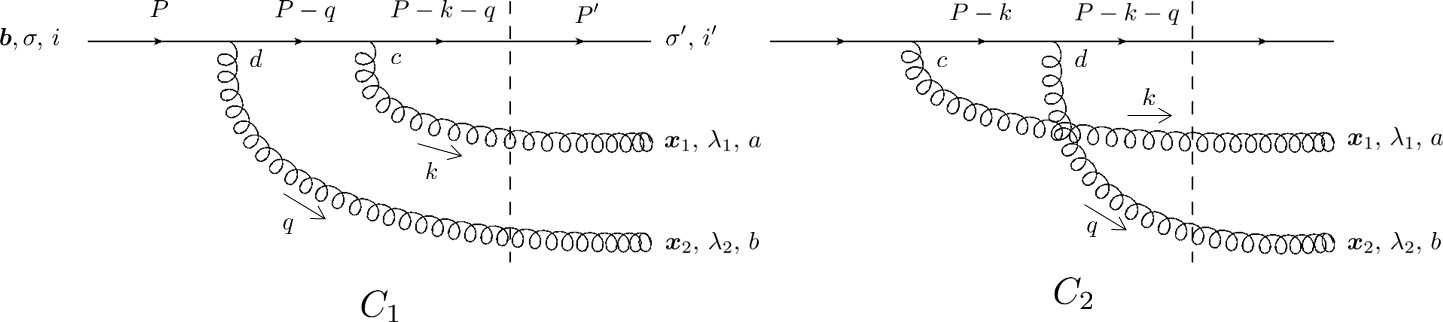}
  \caption{Diagrams $C_1$ and $C_2$. The difference between the two diagrams is the ordering of gluon emissions that happen before the shock wave, represented by a dashed line. The sum of these two diagrams combine into a compact result.}
\label{C1and2} 
\end{figure}
%%%%%%%%%%%%%%%%%%%%%%%%%%%%%%%%%%%%%%%%%%%%%%%%%%%%%%%%%%%%%%%%%%%%%%%%%%%

The procedure to calculate diagrams $C_1$ and $C_2$ is very similar to the one used for the $pA$ case. 
Using the results from last time, we know that the starting and ending points of the shock wave (in the $x^+$ direction) are unimportant to the order of interest.
Thus this dependence is neglected which means it is not included in the Wilson lines and the integrals over the momenta $k^-$ and $q^-$ are no longer Fourier transforms.
The main difference comes from how we treat the color factors.
It is easiest at this point to remember that at the end of the day the amplitude must be squared, so there is going to be another fundamental Wilson line, $\left( V^\dagger_{\bm b} \right)^{i i'}$, from the shock wave interaction on the other side of the cut associated with the projectile quark. 
We include this term while summing over the colors of the projectile quark to simplify the final result.
The expression that needs to be evaluated is
\begin{align}
\left( V^\dagger_{\bm b} \right)^{i i'} & (C_1 + C_2)
\notag \\
  = & \int \frac{d k^-}{2 \pi}
  \frac{d^2 k}{(2 \pi)^2} e^{i \bm k \cdot ( \bm x_1 - \bm b )} \frac{-i}{k^2 + i \epsilon}
  \left( g_{\mu \alpha} - \frac{\eta_\mu k_\alpha +\eta_\alpha k_\mu}{k^+} \right)
  \left(- 2 k^+g^{\mu' \mu}_\perp U^{a c}_{\bm x_1} \right) \epsilon^*_{\lambda_1, \mu'}
\notag \\
  \times & \int \frac{d q^-}{2 \pi}
  \frac{d^2 q}{(2 \pi)^2} e^{i \bm q \cdot ( \bm x_2 - \bm b )} \frac{-i}{q^2 + i \epsilon}
  \left( g_{\nu \beta} - \frac{\eta_\nu q_\beta +\eta_\beta q_\nu}{q^+} \right)
  \left(- 2 q^+g^{\nu' \nu}_\perp U^{b d}_{\bm x_2} \right) \epsilon^*_{\lambda_2, \nu'}
\notag \\
  \times & \frac{1}{2 P^+} \left( V^\dagger_{\bm b} \right)^{i i'} \bar u_{\sigma'} \left( P' \right) \gamma^+
  V^{i'i'''}_{\bm b} \frac{i \left( \slashed{P} - \slashed{k} - \slashed{k}\right)}{(P-k-q)^2 + i \epsilon}
  \left[ i g \, \gamma^\alpha (t^c)^{i''' i''} \frac{i \left( \slashed{P} - \slashed{q} \right)}
  {(P-q)^2 + i \epsilon} i g \, \gamma^\beta (t^d)^{i'' i} \right.
\notag \\
  & \left. + \, i g \, \gamma^\beta (t^d)^{i''' i''} \frac{i \left( \slashed{P} - \slashed{k} \right)}
  {(P-k)^2 + i \epsilon} i g \, \gamma^\alpha (t^c)^{i'' i} \right] u_{\sigma} \left( P \right).
\end{align}

The first step is analyzing the quark line while taking into account the various Wilson lines and generators associated with it.
Using the eikonal limit and evaluating the color object the expression simplifies quite nicely.
\begin{align}
\frac{1}{2 P^+} & \left( V^\dagger_{\bm b} \right)^{i i'} \bar u_{\sigma'}
  \left( P' \right) \cdots u_{\sigma} \left( P \right)
\notag \\
  \approx & \, \frac{-g^2}{2} \delta^{c d} \, g^{- \alpha} \, g^{- \beta} \frac{1}{2 P^+}
  \bar u_{\sigma'} \left( P' \right) \gamma^+ \frac{- i \gamma^- P^+}
  {2 P^+ (k^- +q^-) - i \epsilon} \gamma^+ 
\notag \\
  & \times \left[ \frac{- i \gamma^- P^+}{2 P^+ q^- - i \epsilon}
  + \frac{- i \gamma^- P^+}{2 P^+ k^-   - i \epsilon}\right] \gamma^+ 
  u_{\sigma} \left( P \right)
\notag \\
  = & \, \frac{g^2}{2} \delta^{c d} \, g^{- \alpha} \, g^{- \beta} \frac{1}{2 P^+}
  \bar u_{\sigma'} \left( P' \right) \gamma^+ u_{\sigma} \left( P \right)
  \frac{1}{k^- +q^- - i \epsilon} \left[ \frac{1}{ q^- - i \epsilon}
  + \frac{1}{ k^-   - i \epsilon}\right]
\notag \\
  = & \, \frac{g^2}{2} \, \delta_{\sigma' \sigma} \, \delta^{c d} \, g^{- \alpha} \, g^{- \beta}
  \frac{1}{ k^-   - i \epsilon} \frac{1}{ q^-   - i \epsilon}.
\end{align}
At this point it should be clear why we included the $\left( V^\dagger_{\bm b} \right)^{i i'}$ term.
This allowed us to notice that, despite each fundamental generator being associated with a different observed gluon, the color factor for both diagrams ends up being the same, which allowed us to combine the denominators associated with the eikonal quarks in a convenient manner.

Plugging this term back into the main expression we arrive at
\begin{align}
\left( V^\dagger_{\bm b} \right)^{i i'} (C_1 + C_2)
  = & 2 \, g^2 \, \delta_{\sigma' \sigma} \, \delta^{c d} \,
  \int \frac{d k^-}{2 \pi} \frac{d^2 k}{(2 \pi)^2}
  e^{i \bm k \cdot ( \bm x_1 - \bm b )} \frac{i}{ k^-   - i \epsilon}
  \frac{\bm \epsilon^*_{\lambda_1} \cdot \bm k}{2 k^+ k^- - \bm k^2 + i \epsilon} U^{a c}_{\bm x}
\notag \\
  \times & \int \frac{d q^-}{2 \pi} \frac{d^2 q}{(2 \pi)^2}
  e^{i \bm q \cdot ( \bm x_2 - \bm b )} \frac{i}{ q^-  - i \epsilon}
  \frac{\bm \epsilon^*_{\lambda_2} \cdot \bm q}{2 q^+ q^- - \bm q^2 + i \epsilon}
  U^{b d}_{\bm y}
\end{align}
which is very similar to the result we got for the $pA$ case, \eqref{amp_A1k}.
Using that result we arrive at
\begin{align}
\left( V^\dagger_{\bm b} \right)^{i i'} (C_1 + C_2)
  = & - \frac{g^2}{\pi^2} \delta_{\sigma', \sigma}
  \frac{\bm \epsilon^*_{\lambda_1} \cdot \left( \bm x_1 - \bm b \right)}{ | \bm x_1 - \bm b |^2 }
  \frac{\bm \epsilon^*_{\lambda_2} \cdot \left( \bm x_2 - \bm b \right)}{ | \bm x_2 - \bm b |^2 }
  \left[ U_{\bm x_1} U^\dagger_{\bm x_2} \right]^{a b}.
\end{align}
The other diagrams can be evaluated in a manner similar to this and the $pA$ case.
The final result for each of the $C$ diagrams, now including the averaging of the quark colors and the summing and averaging over the spins, is
\begin{subequations}
\label{Cdiagrams}
\begin{align}
\label{C12}
\frac{1}{2 N_c} \sum_{\sigma', \sigma} \left( V^\dagger_{\bm b} \right)^{i i'} (C_1 + C_2)
  = & - \frac{g^2}{2 N_c \pi^2}
  \frac{\bm \epsilon^*_{\lambda_1} \cdot \left( \bm x_1 - \bm b \right)}{ | \bm x_1 - \bm b |^2 }
  \frac{\bm \epsilon^*_{\lambda_2} \cdot \left( \bm x_2 - \bm b \right)}{ | \bm x_2 - \bm b |^2 }
  \left[ U_{\bm x_1} U^\dagger_{\bm x_2} \right]^{a b}
\\
\label{C3}
\frac{1}{2 N_c} \sum_{\sigma', \sigma} \left( V^\dagger_{\bm b} \right)^{i i'} C_3
  = & \frac{g^2}{2 N_c \pi^2}
  \frac{\bm \epsilon^*_{\lambda_1} \cdot \left( \bm x_1 - \bm b \right)}{ | \bm x_1 - \bm b |^2 }
  \frac{\bm \epsilon^*_{\lambda_2} \cdot \left( \bm x_2 - \bm b \right)}{ | \bm x_2 - \bm b |^2 }
  \left[ U_{\bm b} U^\dagger_{\bm x_2} \right]^{a b}
\\
\label{C4}
\frac{1}{2 N_c} \sum_{\sigma', \sigma} \left( V^\dagger_{\bm b} \right)^{i i'} C_4
  = & \frac{g^2}{2 N_c \pi^2}
  \frac{\bm \epsilon^*_{\lambda_1} \cdot \left( \bm x_1 - \bm b \right)}{ | \bm x_1 - \bm b |^2 }
  \frac{\bm \epsilon^*_{\lambda_2} \cdot \left( \bm x_2 - \bm b \right)}{ | \bm x_2 - \bm b |^2 }
  \left[ U_{\bm x_1} U^\dagger_{\bm b} \right]^{a b}
\\
\label{C56}
\frac{1}{2 N_c} \sum_{\sigma', \sigma} \left( V^\dagger_{\bm b} \right)^{i i'} (C_5 + C_6)
  = & - \frac{g^2}{2 N_c \pi^2}
  \frac{\bm \epsilon^*_{\lambda_1} \cdot \left( \bm x_1 - \bm b \right)}{ | \bm x_1 - \bm b |^2 }
  \frac{\bm \epsilon^*_{\lambda_2} \cdot \left( \bm x_2 - \bm b \right)}{ | \bm x_2 - \bm b |^2 }
  \left[ U_{\bm b} U^\dagger_{\bm b} \right]^{a b}
\end{align}
\end{subequations}

At this point we define the following function
\begin{equation}
C^{a b}_{\lambda_1 \lambda_2} (\bm x_1, \bm x_2, \bm b) = \frac{1}{2 N_c} \sum_{\sigma', \sigma} \left( V^\dagger_{\bm b} \right)^{i i'} \sum^6_{j=1} C_j.
\end{equation}
This is not quite an amplitude, due to the fact we have already considered the other side of the cut when evaluating the diagrams, but it can be used in a similar manner when calculating the cross section.
Evaluating the expression gives the final result for the $C$ diagrams
\begin{equation}
\label{1nucamp}
C^{a b}_{\lambda_1 \lambda_2} (\bm x_1, \bm x_2, \bm b) = - \frac{g^2}{2 N_c \pi^2}
  \frac{\bm \epsilon^*_{\lambda_1} \cdot \left( \bm x_1 - \bm b \right)}{ | \bm x_1 - \bm b |^2 }
  \frac{\bm \epsilon^*_{\lambda_2} \cdot \left( \bm x_2 - \bm b \right)}{ | \bm x_2 - \bm b |^2 }
  \left[ \left(U_{\bm x_1}-U_{\bm b}\right) \left(U^\dagger_{\bm x_2}
  - U^\dagger_{\bm b}\right) \right]^{a b}.
\end{equation}

At this point we have the results necessary to calculate the two-gluon production cross section for the two-nucleon process.
There are numerous ways these amplitudes can combine so it is necessary to split the cross section into two parts: the ``square terms", where a gluon is emitted by the same quark on both sides of the cut; and the ``crossed terms", where a gluon is emitted by different quarks on either side of the cut.

%%%%%%%%%%%%%%%%%%%%%%%%%%%%%%%%%%%%%%%%%%%%%%%%%%%%%%%%%%%%%%%%%%%%%%%%%%%%%%%%%%

\subsection{Two-gluon production with long-range rapidity
  correlations: ``square'' of the single gluon production}
\label{subsec:ggsquare}

\begin{figure}[h]
\centering
  \includegraphics[width= 0.6 \textwidth]{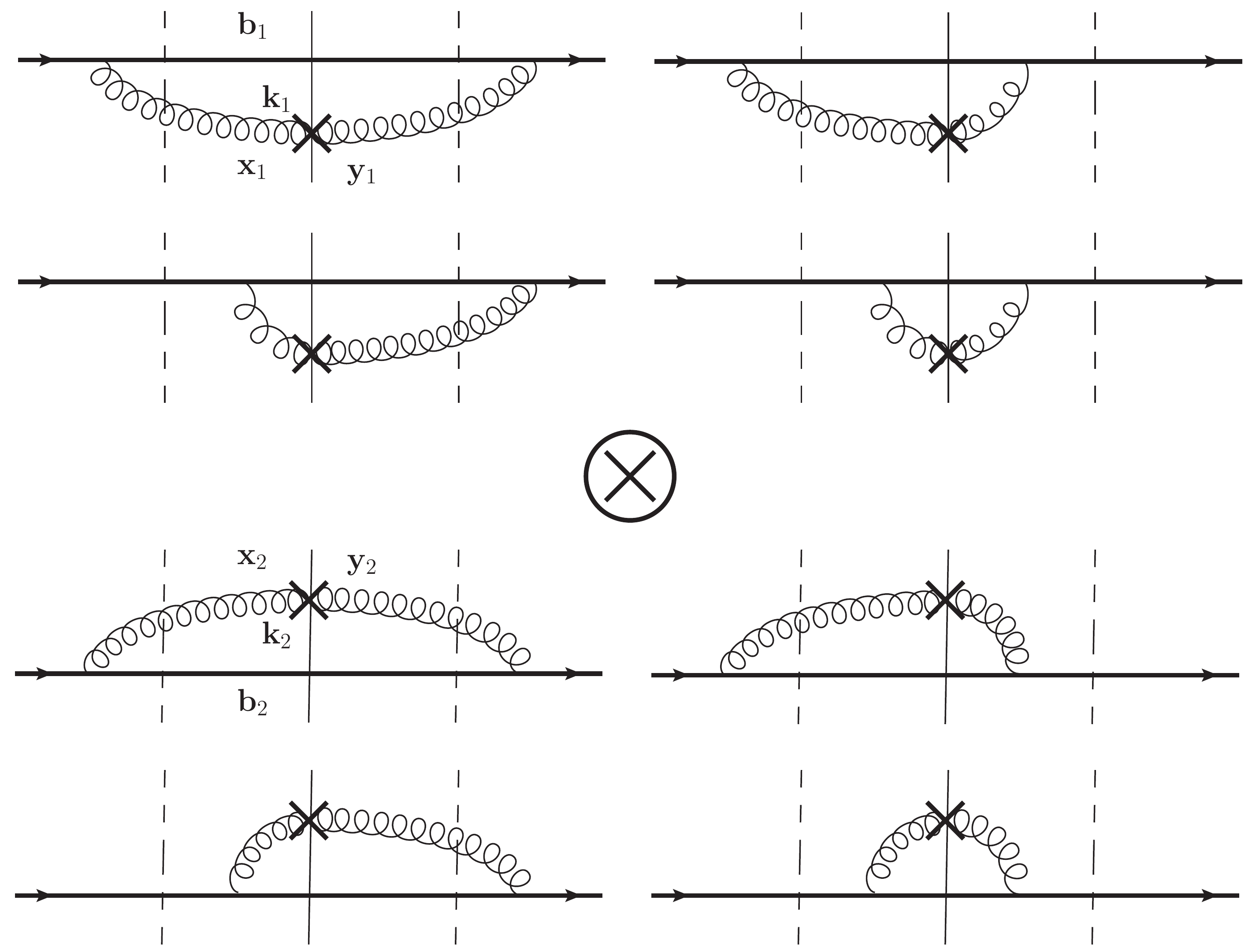}
  \caption{Diagrams contributing to the two-gluon production cross
    section in the heavy--light ion collision. For clarity the
    diagrams are shown as a direct product of gluon production
    processes in collisions of the two interacting nucleons from the
    projectile nucleus with the target nucleus.}
\label{uncrossed} 
\end{figure}

The two-gluon production cross section in heavy--light ion collisions is easily
constructed by analogy to the single-gluon production calculation of
Sec.~\ref{sec:single}. The diagrams contributing to the square of the
scattering amplitude for the double gluon production in heavy--light
ion collisions are shown in Fig.~\fig{uncrossed}, written as a direct
product of the gluon production processes in the interactions of each
of the nucleons from the projectile nucleus with the target. Just as
in Fig.~\fig{1Gincl} the vertical dashed lines in Fig.~\fig{uncrossed} represent
interactions with the target, while the vertical solid line denotes
the final state cut.

%%%%%%%%%%%%%%%%%%%%%%%%%%%%%%%%%%%%%%%%%%%%%%%%%%%%%%%%%%%%%%%%%%%%%%%%%%%%
%%%%%%%%%%%%%%%%%%%%%%%%%%%%%%%%%%%%%%%%%%%%%%%%%%%%%%%%%%%%%%%%%%%%%%%%%%%%

The evaluation of the diagrams in Fig.~\fig{uncrossed} is straightforward.
We follow the same process as we did in the single-gluon production case, \eqref{crosssectionpA1}.
Multiplying the amplitude \eqref{2nucamp} by the corresponding complex conjugated amplitude (making sure the coordinates and spin are contracted in the way that corresponds to the kinematics laid out in Fig.~\fig{uncrossed}), the spins and colors of both quarks are summed and averaged over.
The gluon polarizations are summed over and a Fourier transform over the transverse positions of the gluon at the cut is preformed, leaving the final expression in transverse momentum space.
This is then convoluted with the nuclear profile functions, resulting in
\begin{align}
\label{sigmasquare}
& \frac{d \sigma_{square}}{d^2 k_1 dy_1 d^2 k_2 dy_2} = \frac{1}{[2 (2 \pi)^3]^2}
  \int d^2 B \, d^2 b_1 \, d^2 b_2 \, T_1 ({\bm B} - {\bm
  b}_1) \, T_1 ({\bm B} - {\bm b}_2) \, 
\notag \\
  & \; \; \times  \int d^2 x_1 \, d^2 y_1 \, d^2 x_2 \, d^2 y_2 \;
  e^{- i \; {\bm k}_1 \cdot ({\bm x}_1-{\bm y}_1) -
  i \; {\bm k}_2 \cdot ({\bm x}_2-{\bm y}_2)} \frac{1}{4 N^2_c}
  \sum_{\sigma' \sigma, \rho' \rho} \sum_{\lambda_1, \lambda_2}
\notag \\
  & \; \; \times \left< \left(A^{a,i' i}_{\sigma' \sigma, \lambda_1} (\bm y_1, \bm b_1) 
  A^{b,j' j}_{\rho' \rho, \lambda_2} (\bm y_2, \bm b_2) \right)^\dagger
  A^{a,i' i}_{\sigma' \sigma, \lambda_1} (\bm x_1, \bm b_1) 
  A^{b,j' j}_{\rho' \rho, \lambda_2} (\bm x_2, \bm b_2) \right>.
\end{align}
Using the previous result for the single-gluon case, \ref{A2pA}, we can easily evaluate the amplitude squared term (where we have included all averaging and summations required),
\begin{align}
\label{A2ppAsquare}
\frac{1}{4 N^2_c} & \sum_{\sigma' \sigma, \rho' \rho} \sum_{\lambda_1, \lambda_2}
  \left< \left(A^{a,i' i}_{\sigma' \sigma, \lambda_1} (\bm y_1, \bm b_1) 
  A^{b,j' j}_{\rho' \rho, \lambda_2} (\bm y_2, \bm b_2) \right)^\dagger
  A^{a,i' i}_{\sigma' \sigma, \lambda_1} (\bm x_1, \bm b_1) 
  A^{b,j' j}_{\rho' \rho, \lambda_2} (\bm x_2, \bm b_2) \right>
\notag \\ 
  = & \, \frac{g^4}{4 N^2_c \pi^4}
  \frac{\left( \bm x_1 - \bm b_1 \right)\cdot \left( \bm y_1 - \bm b_1 \right)}
  { | \bm x_1 - \bm b_1 |^2 | \bm y_2 - \bm b_2 |^2 }
    \frac{\left( \bm x_2 - \bm b_2 \right)\cdot \left( \bm y_2 - \bm b_2 \right)}
  { | \bm x_2 - \bm b_2 |^2 | \bm y_2 - \bm b_2 |^2 }
\notag \\
  & \times \left< Tr \left[ \left( U_{\bm x_1} - U_{\bm b_1} \right)
  \left( U^\dagger_{\bm y_1} - U^\dagger_{\bm b_1} \right)\right]
  Tr \left[ \left( U_{\bm x_2} - U_{\bm b_2} \right)
  \left( U^\dagger_{\bm y_2} - U^\dagger_{\bm b_2} \right)\right] \right>.
\end{align}
This is nearly identical to the amplitude due to two independently generated single-gluon emissions, \ref{A2pA} for two different gluons if you will.
Due to the target averaging of the Wilson lines they are not exactly identical, meaning that any dynamically generated correlations are generated though the interaction with the target.

The full ``square" contribution is recovered by inserting  \eqref{A2ppAsquare} into \eqref{sigmasquare}, resulting in (cf. \cite{Kovner:2012jm})
\begin{align} 
\label{eq:2glue_prod_main} 
& \frac{d \sigma_{square}}{d^2 k_1 dy_1 d^2 k_2 dy_2} = \frac{\as^2 \,
  C_F^2}{16 \, \pi^8} \int d^2 B \, d^2 b_1 \, d^2 b_2 \, T_1 ({\bm B}
  - {\bm b}_1) \, T_1 ({\bm B} - {\bm b}_2) \, d^2 x_1 \, d^2 y_1 \, d^2
  x_2 \, d^2 y_2  \notag \\
& \; \; \times \, e^{- i \; {\bm k}_1 \cdot ({\bm x}_1-{\bm y}_1) - i
  \; {\bm k}_2 \cdot ({\bm x}_2-{\bm y}_2)}
  \frac{ {\bm x}_1 - {\bm b}_1}{ |{\bm x}_1 - {\bm b}_1 |^2 } \cdot
  \frac{ {\bm y}_1 - {\bm b}_1}{ |{\bm y}_1 - {\bm b}_1 |^2 } \ \frac{
  {\bm x}_2 - {\bm b}_2}{ |{\bm x}_2 - {\bm b}_2 |^2 } \cdot \frac{
  {\bm y}_2 - {\bm b}_2}{ |{\bm y}_2 - {\bm b}_2 |^2 } \notag \\
& \; \; \times \, \left\langle \left( \frac{1}{N_c^2-1} \; Tr[ U_{{\bm x}_1}
  U_{{\bm y}_1}^\dagger ] \; - \; \frac{1}{N_c^2-1} \; Tr[ U_{{\bm
  x}_1} U_{{\bm b}_1}^\dagger ] \; - \; \frac{1}{N_c^2-1} \;
  Tr[ U_{{\bm b}_1} U_{{\bm y}_1}^\dagger ] \; + \; 1 \right) \right. \notag \\
& \; \; \times \left. \left( \frac{1}{N_c^2-1} \; Tr[ U_{{\bm x}_2} U_{{\bm
  y}_2}^\dagger ] \; - \; \frac{1}{N_c^2-1} \; Tr[ U_{{\bm x}_2}
  U_{{\bm b}_2}^\dagger ] \; - \; \frac{1}{N_c^2-1} \; Tr[ U_{{\bm
  b}_2} U_{{\bm y}_2}^\dagger ] \; + \; 1 \right) \right\rangle.
\end{align}
This is the expression for the two-gluon production cross section
coming from the diagrams in Fig.~\fig{uncrossed}. The interaction with the
target can be evaluated in the quasi-classical multiple rescattering
approximation, as we will show later. The rapidity evolution can be
included using the JIMWLK equation. Note, however, that when the
rapidity difference between the two gluons is sufficiently large,
$|y_1 - y_2| \gtrsim 1/\as$, one has to include the evolution
corrections in the rapidity interval between the produced gluons, such
that the Wilson lines in the two parentheses in
\eqref{eq:2glue_prod_main} should be taken at different rapidities. A
similar effect was included in the two-gluon production cross
section in DIS in \cite{JalilianMarian:2004da}. In such a regime, one
would also need to include evolution corrections in the rapidity
window between the projectile and (at least one of) the produced
gluons. Inclusion of the evolution corrections in terms of the weight
functional $W$ of the JIMWLK evolution equation into the two-gluon
production cross section in nucleus--nucleus collisions was done in
\cite{Dusling:2009ni}. In this chapter we will limit ourselves to the
quasi-classical regime where no evolution corrections are required and
the cross sections are rapidity-independent.

Note that for scattering on a large nuclear target, if one resums
powers of $\as^2 \, A^{1/3}$, as is the case in the Glauber-Mueller
(GM) rescatterings (and in the BK/JIMWLK evolution for which the GM
rescatterings serve as the initial condition), and if one takes the
large-$N_c$ limit, the expectation values of the traces in
\eqref{eq:2glue_prod_main} factorize, such that
\begin{align}
\label{eq:trace_fact}
\left\langle Tr[ U_{{\bm x}_1} U_{{\bm y}_1}^\dagger ] \; Tr[
  U_{{\bm x}_2} U_{{\bm y}_2}^\dagger ] \right\rangle
  \bigg|_{\mbox{large}-N_c, \, \mbox{large}-A_2} \approx \left\langle
  Tr[ U_{{\bm x}_1} U_{{\bm y}_1}^\dagger ] \right\rangle \;
  \left\langle Tr[ U_{{\bm x}_2} U_{{\bm y}_2}^\dagger ]
  \right\rangle.
\end{align}
This leads to factorization of \eqref{int_fact} being valid in the
large-$N_c$ and large-target-nucleus limit. As shown above, even in
this factorized regime one may obtain a non-trivial correlation
function due to the geometric correlations.

%%%%%%%%%%%%%%%%%%%%%%%%%%%%%%%%%%%%%%%%%%%%%%%%%%%%%%%%%%%%%%%%%%%%%%%%%%%%%%%%%%

\subsection{Two-gluon production with long-range rapidity
  correlations: ``crossed'' diagrams}
\label{subsec:ggcrossed}

\begin{figure}[H]
\centering
  \includegraphics[width= 0.77 \textwidth]{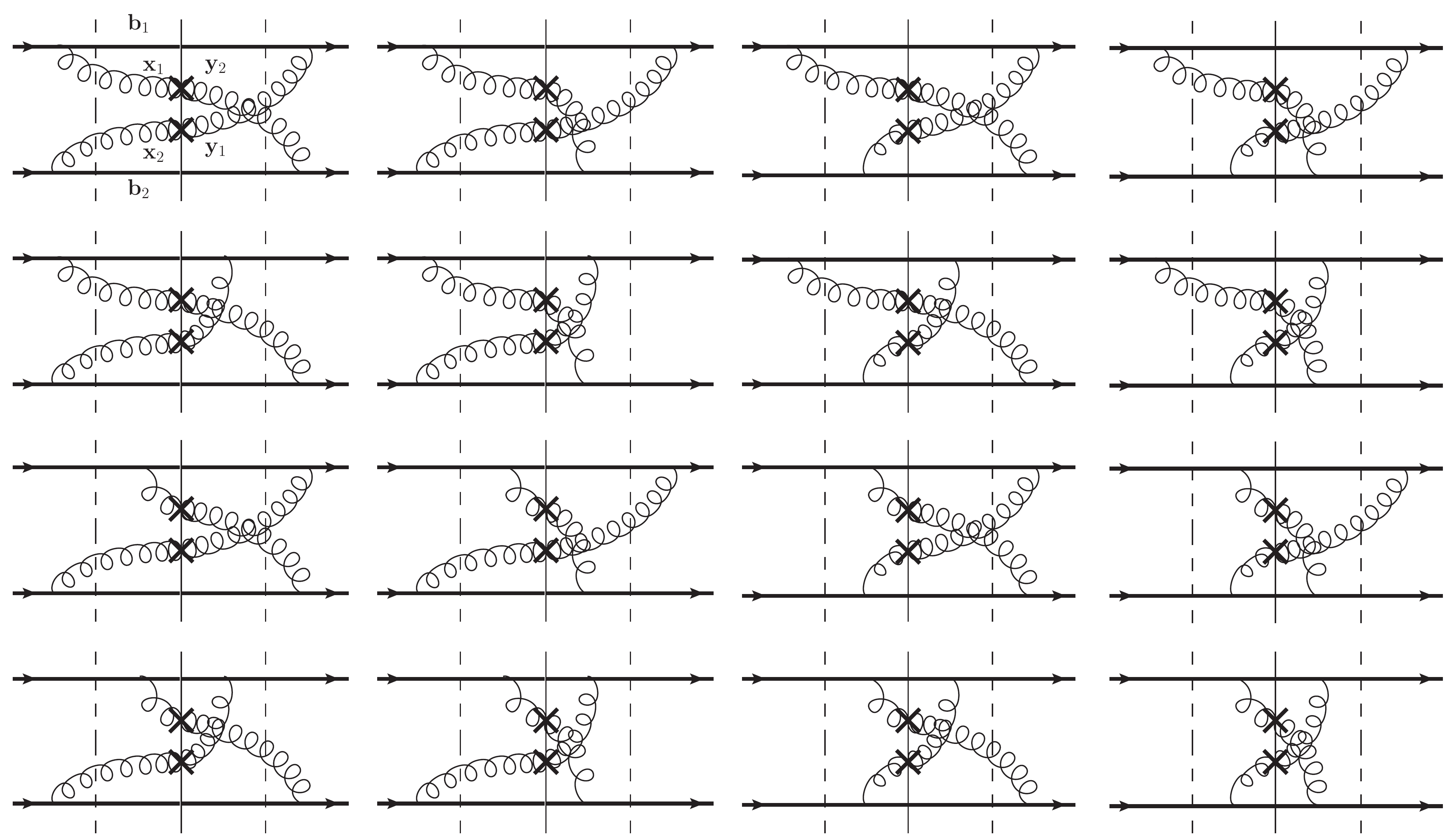}
  \caption{Diagrams contributing to the two-gluon production cross
    section, with the gluon emitted by each nucleon in the amplitude
    absorbed by another nucleon in the complex conjugate
    amplitude. The top cross denotes the gluon with momentum ${\bm
      k}_1$, while the bottom one denotes the gluon with momentum
    ${\bm k}_2$.}
\label{crossed} 
\end{figure}

\begin{figure}[H]
\centering
  \includegraphics[width= 0.77 \textwidth]{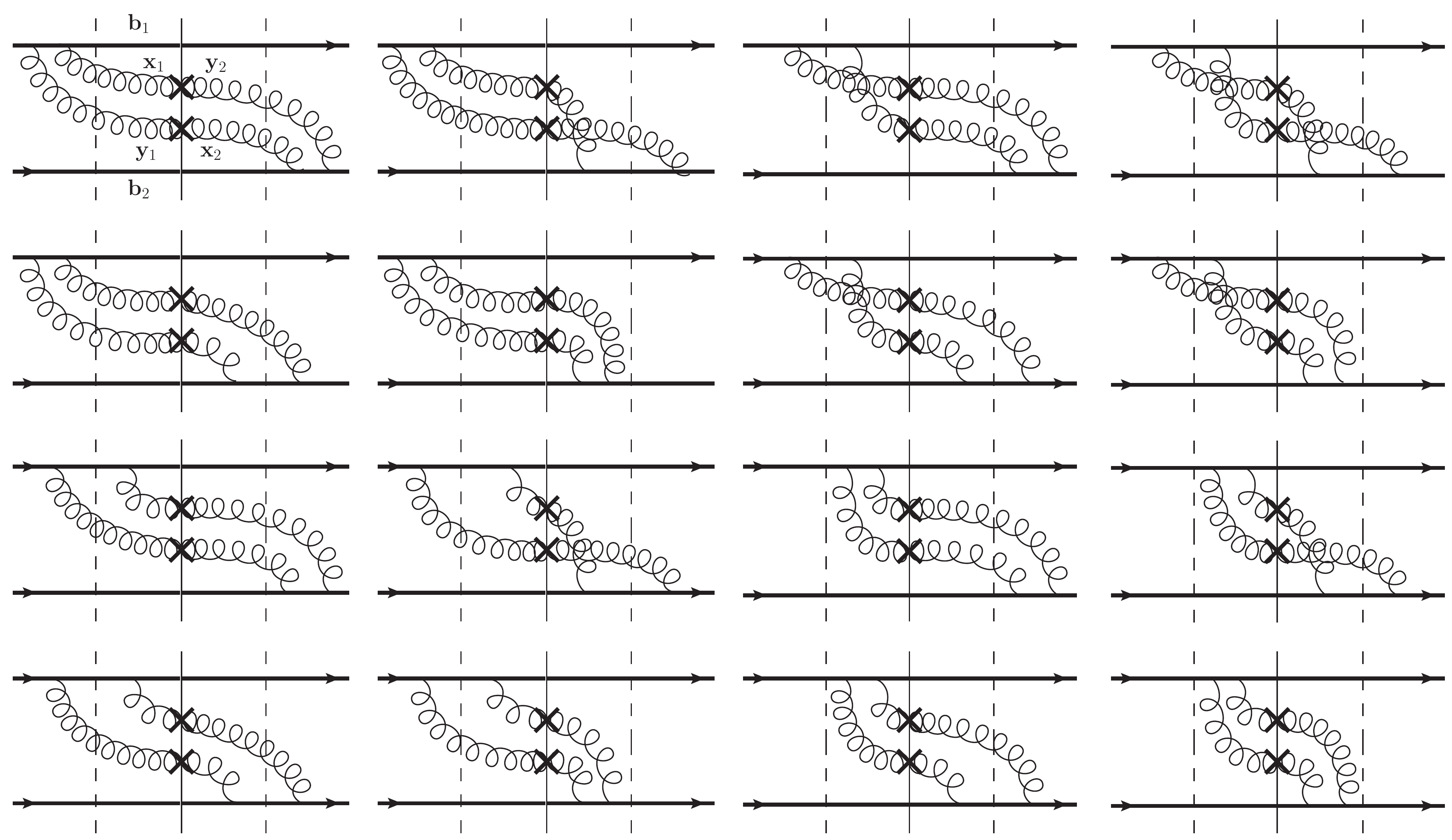}
  \caption{Another set of diagrams contributing to the two-gluon
    production cross section, with the gluon emitted by each nucleon
    in the amplitude absorbed by another nucleon in the complex
    conjugate amplitude. Again the top cross denotes the gluon with
    momentum ${\bm k}_1$, while the bottom one denotes the gluon with
    momentum ${\bm k}_2$.}
\label{crossed2} 
\end{figure}

Before we proceed to evaluating the correlations contained in the cross section \eqref{eq:2glue_prod_main}, let us point out another
contribution to the two-gluon production cross section arising from squaring the sum of the diagrams in Fig.~\fig{ggampB}.
When squaring the diagrams in Fig.~\fig{ggampB} it is possible that the gluon emitted by one nucleon in the amplitude will be absorbed by another nucleon in the complex conjugate amplitude, as shown in Fig.~\ref{crossed}.
We also have the contribution from the diagrams in Fig.~\fig{ggampC} being contracted with diagrams where both gluons are emitted from the other nucleon, shown in Fig.~\ref{crossed2}.
We will refer to these diagrams as the ``crossed'' graphs (it should be noted that in these diagrams crossing gluon lines do not form a vertex).
The diagrams obtained by a mirror reflection with respect to the cut of those in Figs.~\ref{crossed} and \ref{crossed2} correspond to the ${\bm b}_1 \leftrightarrow {\bm b}_2$ interchange, and will be automatically included in the cross section to be calculated below since it will contain integrals over all ${\bm b}_1$ and ${\bm b}_2$.

%%%%%%%%%%%%%%%%%%%%%%%%%%%%%%%%%%%%%%%%%%%%%%%%%%%%%%%%%%%%%%%%%%%%%%%%%%%%
%%%%%%%%%%%%%%%%%%%%%%%%%%%%%%%%%%%%%%%%%%%%%%%%%%%%%%%%%%%%%%%%%%%%%%%%%%%%
%%%%%%%%%%%%%%%%%%%%%%%%%%%%%%%%%%%%%%%%%%%%%%%%%%%%%%%%%%%%%%%%%%%%%%%%%%%%
%%%%%%%%%%%%%%%%%%%%%%%%%%%%%%%%%%%%%%%%%%%%%%%%%%%%%%%%%%%%%%%%%%%%%%%%%%%%

\begin{figure}[H]
\centering
  \includegraphics[width= 0.44 \textwidth]{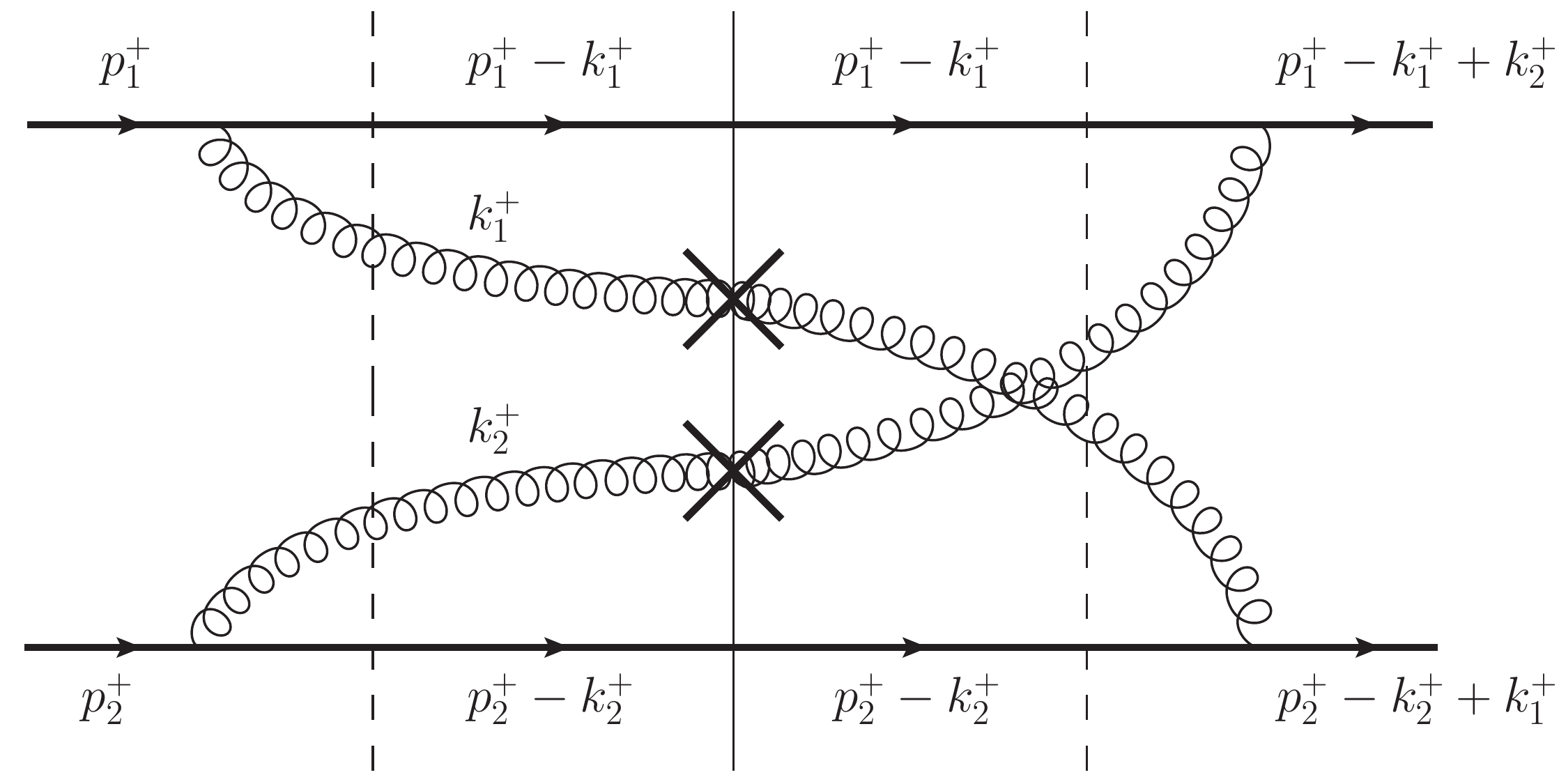}
  \caption{The flow of the ``plus'' momentum component through the
    first diagram in Fig.~\fig{crossed}.}
\label{hbt_rapidities} 
\end{figure}

The longitudinal momentum flow patterns in Figs.~\ref{crossed} and
\ref{crossed2} are different from that in Fig.~\fig{uncrossed}.
This is illustrated in Fig.~\fig{hbt_rapidities}, which shows the flow of
the ``plus'' momentum component through the first diagram in
Fig.~\fig{crossed}. Note that the change in the ``plus'' momentum component
is negligible in the eikonal interactions with the target considered
here. Requiring that the incoming quark lines carry the same ``plus''
momentum both in the amplitude and in the complex conjugate amplitude,
one would obtain $k_1^+ = k_2^+$; however, such requirement is not
correct. The actual scattering happens between two nuclei, and it is
the momenta of the whole incoming nuclei which have to be equal both
in the amplitude and in the complex conjugate amplitude. Hence for
$k_1^+ \neq k_2^+$ the diagrams like that in Fig.~\fig{hbt_rapidities}
would only correspond to different redistributions of the projectile
nucleus momentum between the nucleons in it in the amplitude and in
the complex conjugate amplitude without changing the same ``plus''
momentum of the whole nucleus on both sides of the cut. Hence the
$k_1^+ = k_2^+$ condition is not necessary for the ``crossed''
diagrams.

%%%%%%%%%%%%%%%%%%%%%%%%%%%%%%%%%%%%%%%%%%%%%%%%%%%%%%%%%%%%%%%%%%%%%%%%%%%%
%%%%%%%%%%%%%%%%%%%%%%%%%%%%%%%%%%%%%%%%%%%%%%%%%%%%%%%%%%%%%%%%%%%%%%%%%%%%

The process used here to calculate the cross section is similar to the ``square" diagram case, with the extra complication that we have two sets of diagrams that need to be considered.
To take in to account the terms shown in Fig.~\fig{crossed}, we use the same process used in deriving \eqref{sigmasquare} except the coordinates and spins of the complex conjugated amplitude are different (corresponds to the kinematics shown in Fig.~\fig{uncrossed}).
To take into account the terms in Fig.~\fig{crossed2}, we use \eqref{1nucamp}, multiply it by its complex conjugate (keeping the coordinates such that it matches the kinematics shown in Fig.~\fig{crossed2}) and sum over the gluons' polarizations (the quarks' spins and colors were already taken into account when deriving \eqref{1nucamp}).
The sum of these two terms is Fourier transformed into transverse momentum space and convoluted with the nuclear profile functions, resulting in
\begin{align}
\label{sigmacrossed}
& \frac{d \sigma_{crossed}}{d^2 k_1 dy_1 d^2 k_2 dy_2} = \frac{1}{[2 (2 \pi)^3]^2}
  \int d^2 B \, d^2 b_1 \, d^2 b_2 \, T_1 ({\bm B} - {\bm
  b}_1) \, T_1 ({\bm B} - {\bm b}_2)
\notag \\
  & \; \; \times \, d^2 x_1 \, d^2 y_1 \, d^2 x_2 \, d^2 y_2 \,
  e^{- i \; {\bm k}_1 \cdot ({\bm x}_1-{\bm y}_2) -
  i \; {\bm k}_2 \cdot ({\bm x}_2-{\bm y}_1)}
\notag \\
  & \; \; \times \left< \frac{1}{4 N^2_c}
  \sum_{\sigma' \sigma, \rho' \rho} \sum_{\lambda_1, \lambda_2}
  \left(A^{a,i' i}_{\sigma' \sigma, \lambda_2} (\bm y_2, \bm b_1) 
  A^{b,j' j}_{\rho' \rho, \lambda_1} (\bm y_1, \bm b_2) \right)^\dagger
  A^{a,i' i}_{\sigma' \sigma, \lambda_1} (\bm x_1, \bm b_1) 
  A^{b,j' j}_{\rho' \rho, \lambda_2} (\bm x_2, \bm b_2) \right.
\notag \\
    & \; \; + \left. \sum_{\lambda_1, \lambda_2} \left(
  C^{a b}_{\lambda_1 \lambda_2} (\bm y_2, \bm y_1, \bm b_2) \right)^\dagger
  C^{a b}_{\lambda_1 \lambda_2} (\bm x_1, \bm x_2, \bm b_1) \right>
\end{align}

Evaluating the term associated with Fig.~\ref{crossed} we find
\begin{align}
\label{crossedAA}
\frac{1}{4 N^2_c} &
  \sum_{\sigma' \sigma, \rho' \rho} \sum_{\lambda_1, \lambda_2}
  \left< \left(A^{a,i' i}_{\sigma' \sigma, \lambda_2} (\bm y_1, \bm b_1) 
  A^{b,j' j}_{\rho' \rho, \lambda_1} (\bm y_2, \bm b_2) \right)^\dagger
  A^{a,i' i}_{\sigma' \sigma, \lambda_1} (\bm x_1, \bm b_1) 
  A^{b,j' j}_{\rho' \rho, \lambda_2} (\bm x_2, \bm b_2) \right>
\notag \\
  = & \, \frac{g^4}{4 N^2_c \pi^4} \sum_{\sigma' \sigma, \rho' \rho}
  \delta_{\sigma' \sigma} \delta_{\sigma' \sigma} \delta_{\rho' \rho} \delta_{\rho' \rho}
  \sum_{\lambda_1}
  \frac{\bm \epsilon^*_{\lambda_1} \cdot \left( \bm x_1 - \bm b_1 \right)}{ | \bm x_1 - \bm b_1 |^2 }
  \frac{\bm \epsilon_{\lambda_1} \cdot \left( \bm y_2 - \bm b_2 \right)}{ | \bm y_2 - \bm b_2 |^2 }
\notag \\
  & \times \sum_{\lambda_2}
  \frac{\bm \epsilon^*_{\lambda_2} \cdot \left( \bm x_2 - \bm b_2 \right)}{ | \bm x_2 - \bm b_2 |^2 }
  \frac{\bm \epsilon_{\lambda_2} \cdot \left( \bm y_1 - \bm b_1 \right)}{ | \bm y_1 - \bm b_1 |^2 }
  \, \left< tr[t^e V^\dagger_{b_1} V_{b_1} t^c] \, tr[t^f V^\dagger_{b_2} V_{b_2} t^d] \right.
\notag \\
  & \times \left. \left[ \left( U^\dagger_{\bm y_2} - U^\dagger_{\bm b_2} \right) 
  \left( U_{\bm x_1} - U_{\bm b_1} \right) \right]^{f c}
  \left[ \left( U^\dagger_{\bm y_1} - U^\dagger_{\bm b_1} \right) 
  \left( U_{\bm x_2} - U_{\bm b_2} \right) \right]^{e d} \right>
\notag \\
  = & \, \frac{g^4}{4 N^2_c \pi^4} \,
  \frac{\left( \bm x_1 - \bm b_1 \right)\cdot \left( \bm y_2 - \bm b_2 \right)}
  { | \bm x_1 - \bm b_1 |^2 | \bm y_2 - \bm b_2 |^2 }
  \frac{\left( \bm x_2 - \bm b_2 \right)\cdot \left( \bm y_1 - \bm b_1 \right)}
  { | \bm x_2 - \bm b_2 |^2 | \bm y_1 - \bm b_1 |^2 }
\notag \\
  & \times \left< Tr \left[ \left( U_{\bm x_1} - U_{\bm b_1} \right)
  \left( U^\dagger_{\bm y_1} - U^\dagger_{\bm b_1} \right) 
  \left( U_{\bm x_2} - U_{\bm b_2} \right) 
  \left( U^\dagger_{\bm y_2} - U^\dagger_{\bm b_2} \right) \right] \right>.
\end{align}
Similarly for the term associated with the Fig.~\ref{crossed2},
\begin{align}
\label{crossed2AA}
\sum_{\lambda_1, \lambda_2} &
  \left( C^{a b}_{\lambda_1 \lambda_2} (\bm y_2, \bm y_1, \bm b_2) \right)^\dagger
  C^{a b}_{\lambda_1 \lambda_2} (\bm x_1, \bm x_2, \bm b_1)
\notag \\
  = & \frac{g^4}{4 N^2_c \pi^4} \sum_{\lambda_1 \lambda_2}
  \frac{\bm \epsilon^*_{\lambda_1} \cdot \left( \bm x_1 - \bm b_1 \right)}{ | \bm x_1 - \bm b_1 |^2 }
  \frac{\bm \epsilon_{\lambda_1} \cdot \left( \bm y_2 - \bm b_2 \right)}{ | \bm y_2 - \bm b_2 |^2 }
  \frac{\bm \epsilon^*_{\lambda_2} \cdot \left( \bm x_2 - \bm b_1 \right)}{ | \bm x_2 - \bm b_1 |^2 }
  \frac{\bm \epsilon_{\lambda_2} \cdot \left( \bm y_1 - \bm b_2 \right)}{ | \bm y_1 - \bm b_2 |^2 }
\notag \\
  & \times \left< \left[ \left(U_{\bm y_2}-U_{\bm b_2}\right) \left(U^\dagger_{\bm y_1}
  - U^\dagger_{\bm b_2}\right) \right]^{b a}
  \left[ \left(U_{\bm x_1}-U_{\bm b_1}\right) \left(U^\dagger_{\bm x_2}
  - U^\dagger_{\bm b_1}\right) \right]^{a b} \right>
\notag \\
  = & \, \frac{g^4}{4 N^2_c \pi^4} \,
  \frac{\left( \bm x_1 - \bm b_1 \right)\cdot \left( \bm y_2 - \bm b_2 \right)}
  { | \bm x_1 - \bm b_1 |^2 | \bm y_2 - \bm b_2 |^2 }
  \frac{\left( \bm x_2 - \bm b_1 \right)\cdot \left( \bm y_1 - \bm b_2 \right)}
  { | \bm x_2 - \bm b_1 |^2 | \bm y_1 - \bm b_2 |^2 }
\notag \\
  & \times \left< Tr \left[ \left( U_{\bm x_1} - U_{\bm b_1} \right)
  \left( U^\dagger_{\bm x_2} - U^\dagger_{\bm b_1} \right) 
  \left( U_{\bm y_2} - U_{\bm b_2} \right) 
  \left( U^\dagger_{\bm y_1} - U^\dagger_{\bm b_2} \right) \right] \right>.
\end{align}
Notice this is the same as \eqref{crossedAA} under interchange of $\bm x_2 \leftrightarrow \bm y_1$.
Since in the total expression \eqref{sigmacrossed} the transverse coordinates are all dummy variables, we can interchange $\bm x_2 \leftrightarrow \bm y_1$ in \eqref{crossed2AA} so that it now is the same as \eqref{crossedAA} except it has different coordinates in the exponential.
This means the only difference between the contributions of the diagrams in
Figs.~\ref{crossed} and \ref{crossed2} is in the exponential factors
for the Fourier transform into transverse momentum space.

It is convenient to define $Int_{crossed}$ as the target averaged Wilson line object in both \eqref{crossed2AA} and \eqref{crossed2AA} (normalized by $\frac{1}{N^2_c}$). Defining the color-quadrupole operator
\cite{Chen:1995pa,JalilianMarian:2004da,Dominguez:2011gc,Dumitru:2011vk,Iancu:2011ns}
\begin{align}
\label{quad_def}
Q ( {\bm x}_1 , {\bm x}_2 , {\bm x}_3 , {\bm x}_4 ) \equiv
  \frac{1}{N_c^2-1} \; \left\langle Tr[ U_{{\bm x}_1} U_{{\bm
  x}_2}^\dagger U_{{\bm x}_3} U_{{\bm x}_4}^\dagger ]
  \right\rangle,
\end{align}
we can express $Int_{crossed}$ as a combination of color-quadrupole operators and adjoint (gluon) dipole $S$-matrices, \eqref{gdipole} with the rapidity dependence dropped.
\begin{align}
\label{crossedinteraction}
Int_{crossed} & ( {\bm x}_1, {\bm y}_1 , {\bm b}_1 , {\bm x}_2 ,
  {\bm y}_2, {\bm b}_2)
\notag \\
  \equiv & \, \frac{1}{N^2_c -1} \left< Tr \left[ \left( U_{\bm x_1} - U_{\bm b_1} \right)
  \left( U^\dagger_{\bm x_2} - U^\dagger_{\bm b_1} \right) 
  \left( U_{\bm y_2} - U_{\bm b_2} \right) 
  \left( U^\dagger_{\bm y_1} - U^\dagger_{\bm b_2} \right) \right] \right>
\notag \\
  = & \, Q( {\bm x}_1, {\bm y}_1 , {\bm x}_2 , {\bm
   y}_2 ) - Q( {\bm x}_1, {\bm y}_1 , {\bm x}_2 , {\bm b}_2 ) - Q(
 {\bm x}_1, {\bm y}_1 , {\bm b}_2 , {\bm y}_2 ) + S_G( {\bm x}_1, {\bm y}_1)
\notag \\
  & - Q( {\bm x}_1, {\bm b}_1 , {\bm x}_2 , {\bm y}_2 ) \; + \; Q( {\bm
  x}_1, {\bm b}_1 , {\bm x}_2 , {\bm b}_2 ) + Q( {\bm x}_1, {\bm b}_1
  , {\bm b}_2 , {\bm y}_2 ) - S_G( {\bm x}_1, {\bm b}_1)
\notag \\
  & - Q( {\bm b}_1, {\bm y}_1 , {\bm x}_2 , {\bm y}_2 ) + Q( {\bm b}_1,
  {\bm y}_1 , {\bm x}_2 , {\bm b}_2 ) + Q( {\bm b}_1, {\bm y}_1 , {\bm
   b}_2 , {\bm y}_2 ) - S_G( {\bm b}_1, {\bm y}_1)
\notag \\
  & + S_G( {\bm x}_2, {\bm y}_2) - S_G( {\bm x}_2, {\bm b}_2) - S_G(
  {\bm b}_2, {\bm y}_2) + 1.
\end{align}

Using \eqref{crossedAA}, \eqref{crossed2AA}, \eqref{crossedinteraction} and the fact that the transverse coordinates are all dummy variables we arrive at the expression for the two-gluon production
cross section contribution resulting from the ``crossed'' diagrams
from Figs.~\ref{crossed} and \ref{crossed2}
\begin{align}
\label{crossed_xsect}
& \frac{d \sigma_{crossed}}{d^2 k_1 dy_1 d^2 k_2 dy_2} =
  \frac{1}{[2(2 \pi)^3]^2} \, \int d^2 B \, d^2 b_1 \, d^2 b_2 \, T_1
  ({\bm B} - {\bm b}_1) \, T_1 ({\bm B} - {\bm b}_2) \,
\notag \\
& \times \,  d^2 x_1 \, d^2 y_1 \, d^2 x_2 \, d^2 y_2 \left[ e^{- i \;
  {\bm k}_1 \cdot ({\bm x}_1-{\bm y}_2) - i \; {\bm k}_2 \cdot
  ({\bm x}_2-{\bm y}_1)} + e^{- i \; {\bm k}_1 \cdot ({\bm
  x}_1-{\bm y}_2) + i \; {\bm k}_2 \cdot ({\bm x}_2-{\bm y}_1)}
  \right]
\notag \\
& \times \, \frac{16 \; {\alpha}_s^2}{\pi^2} \, \frac{C_F}{2 N_c} \;
  \frac{ {\bm x}_1 - {\bm b}_1}{|{\bm x}_1 - {\bm b}_1|^2} \cdot
  \frac{{\bm y}_2 - {\bm b}_2 }{ |{\bm y}_2 - {\bm b}_2|^2 } \,
  \frac{{\bm x}_2 - {\bm b}_2}{|{\bm x}_2 - {\bm b}_2|^2} \cdot
  \frac{{\bm y}_1 - {\bm b}_1 }{ |{\bm y}_1 - {\bm b}_1|^2} \notag \\
& \times \, \bigg[ Q( {\bm x}_1, {\bm y}_1 , {\bm x}_2 , {\bm y}_2 )
  - Q( {\bm x}_1, {\bm y}_1 , {\bm x}_2 , {\bm b}_2 ) - Q( {\bm x}_1,
  {\bm y}_1 , {\bm b}_2 , {\bm y}_2 ) + S_G( {\bm x}_1, {\bm y}_1) \notag \\
& - Q( {\bm x}_1, {\bm b}_1 , {\bm x}_2 , {\bm y}_2 ) \; + \; Q( {\bm
  x}_1, {\bm b}_1 , {\bm x}_2 , {\bm b}_2 ) + Q( {\bm
  x}_1, {\bm b}_1 , {\bm b}_2 , {\bm y}_2 ) - S_G( {\bm x}_1, {\bm
  b}_1) \notag \\
& - Q( {\bm b}_1, {\bm y}_1 , {\bm x}_2 , {\bm y}_2 ) + Q(
  {\bm b}_1, {\bm y}_1 , {\bm x}_2 , {\bm b}_2 ) + Q( {\bm b}_1, {\bm
  y}_1 , {\bm b}_2 , {\bm y}_2 ) - S_G( {\bm b}_1, {\bm y}_1) \notag \\
& + S_G(
  {\bm x}_2, {\bm y}_2) - S_G( {\bm x}_2, {\bm b}_2) - S_G( {\bm b}_2, {\bm y}_2) +
  1\bigg].
\end{align}
Just like in \eqref{eq:2glue_prod_main}, the dipole and quadrupole
scattering amplitudes in \eqref{crossed_xsect} can be evaluated either in
the MV model or by using BK and JIMWLK evolution equations. The
quadrupole amplitude evolution equation was derived in the large-$N_c$
limit in \cite{JalilianMarian:2004da}, and beyond the large-$N_c$
limit in \cite{Dominguez:2011gc}. Again, \eqref{crossed_xsect} is valid
only as long as the rapidities $y_1$ and $y_2$ of the two produced
gluons are close to each other, $|y_2 - y_1| \lesssim 1/\as$, such
that no small-$x$ evolution corrections need to be included in the
$[y_1, y_2]$ rapidity interval.

%%%%%%%%%%%%%%%%%%%%%%%%%%%%%%%%%%%%%%%%%%%%%%%%%%%%%%%%%%%%%%%%%%%%%%%%%%%%%%%%%%

\subsection{Two-gluon production with long-range rapidity
  correlations: the net result}

Equations (\ref{eq:2glue_prod_main}) and (\ref{crossed_xsect}), when
combined, give us the two-gluon production cross section in the
heavy--light ion collisions:
\begin{align}\label{eq_all}
  \frac{d \sigma}{d^2 k_1 dy_1 d^2 k_2 dy_2} = \frac{d
    \sigma_{square}}{d^2 k_1 dy_1 d^2 k_2 dy_2} + \frac{d
    \sigma_{crossed}}{d^2 k_1 dy_1 d^2 k_2 dy_2}.
\end{align}

%% file: Section_Angular_results.tex
% !TEX root = WertepnyPhDThesis.tex

\section{Long-range rapidity correlations: away-side and near-side;
  HBT correlations}
\label{sec:lrrc}

Our goal now is to evaluate the correlations resulting from the
two-gluon production cross section \eqref{eq_all}, that is from the
cross sections in Eqs.~(\ref{eq:2glue_prod_main}) and
(\ref{crossed_xsect}). The first step is to evaluate the interaction
with the target. We will be working in the quasi-classical MV/GM
limit, where the interaction and, hence, the production cross sections
\eqref{eq:2glue_prod_main} and \eqref{crossed_xsect}, are
rapidity-independent.

We begin with the cross section in \eqref{eq:2glue_prod_main}. Even in
the quasi-classical limit, the evaluation of the interactions with the target
 is calculationally intensive (though
conceptually rather straightforward). To simplify the calculation we
will also employ the large-$N_c$ expansion: we assume that the
nucleons in the nuclei are made out of an order-$N_c^2$ valence quarks
(or gluons), such that the saturation scale, which in such case is
proportional to $Q_{s0}^2 \sim \alpha_s^2 N_c^2$, is constant in the 't
Hooft's large-$N_c$ limit. In the saturation physics framework such
approximation was used in \cite{JalilianMarian:2004da} for the
quadrupole operator giving a reasonably good approximation to the
exact answer \cite{Dumitru:2011vk}. Similar approximations are
frequently used (albeit, often implicitly) in applications of anti-de
Sitter space/conformal field theory (AdS/CFT) correspondence to
collisions of heavy ions modeled by shock waves
\cite{Grumiller:2008va,Albacete:2008vs,Albacete:2009ji,Chesler:2010bi}.

At the leading order in $1/N_c^2$ expansion of \eqref{eq:2glue_prod_main}
the interaction with the target factorizes, as discussed around
\eqref{eq:trace_fact}. In addition, the cross section in
\eqref{crossed_xsect} is $1/N_c^2$-suppressed (as compared to the leading
term in \eqref{eq:2glue_prod_main}) and can be neglected at the leading
order in $N_c$. The correlation function is then given by
\eqref{corr_fact} with the single gluon production cross section from
\eqref{crosssection12}. In the MV/GM approximation the gluon color dipole
interaction with the target is \cite{Mueller:1989st}
\begin{align}
\label{eq:SG_GM}
S_G ({\bm x}_1, {\bm x}_2, y=0) = \exp \left[ -\frac{1}{4} \, |{\bm
  x}_1 - {\bm x}_2|^2 \, Q_{s0}^2 \left( \frac{{\bm x}_1 + {\bm
  x}_2}{2} \right) \, \ln \left(\frac{1}{|{\bm x}_1 - {\bm
  x}_2| \, \Lambda} \right) \right]
\end{align}
with $Q_{s0}$ the rapidity-independent gluon saturation scale in the
quasi-classical limit evaluated at the dipole center-of-mass $({\bm
  x}_1 + {\bm x}_2)/2$ and $\Lambda$ an infrared (IR) cutoff. We see
that the quasi-classical single gluon production cross section is
rapidity-independent and, for unpolarized nuclei and for
perturbatively large $k_T$ \cite{Teaney:2002kn}, is also independent
of the azimuthal angle $\phi$ of the transverse momentum $\bm k$ of
the outgoing gluon. We conclude that at the leading order in $1/N_c^2$
in the quasi-classical approximation the geometric correlations are
almost absent and the correlation function \eqref{corr_fact} is
approximately zero.

Non-trivial correlations can be obtained from \eqref{eq:2glue_prod_main}
by expanding the interaction term to the first non-trivial order in
$1/N_c^2$. To this end we write a correlator of two Wilson line traces
as
\begin{align}
\label{Ddef}
\frac{1}{(N_c^2 - 1)^2} \, & \langle Tr[ U_{{\bm x}_1} U_{{\bm
  x}_2}^\dagger ] \, Tr[ U_{{\bm x}_3} U_{{\bm x}_4}^\dagger ]
  \rangle = \notag \\
& \frac{1}{(N_c^2 - 1)^2} \, \langle Tr[ U_{{\bm x}_1}
  U_{{\bm x}_2}^\dagger ] \rangle \langle Tr[ U_{{\bm x}_3} U_{{\bm
  x}_4}^\dagger ] \rangle + \Delta ( {\bm x}_1 , {\bm x}_2 , {\bm
  x}_3 , {\bm x}_4 ),
\end{align}
where the correction to the factorized expression
\eqref{eq:trace_fact}, denoted by $\Delta$, is order-$1/N_c^2$. (Note
that each adjoint trace in \eqref{Ddef} is order-$N_c^2$, such that
its left-hand side, as well as the first term on its right-hand side,
are order-one in $N_c$ counting.)

The leading order-$1/N_c^2$ contribution to $\Delta$ is derived in the
Appx.~\ref{Appendix:Wilson}, with the result being
\begin{align}
\label{Delta_exp}
\Delta ( {\bm x}_1 , {\bm x}_2 , {\bm x}_3 , {\bm x}_4 ) & = \notag \\
\frac{(D_3-D_2)^2}{N_c^2} \, & \left[ \frac{e^{D_1}}{D_1-D_2} -
  \frac{2 \, e^{D_1}}{(D_1-D_2)^2} + \frac{e^{D_1}}{D_1-D_3} -
  \frac{2 \, e^{D_1}}{(D_1-D_3)^2} \right. \notag \\
& \left. + \frac{2 \, e^{\frac{1}{2}(D_1+D_2)}}{(D_1-D_2)^2} +
  \frac{2 \, e^{\frac{1}{2}(D_1+D_3)}}{(D_1-D_3)^2} \right] + O
  \left( \frac{1}{N_c^4} \right),
\end{align}
where we have defined
\begin{subequations}\label{Ds}
\begin{align}
D_1 & = - \frac{Q_{s0}^2}{4} \left[ | \bm x_1 - \bm x_2 |^2 \, 
  \ln \left( \frac{1}{| \bm x_1 - \bm x_2 | \Lambda } \right) + | \bm x_3 - \bm x_4 |^2 \,
  \ln \left( \frac{1}{| \bm x_3 - \bm x_4 | \Lambda } \right) \right] \\
D_2 & = - \frac{Q_{s0}^2}{4} \left[ | \bm x_1 - \bm x_3 |^2 \,
  \ln \left( \frac{1}{| \bm x_1 - \bm x_3 | \Lambda } \right) + | \bm x_2 - \bm x_4 |^2 \, 
  \ln \left( \frac{1}{| \bm x_2 - \bm x_4 | \Lambda } \right) \right] \\
D_3 & = - \frac{Q_{s0}^2}{4} \left[ | \bm x_1 - \bm x_4 |^2 \,
  \ln \left( \frac{1}{| \bm x_1 - \bm x_4 | \Lambda } \right) + | \bm x_2 - \bm x_3 |^2 \,
  \ln \left( \frac{1}{| \bm x_2 - \bm x_3 | \Lambda } \right) \right]
\end{align}
\end{subequations}
assuming, for simplicity, that all the saturation scales are evaluated
at the same impact parameter.

Using \eqref{Ddef} in \eqref{eq:2glue_prod_main} we see that the correlated
part of the two-gluon production cross section is
\begin{align} 
\label{eq:2glue_prod_delta} 
& \frac{d \sigma_{square}^{(corr)}}{d^2 k_1 dy_1 d^2 k_2 dy_2} = \frac{\as^2 \,
  C_F^2}{16 \, \pi^8} \int d^2 B \, d^2 b_1 \, d^2 b_2 \, T_1 ({\bm B}
  - {\bm b}_1) \, T_1 ({\bm B} - {\bm b}_2) \, d^2 x_1 \, d^2 y_1 \, d^2
  x_2 \, d^2 y_2 \, \notag \\
& \quad \quad \times \, e^{- i \; {\bm k}_1 \cdot ({\bm x}_1-{\bm y}_1) - i
  \; {\bm k}_2 \cdot ({\bm x}_2-{\bm y}_2)} 
  \frac{ {\bm x}_1 - {\bm b}_1}{ |{\bm x}_1 - {\bm b}_1 |^2 } \cdot
  \frac{ {\bm y}_1 - {\bm b}_1}{ |{\bm y}_1 - {\bm b}_1 |^2 } \ \frac{
  {\bm x}_2 - {\bm b}_2}{ |{\bm x}_2 - {\bm b}_2 |^2 } \cdot \frac{
  {\bm y}_2 - {\bm b}_2}{ |{\bm y}_2 - {\bm b}_2 |^2 } \notag \\
& \quad \quad \times \, \left[ \Delta( {\bm x}_1 , {\bm y}_1 , {\bm x}_2 , {\bm
  y}_2 ) - \Delta ( {\bm x}_1 , {\bm y}_1 , {\bm x}_2 , {\bm b}_2 )
  - \Delta ( {\bm x}_1 , {\bm y}_1 , {\bm b}_2 , {\bm y}_2 ) - \Delta
  ( {\bm x}_1 , {\bm b}_1 , {\bm x}_2 , {\bm y}_2 ) \ \right. \notag \\
& \quad \quad \left. 
  - \Delta ( {\bm b}_1 , {\bm y}_1 , {\bm x}_2 , {\bm y}_2 )+ \Delta ( {\bm x}_1 , {\bm b}_1 , {\bm x}_2 , {\bm b}_2 ) +
  \Delta ( {\bm x}_1 , {\bm b}_1 , {\bm b}_2 , {\bm y}_2 ) + \Delta (
  {\bm b}_1 , {\bm y}_1 , {\bm x}_2 , {\bm b}_2 ) \right. \notag \\
& \quad \quad \left. + \Delta ( {\bm b}_1,
  {\bm y}_1 , {\bm b}_2 , {\bm y}_2 ) \right].
\end{align}
\eqref{eq:2glue_prod_delta} along with the expression for $\Delta$ in
\eqref{Delta_exp} are our most complete results for the contribution to
the two-gluon production cross section coming from
\eqref{eq:2glue_prod_main} in the quasi-classical regime of the
heavy--light ion collisions in the large-$N_c$ limit. The evaluation
of the full \eqref{eq:2glue_prod_delta} appears to be rather involved and
is left for the future work.

Instead we will expand \eqref{Delta_exp} and use the result in
\eqref{eq:2glue_prod_delta} to obtain the correlated gluon production at
the lowest non-trivial order. This result can be used to elucidate the
structure of the long-range rapidity correlations, along with the
comparison to the existing expressions in the literature.

At the lowest non-trivial order in $D_i$'s (that is, since each $D_i$
represents a two-gluon exchange with the target, at the lowest order
in the number of gluon exchanges corresponding to $k_{1T}, k_{2T} \gg
Q_{s0}$), \eqref{Delta_exp} becomes
\begin{align}
\label{eq:Delta_LO}
\Delta ( {\bm x}_1 , {\bm x}_2 , {\bm x}_3 , {\bm x}_4 ) \approx
  \frac{(D_3 - D_2)^2}{2 \, N_c^2}.
\end{align}
Substituting this into \eqref{eq:2glue_prod_delta} we can further
simplify the expression by assuming that for the connected diagrams
that contribute to the $\Delta$'s one has ${\bm b}_1$ and ${\bm b}_2$
perturbatively close to each other, with the typical separation
between these two impact parameters much smaller than the nucleon
radius. Since the nuclear profile functions for a large nucleus do not
vary much over perturbatively short distances we can put ${\bm b}_1
\approx {\bm b}_2 \approx {\bm b}$ in the arguments of $T_1$'s and
$Q_{s0}$, where ${\bm b} \equiv ({\bm b}_1 + {\bm b}_2)/2$. Defining
$\Delta {\bm b} \equiv {\bm b}_1 - {\bm b}_2$ we can write $d^2 b_1 \,
d^2 b_2 = d^2 b \, d^2 \Delta b$. The integral over $\Delta {\bm b}$
can be then carried out with the help of a Fourier transform
\begin{align}
\label{eq:Fourier}
\frac{1}{4} \, |{\bm x}|^2 \, \ln \left( \frac{1}{|{\bm x}| \,
  \Lambda} \right) = \int \frac{d^2 l}{2 \, \pi} \, ( 1 - e^{i \,
  {\bm l} \cdot {\bm x}}) \, \frac{1}{({\bm l}^2)^2}
\end{align}
used to replace $(D_3 - D_2)^2$ in \eqref{eq:Delta_LO} and, hence, all
$\Delta$'s in \eqref{eq:2glue_prod_delta} as a double Fourier-integral:
for instance
\begin{align}
\label{eq:Delta_sample}
\Delta & ( {\bm x}_1 , {\bm y}_1 , {\bm x}_2 , {\bm y}_2 ) \notag \\
& = \frac{Q_{s0}^4}{2 \, N_c^2} \int \frac{d^2 l \, d^2 l'}{(2 \pi)^2}
  \, \, \frac{1}{({\bm l}^2)^2} \, \frac{1}{({\bm l}'^2)^2} \, \left[
  e^{i \, {\bm l} \cdot ({\bm x}_1 - {\bm x}_2)} + e^{i \, {\bm l}
  \cdot ({\bm y}_1 - {\bm y}_2)} - e^{i \, {\bm l} \cdot ({\bm
  x}_1 - {\bm y}_2)} - e^{i \, {\bm l} \cdot ({\bm y}_1 - {\bm
  x}_2)} \right] \notag \\
& \quad \times \, \left[ e^{- i \, {\bm l}' \cdot ({\bm x}_1 - {\bm x}_2)} +
  e^{-i \, {\bm l}' \cdot ({\bm y}_1 - {\bm y}_2)} - e^{- i \, {\bm
  l}' \cdot ({\bm x}_1 - {\bm y}_2)} - e^{- i \, {\bm l}' \cdot
  ({\bm y}_1 - {\bm x}_2)} \right] \notag \\
& = \frac{Q_{s0}^4}{2 \, N_c^2} \int \frac{d^2 l \, d^2 l'}{(2 \pi)^2}
  \, \, \frac{1}{({\bm l}^2)^2} \, \frac{1}{({\bm l}'^2)^2} \, \left[
  e^{i \, {\bm l} \cdot ({\tilde {\bm x}}_1 - {\tilde {\bm x}}_2)} +
  e^{i \, {\bm l} \cdot ({\tilde {\bm y}}_1 - {\tilde {\bm y}}_2)} -
  e^{i \, {\bm l} \cdot ({\tilde {\bm x}}_1 - {\tilde {\bm y}}_2)} -
  e^{i \, {\bm l} \cdot ({\tilde {\bm y}}_1 - {\tilde {\bm
  x}_2)}} \right] \notag \\
& \quad \times \, \left[ e^{- i \, {\bm l}' \cdot ({\tilde {\bm x}}_1 -
  {\tilde {\bm x}}_2)} + e^{-i \, {\bm l}' \cdot ({\tilde {\bm
  y}}_1 - {\tilde {\bm y}}_2)} - e^{- i \, {\bm l}' \cdot
  ({\tilde {\bm x}}_1 - {\tilde {\bm y}}_2)} - e^{- i \, {\bm l}'
  \cdot ({\tilde {\bm y}}_1 - {\tilde {\bm x}}_2)} \right] \, e^{i
  \, ({\bm l} - {\bm l}') \cdot \Delta {\bm b}}
\end{align}
where we employed the substitution 
\begin{align}
\label{eq:change}
{\bm {\tilde x}}_1 = {\bm x}_1 - {\bm b}_1, \ \ {\bm {\tilde y}}_1 =
  {\bm y}_1 - {\bm b}_1, \ \ {\bm {\tilde x}}_2 = {\bm x}_2 - {\bm
  b}_2, \ \ {\bm {\tilde y}}_2 = {\bm y}_2 - {\bm b}_1.
\end{align}
Performing similar substitutions for all $\Delta$'s in
\eqref{eq:2glue_prod_delta}, we can integrate over $\Delta {\bm b}$,
${\tilde {\bm x}}_1$, ${\tilde {\bm x}}_2$, ${\tilde {\bm y}}_1$,
${\tilde {\bm y}}_2$, and ${\bm l}'$. After some algebra one arrives
at (cf. \cite{Dusling:2009ni,Dumitru:2010iy})
\begin{align} 
\label{eq:2glue_prod_LO} 
\frac{d \sigma_{square}^{(corr)}}{d^2 k_1 dy_1 d^2 k_2 dy_2}\bigg|_{LO} & =
  \frac{\as^2}{4 \, \pi^4} \, \int d^2 B \, d^2 b \, [T_1 ({\bm B} -
  {\bm b})]^2 \, \frac{Q_{s0}^4 ({\bm b})}{{\bm k}_1^2 \, {\bm k}_2^2}
  \, \notag \\
& \times \, \int\limits_\Lambda \frac{d^2 l}{({\bm l}^2)^2} \, \left[
  \frac{1}{({\bm k}_1 - {\bm l})^2 \, ({\bm k}_2 + {\bm l})^2} +
  \frac{1}{({\bm k}_1 - {\bm l})^2 \, ({\bm k}_2 - {\bm l})^2}
\right],
\end{align}
where the subscript $LO$ denotes the lowest-order cross section.

\begin{figure}[H]
\centering
  \includegraphics[width= 0.9 \textwidth]{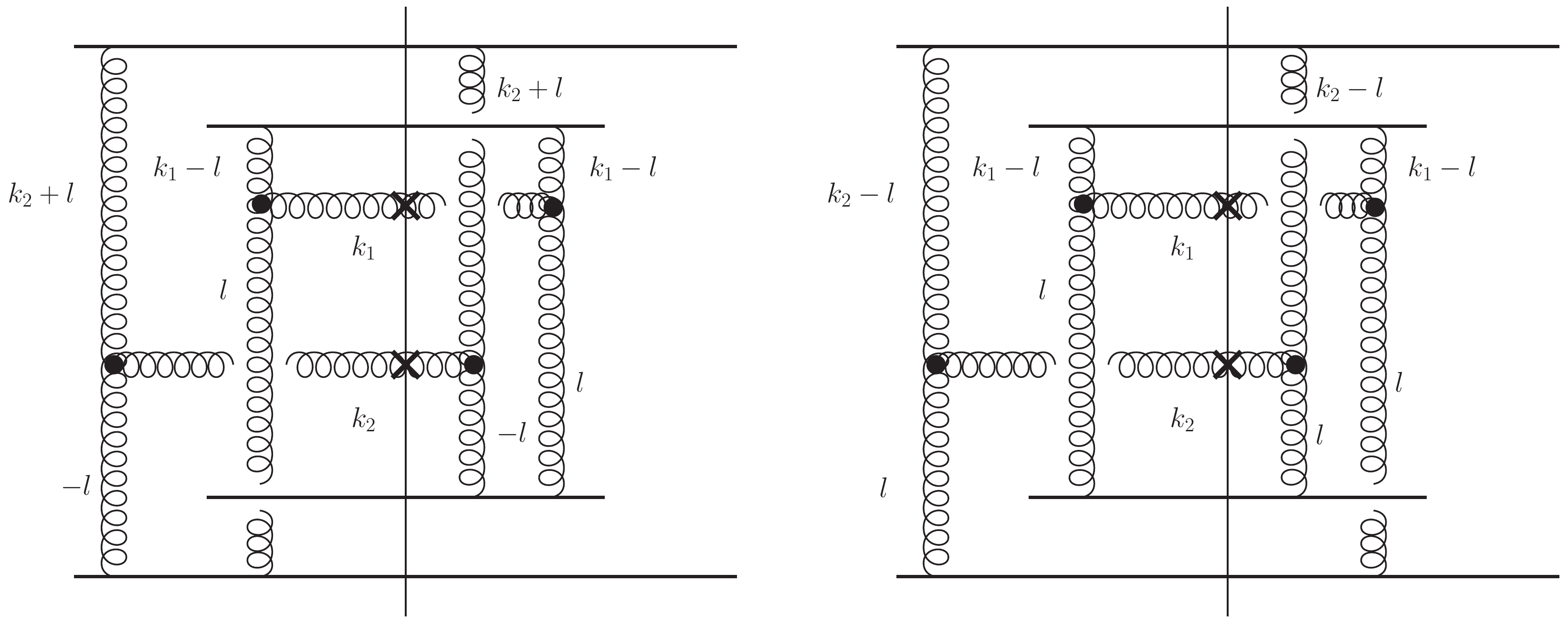}
  \caption{Examples of diagrams generating contributions to
    \eqref{eq:2glue_prod_LO}: the left panel represent the away-side
    correlations (the first term in \eqref{eq:2glue_prod_LO}), while the
    right panel contributes near-side correlations (the second term in
    \eqref{eq:2glue_prod_LO}). The $t$-channel gluon momenta flow toward
    the triple-gluon vertices to the left of the cut, and away from
    those vertices to the right of the cut.}
\label{glgraph1} 
\end{figure}

Equation \eqref{eq:2glue_prod_LO} is illustrated in \fig{glgraph1} by
regular Feynman diagrams that contribute to its right-hand-side
(cf. \cite{Dusling:2009ni,Dumitru:2010iy}): these diagrams are
referred to as the ``glasma'' graphs in the literature. The momenta of
the gluon lines are labeled in \fig{glgraph1}, and the triple gluon
vertices, marked by the dark circles, are the effective Lipatov
vertices \cite{Kuraev:1977fs,Fadin:1975cb}. These particular graphs contribute to the first and the
second terms in the square brackets of \eqref{eq:2glue_prod_LO}
correspondingly, and can be calculated by taking two Lipatov vertices
squared.

%%%%%%%%%%%%%%%%%%%%%%%%%%%%%%%%%%%%%%%%%%%%%%%%%%%%%%%%%%%%%%%%%%%%%%%%%%%%
%%%%%%%%%%%%%%%%%%%%%%%%%%%%%%%%%%%%%%%%%%%%%%%%%%%%%%%%%%%%%%%%%%%%%%%%%%%%

The obtained expression \eqref{eq:2glue_prod_LO} contains both the
near-side and away-side azimuthal correlations
\cite{Dusling:2009ni,Dumitru:2010iy}: clearly the first term in the
square brackets of \eqref{eq:2glue_prod_LO} contains poles at ${\bm l} =
{\bm k}_1$ and ${\bm l} = - {\bm k}_2$, which, after integration over
$l$, lead to a contribution\footnote{Note that both terms in
  \eqref{eq:2glue_prod_LO} also contain a pole at ${\bm l} =0$, which
  leads to a correlated contribution independent of the azimuthal
  angle between ${\bm k}_1$ and ${\bm k}_2$.}
\begin{align}
\label{eq:b2b}
\sim \frac{1}{({\bm k}_1 + {\bm k}_2)^2},
\end{align}
characteristic of the away-side correlations.

The second term in the square brackets of \eqref{eq:2glue_prod_LO} has
poles at ${\bm l} = {\bm k}_1$ and ${\bm l} = {\bm k}_2$, yielding a
contribution
\begin{align}
\label{eq:near_side}
\sim \frac{1}{({\bm k}_1 - {\bm k}_2)^2},
\end{align}
indicating near-side correlations. Note that all correlations are
long-range in rapidity since the cross section
\eqref{eq:2glue_prod_LO} is rapidity-independent.

Now we turn our attention to \eqref{crossed_xsect}. There the cross
section contribution itself is $1/N_c^2$-suppressed as compared to the
leading (uncorrelated) part of \eqref{eq:2glue_prod_main}: hence we need
to evaluate the interaction with the target in \eqref{crossed_xsect}
using the large-$N_c$ limit. We work in the same quasi-classical MV/GM
approximation, along with the large-$N_c$ limit. The fundamental
(quark) quadrupole amplitude was evaluated in this approximation in
\cite{JalilianMarian:2004da} (see Eq. (14) there), yielding
\begin{align}
\label{eq:quad_quark}
Q_{quark} ({\bm x}_1, {\bm x}_2, {\bm x}_3, {\bm x}_4) = e^{D_1/2} +
  \frac{D_3 - D_2}{D_1 - D_3} \, \left[ e^{D_1/2} - e^{D_3/2} \right].
\end{align}
Since the adjoint (gluon) quadrupole \eqref{quad_def} in the
large-$N_c$ limit is simply
\begin{align}
\label{eq:quad_rel}
Q ({\bm x}_1, {\bm x}_2, {\bm x}_3, {\bm x}_4) = \left[ Q_{quark}
  ({\bm x}_1, {\bm x}_2, {\bm x}_3, {\bm x}_4) \right]^2
\end{align}
we get
\begin{align}
\label{eq:quad_MV}
Q ({\bm x}_1, {\bm x}_2, {\bm x}_3, {\bm x}_4) = \left[e^{D_1/2} +
  \frac{D_3 - D_2}{D_1 - D_3} \, \left( e^{D_1/2} - e^{D_3/2}
  \right) \right]^2.
\end{align}
Equations \eqref{eq:quad_MV} and \eqref{eq:SG_GM}, when used in
\eqref{crossed_xsect}, give us the remaining contribution to the
two-gluon production cross section in the quasi-classical
approximation, which has to be added to \eqref{eq:2glue_prod_delta}.

Again the full expression \eqref{crossed_xsect} is hard to evaluate:
instead, similar to \eqref{eq:2glue_prod_LO}, we will consider the limit
of large $k_{1T} = |{\bm k}_1|$ and $k_{2T} = |{\bm k}_2|$: $k_1, k_2
\gg Q_{s0}$. Expanding \eqref{eq:quad_MV} in the powers of $D_i$'s yields
\begin{align}
\label{eq:quad_exp}
Q & ({\bm x}_1, {\bm x}_2, {\bm x}_3, {\bm x}_4) = 1 + D_1 - D_2 + D_3 \notag \\
& + \frac{1}{4} \, \left[ 2 \, D_1^2 + D_2^2 + 2 \, D_3^2 - 3 \, D_1
  \, D_2 - 3 \, D_2 \, D_3 + 3 \, D_1 \, D_3 \right] + O \left(
  D_i^3 \right).
\end{align}
Using this result along with a similar expansion for \eqref{eq:SG_GM} in
\eqref{crossedinteraction}, and employing again the Fourier transform
\eqref{eq:Fourier} along with the substitution \eqref{eq:change} one
can write
\begin{align}
\label{eq:intQ}
Int_{crossed} & ( {\bm x}_1, {\bm y}_1 , {\bm b}_1 , {\bm x}_2 , {\bm
  y}_2, {\bm b}_2) \notag \\
& = Q_{s0}^4 \, \int \frac{d^2 l \, d^2 l'}{(2
  \pi)^2} \, \, \frac{1}{({\bm l}^2)^2} \, \frac{1}{({\bm l}'^2)^2}
  \, \bigg\{ e^{i \, ({\bm l} - {\bm l}') \cdot \Delta {\bm b}} \,
  \left( 1 - e^{i \, {\bm l}' \cdot {\bm {\tilde x}}_2} \right) \,
  \left( 1 - e^{- i \, {\bm l} \cdot {\bm {\tilde y}}_2} \right)
  \notag \\
& \times \, \bigg[ \frac{1}{2} \, \left( 1 - e^{- i \, {\bm
  l}' \cdot {\bm {\tilde x}}_1} \right) \, \left( 1 - e^{i \,
  {\bm l} \cdot {\bm {\tilde y}}_1} \right) + \left( 1 - e^{i \,
  {\bm l} \cdot {\bm {\tilde x}}_1} \right) \, \left( 1 - e^{- i
  \, {\bm l}' \cdot {\bm {\tilde y}}_1} \right) \bigg] \notag \\
& + \left( 1 - e^{i \, {\bm l} \cdot {\bm {\tilde x}}_1} \right) \,
  \left( 1 - e^{- i \, {\bm l}' \cdot {\bm {\tilde x}}_2} \right) \,
  \left( 1 - e^{- i \, {\bm l} \cdot {\bm {\tilde y}}_1} \right) \,
  \left( 1 - e^{i \, {\bm l}' \cdot {\bm {\tilde y}}_2} \right)
  \bigg\}
\end{align}
where the ${\bm l} \leftrightarrow {\bm l}'$, ${\bm l} \leftrightarrow
- {\bm l}$, and ${\bm l}' \leftrightarrow - {\bm l}'$ symmetries of
the integrand were utilized to cast the expression in its present
form. Using \eqref{eq:intQ} to replace the interaction with the target in
\eqref{crossed_xsect} and integrating over ${\bm {\tilde x}}_1$, ${\bm
  {\tilde x}}_2$, ${\bm {\tilde y}}_1$, ${\bm {\tilde y}}_2$, and, in
some terms, over ${\bm l}'$, leads to the following result
\begin{align}
\label{eq:rap_corr2}
\frac{d \sigma_{crossed}}{d^2 k_1 dy_1 d^2 k_2 dy_2}\bigg|_{LO} =
  \frac{d \sigma_{crossed}^{(corr)}}{d^2 k_1 dy_1 d^2 k_2
  dy_2}\bigg|_{LO} + \frac{d \sigma_{HBT}}{d^2 k_1 dy_1 d^2 k_2
  dy_2}\bigg|_{LO}
\end{align}
where
\begin{align}\label{eq:rap_corr3}
& \frac{d \sigma^{(corr)}_{crossed}}{d^2 k_1 dy_1 d^2 k_2
  dy_2}\bigg|_{LO} \notag \\
& \quad = \frac{\as^2}{32 \, \pi^4} \, \int d^2 B \,
  d^2 b \, [T_1 ({\bm B} - {\bm b})]^2 \, \frac{Q_{s0}^4 ({\bm
  b})}{{\bm k}_1^2 \, {\bm k}_2^2} \, \notag \\
& \quad \times \int \frac{d^2 l}{({\bm
  l}^2)^2 \, (({\bm l} - {\bm k}_1 + {\bm k}_2)^2)^2 \, (({\bm
  k}_1 - {\bm l})^2)^2 \, (({\bm k}_2 + {\bm l})^2)^2} \notag \\
& \quad \times \big\{ \left[ {\bm l}^2 \, ({\bm k}_2 + {\bm l})^2 + ({\bm
  k}_1 - {\bm l})^2 \, ({\bm l} - {\bm k}_1 + {\bm k}_2)^2 - {\bm
  k}_1^2 \, ({\bm k}_2 - {\bm k}_1 + 2 \, {\bm l})^2 \right]
  \notag \\
& \quad \times \, \left[ {\bm l}^2 \, ({\bm k}_1 - {\bm l})^2 +
  ({\bm k}_2 + {\bm l})^2 \, ({\bm l} - {\bm k}_1 + {\bm k}_2)^2 -
  {\bm k}_2^2 \, ({\bm k}_2 - {\bm k}_1 + 2 \, {\bm l})^2 \right]
  \notag \\
& \quad  + 4 \, {\bm l}^2 \, ({\bm l} - {\bm k}_1 + {\bm k}_2)^2
  \, \left[ (({\bm k}_1 - {\bm l})^2)^2 + (({\bm k}_2 + {\bm l})^2)^2
  \right] \big\} + ({\bm k}_2 \rightarrow - {\bm k}_2)
\end{align}
and
\begin{align} 
\label{eq:HBT_LO} 
& \frac{d \sigma_{HBT}}{d^2 k_1 dy_1 d^2 k_2 dy_2}\bigg|_{LO} \notag \\
& = \frac{\as^2}{16 \, \pi^4} \, \int d^2 B \, d^2 b \, [T_1 ({\bm B} -
  {\bm b})]^2 \, \frac{Q_{s0}^4 ({\bm b})}{{\bm k}_1^2 \, {\bm k}_2^2}
  \, \left[ \delta^2 ({\bm k}_1 - {\bm k}_2) + \delta^2 ({\bm k}_1 +
  {\bm k}_2) \right] \, \notag \\
& \times \, \int\limits_\Lambda
  \frac{d^2 l \, d^2 l'}{({\bm l}^2)^2 \, ({\bm l '}^2)^2 \, (({\bm k}_1
  - {\bm l})^2)^2 \, (({\bm k}_1 + {\bm l'})^2)^2} \, \left[ {\bm l}^2
  \, ({\bm k}_1 + {\bm l'})^2 + {\bm l'}^2 \, ({\bm k}_1 - {\bm l})^2
  - {\bm k}_1^2 \, ({\bm l} + {\bm l'})^2 \right]^2.
\end{align}
We will explain the origin of the ``HBT'' label on the cross section
in \eqref{eq:HBT_LO} in a little while below: we defer the analysis of
\eqref{eq:HBT_LO} until then.

First we concentrate on the expression \eqref{eq:rap_corr3}. We note
that by shifting the integration momentum $\bm l$ one could reduce the
second term in the curly brackets (together with the ${\bm k}_2
\rightarrow - {\bm k}_2$ term) to that in \eqref{eq:2glue_prod_LO}, thus
doubling \eqref{eq:2glue_prod_LO}. That term is still described by the
diagrams of the type shown in \fig{glgraph1} and gives the near- and
away-side correlations in Eqs. \eqref{eq:b2b} and
\eqref{eq:near_side}.

%%%%%%%%%%%%%%%%%%%%%%%%%%%%%%%%%%%%%%%%%%%%%%%%%%%%%%%%%%%%%%%%%%%%%%%%%%%%
\begin{figure}[H]
\centering
  \includegraphics[width= 0.45 \textwidth]{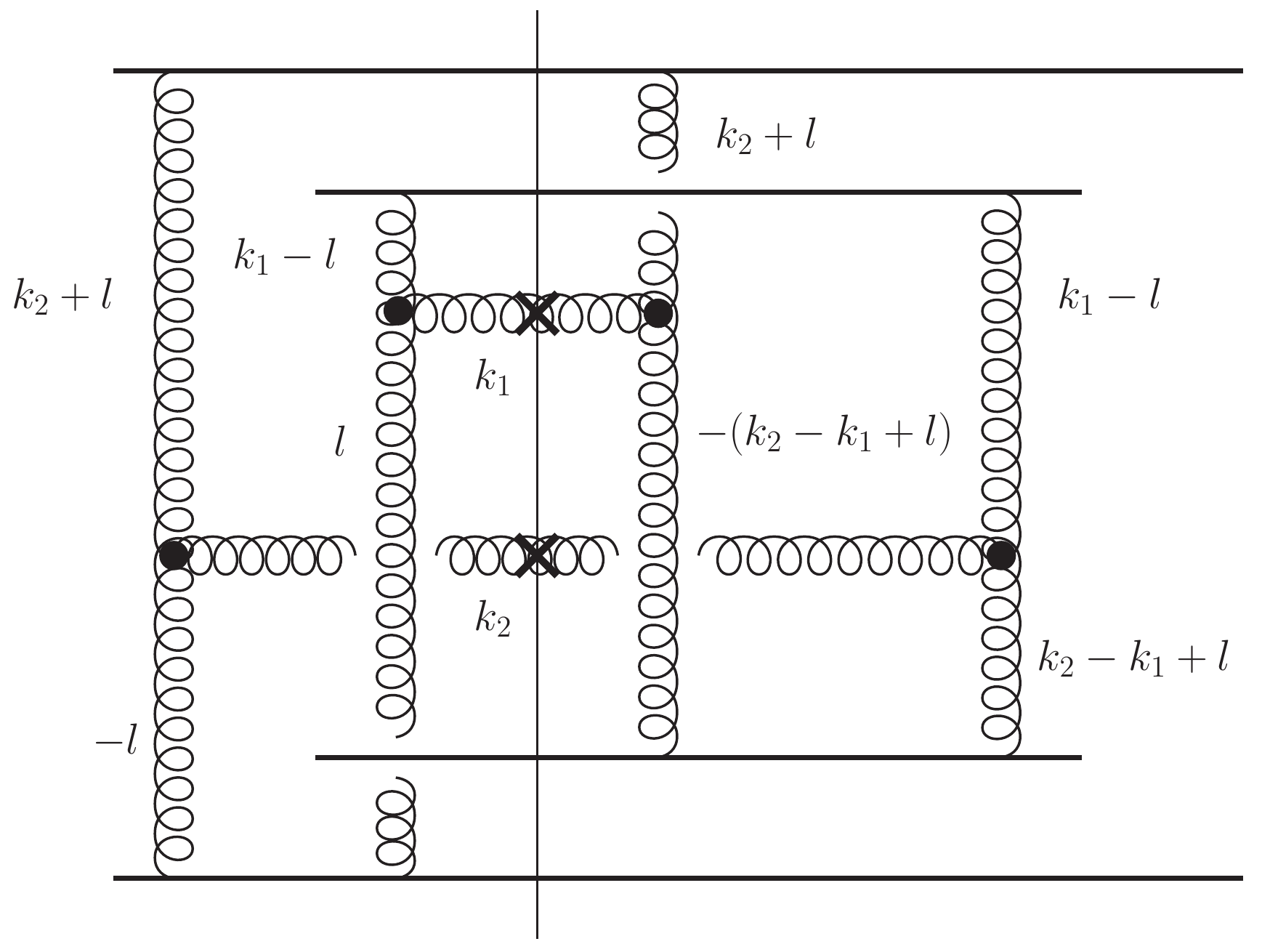}
  \caption{An example of a diagram giving a contribution to the first
    term in the curly brackets of \eqref{eq:rap_corr3}.}
\label{glgraph2} 
\end{figure}
%%%%%%%%%%%%%%%%%%%%%%%%%%%%%%%%%%%%%%%%%%%%%%%%%%%%%%%%%%%%%%%%%%%%%%%%%%%%

The first term in the curly brackets of \eqref{eq:rap_corr3} corresponds
to a different class of diagrams, one of which is shown in
\fig{glgraph2}. One can see that the diagram in \fig{glgraph2}
contributes two non-forward ``squares'' of the effective Lipatov
vertices. An analysis of the poles of the integral in the first term
of \eqref{eq:rap_corr3} shows that it contains similar near- and
away-side correlations to that in Eqs. \eqref{eq:b2b} and
\eqref{eq:near_side}, though with an additional enhancement due to a
prefactor, 
\begin{align}
\label{eq:near_side2}
\sim [2 ({\bm k}_1 \cdot {\bm k}_2)^2 - {\bm k}_1^2 \, {\bm k}_2^2]
  \, \left[\frac{1}{({\bm k}_1 - {\bm k}_2)^2} + \frac{1}{({\bm k}_1 +
  {\bm k}_2)^2} \right].
\end{align}
The origin of this contribution to correlations is in the non-forward
squares of the Lipatov vertices, which is similar to the mechanism for
generating long-range rapidity correlations proposed in
\cite{Levin:2011fb} using two BFKL ladders with non-zero momentum
transfer. Note, however, that we do not obtain the correlation
proportional to the first power of ${\bm k}_1 \cdot {\bm k}_2$
advocated in \cite{Levin:2011fb}, perhaps due to the lowest-order
nature of the result \eqref{eq:rap_corr3}. It appears that the
correlations \eqref{eq:near_side2} resulting from the lowest-order
diagrams like the one depicted in \fig{glgraph2} have not been
considered in the literature yet.

Note that the away- and the near-side correlations enter
Eqs. \eqref{eq:2glue_prod_LO} and \eqref{eq:rap_corr3} on equal
footing: in fact, one could be obtained from another by a simple ${\bm
  k}_2 \rightarrow - {\bm k}_2$ substitution in either of those
expressions. While evaluating the integrals over $l$ in
Eqs. \eqref{eq:2glue_prod_LO} and \eqref{eq:rap_corr3} analytically
appears to be rather algebra-intensive, we do not need to do this to
observe that, once the $l$-integral is carried out for one of the
terms in the square brackets of \eqref{eq:2glue_prod_LO}, the answer for
the other term is obtained by substituting ${\bm k}_2 \rightarrow -
{\bm k}_2$ in the result. Equation \eqref{eq:rap_corr3} simply
contains an additive ${\bm k}_2 \rightarrow - {\bm k}_2$ term.

Since the correlated cross sections \eqref{eq:2glue_prod_LO} and
\eqref{eq:rap_corr3} are sums of two terms related by the ${\bm k}_2
\rightarrow - {\bm k}_2$ substitution and are symmetric under the
${\bm k}_1 \leftrightarrow {\bm k}_2$ interchange, we conclude that
they are functions only of even powers of ${\bm k}_1$ and ${\bm k}_2$,
that is functions of ${\bm k}_1^2$, ${\bm k}_2^2$, $({\bm k}_1 \cdot
{\bm k}_2)^2$, and possibly $({\bm k}_1 \times {\bm k}_2)^2$. Clearly
this implies that the Fourier series representation of
Eqs. \eqref{eq:2glue_prod_LO} and \eqref{eq:rap_corr3} would only
contain even cosine harmonics of the azimuthal angle, that is
\begin{align}
\label{eq:series}
\frac{d \sigma^{(corr)}}{d^2 k_1 dy_1 d^2 k_2 dy_2}\bigg|_{LO}
  & \equiv \frac{d \sigma^{(corr)}_{square}}{d^2 k_1 dy_1 d^2 k_2
  dy_2}\bigg|_{LO} + \frac{d \sigma^{(corr)}_{crossed}}{d^2 k_1 dy_1
  d^2 k_2 dy_2}\bigg|_{LO} \notag \\
& \sim \sum_{n=0}^\infty c_n (k_{1T},
  k_{2T}) \, \cos (2 \, n \, \Delta \phi)
\end{align}
where $\Delta \phi = \phi_1 - \phi_2$ is the angle between momenta
${\bm k}_1$ and ${\bm k}_2$ while $k_{1T} = |{\bm k}_1|$ and $k_{2T} =
|{\bm k}_2|$. Here $c_n (k_1, k_2)$ are some coefficient to be
determined by an exact calculation. It is quite interesting that only
even harmonics contribute to the correlated cross section in
\eqref{eq:series}. Let us stress here that we have not made any
assumptions about the centrality of the collision: we do not have an
almond-shaped overlap of the two nuclei. In fact we integrate over all
impact parameters $\bm B$. (Also the impact parameter dependence
factorizes from the rest of the expression.) Hence the correlation in
\eqref{eq:series} is not caused by the geometry of the collision.

To construct the correlation function one has to use
Eqs. \eqref{eq:2glue_prod_LO} and \eqref{eq:rap_corr3} in
\eqref{corr_main}. In the latter, the uncorrelated two-gluon production
would dominate in the denominator of the normalization factor, that
is, in the denominator of the first factor on its right-hand side. The
single-gluon production cross-section \eqref{crosssection12} at the
lowest order is equal to (see \cite{Jalilian-Marian:2005jf} and
references therein)
\begin{align}
\label{eq:1GLO}
\left\langle \frac{d \sigma^{pA_2}}{d^2 k \, dy \, d^2 b}
  \right\rangle = \frac{\as \, C_F}{\pi^2} \, \frac{Q_{s0}^2 ({\bm
  b})}{k_T^4} \, \ln \frac{k_T^2}{\Lambda^2}.
\end{align}
Using this along with the sum of Eqs. \eqref{eq:2glue_prod_LO} and
\eqref{eq:rap_corr3} in \eqref{corr_main} and taking the large-$N_c$
limit in \eqref{eq:1GLO} we get
\begin{align}
\label{eq:corr_LO}
& C ({\bm k}_1, y_1, {\bm k}_2, y_2)\big|_{LO} = \frac{1}{N_c^2} \,
  \frac{\int d^2 B \, d^2 b \, [T_1 ({\bm B} - {\bm b})]^2 \, Q_{s0}^4
  ({\bm b})}{\int d^2 B \, d^2 b_1 \, d^2 b_2 \, T_1 ({\bm B} - {\bm
  b}_1) \, T_1 ({\bm B} - {\bm b}_2) \, Q_{s0}^2 ({\bm b}_1) \,
  Q_{s0}^2 ({\bm b}_2) } \notag \\
& \times \, \frac{{\bm k}_1^2 \,
  {\bm k}_2^2}{\ln \frac{k_1^2}{\Lambda^2} \, \ln
  \frac{k_2^2}{\Lambda^2}} \, \bigg\{ 2 \, \int\limits_\Lambda
  \frac{d^2 l}{({\bm l}^2)^2} \, \left[ \frac{1}{({\bm k}_1 - {\bm
  l})^2 \, ({\bm k}_2 + {\bm l})^2} + \frac{1}{({\bm k}_1 - {\bm
  l})^2 \, ({\bm k}_2 - {\bm l})^2} \right] \notag \\
& + \frac{1}{8} \, \bigg[ \int \frac{d^2 l}{({\bm l}^2)^2 \, (({\bm l} -
  {\bm k}_1 + {\bm k}_2)^2)^2 \, (({\bm k}_1 - {\bm l})^2)^2 \,
  (({\bm k}_2 + {\bm l})^2)^2} \, \notag \\
& \times \, \left[ {\bm l}^2 \, ({\bm k}_2 +
  {\bm l})^2 + ({\bm k}_1 - {\bm l})^2 \, ({\bm l} - {\bm k}_1 +
  {\bm k}_2)^2 - {\bm k}_1^2 \, ({\bm k}_2 - {\bm k}_1 + 2 \, {\bm
  l})^2 \right] \notag \\
& \times \, \left[ {\bm l}^2 \, ({\bm
  k}_1 - {\bm l})^2 + ({\bm k}_2 + {\bm l})^2 \, ({\bm l} - {\bm
  k}_1 + {\bm k}_2)^2 - {\bm k}_2^2 \, ({\bm k}_2 - {\bm k}_1 + 2
  \, {\bm l})^2 \right] + ({\bm k}_2 \rightarrow - {\bm k}_2) \bigg]
  \bigg\}.
\end{align}

To evaluate this further let us assume for a moment that both
colliding nuclei are simply cylinders oriented along the collision
axis, such that the nuclear profile functions of the nuclei are $T_i
({\bm b}) = 2 \, \rho \, R_i \, \theta (R_i - b)$ where $i = 1,2$
labels the nuclei, $\rho$ is the (constant) nucleon number density in
the nucleus, $R_1$ and $R_2$ are the radii of the projectile and
target nuclei, and the cylindrical nucleus is assumed to have length
$2 R_i$ along its axis. Assuming that both nuclei are large and
neglecting the edge effects, we can neglect the ${\bm b}_1$ and ${\bm
  b}_2$ dependence in these nuclear profile functions of both
nuclei. Since the gluon saturation scale in the MV model is $Q_{s0}^2
= 4 \, \pi \, \as^2 \, T_2 ({\bm b})$ we obtain in the $R_1 \ll R_2$
limit
\begin{align}
\label{eq:corr_LO_cyl}
& C ({\bm k}_1, y_1, {\bm k}_2, y_2)\big|_{LO} = \frac{1}{N_c^2 \,
  \pi R_1^2} \, \frac{{\bm k}_1^2 \, {\bm k}_2^2}{\ln
  \frac{k_1^2}{\Lambda^2} \, \ln \frac{k_2^2}{\Lambda^2}} \notag \\
& \times \, \bigg\{
  2 \, \int\limits_\Lambda \frac{d^2 l}{({\bm l}^2)^2} \, \left[
  \frac{1}{({\bm k}_1 - {\bm l})^2 \, ({\bm k}_2 + {\bm l})^2} +
  \frac{1}{({\bm k}_1 - {\bm l})^2 \, ({\bm k}_2 - {\bm l})^2}
  \right] \notag \\
& + \frac{1}{8} \, \bigg[ \int \frac{d^2 l}{({\bm
  l}^2)^2 \, (({\bm l} - {\bm k}_1 + {\bm k}_2)^2)^2 \, (({\bm
  k}_1 - {\bm l})^2)^2 \, (({\bm k}_2 + {\bm l})^2)^2} \, \notag \\
& \times \, \left[
  {\bm l}^2 \, ({\bm k}_2 + {\bm l})^2 + ({\bm k}_1 - {\bm l})^2 \,
  ({\bm l} - {\bm k}_1 + {\bm k}_2)^2 - {\bm k}_1^2 \, ({\bm k}_2 -
  {\bm k}_1 + 2 \, {\bm l})^2 \right] \notag \\
& \times \, \left[
  {\bm l}^2 \, ({\bm k}_1 - {\bm l})^2 + ({\bm k}_2 + {\bm l})^2 \,
  ({\bm l} - {\bm k}_1 + {\bm k}_2)^2 - {\bm k}_2^2 \, ({\bm k}_2 -
  {\bm k}_1 + 2 \, {\bm l})^2 \right] + ({\bm k}_2 \rightarrow -
  {\bm k}_2) \bigg] \bigg\}.
\end{align}
Indeed the correlator is suppressed by a power of $N_c^2$ and a power
of the cross sectional area of the projectile nucleus $\pi R_1^2$, as
discussed in the literature
\cite{Dusling:2009ni,Dumitru:2010iy,Dumitru:2010mv,Kovner:2010xk,Kovner:2012jm}.

While our conclusion about the correlator \eqref{eq:corr_LO_cyl} and
the cross section \eqref{eq:series} contributing only to even Fourier
harmonics has been verified above at the lowest order only, it is true
for the full two-gluon production cross section in heavy-light ion
collisions \eqref{eq_all}. This can be seen by noticing that ${\bm
  k}_2 \rightarrow - {\bm k}_2$ substitution in
\eqref{eq:2glue_prod_delta} does not change the cross section, as it is
equivalent to the ${\bm x}_2 \leftrightarrow {\bm y}_2$ interchange of
the integration variables. The integrand of \eqref{eq:2glue_prod_delta}
(or, equivalently, of \eqref{eq:2glue_prod_main}) is invariant under
${\bm x}_2 \leftrightarrow {\bm y}_2$ interchange since the gluon is
its own anti-particle such that $Tr \left[ U_{\bm x} \, U^\dagger_{\bm
    y} \right] = Tr \left[ U_{\bm y} \, U^\dagger_{\bm x}
\right]$. The expression in \eqref{eq:rap_corr3} is explicitly invariant
under ${\bm k}_2 \rightarrow - {\bm k}_2$ substitution. Hence the net
correlated cross section \eqref{eq_all} is an even function of ${\bm
  k}_2$. Note also that Eqs. \eqref{eq:2glue_prod_main} and
\eqref{crossed_xsect} are ${\bm k}_1 \leftrightarrow {\bm
  k}_2$-symmetric: in \eqref{eq:2glue_prod_main} and in the term arising
from the first exponential in \eqref{crossed_xsect} the symmetry is a
consequence of the symmetry of the integrand under the simultaneous
${\bm x}_1 \leftrightarrow {\bm x}_2$, ${\bm y}_1 \leftrightarrow {\bm
  y}_2$, and ${\bm b}_1 \leftrightarrow {\bm b}_2$ interchanges. The
term multiplying the second exponential in \eqref{crossed_xsect} is ${\bm
  k}_1 \leftrightarrow {\bm k}_2$-symmetric due to the ${\bm x}_1
\leftrightarrow {\bm y}_1$, ${\bm x}_2 \leftrightarrow {\bm y}_2$
symmetry of the integrand, which follows from the following property
of the quadrupole operator: $Tr \left[ U_{{\bm x}_1} \,
  U^\dagger_{{\bm y}_1} \, U_{{\bm x}_2} \, U^\dagger_{{\bm y}_2}
\right] = Tr \left[ U_{{\bm y}_2} \, U^\dagger_{{\bm x}_2} \, U_{{\bm
      y}_1} \, U^\dagger_{{\bm x}_1} \right]$. We see that the net
cross section \eqref{eq_all} is decomposable into a Fourier series
with even harmonics only. Therefore, the correlation function in the
heavy-light ion collision can be also written as an even-harmonics
series
\begin{align}
\label{eq:Cseries}
C ({\bm k}_1, y_1, {\bm k}_2, y_2)\big|_{A_2 \gg A_1 \gg 1} \sim
  \sum_{n=0}^\infty d_n (k_{1T}, k_{2T}) \, \cos (2 \, n \, \Delta
  \phi)
\end{align}
with some coefficients $d_n$. This conclusion also seems to hold in
the case of classical gluon fields produced in a collision of two
heavy ions, as can be seen from the result of the full numerical
simulation of two-gluon production in heavy ion collisions due to
classical gluon field carried out in \cite{Lappi:2009xa}. The
correlators in Figs.~9 and 10 of \cite{Lappi:2009xa} do appear to have
similar-looking maxima at $\Delta \phi =0$ and $\Delta \phi =\pi$
(within the accuracy of the numerical error bars), though a more
careful analysis is needed to figure out if our conclusion is true for
two colliding heavy ions.

\begin{figure}[H]
\centering
  \includegraphics[width= 0.7 \textwidth]{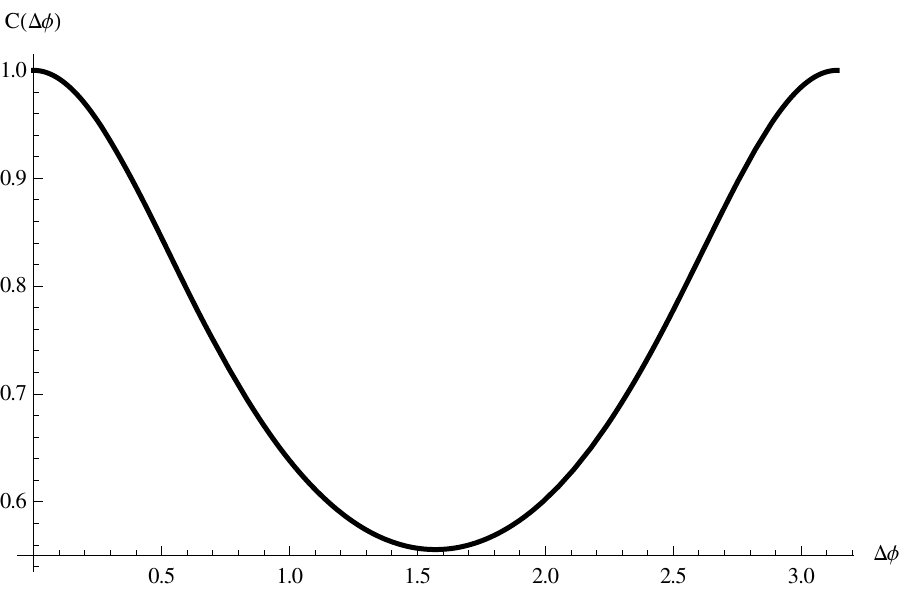}
  \caption{A toy azimuthal two-gluon correlation function motivated by
    the calculation in this section. Vertical scale is arbitrary.}
\label{corr_toy} 
\end{figure}

To better visualize the correlator let us envision a toy model in
which multiple rescatterings (and other saturation effects) regulate
the singularities at ${\bm k}_1 = \pm {\bm k}_2$ by the saturation
scale $Q_{s0}$ in such a way that the correlator can be modeled as
proportional to
\begin{align}
\label{eq:corr_model}
C_{\mbox{toy} \ \mbox{model}} ({\bm k}_1, y_1, {\bm k}_2, y_2) \sim
  \frac{1}{({\bm k}_1 - {\bm k}_2)^2 + Q_{s0}^2} + \frac{1}{({\bm k}_1
  + {\bm k}_2)^2 + Q_{s0}^2}.
\end{align}
This correlator is plotted in \fig{corr_toy} using arbitrary units
along the vertical axis as a function of the azimuthal angle $\Delta
\phi$ between ${\bm k}_1$ and ${\bm k}_2$ for $k_1 = k_2 =
Q_{s0}$. The shape illustrates what the full correlation function may
look like, having identical near- and away-side correlation peaks.

%%%%%%%%%%%%%%%%%%%%%%%%%%%%%%%%%%%%%%%%%%%%%%%%%%%%%%%%%%%%%%%%%%%%%%%%%%%%
%%%%%%%%%%%%%%%%%%%%%%%%%%%%%%%%%%%%%%%%%%%%%%%%%%%%%%%%%%%%%%%%%%%%%%%%%%%%

Our conclusion in this section is that the saturation/CGC dynamics in
nuclear collisions appears to generate the long-range rapidity
correlations which have identical maxima at both $\Delta \phi =0$ and
$\Delta \phi =\pi$ in the azimuthal angle. Such correlations are
non-flow in nature, since they do not arise due to almond-shaped
geometry of the collisions. (In fact the correlations should persist
for the most central heavy ion collisions, though they should be
suppressed by an inverse power of the overlap area as discussed above
and in
\cite{Dusling:2009ni,Dumitru:2010iy,Dumitru:2010mv,Kovner:2010xk,Kovner:2012jm}.)
However, since the elliptic flow observable $v_2$ (even in the
reaction-plane method) is determined from two-particle correlations,
it is possible that the correlations discussed here may contribute to
the elliptic flow (and to higher-order even-harmonics flow observables
$v_{2n}$) measured experimentally. It is, therefore, very important to
experimentally separate our initial-state correlations from the
late-time QGP effects.

Naturally the cumulant analysis \cite{Borghini:2000sa,Borghini:2001vi}
is likely to remove these two-particle non-flow correlations from the
flow observables. However, the effectiveness of the cumulant method
may again depend on the collision geometry. Let us illustrate this by
considering the fourth order cumulant for elliptic flow, defined by
\cite{Borghini:2000sa,Borghini:2001vi}
\begin{align}
\label{eq:4cumulant}
c_2 \{ 4 \} \equiv \left\langle e^{2 \, i \, (\phi_1 + \phi_2 -
  \phi_3 - \phi_4)} \right\rangle - \left\langle e^{2 \, i \,
  (\phi_1 - \phi_3)} \right\rangle \, \left\langle e^{2 \, i \,
  (\phi_2 - \phi_4)} \right\rangle - \left\langle e^{2 \, i \,
  (\phi_1 - \phi_4)} \right\rangle \, \left\langle e^{2 \, i \,
  (\phi_2 - \phi_3)} \right\rangle
\end{align}
where the angle brackets denote event averages along with the
averaging over all angles $\phi_1, \phi_2, \phi_3$, and $\phi_4$ of
the four particles employed in the definition. Using the lowest-order
correlations employed in arriving at \eqref{eq:corr_LO} one can
straightforwardly show that the cumulant due to these correlations
only is proportional to
\begin{align}
\label{eq:LOcumulant}
& c_2 \{ 4 \} \big|_{LO} \propto \left( \int d^2 B \, d^2 b_1 \, d^2
  b_2 \, [T_1 ({\bm B} - {\bm b}_1)]^2 \, [T_1 ({\bm B} - {\bm
  b}_2)]^2 \, Q_{s0}^4 ({\bm b}_1) \, Q_{s0}^4 ({\bm b}_2) \right)
\notag \\
  & \; \times \left( \int d^2 B \, d^2 b_1 \, d^2 b_2 \, d^2 b_3 \, d^2 b_4 \, T_1 ({\bm B}
  - {\bm b}_1) \, T_1 ({\bm B} - {\bm b}_2) \, T_1 ({\bm B} - {\bm
  b}_3) \, T_1 ({\bm B} - {\bm b}_4) \right.
\notag \\
  & \qquad \times Q_{s0}^2 ({\bm b}_1) \,
  Q_{s0}^2 ({\bm b}_2) \, Q_{s0}^2 ({\bm b}_3) \, Q_{s0}^2 ({\bm
  b}_4) \bigg)^{-1}
\notag \\
  & \; - \left[ \frac{\int d^2 B \, d^2 b \, [T_1 ({\bm B} - {\bm b})]^2
  \, Q_{s0}^4 ({\bm b})}{\int d^2 B \, d^2 b_1 \, d^2 b_2 \, T_1
  ({\bm B} - {\bm b}_1) \, T_1 ({\bm B} - {\bm b}_2) \, Q_{s0}^2
  ({\bm b}_1) \, Q_{s0}^2 ({\bm b}_2) } \right]^2
\end{align}
\iffalse
\begin{align}
\label{eq:LOcumulant}
c_2 \{ 4 \} \big|_{LO} \propto & \frac{\int d^2 B \, d^2 b_1 \, d^2
  b_2 \, [T_1 ({\bm B} - {\bm b}_1)]^2 \, [T_1 ({\bm B} - {\bm
  b}_2)]^2 \, Q_{s0}^4 ({\bm b}_1) \, Q_{s0}^4 ({\bm b}_2)}{\int
  d^2 B \, d^2 b_1 \, d^2 b_2 \, d^2 b_3 \, d^2 b_4 \, T_1 ({\bm B}
  - {\bm b}_1) \, T_1 ({\bm B} - {\bm b}_2) \, T_1 ({\bm B} - {\bm
  b}_3) \, T_1 ({\bm B} - {\bm b}_4) \, Q_{s0}^2 ({\bm b}_1) \,
  Q_{s0}^2 ({\bm b}_2) \, Q_{s0}^2 ({\bm b}_3) \, Q_{s0}^2 ({\bm
  b}_4)} \notag \\
& - \left[ \frac{\int d^2 B \, d^2 b \, [T_1 ({\bm B} - {\bm b})]^2
  \, Q_{s0}^4 ({\bm b})}{\int d^2 B \, d^2 b_1 \, d^2 b_2 \, T_1
  ({\bm B} - {\bm b}_1) \, T_1 ({\bm B} - {\bm b}_2) \, Q_{s0}^2
  ({\bm b}_1) \, Q_{s0}^2 ({\bm b}_2) } \right]^2
\end{align}
\fi
if the 4-particle correlator in \eqref{eq:4cumulant} (the correlated part
of the first term on its right-hand-side) is due to the pairwise
correlations of particles $1, 3$ and $2, 4$ or $1, 4$ and
$2,3$. Clearly, in the general case, $c_2 \{ 4 \} \big|_{LO}$ from
\eqref{eq:LOcumulant} is non-zero.\footnote{Note, however, that for the
  heavy-light nuclear collision case considered here, $A_1 \ll A_2$,
  and for cylindrical nuclei, one gets $c_2 \{ 4 \} \big|_{LO} =0$.}
Moreover, the fourth order cumulant would contain the azimuthal angles
dependence resulting from \eqref{eq:corr_LO} but not shown explicitly in
\eqref{eq:LOcumulant}. We see that the geometric correlations from
Sec.~\ref{sub:geocor} may prevent complete removal of these non-flow
correlations in the cumulants. Just like with the correlator of
\eqref{eq:CfixedB}, the above non-flow correlations can be completely
removed from the cumulants for the fixed impact parameter $\bm B$: if
we fix $\bm B$ in \eqref{eq:LOcumulant} (that is, remove all the $d^2 B$
integrations from it), we get $c_2 \{ 4 \} \big|_{LO} =0$, which is
exactly what the cumulant is designed to do
\cite{Borghini:2000sa,Borghini:2001vi} --- completely cancel for
non-flow correlations. Note that since, in the actual experimental
analyses one effectively integrates over $\bm B$ in a given centrality
bin, it is possible that some non-flow correlations \eqref{eq:corr_LO}
would remain in the cumulant \eqref{eq:LOcumulant}. Even fixing $|{\bm
  B}|$ precisely and integrating over the angles of $\bm B$ may
generate a non-zero contribution of these correlations to the
cumulant. The question of the interplay of the true QGP flow and the
non-flow correlations discussed here has to be resolved by a more
detailed numerical study.

%%%%%%%%%%%%%%%%%%%%%%%%%%%%%%%%%%%%%%%%%%%%%%%%%%%

To clarify the physical meaning of the cross section obtained in
\eqref{eq:HBT_LO} we again consider a collisions of cylindrical
nuclei. With this simplification we can consider one of the terms in
the interaction with the target, say the $Q( {\bm x}_1, {\bm y}_1 ,
{\bm x}_2 , {\bm y}_2 )$ in \eqref{crossedinteraction}.
Among many terms which contribute to the quadrupole amplitude in
\eqref{eq:quad_exp}, there is a term 
\begin{align}
\label{eq:quad}
Q & ( {\bm x}_1, {\bm y}_1 , {\bm x}_2 , {\bm y}_2 )
  \bigg|_{\mbox{order}-Q_{s0}^4} \notag \\
& \sim ({\bm x}_1 - {\bm y}_1)^2 \,
  Q_{s0}^2 \, \ln \left( \frac{1}{|{\bm x}_1 - {\bm y}_1| \, \Lambda}
  \right) \ ({\bm x}_2 - {\bm y}_2)^2 \, Q_{s0}^2 \, \ln \left(
  \frac{1}{|{\bm x}_2 - {\bm y}_2| \, \Lambda} \right) + \ldots \, .
\end{align}
Using \eqref{eq:quad} in \eqref{crossed_xsect} and changing the interaction
variables to those defined in \eqref{eq:change} we see that the remaining
${\bm b}_1$ and ${\bm b}_2$ integrals, after shifting those variables
by $\bm B$, become simple Fourier transforms of the projectile nucleus
profile function $T_1 ({\bm b})$,
\begin{align}
\label{eq:fourier}
\int d^2 b_1 \, d^2 b_2 \, T_1 ({\bm b}_1) \, T_1 ({\bm b}_2) \,
  e^{- i \, ({\bm k}_1 - {\bm k}_2) \cdot ({\bm b}_1-{\bm b}_2)} +
  ({\bm k}_2 \rightarrow - {\bm k}_2).
\end{align}
In the limit of a sufficiently large projectile nucleus the Fourier
transforms give a delta-function, yielding
\begin{align}
\label{eq:hbt1}
\frac{d \sigma_{HBT}}{d^2 k_1 dy_1 d^2 k_2 dy_2} \sim \delta^2 ({\bm
  k}_1 - {\bm k}_2) + \delta^2 ({\bm k}_1 + {\bm k}_2),
\end{align}
in agreement with the result in \eqref{eq:HBT_LO}. Our present estimate
shows that a more careful evaluation of \eqref{eq:fourier} would give a
smoother peak in $|{\bm k}_1 - {\bm k}_2|$ at ${\bm k}_1 = {\bm k}_2$
with the width determined by the inverse radius of the projectile
nucleus $R_1$.

%%%%%%%%%%%%%%%%%%%%%%%%%%%%%%%%%%%%%%%%%%%%%%%%%%%%%%%%%%%%%%%%%%%%%%%%%%%%
\begin{figure}[H]
\centering
  \includegraphics[width= 0.45 \textwidth]{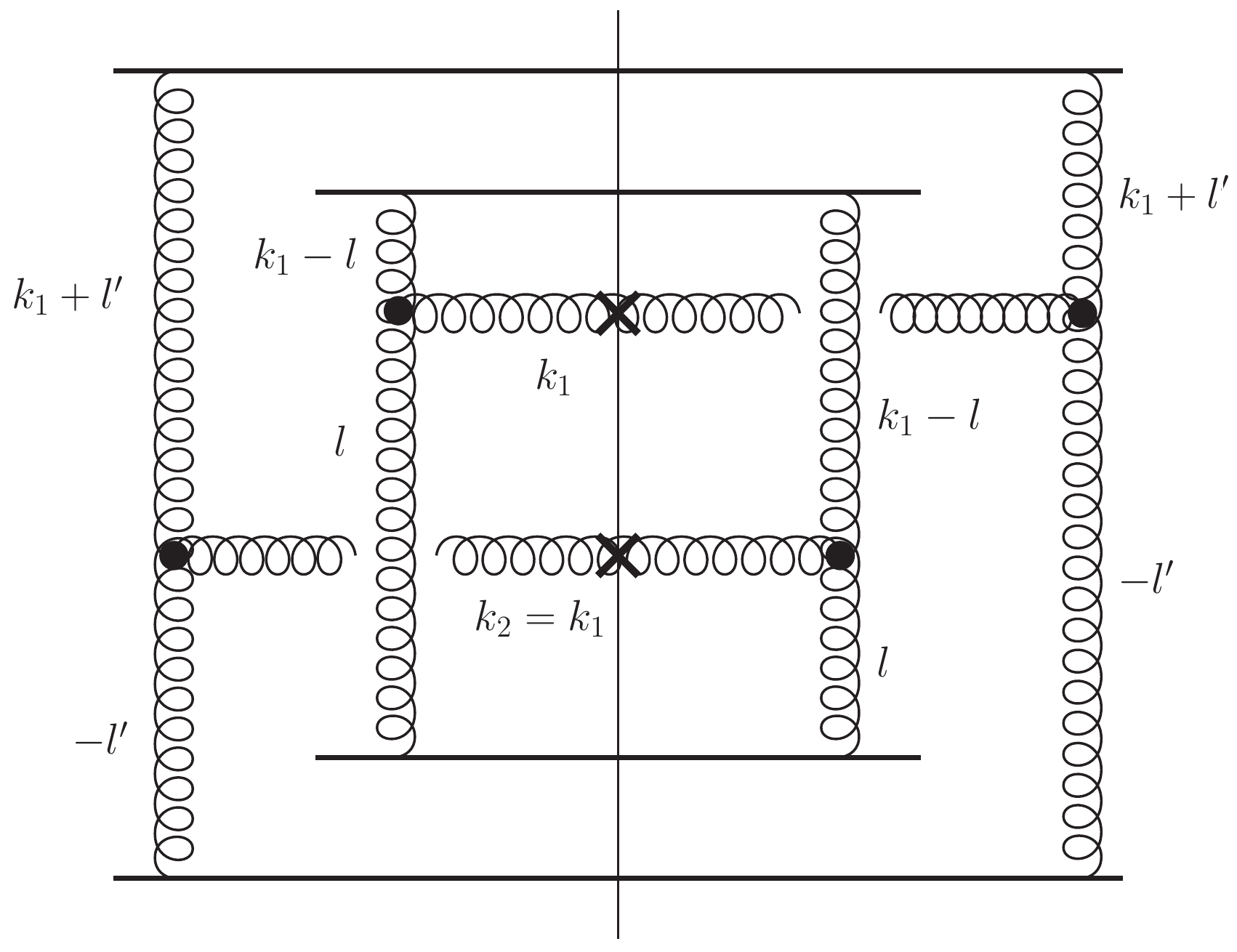}
  \caption{An example of a lowest-order diagram generating HBT-type
    correlations.}
\label{hbt_graph} 
\end{figure}
%%%%%%%%%%%%%%%%%%%%%%%%%%%%%%%%%%%%%%%%%%%%%%%%%%%%%%%%%%%%%%%%%%%%%%%%%%%%

The first term on the right-hand-side of \eqref{eq:hbt1} (or
\eqref{eq:fourier}) has the trademark form of the Hanbury Brown--Twiss
(HBT) correlations \cite{HanburyBrown:1956pf}, which are widely
studied in heavy ion physics
\cite{Heinz:1999rw,Adams:2003ra,Kopylov:1972qw}. These correlations
are normally local both in rapidity ($y_1 = y_2$) and in the
transverse momentum ${\bm k}_1 = {\bm k}_2$: however, in
\eqref{eq:HBT_LO} (or, in \eqref{eq:hbt1}) we only have locality in
transverse momenta. This can be explained by the fact that the extent
of the interaction region in the longitudinal direction, commonly
labeled $R_{long}$, is very small in our case due to the extreme
Lorentz-contraction of the two colliding nuclei leading to smearing of
the longitudinal HBT correlations.

We conclude that the correlations resulting from \eqref{crossed_xsect}
include HBT. Diagrammatic representation of the HBT correlations was
discussed earlier in \cite{Capella:1991mp}. An example of the diagram
giving rise to the correlations in \eqref{eq:HBT_LO} is shown in
\fig{hbt_graph}.

Interestingly though, due to the ${\bm k}_2 \leftrightarrow - {\bm
  k}_2$ symmetry of the two-gluon production cross section
\eqref{crossed_xsect}, the HBT peak at ${\bm k}_1 = {\bm k}_2$ in
\eqref{eq:HBT_LO} is accompanied by an identical peak at ${\bm k}_1 = -
{\bm k}_2$ resulting from $\delta^2 ({\bm k}_1 + {\bm k}_2)$. We thus
obtain a back-to-back HBT correlation resulting from multiple
rescatterings in nuclei. (The origin of our back-to-back HBT
correlations is different from that of the back-to-back HBT-like
correlations proposed in \cite{Asakawa:1998cx,Csorgo:2007iv}.)

Note that it is possible that the process of hadronization would
affect the phases of the produced gluons, possibly destroying the
perturbative HBT correlations of \eqref{crossed_xsect} and replacing them
with the HBT correlations of the non-perturbative origin. The same
hadronization process may also destroy the back-to-back HBT
correlations from \eqref{eq:HBT_LO}.

%% file: Chapter_Properties.tex
% !TEX root = WertepnyPhDThesis.tex
\cleardoublepage
\chapter{Geometric effects and $k_T$-factorization}
\label{ch:Properties} 

After calculating the two-gluon production cross-section and using this to look at the angular dependence of the two-gluon correlation functions in Ch.~\ref{ch:Calculation} we now examine a few specific properties of the two-gluon correlation function.
We make a prediction for the collision of asymmetric ions (specifically Uranium-Uranium collisions) and show that the saturation framework
gives a different result than hydrodynamical simulations.
We also derive the $k_T$-factorized form for the two-gluon production cross section, which allows us to isolate the ``HBT" like contributions.
This chapter borrows heavily from \cite{Kovchegov:2013ewa}.

\input{Section_UU_Collisions}
\input{Section_kT_factorization}

\section{Summary}

In this chapter we looked at two different topics.
The first topic was the effect different collisional geometries had on the two-gluon correlations of interest.
We found that in U+U collisions there is an enhancement of the correlations for tip-on-tip collisions as opposed to side-on-side.
This effect is opposite of what one would see for correlations due to elliptic flow, which gives a possible way to distinguish the two mechanisms for generating two-particle correlations.
Secondly we found the $k_T$-factorized form for the two-gluon production cross section.
This is drastically different from the form associated with single-gluon production and shows that one has to be careful when using $k_T$-factorization to calculate multi-gluon production.
We were also able to isolate the ``HBT" correlations and showed that these correlations exist to all orders in saturation corrections of the target.
This implies that the ``HBT" correlations give information about the geometry of the collision immediately after the multiple rescatterings. 

%% file: Section_UU_Collisions.tex
% !TEX root = WertepnyPhDThesis.tex
\section{Geometry-dependent correlations}
\label{sec:Geo}

In \cite{Kovchegov:2012nd} we pointed out that the geometry of the
collision can have an effect on the correlation function, both through
the so-called geometric correlations introduced in
\cite{Kovchegov:2012nd} (Sec.~\ref{sub:geocor}) (see also
\cite{Frankfurt:2003td,Frankfurt:2010ea} for a discussion of the role
of geometry in di-jet production in $p+p$ collisions) and through a
collision geometry-dependent prefactor of the correlator, like that in
\eqref{eq:corr_LO}. Note also that in the approximation considered, the
two-gluon production cross section contains only the even Fourier
harmonics in the azimuthal opening angle $\Delta \phi$: it would be
important to better understand the effect of geometry on the Fourier
expansion coefficients. We know that even Fourier harmonics in the
di-hadron correlators are also generated by the event-averaged
hydrodynamics, describing the flow of the quark-gluon plasma. (The odd
harmonics are generated by the event-by-event hydrodynamic
simulations, including geometry fluctuations \cite{Alver:2010gr}.) It
would be interesting to understand the differences and similarities of
the two types of correlations.

Let us concentrate specifically on the elliptic flow observable $v_2$,
resulting from the 2nd Fourier harmonic of the correlation
function. The value of $v_2$ in the event-averaged hydrodynamics is
driven by the ellipticity of the overlap region of the colliding
nuclei: the larger the ellipticity, the larger is $v_2$. In contrast
to this behavior, the non-flow correlations in
Eqs.~\eqref{eq:2glue_prod_main} and \eqref{crossed_xsect} do not seem
to require any ellipticity at all to produce a second harmonic (and
other even harmonics) in the correlator, resulting in the
geometry-dependent non-flow contribution to $v_2$ which is not
ellipticity-driven. This can be seen from the lowest-order correlator
in \eqref{eq:corr_LO}: there the geometry-dependent factor factorizes
from the momentum-dependent term which contains the azimuthal angle
dependence of the correlations. The strength of the correlations in
\eqref{eq:corr_LO} is indeed dependent on the geometry-dependent
prefactor: however, it is not {\sl a priori} clear whether this factor
depends on the ellipticity of the overlap region.

\begin{figure}[h]
\centering
  \includegraphics[width=10cm]{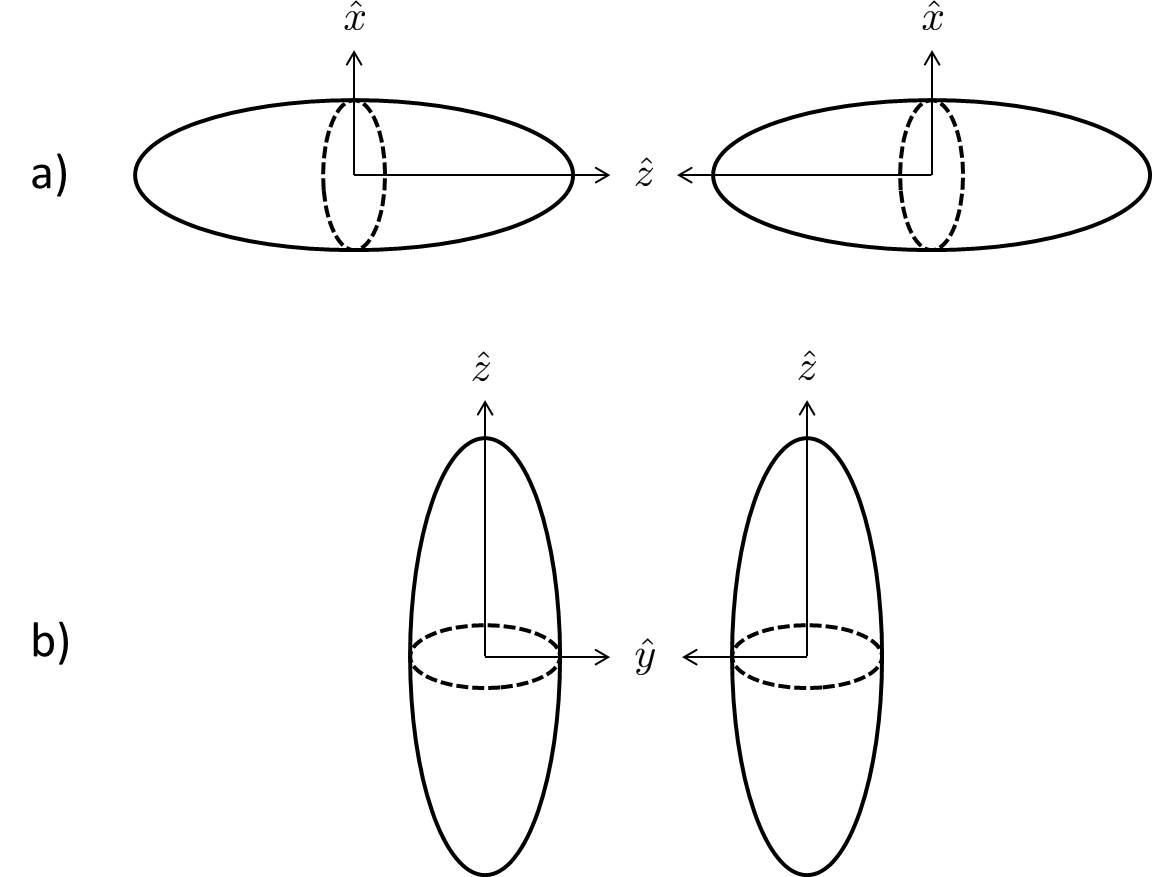}
  \caption{The layout of two possible geometries for the $U+U$
    collisions. The top panel is the tip-on-tip collision, which has
    the z-axis of the two nuclei anti-parallel to each other and
    (anti-)parallel to the collision axis. The bottom diagram is the
    side-on collision, which has the z-axis of the two nuclei parallel
    to each other and perpendicular to the collision axis. }
\label{collision_geometry} 
\end{figure}

To elucidate this issue let us consider uranium-uranium ($U+U$)
collisions. Data from such collisions have been collected at RHIC, in
order to study the properties of hydrodynamic evolution, which
predicts stronger elliptic flow (larger $v_2$) in the side-on-side
collisions (bottom panel in \fig{collision_geometry}) than in the
tip-on-tip collisions (top panel in \fig{collision_geometry}), since
the ellipticity in the former case is much larger than that in the
latter case \cite{Heinz:2004ir,Kuhlman:2005ts,Kuhlman:2006qp}.

To compare this with the behavior of the correlations in the CGC
dynamics we will employ the lowest-order correlator
\eqref{eq:corr_LO}. Note that the higher-order corrections to this
correlator, which are contained in Eqs.~\eqref{eq:2glue_prod_main} and
\eqref{crossed_xsect}, are likely to regulate some of the IR
singularities present in \eqref{eq:corr_LO}, introducing new factors
of the saturation scale $Q_{s2} ({\bm b})$, which may modify the
geometry-dependence of the lowest-order correlator
\eqref{eq:corr_LO}. However, as we will see below, the power-law IR
divergences in \eqref{eq:corr_LO} do not affect the azimuthal
angle-dependent correlations; hence our estimate of the magnitude of
the Fourier harmonics with index $n \ge 2$ should not be affected
qualitatively by higher-order corrections.

To see how the geometry of the collision affects the correlation we
take the ratio of two correlation functions which have different
geometries associated with the $U+U$ collision illustrated in
\fig{collision_geometry}: tip-on-tip (top panel) and side-on-side
(bottom panel). This requires fixing the impact parameter between the
two nuclei, ${\bm B}$, which, in this case, is fixed to ${\bm 0}$ for
both correlations. In the MV model which we have used here $Q_{s2}^2 =
4 \pi \alpha_s^2 T_2 (\bm b)$. In our case the two nuclei involved in
a collision are identical and, hence, have the same nuclear profile
functions, $T_1 (\bm b) = T_2 (\bm b)$. (Note that, while the gluon
production cross section in Eqs.~\eqref{eq:2glue_prod_main} and
\eqref{crossed_xsect} was derived in the $A_2 \gg A_1 \gg 1$ limit
with $k_1, k_2 \gtrsim Q_{s1}$, the lowest-order correlator
\eqref{eq:corr_LO} is valid for $k_1, k_2 \gg Q_{s1}, Q_{s2}$ with the
ordering condition relaxed on $A_1, A_2 \gg 1$.) The difference
between the two geometries in \fig{collision_geometry} is governed by
the nuclear profile function. The ratio between the tip-on-tip and
side-on-side correlation functions \eqref{eq:corr_LO} can be written
as
\begin{align}
\label{georatio}
\frac{C_{tip-on-tip} ({\bm k}_1, y_1, {\bm k}_2,
  y_2)\big|_{LO}}{C_{side-on-side} ({\bm k}_1, y_1, {\bm k}_2,
  y_2)\big|_{LO}} = \frac{\int d^2 b \, [T_{tip-on-tip} ({\bm
    b})]^4}{\left[ \int d^2 b \, [T_{tip-on-tip} ({\bm b})]^2
  \right]^2} \frac{\left[ \int d^2 b \, [T_{side-on-side} ({\bm b})]^2
  \right]^2}{\int d^2 b \, [T_{side-on-side} ({\bm b})]^4}. 
\end{align}
Note that the momentum dependence cancels out in the ratio of two
lowest-order correlators.

%%%%%%%%%%%%%%%%%%%%%%%%%%%%%%%%%%%%%%%%%%%%%%%%%%%%%%%%%%%%%%%%%%%%%%%%%%%%
%%%%%%%%%%%%%%%%%%%%%%%%%%%%%%%%%%%%%%%%%%%%%%%%%%%%%%%%%%%%%%%%%%%%%%%%%%%

For the analytical estimate we are about to perform here we employ a
toy model of a uranium nucleus as a prolate spheroid with the Gaussian
distribution of the nucleon number density
\begin{align}
\label{density}
  \rho (\vec{\bm r}) = \rho_0 \; e^{-\frac{x^2}{R^2}-\frac{y^2}{R^2}-\frac{\lambda^2}{R^2}z^2}
\end{align}
where $\lambda \approx 0.79$ is related to the ellipticity $\epsilon$
of the spheroid by $\lambda = \sqrt{1 - \epsilon^2}$. To translate
this into a nuclear profile function we integrate over one of the
spatial coordinates: $z$ for the tip-on-tip collisions and $y$ for the
side-on-side collisions (see \fig{collision_geometry}). Thus we have
\begin{align}
\label{nuc_thick}
& T_{tip-on-tip}({\bm b} = (x,y)) = \int\limits_{-\infty}^\infty d z
\, \rho (\vec{\bm r}) = \sqrt{\pi} \; \frac{R}{\lambda} \; \rho_0 \;
e^{-\frac{b^2}{R^2}} \notag \\ & T_{side-on-side}({\bm b} = (z,x)) =
\int\limits_{-\infty}^\infty d y \, \rho (\vec{\bm r}) = \sqrt{\pi} \;
R \; \rho_0 \; e^{-\frac{x^2}{R^2}-\frac{\lambda^2}{R^2}z^2}.
\end{align}

Plugging these results into \eqref{georatio} and integrating we arrive at
\begin{align}
  \frac{C_{tip-on-tip} ({\bm k}_1, y_1, {\bm k}_2,
    y_2)\big|_{LO}}{C_{side-on-side} ({\bm k}_1, y_1, {\bm k}_2,
    y_2)\big|_{LO}} = \frac{1}{\lambda} \approx 1.26 \ \ \ (\mbox{for}
  \ U + U).
\end{align}
Thus a tip-on-tip collision enhances the initial-state (CGC)
correlation between two gluons as compared to the side-on-side
collision. We have checked this conclusion numerically by using more
realistic nuclear density profiles in \eqref{georatio}, invariably
getting stronger correlations in the tip-on-tip versus side-on-side
collisions.

We conclude that, at least at the lowest order, the two-gluon
correlations behave in an exactly opposite way from hydrodynamics:
while hydrodynamic contribution to $v_2$ is ellipticity-driven, and is
hence larger in the side-on-side $U+U$ collisions, the CGC
correlations considered here give stronger correlations for the
tip-on-tip $U+U$ collisions. This difference in geometry dependence
should allow these two effects to be experimentally
distinguishable. Further work is needed to understand the geometry
dependence of the full correlator resulting from the two-gluon
production cross section in Eqs.~\eqref{eq:2glue_prod_main} and
\eqref{crossed_xsect}.

%% file: Section_kT_factorization.tex
% !TEX root = WertepnyPhDThesis.tex

\section{$k_T$-factorization}
\label{sec:fact}

As discussed in \ref{sec:pAfact}, there exists a $k_T$-factorized from of the single-gluon production for heavy-light ion collisions to leading order in saturation of the projectile.
This single-gluon production process is slightly different
from the two-gluon production process in heavy-light ion collisions at hand:
in obtaining Eqs.~\eqref{eq:2glue_prod_main} and \eqref{crossed_xsect}
we considered two gluons originating in the projectile wave function,
which could be deemed a ``saturation effect'' compared to the single
gluon needed for quasi-classical gluon production in $pA$
collisions. It appears to be interesting to investigate whether the
two-gluon production cross section \eqref{eq_all} could also be
written in a $k_T$-factorized form.

The distribution functions \eqref{eq:dipole_wave_int} and
\eqref{eq:trace_wave} defined in \ref{sec:pAfact} are needed for the $k_T$
factorization expression \eqref{eq:singlegluon} of the single gluon
production cross-section in $pA$ (or heavy-light ion)
collisions. However, when we are dealing with the two-gluon production
cross-section \eqref{eq_all}, these distribution functions are likely
not to be adequate. First we notice that the only Wilson line operator
in the single-gluon production case is the gluon dipole
\eqref{ngdipole} . In the expression for the two-gluon production
cross section \eqref{eq_all} we have both the quadruple operator
\eqref{quad_def}, and the double trace operator \eqref{Ddef}, which
would lead to different distribution functions.

Secondly, the two gluon production cross-section has geometric
correlations \cite{Kovchegov:2012nd} (Sec.~\ref{sub:geocor}), which arise purely from the
integration over the impact parameters $B$, $b_1$ and $b_2$ in
Eqs.~\eqref{eq:2glue_prod_main} and \eqref{crossed_xsect}. This
prevents the integrals over the impact parameters from being contained
within the distribution functions themselves. This will end up
drastically changing the nature of the distribution functions and thus
the final factorized from.

The last major difference comes from the 'crossed' diagrams. These
diagrams contain the interference of the wave functions of the
incoming nucleons, which generates a significant ``cross-talk''
between different parts of the diagram; it is, therefore, {\it a
  priori} unlikely that factorization would take place. As we will see
below, the factorized form of the expression cannot be written purely
as a convolution of distribution functions without additional factors,
like in \eqref{eq:singlegluon}. While factorized form can be achieved, it
would also contain an extra factor (a ``coefficient function'') in the
final result for the convolution.

With these considerations in mind we first should take a look at the
nature of the distribution functions needed for the $k_T$-factorized
expression for the two-gluon production.

%%%%%%%%%%%%%%%%%%%%%%%%%%%%%%%%%%%%%%%%%%%%%%%%%%%%%%%%%%%%%%%%%%%%%%%%%%%%%

\subsection{One- and two-gluon distribution functions}
\label{sec:fact-distribution}

As we mentioned above, the impact parameter convolutions in
Eqs.~\eqref{eq:2glue_prod_main} and \eqref{crossed_xsect} do not
appear to be factorizable into the integral over the distances between
the gluons and the projectile and a separate integral over the
distances between the gluons and the target, in stark contrast to the
single-gluon production case
\cite{Braun:2000bh,Kovchegov:2001sc}. Therefore, any factorization
expression we could obtain for the two-gluon production has to have an
explicit convolution over the impact parameters. Therefore, we first
need to rewrite the single-gluon distribution functions introduced
above for the fixed impact parameter. We can easily recast
Eqs.~\eqref{eq:dipole_wave_int} and \eqref{eq:trace_wave} as
\begin{align} 
\label{eq:dipole_wave} 
\left\langle \frac{d \phi_{A_1} ({\bm q}, y)}{d^2 b} \right\rangle_{A_1} =
  \frac{C_F}{\alpha_s ( 2 \pi)^3} \int d^2 r \; 
  e^{-i {\bm q} \cdot {\bm r}} \; \nabla_{{\bm r}}^2 \; n_G ({\bm b} + {\bm r}, {\bm b}, y)
\end{align}
and
\begin{align} 
\label{eq:dipole_dist} 
\left\langle \frac{d \phi_{A_2} ({\bm q}, y)}{d^2 b}
  \right\rangle_{A_2} = \frac{C_F}{\alpha_s ( 2 \pi)^3} \int d^2 r \; e^{-i
  {\bm q} \cdot {\bm r}} \; \nabla_{{\bm r}}^2 \; N_G ({\bm b} + {\bm
  r}, {\bm b}, y).
\end{align}
Since now these distribution functions fix both the momentum of the
gluon $\bm q$ and its (approximate) position in the transverse
coordinate space $\bm b$, along with its rapidity $y$ specifying the
value of Bjorken-$x$ variable, we identify the differential
unintegrated gluon distribution functions in
Eqs.~\eqref{eq:dipole_wave} and \eqref{eq:dipole_dist} with the Wigner
distribution \cite{Wigner:1932eb} for gluons (see
\cite{Belitsky:2002sm,Accardi:2012qut} and references therein for
applications of Wigner distributions in perturbative QCD).

\begin{figure}[H]
  \centering
  \includegraphics[width=10cm]{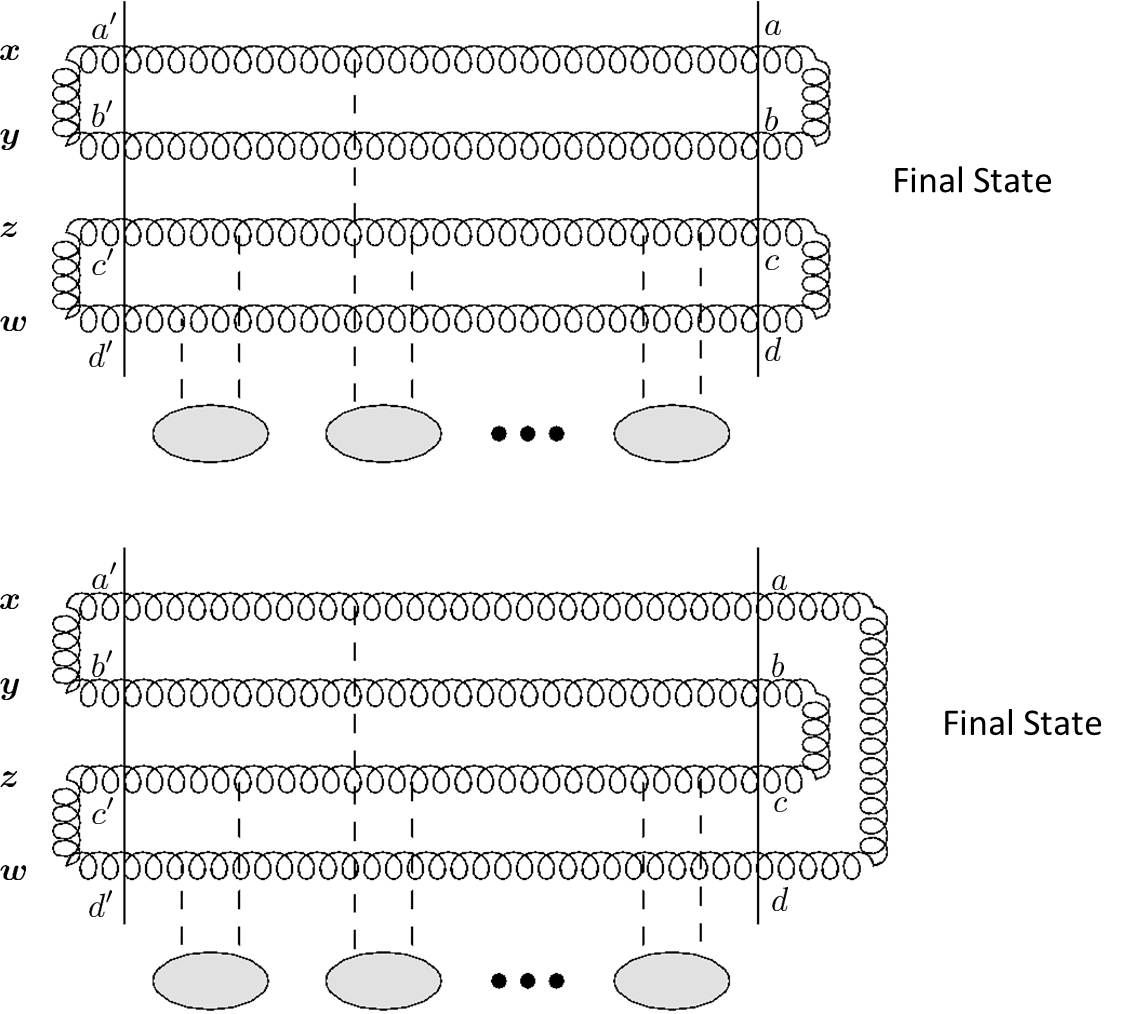}
  \caption{The top panel represents the forward amplitude for the
    scattering of two gluon dipoles on a target nucleus in the
    quasi-classical approximation: this is an essential contribution
    to the definition of the two-gluon distribution in
    \eqref{eq:doubletrace_dist}. The bottom panel represents the
    quadrupole scattering on the target, as in the definition of the
    two-gluon distribution in \eqref{eq:quad_dist}. The vectors ${\bm
      x}$, ${\bm y}$, ${\bm z}$, and ${\bm w}$ label the positions of
    the gluon Wilson lines. Vertical solid lines denote the initial
    (left) and final (right) states. The final state of the gluons is
    labeled to stress that the difference between the two panels is in
    the color configurations of the final state.}
\label{distributions} 
\end{figure}

Here we introduce two different distribution functions which are
associated with the two Wilson line operators entering the two-gluon
production cross-section \eqref{eq_all}, the gluon quadrupole and the
double-trace operators. The two-gluon distribution function associated
with the gluon double-trace operator is
\begin{align} 
\label{eq:doubletrace_dist} 
\left\langle \frac{d \phi_{A_2}^{D} ({\bm q}_1, {\bm q}_2, y)}{d^2 b_1
  \; d^2 b_2} \right\rangle_{A_2} & = \left( \frac{C_F}{\alpha_s ( 2
  \pi)^3} \right)^2 \int d^2 r_1 \; d^2 r_2 \, e^{-i {\bm q}_1 \cdot
  {\bm r}_1 -i {\bm q}_2 \cdot {\bm r}_2}
\\ \notag
& \times \nabla_{{\bm r}_1}^2 \;
\nabla_{{\bm r}_2}^2 \; N_D ({\bm b}_1 + {\bm r}_1, {\bm b}_1, {\bm
  b}_2 + {\bm r}_2, {\bm b}_2, y),
\end{align} 
where
\begin{align} 
\label{eq:doubletrace_amp} 
N_D ({\bm x}, {\bm y}, {\bm z}, {\bm w}, Y) = \; \frac{1}{(N_c^2-1)^2}
  \; \left\langle \mbox{Tr} \left[ \mathbb{1} - U_{{\bm x}} U_{{\bm
  y}}^\dagger \right] \mbox{Tr} \left[ \mathbb{1} - U_{{\bm z}}
  U_{{\bm w}}^\dagger \right] \right\rangle_{A_2} (Y).
\end{align} 
The correlator $N_D$ is illustrated diagrammatically in the top panel
of Fig.~\fig{distributions} for the quasi-classical approximation. The
distribution function \eqref{eq:doubletrace_dist} gives us the number
density for pairs of gluons, with the transverse momenta ${\bm q}_1,
{\bm q}_2$ and positions ${\bm b}_1, {\bm b}_2$ of the gluons fixed
and with the rapidity of both gluons being close to $y$ (up to $\ll
1/\alpha_s$ variations): we can think of this distribution function as a
two-gluon Wigner distribution.

The distribution function associated with the gluon quadrupole
operator is 
\begin{align} 
\label{eq:quad_dist} 
\left\langle \frac{d \phi_{A_2}^{Q} ({\bm q}_1, {\bm q}_2, y)}{d^2 b_1
  \; d^2 b_2} \right\rangle_{A_2} & = \left( \frac{C_F}{\alpha_s ( 2
  \pi)^3} \right)^2 \int d^2 r_1 \; d^2 r_2 \, e^{-i {\bm q}_1 \cdot
  {\bm r}_1 -i {\bm q}_2 \cdot {\bm r}_2} \\ \notag
& \times \nabla_{{\bm r}_1}^2 \;
  \nabla_{{\bm r}_2}^2 \; N_{Q} ({\bm b}_1 + {\bm r}_1, {\bm b}_1, {\bm
  b}_2 + {\bm r}_2, {\bm b}_2, y)
\end{align}
with
\begin{align} 
\label{eq:quad_amp} 
N_{Q} ({\bm x}, {\bm y}, {\bm z}, {\bm w}, Y) = \; \frac{1}{N_c^2-1}
  \; \left\langle \mbox{Tr} \left[ \left( \mathbb{1} - U_{{\bm x}}
  U_{{\bm y}}^\dagger \right) \left( \mathbb{1} - U_{{\bm z}}
  U_{{\bm w}}^\dagger \right) \right] \right\rangle_{A_2} (Y).
\end{align} 
The definition \eqref{eq:quad_dist} is illustrated diagrammatically in
the lower panel of Fig.~\fig{distributions} in the quasi-classical
approximation. The object defined in \eqref{eq:quad_dist} can, similar
to \eqref{eq:doubletrace_dist}, be thought of as a (different)
two-gluon Wigner distribution.

Notice how both the two-dipole \eqref{eq:doubletrace_dist} and
quadrupole \eqref{eq:quad_dist} two-gluon distribution are composed of
Wilson line operators. This is natural for distribution functions
entering production cross section, since in high energy scattering all
cross sections are expressed in terms of Wilson lines. This is in
exact parallel to the single-gluon distribution
\eqref{eq:dipole_dist}, which is related to the adjoint dipole
operator. Note that since the single-gluon production cross section
depends only on the adjoint dipole operator, one can express it only
in terms of the single-gluon distribution \eqref{eq:trace_wave}. For
the two-gluon production \eqref{eq_all}, which contains both the
double-trace and quadrupole operators, we end up with two different
two-gluon distributions \eqref{eq:doubletrace_dist} and
\eqref{eq:quad_dist}.

%%%%%%%%%%%%%%%%%%%%%%%%%%%%%%%%%%%%%%%%%%%%%%%%%%%%%%%%%%%%%%%%%%%%%%%%%%%%
%%%%%%%%%%%%%%%%%%%%%%%%%%%%%%%%%%%%%%%%%%%%%%%%%%%%%%%%%%%%%%%%%%%%%%%%%%%

There is also an alternative single-gluon distribution, the so-called
Weizs\"{a}cker-Williams (WW) distribution
\cite{Jalilian-Marian:1997xn,Kovchegov:1998bi,Kovchegov:2001sc,Kharzeev:2003wz},
which was found to be related to the $q\bar q$ back-to-back jet
production in DIS \cite{Dominguez:2011wm}. In the quasi-classical MV
picture the Weizs\"{a}cker-Williams two-gluon distribution, given by
the correlator of four different gluon fields, would simply factorize
into a product of two single-gluon WW distributions. It is possible,
however, that beyond the quasi-classical limit the two-gluon WW
distribution (properly defined in terms of Wilson line operators along
the lines of the single-gluon WW distribution from
\cite{Dominguez:2011wm}) would constitute an independent new object,
related to some observables. Investigating this possibility further
is beyond the scope of this work.

Since there exists more experience in the field with dipole
distribution functions \eqref{eq:dipole_dist}, it would be nice to be
able to write the two-gluon distributions \eqref{eq:doubletrace_dist}
and \eqref{eq:quad_dist} as combinations of dipole
distributions. Unfortunately this is not possible in general; however
each distribution does contain a piece that can be written in terms of
dipole distributions.

The most obvious is the double-trace two-gluon distribution function
\eqref{eq:doubletrace_dist}. Since in the large-$N_c$ limit
\begin{align} 
\label{eq:singlet_amp_dipole}
  N_{D} ({\bm x}, {\bm y}, {\bm z}, {\bm w}) \bigg|_{\mbox{large}-N_c} = \;
  N_{G} ({\bm x}, {\bm y}) \; N_{G} ({\bm z}, {\bm w})
\end{align}
with $N_G = 1 - S_G$, we can see by plugging this result into
\eqref{eq:doubletrace_dist} and comparing to \eqref{eq:dipole_dist} that
\begin{align} 
\label{eq:singlet_app} 
\left\langle \frac{d \phi_{A_2}^{D} ({\bm q}_1, {\bm q}_2, y)}{d^2 b_1
  \; d^2 b_2} \right\rangle_{\! \! A_2} \Bigg|_{\mbox{large}-N_c} =
  \left\langle \frac{d \phi_{A_2} ({\bm q}_1, y)}{d^2 b_1}
  \right\rangle_{\! \! A_2} \left\langle \frac{d \phi_{A_2} ({\bm q}_2,
  y)}{d^2 b_2} \right\rangle_{\! \! A_2}.
\end{align}
The double-trace two-gluon distribution function factorizes into two
dipole distribution functions only in the large-$N_c$
limit. Unfortunately, the only correlations left in the two-gluon
production cross section \eqref{eq_all} evaluated in the large-$N_c$
limit are the geometric correlations \cite{Kovchegov:2012nd} (Sec.~\ref{sub:geocor}). All of
the other correlations contained in \eqref{eq_all} are subleading in
$N_c$; for instance, the correlations \eqref{eq:corr_LO} are
explicitly $\mathcal{O}\left(1/N_c^2\right)$. 

In order to isolate the dipole contribution to the two-gluon
quadrupole distribution \eqref{eq:quad_dist} we cannot just take the
large-$N_c$ limit like we did for the singlet distribution. (In
addition the whole corresponding contribution to the cross section
\eqref{crossed_xsect} is $\mathcal{O}\left(1/N_c^2\right)$ when compared to
\eqref{eq:2glue_prod_main}.) Instead we can single out the part of the
two-gluon quadrupole distribution which is expressible in terms of
single-gluon dipole distributions: we will show later that this is
exactly the part that gives rise to the early-time
Hanbury-Brown--Twiss (HBT) correlations \cite{HanburyBrown:1956pf}
discussed in \cite{Kovchegov:2012nd} (Sec.~\ref{sec:lrrc}).

First let us analyze the quadrupole operator (cf. \eqref{quad_def})
\begin{align} 
\label{eq:quad_color1} 
Q({\bm x}, {\bm y}, {\bm z}, {\bm w}, Y) & = \frac{1}{N_c^2-1} \;
  \left\langle \mbox{Tr} \left[ U_{{\bm x}} U_{{\bm y}}^\dagger U_{{\bm
  z}} U_{{\bm w}}^\dagger \right] \right\rangle_{A_2} (Y) \notag \\
& = \frac{1}{N_c^2-1} \; \left\langle \delta^{ad} \delta^{bc} \; U_{{\bm
  x}}^{a a'} U_{{\bm y}}^{b b'} U_{{\bm z}}^{c c'} U_{{\bm w}}^{d
  d'} \; \delta^{a'b'} \delta^{c'd'} \right\rangle_{\! A_2} (Y). 
\end{align}
Here we have written out the color structure implied by the trace
notation in terms of the adjoint color indices $a$, $b$, $c$, $d$ (and
the corresponding primed variables) shown in the lower panel of
Fig.~\fig{distributions}. The four gluon lines in the final state in
Fig.~\fig{distributions} carrying indices $a$, $b$, $c$, $d$ are in a net
color-neutral state. This allows us to classify the color states in
the quadrupole operator by the color states of the two gluons with
indices $a$ and $b$. Choosing the color state of gluons $a$ and $b$
sets the color state of gluons $c$ and $d$ due to the color neutrality
of all four gluons in the final state. The same applies to the initial
state gluons with the color indices $a'$, $b'$, $c'$, and $d'$.

A pair of gluons may be found in either of the following irreducible
representations of SU($N_c$)
\begin{align}
\label{8x8}
& {(N_c^2 - 1)} \otimes {(N_c^2 - 1)} = V_1 \oplus V_2 \oplus V_3
  \oplus V_4 \oplus V_5 \oplus V_6 \oplus V_7 \notag \\
& = {\bm 1}
  \oplus {(N_c^2 -1)} \oplus \frac{N_c^2 (N_c -3) (N_c +1)}{4} \oplus
  \frac{N_c^2 (N_c + 3) (N_c - 1)}{4} \notag \\
& \oplus {(N_c^2 -1)} \oplus
  \frac{(N_c^2 -1) (N_c^2 - 4)}{4} \oplus \frac{(N_c^2 -1) (N_c^2 -
  4)}{4}.
\end{align}
In \eqref{8x8} we follow the notation for the irreducible representations
introduced in \cite{Cvitanovic:2008zz}, see page 120 there. We will,
however, use a different normalization scheme from the projection
operators $P_i^{abcd}$'s. We normalize the states such that
$P_i^{abcd} \; P_i^{abcd} = 1$ (summation over repeated indices is
implied), which implies, due to the orthonormality of the projection
operators,
\begin{align} 
\label{eq:unit_op}
\mathbb{1}^{abcd, \; a'b'c'd'} = \sum_{i=1}^7 P_i^{abcd} \;
P_i^{a'b'c'd'}.
\end{align}

The only projection operator we need to know explicitly for the
following calculation is the singlet projector,
\begin{align} 
\label{eq:blank1} 
P_1^{abcd} = \frac{1}{N_c^2-1} \delta^{ab} \delta^{cd}.
\end{align}
Using the singlet projection and the unit operator \eqref{eq:unit_op}
we can rewrite \eqref{eq:quad_color1} as (dropping the $A_2$ subscript
and not showing rapidity dependence for brevity)
\begin{align} 
\label{eq:quad_color2} 
\frac{1}{N_c^2-1} \; \left\langle \mbox{Tr} \left[ U_{{\bm x}} U_{{\bm
  y}}^\dagger U_{{\bm z}} U_{{\bm w}}^\dagger \right]
  \right\rangle = \sum_{i=1}^7 P_i^{a''b''b''a''} \; \left\langle
  P_i^{abcd} \; U_{{\bm x}}^{a a'} U_{{\bm y}}^{b b'} U_{{\bm z}}^{c
  c'} U_{{\bm w}}^{d d'} \; P_1^{a'b'c'd'} \right\rangle .
\end{align}
We can isolate the part that gives the factorized dipole contribution
in the sum of \eqref{eq:quad_color2}. This contribution comes from the
large-$N_c$ part of the double-dipole operator, which, in turn,
originates in the $P_1$-term in the sum in
\eqref{eq:quad_color2}. Isolating the double trace operator from the
rest of the expression in \eqref{eq:quad_color2} we arrive at
\begin{align} 
\label{eq:quad_color3} 
\frac{1}{N_c^2-1} \; \left\langle \mbox{Tr} \left[ U_{{\bm x}} U_{{\bm
        y}}^\dagger U_{{\bm z}} U_{{\bm w}}^\dagger \right]
\right\rangle & = \frac{1}{(N_c^2-1)^2} \; \left\langle \mbox{Tr}
  \left[ U_{{\bm x}} U_{{\bm y}}^\dagger \right] \mbox{Tr} \left[
    U_{{\bm z}} U_{{\bm w}}^\dagger \right] \right\rangle \notag \\ &
+ \, \sum_{i=2}^7 P_i^{a''b''b''a''} \; \left\langle P_i^{abcd} \;
  U_{{\bm x}}^{a a'} U_{{\bm y}}^{b b'} U_{{\bm z}}^{c c'} U_{{\bm
      w}}^{d d'} \; P_1^{a'b'c'd'} \right\rangle .
\end{align}
The double trace operator comes with a prefactor of $\frac{1}{(N_c^2 -
  1)^2}$, which means that when we combine \eqref{eq:quad_color3} with
\eqref{eq:quad_amp} we arrive at
\begin{align} 
\label{eq:quad_amp_dipole}
  N_{Q} ({\bm x}, {\bm y}, {\bm z}, {\bm w}) = \;
  N_{G} ({\bm x}, {\bm y}) \; N_{G} ({\bm z}, {\bm w})
  + \cdots .
\end{align}
The ellipses in \eqref{eq:quad_amp_dipole} represent the remaining
contributions which are not contained in the factorized gluon dipoles,
the $P_2$ through $P_7$ terms and the sub-leading in $N_c$ terms from
the double trace operator in \eqref{eq:quad_color3}. Plugging
\eqref{eq:quad_amp_dipole} into \eqref{eq:quad_dist} we arrive at
\begin{align} 
\label{eq:quad_approx} 
\left\langle \frac{d \phi_{A_2}^{Q} ({\bm q}_1, {\bm q}_2, y)}{d^2 b_1
    \; d^2 b_2} \right\rangle_{\! \! A_2} = \left\langle \frac{d
    \phi_{A_2} ({\bm q}_1, y)}{d^2 b_1} \right\rangle_{\! \! A_2}
\left\langle \frac{d \phi_{A_2} ({\bm q}_2, y)}{d^2 b_2}
\right\rangle_{\! \! A_2} + \cdots,
\end{align}
where we have isolated the factorized dipole distributions from the
rest of the expression. Let us stress that the terms represented by
ellipsis in \eqref{eq:quad_approx} are not suppressed by any parameter
involved in the problem: these corrections are comparable to the term
shown explicitly on the right of \eqref{eq:quad_approx}. Hence even if
we took the leading-$N_c$ limit of the two-gluon quadrupole
distribution there would still be terms that would not be contained
inside the two factorized gluon distributions of
\eqref{eq:quad_approx}.

%%%%%%%%%%%%%%%%%%%%%%%%%%%%%%%%%%%%%%%%%%%%%%%%%%%%%%%%%%%%%%%%%%%%%%%%%%%%%

\subsection{Derivation of the factorized forms}
\label{sec:fact-main}

Now that we have defined the necessary distribution functions we can
start constructing the factorized form of the two-gluon production
cross-section. Each of the parts of the cross section \eqref{eq_all}
given by Eqs.~\eqref{eq:2glue_prod_main} and \eqref{crossed_xsect}
factorizes differently.

The easiest case to factorize, and thus the first one we will cover,
is the 'square' diagram component
\eqref{eq:2glue_prod_main}. Separating the transverse vectors
associated with either one of the valence quarks and emitted gluons,
we can write \eqref{eq:2glue_prod_main} in the following form,
\begin{align} 
\label{eq:square_cross_form2} 
& \frac{d \sigma_{square}}{d^2 k_1 dy_1 d^2 k_2 dy_2} = \frac{\alpha_s^2 \,
  C_F^2}{16 \, \pi^8} \int d^2 B \, \notag \\
& \times \left\langle \int \, d^2 x_1 \,
  d^2 y_1 \, d^2 b_1 \, T_1 ({\bm B} - {\bm b}_1) e^{- i \; {\bm k}_1
  \cdot ({\bm x}_1-{\bm y}_1)} \frac{ {\bm x}_1 - {\bm b}_1}{ |{\bm
  x}_1 - {\bm b}_1 |^2 } \cdot \frac{ {\bm y}_1 - {\bm b}_1}{
  |{\bm y}_1 - {\bm b}_1 |^2 } \right.  \notag \\
& \times \, \left(
  \frac{1}{N_c^2-1} \; \mbox{Tr}[ U_{{\bm x}_1} U_{{\bm y}_1}^\dagger
  ] \; - \; \frac{1}{N_c^2-1} \; \mbox{Tr}[ U_{{\bm x}_1} U_{{\bm
  b}_1}^\dagger ] \; - \; \frac{1}{N_c^2-1} \; \mbox{Tr}[ U_{{\bm
  b}_1} U_{{\bm y}_1}^\dagger ] \; + \; 1 \right) \notag \\
& \times \, \int \, d^2 x_2 \, d^2 y_2 \, d^2 b_2 \, T_1 ({\bm B} - {\bm
  b}_2) e^{- i \; {\bm k}_2 \cdot ({\bm x}_2-{\bm y}_2)} \frac{ {\bm
  x}_2 - {\bm b}_2}{ |{\bm x}_2 - {\bm b}_2 |^2 } \cdot \frac{ {\bm
  y}_2 - {\bm b}_2}{ |{\bm y}_2 - {\bm b}_2 |^2 } \notag \\ & \times
  \, \left. \left( \frac{1}{N_c^2-1} \; \mbox{Tr}[ U_{{\bm x}_2} U_{{\bm
  y}_2}^\dagger ] \; - \; \frac{1}{N_c^2-1} \; \mbox{Tr}[
  U_{{\bm x}_2} U_{{\bm b}_2}^\dagger ] \; - \; \frac{1}{N_c^2-1} \;
  \mbox{Tr}[ U_{{\bm b}_2} U_{{\bm y}_2}^\dagger ] \; + \; 1 \right)
\right\rangle_{A_2}. 
\end{align}
Notice that the first two lines in \eqref{eq:square_cross_form2} are the
only two lines that contain the variables ${\bm x}_1, \; {\bm y}_1, \;
{\bm b}_1$, while the next two lines are the only ones that contain
the variables ${\bm x}_2, \; {\bm y}_2, \; {\bm b}_2$. In the limit we
are dealing with ${\bm x}_1, \; {\bm y}_1, \; {\bm b}_1$ are
perturbatively close to each other. Since $T_1({\bm b})$ is slowly
varying it is approximately constant over perturbatively short
scales. Thus we can make the approximation
\begin{align}
\label{Tapprox}
T_1({\bm B} - {\bm b}_1) \; \approx \; T_1({\bm B} - {\bm x}_1)
  \; \approx \; T_1({\bm B} - {\bm y}_1).
\end{align}
This same approximation also applies to ${\bm x}_2, \; {\bm y}_2, \;
{\bm b}_2$. Notice that the second line of
\eqref{eq:square_cross_form2} has four different terms in the
parentheses, each of which is at most a function of two of the three
variables ${\bm x}_1, \; {\bm y}_1, \; {\bm b}_1$. Combining this fact
with the approximation \eqref{Tapprox} we can perform one of the ${\bm
  x}_1, \; {\bm y}_1, \; {\bm b}_1$ integrals over a different
variable for each term in the second line depending on which variable
is not in the trace. A similar thing is done with the ${\bm x}_2, \;
{\bm y}_2, \; {\bm b}_2$ integral. After doing this and integrating by
parts we arrive at
\begin{align} 
\label{eq:square_cross_form3} 
\frac{d \sigma_{square}}{d^2 k_1 dy_1 d^2 k_2 dy_2} & = \frac{ \alpha_s^2
  \, C_F^2 }{4 \, \pi^6} \frac{1}{{\bm k}_1^2 \; {\bm k}_2^2} \int d^2
  B \, d^2 b_1 \, d^2 b_2 \, d^2 x_1 \, d^2 x_2 \, T_1 ({\bm B} - {\bm
  b}_1) \; T_1 ({\bm B} - {\bm b}_2) \notag \\
& \times \, e^{- i \; {\bm k}_1
  \cdot ({\bm x}_1-{\bm b}_1)- i \; {\bm k}_2 \cdot ({\bm x}_2-{\bm
  b}_2)} \ln \left( \frac{1}{|{\bm x}_1 -
  {\bm b}_1| \Lambda} \right) \ln \left( \frac{1}{|{\bm x}_2 - {\bm
  b}_2| \Lambda} \right) \notag \\
& \times  \; \nabla_{{\bm x}_1}^2 \; \nabla_{{\bm x}_2}^2 \;
  \frac{1}{(N_c^2-1)^2} \left\langle \mbox{Tr} \left[ \mathbb{1} -
  U_{{\bm x}_1} U_{{\bm b}_1}^\dagger \right] \mbox{Tr} \left[
  \mathbb{1} - U_{{\bm x}_2} U_{{\bm b}_2}^\dagger \right]
\right\rangle_{\! A_2}.
\end{align}

From here we can manipulate this expression into a form reminiscent of
\eqref{eq:singlegluon} but not quite the same. As mentioned in the
discussion at the beginning of Sec.~\ref{sec:fact}, the integrals over
the impact parameters cannot be absorbed into the distribution
functions. This is now manifest in \eqref{eq:square_cross_form3}: we have
three integrals (over ${\bm B}$, ${\bm b}_1$ and ${\bm b}_2$) and four
impact parameter-related distances (${\bm B} - {\bm b}_1$, ${\bm B} -
{\bm b}_2$, ${\bm b}_1$ and ${\bm b}_2$). We conclude that we must use
the new distribution functions defined in Eqs.~\eqref{eq:dipole_wave}
and \eqref{eq:doubletrace_dist} while convoluting them over the impact
parameters ${\bm B}$, ${\bm b}_1$ and ${\bm b}_2$. Employing
Eqs.~\eqref{eq:dipole_wave} and \eqref{eq:doubletrace_dist} we can
rewrite the 'square' diagrams contribution to the two-gluon production
cross section \eqref{eq:2glue_prod_main} in the factorized form
\begin{align} 
\label{eq:factorized_square} 
\frac{d \sigma_{square}}{d^2 k_1 dy_1 d^2 k_2 dy_2} & = \left( \frac{2
  \; \alpha_s}{C_F} \right)^2 \frac{1}{k_1^2 \; k_2^2} \int d^2 B \, d^2
  b_1 \, d^2 b_2 \int d^2 q_1 \, d^2 q_2 \, \notag \\
& \times \;
  \left\langle \frac{d \phi_{A_1} ({\bm q}_1, y=0)}{d^2 ({\bm B}-{\bm
  b}_1)} \right\rangle_{\! \! A_1} \left\langle \frac{d \phi_{A_1}
  ({\bm q}_2, y=0)}{d^2 ({\bm B}-{\bm b}_2)} \right\rangle_{\! \!
  A_1} \notag \\
& \times \; \left\langle \frac{d \phi_{A_2}^{D} ({\bm q}_1 - {\bm k}_1,
  {\bm q}_2 - {\bm k}_2, y \approx y_1 \approx y_2)}{d^2 b_1 \; d^2
  b_2} \right\rangle_{\! \! A_2}.
\end{align}
The asymmetry in rapidity arguments of the distribution entering
\eqref{eq:factorized_square} is due to the fact that the projectile in
the original \eqref{eq:2glue_prod_main} was treated in the lowest-order
quasi-classical approximation, while the whole non-linear evolution
\cite{Balitsky:1996ub,Balitsky:1998ya,Kovchegov:1999yj,Kovchegov:1999ua,Jalilian-Marian:1997dw,Jalilian-Marian:1997gr,Iancu:2001ad,Iancu:2000hn}
is included in the rapidity interval between the produced gluons and
the target by the use of the Wilson lines.  As mentioned previously,
\eqref{eq:factorized_square} is similar to \eqref{eq:singlegluon} but has a
few key differences. \eqref{eq:singlegluon} employs unintegrated gluon
distributions (gluon transverse momentum distributions (TMDs)), while
\eqref{eq:factorized_square} uses one- and two-gluon Wigner
distributions. Related to that, in \eqref{eq:singlegluon} the convolution
happens only over transverse momentum, while \eqref{eq:factorized_square}
also contains integrals over impact parameters ${\bm B}, \; {\bm
  b}_1$, and ${\bm b}_2$.

One may also note that \eqref{eq:factorized_square} is not
target-projectile symmetric: the target is described by a single
two-gluon distribution, while the projectile is represented by two
single-gluon distributions. In contrast, \eqref{eq:singlegluon} is
completely target-projectile symmetric. In fact, \eqref{eq:singlegluon}
is often generalized to the case of nucleus--nucleus ($AA$) collisions
by using \eqref{eq:trace_wave} for both unintegrated gluon distributions
in it. While such generalization allows for successful phenomenology
(see e.g. \cite{ALbacete:2010ad}), it is theoretically not justified
below the saturation scales of both nuclei. Moreover, there is
numerical evidence \cite{Blaizot:2010kh} demonstrating that the
$k_T$-factorization formula \eqref{eq:singlegluon} is not valid in
$AA$ collisions. Therefore, it appears that the apparent
target-projectile symmetry of \eqref{eq:singlegluon} is, in fact,
somewhat misleading: the equation was derived in the limit where the
projectile is dilute, while the target may or may not be dense,
leading to the difference in the definitions of the unintegrated gluon
distributions of the target and the projectile in
Eqs.~\eqref{eq:dipole_wave_int} and \eqref{eq:trace_wave}. It is
likely that \eqref{eq:singlegluon} is not valid for dense-dense
scattering \cite{Blaizot:2010kh}, and is thus not truly
target-projectile symmetric due to the underlying assumptions.

With the 'square' diagrams contribution to the cross section cast in a
factorized form we now turn our attention to the 'crossed' diagrams
contribution \eqref{crossed_xsect}. It is helpful to write out the
crossed diagrams part of the cross section, \eqref{crossed_xsect}, in the
following form,
\begin{align} 
\label{crossed_xsect_fac} 
& \frac{d \sigma_{crossed}}{d^2 k_1 dy_1 d^2 k_2 dy_2} = \frac{\alpha_s^2
  \, C_F^2}{16 \, \pi^8} \int d^2 B \, \left\langle \int \; d^2 x_1 \;
  d^2 y_1 \; d^2 b_1 \; T_1 ({\bm B} - {\bm b}_1) \; e^{- i \; {\bm
      k}_1 \cdot {\bm x}_1 + i \; {\bm k}_2 \cdot {\bm y}_1} \right.  \notag \\
& \times \left[
    \frac{ {\bm x}_1 - {\bm b}_1}{ |{\bm x}_1 - {\bm b}_1 |^2 }
  \right]_i \left[ \frac{ {\bm y}_1 - {\bm b}_1}{ |{\bm y}_1 - {\bm b}_1 |^2 } \right]_j
  \frac{1}{N_c^2-1} \left[ U_{{\bm x}_1} U_{{\bm y}_1}^\dagger \; - \;
  U_{{\bm x}_1} U_{{\bm b}_1}^\dagger \; - \; U_{{\bm b}_1} U_{{\bm
  y}_1}^\dagger \; + \; \mathbb{1} \right]^{ab} \notag \\
& \times
  \; \int \; d^2 x_2 \; d^2 y_2 \; d^2 b_2 \; T_1 ({\bm B} - {\bm b}_2) 
  \; e^{- i \; {\bm k}_2 \cdot {\bm x}_2 + i \; {\bm k}_1 \cdot {\bm
  y}_2} \left[ \frac{ {\bm x}_2 - {\bm b}_2}{ |{\bm x}_2 - {\bm b}_2
  |^2 } \right]_j \left[ \frac{ {\bm y}_2 - {\bm b}_2}{ |{\bm y}_2 -
  {\bm b}_2 |^2 } \right]_i \notag \\
& \times \; \frac{1}{N_c^2-1}
  \left. \left[ U_{{\bm x}_2} U_{{\bm y}_2}^\dagger \; - \; U_{{\bm
  x}_2} U_{{\bm b}_2}^\dagger \; - \; U_{{\bm b}_2} U_{{\bm
  y}_2}^\dagger \; + \; \mathbb{1} \right]^{ba}
  \right\rangle_{A_2} \; + \; ({\bm k}_2 \rightarrow - {\bm k}_2),
\end{align}
where $i,j = 1,2$ are transverse vector indices and $a,b = 1, \ldots ,
N_c^2 -1$ are adjoint color indices, with summation assumed over
repeated indices. Here we have again separated the terms that depend
on ${\bm x}_1, \; {\bm y}_1, \; {\bm b}_1$ from the terms that depend
on ${\bm x}_2, \; {\bm y}_2, \; {\bm b}_2$. Using the same trick we
employed when factorizing the 'square' diagrams contribution, we
evaluate the ${\bm x}_1, \; {\bm y}_1, \; {\bm b}_1$, and ${\bm x}_2,
\; {\bm y}_2, \; {\bm b}_2$ integrals piece by piece arriving at
(after transverse vector relabeling)
\begin{align} 
\label{crossed_xsect_fac2}
& \frac{d \sigma_{crossed}}{d^2 k_1 dy_1 d^2 k_2 dy_2} = \frac{
  \alpha_s^2 \; C_F^2 }{4 \; \pi^6} \int d^2 B \; d^2 b_1 \; d^2 b_2 \;
  d^2 x_1 \; d^2 x_2 \; T_1 ({\bm B} - {\bm b}_1) \; T_1 ({\bm B} -
  {\bm b}_2) \notag \\ & \times \; \left\{ \frac{1}{2} \delta_{ij} \ln
  \left( \frac{1}{|{\bm x}_1 - {\bm b}_1| \Lambda} \right) \; - \;
  \frac{ \left[ {\bm x}_1 - {\bm b}_1 \right]_i \; \left[ {\bm x}_1
  - {\bm b}_1 \right]_j}{2 \, |{\bm x}_1 - {\bm b}_1|^2} \; - i
  \left[ \frac{{\bm k}_1}{k_1^2} \right]_i \left[ \frac{ {\bm x}_1 -
  {\bm b}_1 }{ |{\bm x}_1 - {\bm b}_1|^2} \right]_j \right. \notag \\
& \left. - i
  \left[ \frac{ {\bm x}_1 - {\bm b}_1 }{ |{\bm x}_1 - {\bm b}_1|^2}
   \right]_i \left[ \frac{{\bm k}_2}{k_2^2} \right]_j \right\} \; \left\{ \frac{1}{2} \delta_{ij} \ln \left(
  \frac{1}{|{\bm x}_2 - {\bm b}_2| \Lambda} \right) \; - \; \frac{
  \left[ {\bm x}_2 - {\bm b}_2 \right]_i \; \left[ {\bm x}_2 -
  {\bm b}_2 \right]_j}{2 \, |{\bm x}_2 - {\bm b}_2|^2} \right. \notag \\
& \left. - i
  \left[ \frac{{\bm k}_1}{k_1^2} \right]_i \left[ \frac{ {\bm x}_2 -
  {\bm b}_2 }{ |{\bm x}_2 - {\bm b}_2|^2} \right]_j \; - i
  \left[ \frac{ {\bm x}_2 - {\bm b}_2 }{ |{\bm x}_2 - {\bm b}_2|^2}
  \right]_i \left[ \frac{{\bm k}_2}{k_2^2} \right]_j \right\} \; e^{- i \; {\bm k}_1 \cdot ({\bm x}_1-{\bm b}_2)- i \;
  {\bm k}_2 \cdot ({\bm x}_2-{\bm b}_1)} \notag \\
& \times \; \frac{1}{(N_c^2-1)^2}
  \left\langle \mbox{Tr} \left[ \left( \mathbb{1} - U_{{\bm x}_1}
  U_{{\bm b}_1}^\dagger \right) \left( \mathbb{1} - U_{{\bm
  x}_2} U_{{\bm b}_2}^\dagger \right) \right]
  \right\rangle_{A_2} \; + \; ({\bm k}_2 \rightarrow - {\bm k}_2),
\end{align}
where we have employed
\begin{align}
  \label{formula1}
  \int d^2 b \, \left[ \frac{ {\bm x} - {\bm b}}{ |{\bm x} - {\bm
        b}|^2 } \right]_i \left[ \frac{ {\bm y} - {\bm b}}{ |{\bm y} -
      {\bm b}|^2 } \right]_j = \pi \left\{ \delta_{ij} \, \ln \left(
      \frac{1}{|{\bm x} - {\bm y}| \, \Lambda} \right) - \frac{\left[
        {\bm x} - {\bm y} \right]_i \; \left[ {\bm x} - {\bm y}
      \right]_j}{|{\bm x} - {\bm y}|^2} \right\}
\end{align}
along with other, more common, two-dimensional integrals (see, e.g.,
Appendix A.2 of \cite{KovchegovLevin} for a list of useful integrals).

To proceed we rewrite \eqref{crossed_xsect_fac2} as
\begin{align} 
\label{crossed_xsect_fac3}
& \frac{d \sigma_{crossed}}{d^2 k_1 dy_1 d^2 k_2 dy_2} = \frac{
  \alpha_s^2 \; C_F^2 }{4^3 \; \pi^6} \frac{1}{k_1^4 \, k_2^4} \int d^2 B
  \; d^2 b_1 \; d^2 b_2 \; d^2 x_1 \; d^2 x_2 \, e^{- i \; {\bm k}_1
  \cdot ({\bm x}_1-{\bm b}_2)- i \; {\bm k}_2 \cdot ({\bm x}_2-{\bm
   b}_1)} \notag \\
& \times \; \left\{ \left[
  \overleftarrow{\nabla}_{x_1}^2 \, \overleftarrow{\nabla}_{b_1}^2
  \, \nabla^i_{x_1} \, \nabla^j_{x_1} +
  \overleftarrow{\nabla}_{b_1}^2 \, \overleftarrow{\nabla}_{x_1}^i
  \, \nabla^j_{x_1} \, \nabla^2_{x_1} -
  \overleftarrow{\nabla}_{x_1}^2 \, \overleftarrow{\nabla}^j_{b_1}
  \, \nabla^i_{x_1} \, \nabla^2_{x_1} \right] \right. \notag \\
& \left. \qquad \times \; ({\bm x}_1 - {\bm
  b}_1)^2 \, \ln \left( \frac{1}{|{\bm x}_1 - {\bm b}_1| \Lambda}
  \right) \, T_1 ({\bm B} - {\bm b}_1) \right\} \notag \\
& \times
  \; \left\{ \left[ \overleftarrow{\nabla}_{x_2}^2 \,
  \overleftarrow{\nabla}_{b_2}^2 \, \nabla^i_{x_2} \,
  \nabla^j_{x_2} - \overleftarrow{\nabla}_{x_2}^2 \,
  \overleftarrow{\nabla}_{b_2}^i \, \nabla^j_{x_2} \,
  \nabla^2_{x_2} + \overleftarrow{\nabla}_{b_2}^2 \,
  \overleftarrow{\nabla}^j_{x_2} \, \nabla^i_{x_2} \,
  \nabla^2_{x_2} \right] \right. \notag \\
& \left. \qquad \times \; ({\bm x}_2 - {\bm b}_2)^2 \, \ln \left(
  \frac{1}{|{\bm x}_2 - {\bm b}_2| \Lambda} \right) \; T_1 ({\bm
  B} - {\bm b}_2) \right\} \notag \\
& \times \;
  \frac{1}{(N_c^2-1)^2} \left\langle \mbox{Tr} \left[ \left(
  \mathbb{1} - U_{{\bm x}_1} U_{{\bm b}_1}^\dagger \right)
  \left( \mathbb{1} - U_{{\bm x}_2} U_{{\bm b}_2}^\dagger \right)
  \right] \right\rangle_{\! \! A_2} + ({\bm k}_2 \rightarrow - {\bm
  k}_2),
\end{align}
where $\nabla$'s denote transverse coordinate derivatives and the left
arrow over $\nabla$ indicates that the derivative is acting on the
exponential to the left of the curly brackets.

Notice the non-trivial transverse index structure in
\eqref{crossed_xsect_fac3}: this drastically alters the factorized form
of the expression, as compared to, say, \eqref{eq:factorized_square}.
Inverting Fourier transforms in Eqs.~\eqref{eq:dipole_wave} and
\eqref{eq:quad_dist}, employing \eqref{eq:dipole_amp}, and substituting
the results into \eqref{crossed_xsect_fac3} yields, after a fair bit of
algebra,
\begin{align} 
\label{eq:factorized_crossed} 
& \frac{d \sigma_{crossed}}{d^2 k_1 dy_1 d^2 k_2 dy_2} = \left(
  \frac{2 \; \alpha_s}{C_F} \right)^2 \frac{1}{k_1^2 \; k_2^2} \int d^2 B
  \; d^2 b_1 \; d^2 b_2 \int d^2 q_1 \; d^2 q_2 \;
\notag \\  
& \times \; \frac{\mathcal{K} (
  {\bm b}_1, {\bm b}_2, {\bm k}_1, {\bm k}_2, {\bm q}_1, {\bm
  q}_2)}{N_c^2-1} \;
  \left\langle \frac{d
  \phi_{A_1} ({\bm q}_1, y=0)}{d^2 ({\bm B}-{\bm b}_1)}
  \right\rangle_{\! \! A_1} \left\langle \frac{d \phi_{A_1} ({\bm q}_2,
  y=0)}{d^2 ({\bm B}-{\bm b}_2)} \right\rangle_{\! \! A_1}
\notag \\  
& \times \;
  \left\langle \frac{d \phi_{A_2}^{Q} ({\bm k}_1 - {\bm q}_1, {\bm k}_2
  - {\bm q}_2, y \approx y_1 \approx y_2)}{d^2 b_1 \; d^2 b_2}
  \right\rangle_{\! \! A_2}
  + \; ({\bm k}_2 \rightarrow - {\bm k}_2),
\end{align}
where the ``coefficient function'' is defined as
\begin{align} 
\label{eq:kernel_crossed}
\mathcal{K} ( {\bm b}_1, {\bm b}_2, {\bm k}_1, {\bm k}_2, {\bm q}_1,
  {\bm q}_2) & = \frac{1}{q_1^2 \; q_2^2 \; ({\bm k}_1-{\bm q}_1)^2
  ({\bm k}_2-{\bm q}_2)^2} \; e^{-i \, ( {\bm k}_1 - {\bm k}_2 )
  \cdot ( {\bm b}_1 - {\bm b}_2 )} \; \left\{ k_1^2 \; k_2^2 ({\bm
  q}_1 \cdot {\bm q}_2)^2 \right.  \notag \\
& - \; k_1^2 \; ({\bm
  q}_1 \cdot {\bm q}_2) \left[ ({\bm k}_2 \cdot {\bm q}_1) \; q_2^2
  \; + \; ({\bm k}_2 \cdot {\bm q}_2) \; q_1^2 \; - \; q_1^2 \;
  q_2^2 \right] \notag \\ & - \; k_2^2 \; ({\bm q}_1 \cdot {\bm
  q}_2) \left[ ({\bm k}_1 \cdot {\bm q}_1) \; q_2^2 \; + \; ({\bm
  k}_1 \cdot {\bm q}_2) \; q_1^2 \; - \; q_1^2 \; q_2^2 \right]
  \notag \\
& \left.  + \; q_1^2 \; q_2^2 \; \left[ ({\bm k}_1 \cdot
  {\bm q}_1) ({\bm k}_2 \cdot {\bm q}_2) \; + \; ({\bm k}_1 \cdot
  {\bm q}_2) ({\bm k}_2 \cdot {\bm q}_1) \right] \right\}
\end{align}
with $q_i = |{\bm q}_i|$, $k_i = |{\bm k}_i|$.

Inserting \eqref{eq:factorized_square} and \eqref{eq:factorized_crossed}
into \eqref{eq_all} we arrive at the $k_T$-factorized form for the two
gluon production cross section in heavy-light ion collisions
\begin{align} 
\label{eq:factorized_final} 
& \frac{d \sigma}{d^2 k_1 dy_1 d^2 k_2 dy_2} = \left( \frac{2 \;
  \alpha_s}{C_F} \right)^2 \frac{1}{k_1^2 \; k_2^2} \int d^2 B \; d^2 b_1
  \; d^2 b_2 \int d^2 q_1 \; d^2 q_2 \notag \\
& \times \left\langle \frac{d \phi_{A_1}
  ({\bm q}_1, y=0)}{d^2 ({\bm B}-{\bm b}_1)} \right\rangle_{\! \!
  A_1} \left\langle \frac{d \phi_{A_1} ({\bm q}_2, y=0)}{d^2 ({\bm
  B}-{\bm b}_2)} \right\rangle_{\! \! A_1}
  \left\{ \left\langle \frac{d \phi_{A_2}^{D} ({\bm q}_1 - {\bm k}_1,
  {\bm q}_2 - {\bm k}_2, y)}{d^2 b_1 \; d^2 b_2} \right\rangle_{\!
  \! A_2} \! \! \right. \notag \\
& .\left. + \left[ \frac{\mathcal{K} ( {\bm b}_1, {\bm b}_2,
  {\bm k}_1, {\bm k}_2, {\bm q}_1, {\bm q}_2)}{N_c^2-1}
  \left\langle \frac{d \phi_{A_2}^{Q} ({\bm q}_1 - {\bm k}_1, {\bm
  q}_2 - {\bm k}_2, y)}{d^2 b_1 \; d^2 b_2} \right\rangle_{\!
  \! A_2} \! \! + ({\bm k}_2 \rightarrow - {\bm k}_2) \right]
\right\}
\end{align}
with $y \approx y_1 \approx y_2$ in the curly
brackets. \eqref{eq:factorized_final} is the main result of this section.

Notice that \eqref{eq:factorized_final} has all of the properties we
expected: it contains the convolution over the impact parameters ${\bm
  B}$, ${\bm b}_1$, and ${\bm b}_2$ along with different two-gluon
distribution functions. As advertised, \eqref{eq:factorized_final} also
contains a ``coefficient function'' associated with the factorized
form of 'crossed' diagrams.

The convolution over impact parameters in \eqref{eq:factorized_final}
appears to imply that the 2-gluon production cross section is
sensitive to the $b$-dependence of the one- and two-gluon
distributions $\phi$, $\phi^D$, $\phi^Q$. From
Eqs.~\eqref{eq:dipole_dist}, \eqref{eq:doubletrace_dist}, and
\eqref{eq:quad_dist} we see that the $b$-dependence of those gluon
distributions is related to that of the dipole, double-trace and
quadrupole operators. It is known that any perturbative approach, such
as the CGC formalism employed here, cannot describe correctly the
$b$-dependence of scattering amplitudes in peripheral collisions due
to the importance of non-perturbative effects
\cite{Kovner:2001bh}. It, therefore, appears that the two-gluon
production cross-section is also sensitive to the non-perturbative
large-$b$ physics. Note, however, that this conclusion also applies to
the single-gluon production in \eqref{eq:singlegluon}, since the impact
parameter integral in \eqref{eq:dipole_wave_int} is also sensitive to
large-$b$ physics. Recent studies \cite{Levin:2014bwa} appear to
indicate that this sensitivity to non-perturbative effects at the
periphery is not very strong, and may be negligible at high energies.

Unfortunately the factorization expression \eqref{eq:factorized_final}
is different from that used in
\cite{Dusling:2012cg,Dusling:2012wy,Dusling:2012iga}. The expression
in those references was motivated by extrapolation of the
dilute--dilute scattering case to the dense--dense scattering by
analogy with the single-gluon production \eqref{eq:singlegluon}. While
our result is valid only for the dense-dilute scattering, we can
conclude that the extrapolation suggested in
\cite{Dumitru:2010mv,Dumitru:2010iy,Dusling:2012cg,Dusling:2012wy,Dusling:2012iga}
does not work in the dense-dilute case, and is, therefore, unlikely to
be valid in the dense-dense scattering case either.

Just like \eqref{eq:factorized_square}, the expression
\eqref{eq:factorized_final} is not projectile-target symmetric. While
this is natural due to the asymmetric treatment of the target and
projectile in our dense-dilute scattering approximation, this
asymmetry also means that a simple generalization to the case of
nucleus--nucleus scattering along the lines of what was done with
\eqref{eq:singlegluon} in
\cite{Kharzeev:2001gp,Kharzeev:2001yq,ALbacete:2010ad,Albacete:2007sm}
appears to be impossible for \eqref{eq:factorized_final}.

The factorized form of the two-gluon production cross section
\eqref{eq:factorized_final} contains a few interesting properties. If
we look at the large-$N_c$ limit the 'crossed' diagrams contribution
can be neglected and, using \eqref{eq:singlet_app}, the two-gluon singlet
distribution function factorizes into two single gluon distribution
functions,
\begin{align} 
\label{eq:factorized_largeNc} 
& \frac{d \sigma}{d^2 k_1 dy_1 d^2 k_2 dy_2} \
  \bigg|_{\mbox{large}-N_c} = \left( \frac{2 \; \alpha_s}{C_F} \right)^2
  \frac{1}{k_1^2 \; k_2^2} \int d^2 B \, d^2 b_1 \, d^2 b_2 \int d^2 q_1
  \, d^2 q_2
\notag \\
& \times \; \left\langle \frac{d \phi_{A_1} ({\bm
  q}_1, y=0)}{d^2 ({\bm B}-{\bm b}_1)} \right\rangle_{\! \!  A_1}
  \left\langle \frac{d \phi_{A_1} ({\bm q}_2, y=0)}{d^2 ({\bm B}-{\bm
  b}_2)} \right\rangle_{\!  \!  A_1}
\notag \\
& \times \; \left\langle \frac{d
  \phi_{A_2} ({\bm q}_1 - {\bm k}_1, y)}{d^2 b_1} \right\rangle_{\!
  \!  A_2} \left\langle \frac{d \phi_{A_2} ({\bm q}_2 - {\bm k}_2,
  y)}{d^2 b_2} \right\rangle_{\! \!  A_2}.
\end{align}
This equation can only generate correlations between the two gluons
through the convolution over the impact parameters, which are
geometric correlations \cite{Kovchegov:2012nd} (Sec.~\ref{sub:geocor}). This form does not
contain the information needed to, say, calculate the correlation
function \eqref{eq:corr_LO}, since for that one needs terms that are
subleading in the large-$N_c$ limit.

Another interesting property is that we can isolate the part of the
cross-section that gives rise to HBT correlations
\cite{HanburyBrown:1956pf}. Due to the nature of HBT correlations, the
only way the correlations can be generated is through interference
effects. Thus only the 'crossed' diagrams contribute. In addition, for
the correlation to be pure HBT, the two produced gluons should have
the same colors (to be identical particles): imposing same-color
requirement on the 'crossed' diagrams is equivalent to the projection
employed in arriving at \eqref{eq:quad_approx}. We conclude that the only
part of the quadrupole two-gluon distribution function
\eqref{eq:quad_dist} that contributes to HBT correlations is the
portion that can be factorized into two single-gluon distributions
shown in \eqref{eq:quad_approx}. With the help of \eqref{eq:quad_approx}
the HBT part of the two-gluon production cross-section can be written
as
\begin{align} 
\label{eq:factorized_HBT} 
& \frac{d \sigma_{HBT}}{d^2 k_1 dy_1 d^2 k_2 dy_2} = \left( \frac{2 \;
  \alpha_s}{C_F} \right)^2 \frac{1}{k_1^2 \; k_2^2} \int d^2 B \; d^2 b_1
  \; d^2 b_2 \int d^2 q_1 \; d^2 q_2 \;
\notag \\
& \times \; \frac{\mathcal{K} ( {\bm b}_1,
  {\bm b}_2, {\bm k}_1, {\bm k}_2, {\bm q}_1, {\bm q}_2)}{N_c^2-1}
  \; \left\langle \frac{d \phi_{A_1} ({\bm q}_1,
  y=0)}{d^2 ({\bm B}-{\bm b}_1)} \right\rangle_{\! \!  A_1}
  \left\langle \frac{d \phi_{A_1} ({\bm q}_2, y=0)}{d^2 ({\bm B}-{\bm
  b}_2)} \right\rangle_{\! \!  A_1}
\notag \\
& \times \;  \left\langle \frac{d \phi_{A_2} ({\bm q}_1 - {\bm k}_1, y)}
  {d^2 b_1} \right\rangle_{\! \!  A_2}
  \left\langle \frac{d \phi_{A_2} ({\bm q}_2 - {\bm k}_2,
  y)}{d^2 b_2} \right\rangle_{\! \!  A_2} \; + \; ({\bm k}_2
  \rightarrow - {\bm k}_2).
\end{align}

To summarize this section let us stress that we were able to find a
factorized form for the two-gluon production cross section in
heavy-light ion collisions, given by \eqref{eq:factorized_final}. It had
to be written in a different form than that of the single gluon
production cross-section \eqref{eq:singlegluon}. In particular, in
\eqref{eq:factorized_final} we have a convolution over the impact
parameters which requires that the distribution functions have to be
written as differentials with respect to impact parameters, that is as
gluon Wigner distributions. There was also a ``coefficient function''
factor \eqref{eq:kernel_crossed} that was associated with the
interference effects included in the 'crossed' diagrams. These facts
may have important implications for $k_T$-factorization when it comes
to multi-gluon cross-sections and could possibly give insight into the
nature of $k_T$-factorization in general.

%% file: Chapter_Conclusions.tex
% !TEX root = WertepnyPhDThesis.tex
\cleardoublepage

\chapter{Conclusions}
\label{ch:Conclusions}

Starting out with a general introduction into the physics involved we briefly covered QCD, heavy-ion collisions and saturation physics in Ch.~\ref{ch:Introduction}.
This laid the ground work for looking at the problem of interest, two-particle correlations in heavy-light ion collisions.
A quantitative overview of QCD (needed for the calculation), an explanation of the heavy-light ion regime and its implication on the two-particle correlations was given in Ch.~\ref{ch:Setup}.

Ch.~\ref{ch:pAreview} consisted of a review of the known result for single-gluon production in $pA$ collisions within the saturation framework in the Classical MV limit.
This served as a pedagogical way of seeing how exactly the saturation framework is used in practical situations.
Not only that but we were able to extend calculations here to a more complicated situation, the two-gluon correlations in heavy-light ion collisions of interest.
The topics reviewed included how the target nucleus produces a shock wave (Sec.~\ref{sec:target}), how exactly the resulting shock wave can be analyzed (Sec.~\ref{sec:Dipole}), the $pA$ single-gluon production cross section (Sec.~\ref{sec:single}) and its $k_T$-factorized form (Sec.~\ref{sec:pAfact}).

With Ch.~\ref{ch:Calculation} we covered the result of the work presented in \cite{Kovchegov:2012nd}.
This included the calculation of the two-gluon production cross section in the heavy-light ion regime.
There were three distinct contributions which contributed to the cross section, one being the ``square" terms (diagrams), Sec.~\ref{subsec:ggsquare}.
The other two contribution were combined and formed a single contribution we call the ``crossed" terms (diagrams), Sec.~\ref{subsec:ggcrossed}.
In Sec.~\ref{sec:lrrc}, we saw three distinct types of long-range in rapidity correlations:
a correlation which is peaked in azimuthal angle on the near-side ($\Delta \phi = 0$),
an equal-and-opposite correlation on the away-side ($\Delta \phi = \pi$),
and a correlation that seems similar to a HBT correlation which are peaked when the transverse momenta of the particles are the same and when they are equal and opposite ($\bm k_1 = \bm k_2$ and $\bm k_1 = - \bm k_2$).
All of the two-gluon correlations produced in this heavy-light ion regime end up being symmetric in azimuthal angle ($\Delta \phi$).
It should be noted that the theoretical results presented here were posted on arXiv before the away-side ridge was discovered at ALICE in p-Pb collisions \cite{Abelev:2012aa}.

In Ch.~\ref{ch:Properties} we covered some of the work presented in \cite{Kovchegov:2013ewa}, two more implications of the two-gluon production cross section and the correlation function derived in Ch.~\ref{ch:Calculation}.
It was shown, Sec.~\ref{sec:Geo} that in the collisions of asymmetric ions (examples being Uranium and Gold) the effects of saturation give rise to correlations that behave differently than those due to hydrodynamical flow.
We focused on U+U collisions and showed that there is an enhancement of the two-particle correlations in tip-on-tip collisions when compared with side-on-side collisions, which is opposite of that predicted by hydrodynamics.
This gives a way to distinguish correlations due to saturation effects and correlations dues to hydrodynamics.
In Sec.~\ref{sec:fact} we derived the $k_T$-factorized form of the two-gluon production cross section in the heavy-light ion limit.
We found that, unlike the $pA$ case the cross section cannot be written in terms of the usual unintegrated gluon distributions.
Instead we had to introduce our own gluon distribution which end up being Wigner distributions, functions of both transverse momentum and transverse position.
This gives insight into how $k_T$-factorization works in general.

There are still a few outstanding issues with two-gluon correlations in heavy-light ion collisions that need to be address.
One such issue is rapidity dependence of the correlations. 
This was partially addressed in \cite{Kovchegov:2013ewa}, where we found that to lowest order in saturation effects the correlations are independent of the center-of-mass energy of the collision.
However, this did not include rapidity evolution between the two gluons, which needs to be addressed to understand the full correlation function.

We would like to know the full heavy ion collision case.
In order to approach this issue a basic starting point would be finding the single-gluon production cross section in heavy ion collisions.
Work in this direction has been done in \cite{Chirilli:2015tea}, where we calculate part of the single-gluon production cross section for the heavy-light ion case, taking into account two nucleons in the projectile and all saturation corrections in the target (beyond the leading order result presented in Ch.~\ref{ch:pAreview}). 
This is not a complete calculation though and the continuation of this project is a work in progress.

The correlations presented in this work were leading-order in saturation scale in the heavy-light ion regime: however they were sub-leading in $N_c$. 
It could be possible that this suppression makes it so that the sub-leading in projectile saturation corrections two-gluon production process, where we have a single nucleon in the projectile emitting two gluons, could be comparable in magnitude in the actual experiment.
The cross section for this process needs to be calculated in order to see what types of additional correlations arise, which is left for future work.

%% file: Appendix_Wilson.tex
% !TEX root = WertepnyPhDThesis.tex
% !TEX encoding = UTF-8 Unicode
% !TEX spellcheck = en_US
\cleardoublepage
\chapter{Double-trace operator}
\label{Appendix:Wilson}
 
Our goal here is to calculate the following object
\begin{align}
\label{corr1}
  \frac{1}{(N_c^2 -1)^2} \, \langle Tr[ U_{{\bm x}_1} U_{{\bm
      x}_2}^\dagger ] \, Tr[ U_{{\bm x}_3} U_{{\bm x}_4}^\dagger ]
  \rangle
\end{align}
in the quasi-classical MV/GM approximation at the lowest non-trivial
order in $1/N_c^2$ expansion in the 't Hooft's large-$N_c$ limit.  We
assume that the saturation scale is $N_c$-independent, that is
$Q_{s0}^2 \sim (N_c)^0$, due to an order-$N_c^2$ number of ``valence''
partons in each nucleon in the target nucleus.

%%%%%%%%%%%%%%%%%%%%%%%%%%%%%%%%%%%%%%%%%%%%%%%%%%%%%%%%%%%%%%%%%%%%%%%%%%%%
\begin{figure}[h]
\centering
  \includegraphics[width= 0.6 \textwidth]{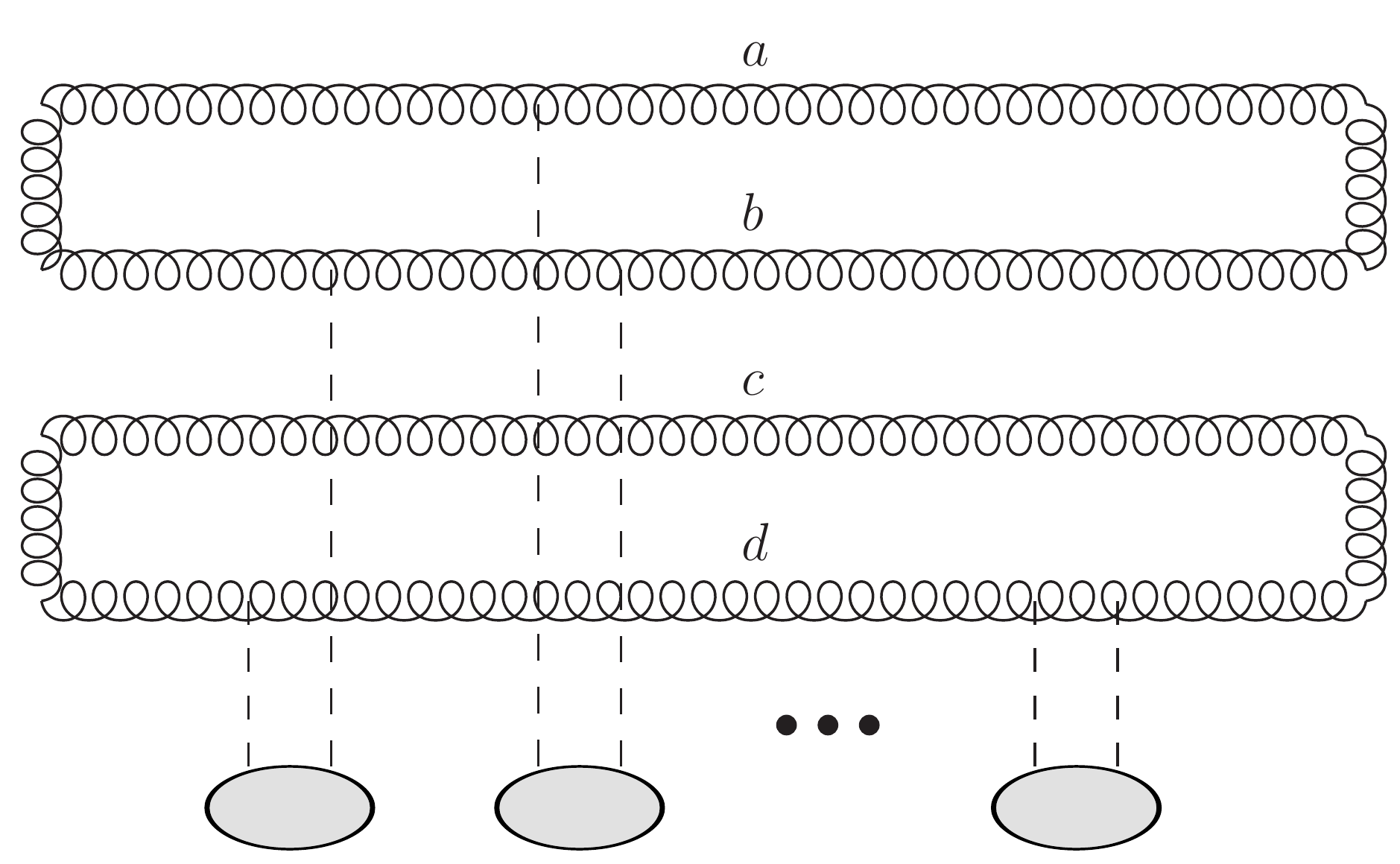}
  \caption{Diagrams contributing to the correlator in \eqref{corr1} in
    the quasi-classical approximation. Vertical dashed lines denote
    $t$-channel gluons, while the shaded ovals represent the nucleons
    in the target nucleus.}
\label{trtr} 
\end{figure}
%%%%%%%%%%%%%%%%%%%%%%%%%%%%%%%%%%%%%%%%%%%%%%%%%%%%%%%%%%%%%%%%%%%%%%%%%%%%

A sample of the diagrams contributing to \eqref{corr1} is shown in
\fig{trtr}, where the $t$-channel gluons are denoted by dashed lines
to simplify the picture. The calculation is straightforward
\cite{Kovchegov:2001ni}: one simply has to exponentiate the two-gluon
exchange interaction with a single nucleon (the nucleons are denoted
by shaded ovals in \fig{trtr}). The only complication is that, unlike
the dipole amplitude calculated in \cite{Mueller:1989st}, the
interaction now, for the double-trace operator \eqref{corr1}, is a
matrix in the color space. A double trace operator like \eqref{corr1}
with the Wilson lines in the fundamental representation was calculated
earlier in \cite{Marquet:2010cf} (for similar calculations see also
\cite{Nikolaev:2005zj,Dominguez:2011gc,Iancu:2011ns}).

To exponentiate the matrix we have to choose a basis in the color
space of four $s$-channel gluons in \fig{trtr}: clearly the net color
of the four gluons is always zero. The color states of the four gluons
can be classified according to the color states of the top two
$s$-channel gluons, since the color of the bottom pair of $s$-channel
gluons is determined by requiring net color-neutrality of the four
$s$-channel gluon system. The colors of a gluon pair can be decomposed
in the following irreducible representations
\begin{align}
 \label{8x8}
  & {(N_c^2 - 1)} \otimes {(N_c^2 - 1)} = V_1 \oplus V_2 \oplus V_3
  \oplus V_4 \oplus V_5 \oplus V_6 \oplus V_7
\notag \\
  & = {\bm 1} \oplus {(N_c^2 -1)} \oplus \frac{N_c^2 (N_c -3) (N_c +1)}{4} \oplus
  \frac{N_c^2 (N_c + 3) (N_c - 1)}{4} \oplus {(N_c^2 -1)}
\notag \\ & \oplus \frac{(N_c^2 -1) (N_c^2 - 4)}{4} \oplus
  \frac{(N_c^2 -1) (N_c^2 - 4)}{4}.
\end{align}
Here we are following the notations introduced in
\cite{Cvitanovic:2008zz}, see page 120. Labeling the colors of
the four $s$-channel gluons in an arbitrary color state by $a, b, c$,
and $d$ as shown in \fig{trtr} we define the color states
corresponding to representations $V_1$, $V_2$, and $V_5$ by
\begin{align}
  \label{eq:color_states}
  |P_1 \rangle = \frac{1}{N_c^2 -1} & \, \delta^{ab} \, \delta^{cd} , \
  \ \ |P_2 \rangle = \frac{1}{\sqrt{N_c^2 -1}} \, \frac{N_c}{N_c^2 -4}
  \, d^{abe} \, d^{cde}, \ \ \ 
\notag \\
& |P_5 \rangle = \frac{1}{\sqrt{N_c^2
      -1}} \, \frac{1}{N_c} \, f^{abe} \, f^{cde},
\end{align}
where we differ from $P_i$'s in \cite{Cvitanovic:2008zz} by
prefactors, since here we demanded that our color states are
normalized to one, $\langle P_i | P_j \rangle = \delta^{ij} $. Other
color states can be constructed as well \cite{Cvitanovic:2008zz}, but
we will only need the states in \eqref{eq:color_states} for the
calculation below.

\begin{figure}[H]
\centering
  \includegraphics[width= 0.5 \textwidth]{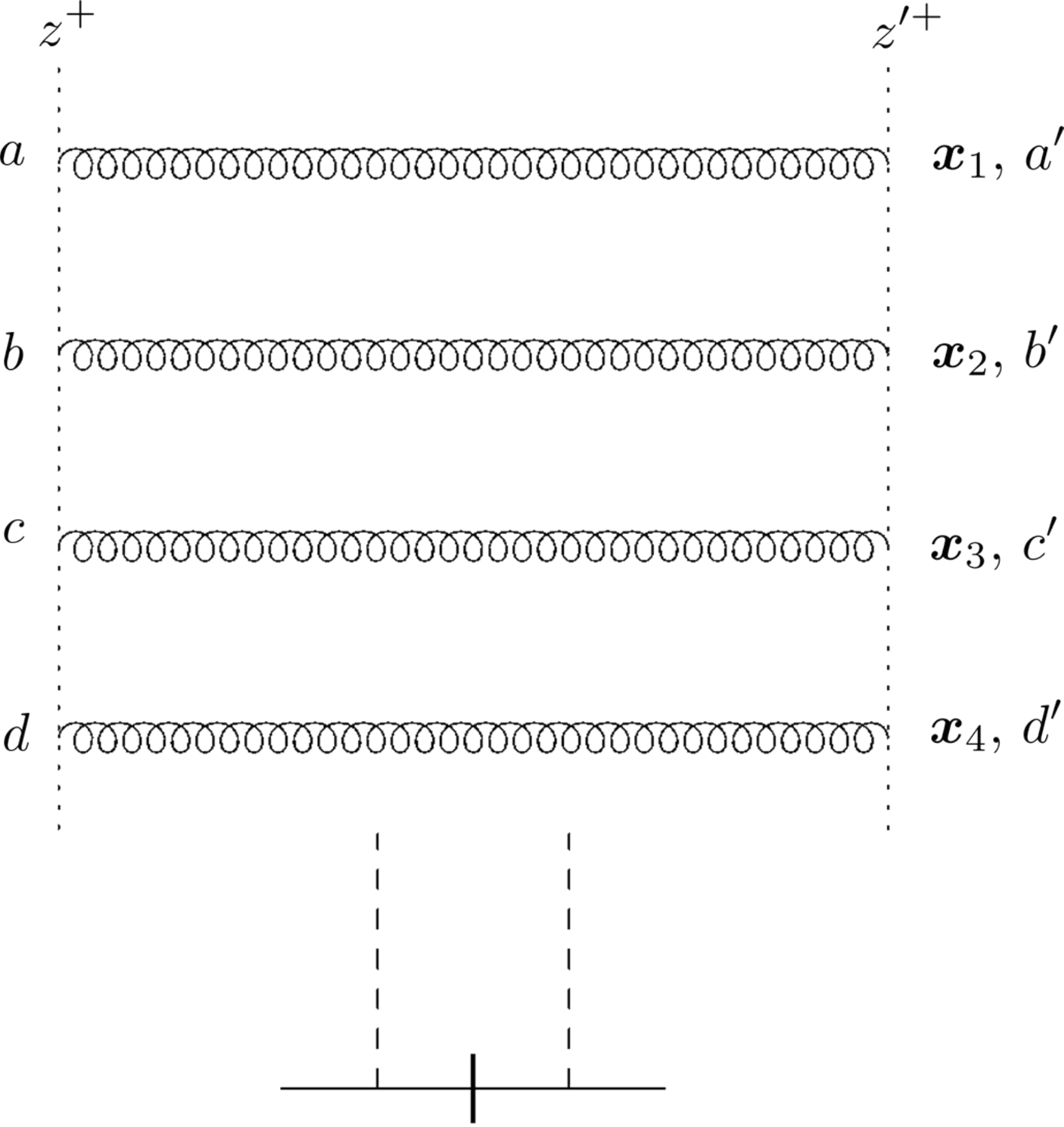}
  \caption{The diagramatic representation of the matrix element associated with a nucleon interacting with 4 adjoint Wilson lines, $\hat M^{a'a,b'b,c'c,d'd} \left(\bm x_1, \bm x_2, \bm x_3, \bm x_4 \right)$. The dotted line represents the gluon fields and it interacts in all possible ways with the Wilson lines.}
\label{WilsonMf} 
\end{figure}

We denote the interaction with a single nucleon by a two-gluon exchange as $\hat M^{a'a,b'b,c'c,d'd} \left(\bm x_1, \bm x_2, \bm x_3, \bm x_4 \right)$: it is a matrix in the color space of the $s$-channel gluons.
Often we suppress the color indices and transverse coordinates and notate this as $\hat M$.
It is defined by taking averaging over the transverse positions of a single nucleon source with 4 wilson line operators this time
\begin{equation}
\label{MhatDef}
\hat M^{a'a,b'b,c'c,d'd} \left(\bm x_1, \bm x_2, \bm x_3, \bm x_4 \right) =
  \left< U^{a' a}_{\bm x_1} [z'^+, z^+] U^{b' b}_{\bm x_2} [z'^+, z^+]
  U^{c' c}_{\bm x_3} [z'^+, z^+] U^{d' d}_{\bm x_4} [z'^+, z^+]\right>_1 .
\end{equation}
The corresponding diagram is shown in Fig. \ref{WilsonMf}.

%%%%%%%%%%%%%%%%%%%%%%%%%%%%%%%%%%%%%%%%%%%%%%%%%%%%%%%%%%%%%%%%%%%%%%%%%%%%
%%%%%%%%%%%%%%%%%%%%%%%%%%%%%%%%%%%%%%%%%%%%%%%%%%%%%%%%%%%%%%%%%%%%%%%%%%%%

The procedure to calculate this follows the same lines as the dipole calculation did, the only difference being the number of possible interactions have increased.
It should be no surprise that it bares a striking resemblance to \ref{1nucleonint}
\begin{align}
\label{Moperator}
\hat M^{a'a,b'b,c'c,d'd} & \left(\bm x_1, \bm x_2, \bm x_3, \bm x_4 \right)
\notag \\
  = & - \, Q^2_{s,2}
  \int \frac{d^2 k}{2 \pi} \frac{1}{|\bm k^2|^2}
  \left[ 2 \, \delta^{a' a}\delta^{b' b} \delta^{c' c} \delta^{d' d}
  + \frac{1}{N_c} [T^e]^{a' a} [T^e]^{b' b} \delta^{c' c} \delta^{d' d}
  e^{-i \bm k \cdot (\bm x_1 -\bm x_2)} \right.
\notag \\
 & + \frac{1}{N_c} [T^e]^{a' a}\delta^{b' b}
  [T^e]^{c' c} \delta^{d' d} e^{-i \bm k \cdot (\bm x_1 -\bm x_3)} 
  + \frac{1}{N_c} [T^e]^{a' a} \delta^{b' b} \delta^{c' c}
   [T^e]^{d' d} e^{-i \bm k \cdot (\bm x_1 -\bm x_4)} 
\notag \\
 & + \frac{1}{N_c} \delta^{a' a} [T^e]^{b' b}
  [T^e]^{c' c} \delta^{d' d} e^{-i \bm k \cdot (\bm x_2 -\bm x_3)} 
  + \frac{1}{N_c} \delta^{a' a} [T^e]^{b' b} \delta^{c' c}
   [T^e]^{d' d} e^{-i \bm k \cdot (\bm x_2 -\bm x_4)}
\notag \\
 & \left. + \frac{1}{N_c} \delta^{a' a} \delta^{b' b}
  [T^e]^{c' c} [T^e]^{d' d} e^{-i \bm k \cdot (\bm x_3 -\bm x_4)} 
  \right]
\end{align}

Since in the correlator \eqref{corr1} the top
(bottom) two $s$-channel gluons are in the color-singlet state both
before and after the interaction, we write
\begin{align}
  \label{corr2}
  \frac{1}{(N_c^2 -1)^2} \, \langle Tr[ U_{{\bm x}_1} U_{{\bm
      x}_2}^\dagger ] \, Tr[ U_{{\bm x}_3} U_{{\bm x}_4}^\dagger ]
  \rangle = \langle P_1 | e^{\hat M} | P_1 \rangle. 
\end{align}
Expanding the exponential in a power series and inserting unit
operators ${\bm 1} = \sum_i |P_i \rangle \, \langle P_i |$ between all
the $\hat M$'s yields
\begin{align}
  \label{corr3}
  \frac{1}{(N_c^2 -1)^2} \, \langle Tr[ U_{{\bm x}_1} U_{{\bm
      x}_2}^\dagger ] \, Tr[ U_{{\bm x}_3} U_{{\bm x}_4}^\dagger ]
  \rangle = \left( e^M \right)_{11}
\end{align}
where the $7 \times 7$ matrix $M$ is defined by its elements,
\begin{align}
\label{Mdef}
M_{ij} = \langle P_i | \hat M | P_j \rangle .  
\end{align}
As an example we calculate $M_{11}$,
\begin{align}
\label{M11}
M_{11} = & \frac{1}{N^2_c -1} \delta^{a' b'} \delta^{c' d'}
  \hat M^{a'a,b'b,c'c,d'd} \left(\bm x_1, \bm x_2, \bm x_3, \bm x_4 \right) 
  \frac{1}{N^2_c -1} \delta^{ab} \delta^{cd}
\notag \\
  = & \frac{-1}{(N^2_c -1)^2} \delta^{a' b'} \delta^{c' d'}
  \, Q^2_{s,2} \int \frac{d^2 k}{2 \pi} \frac{1}{|\bm k^2|^2}
\notag \\
  & \times \left[ 2 \, \delta^{a' b'}\delta^{c' d'}
  - \frac{1}{N_c} [T^e T^e]^{a' b'} \delta^{c' d'}
  e^{-i \bm k \cdot (\bm x_1 -\bm x_2)} \right.
  + \frac{1}{N_c} [T^e]^{a' b'}
  [T^e]^{c' d'}  e^{-i \bm k \cdot (\bm x_1 -\bm x_3)} 
\notag \\
  & - \frac{1}{N_c} [T^e]^{a' b'}
   [T^e]^{c' d'} e^{-i \bm k \cdot (\bm x_1 -\bm x_4)} 
  - \frac{1}{N_c} [T^e]^{a' b'}
  [T^e]^{c' d'} e^{-i \bm k \cdot (\bm x_2 -\bm x_3)} 
\notag \\
 & \left. + \frac{1}{N_c} [T^e]^{a' b'}
   [T^e]^{c' d'} e^{-i \bm k \cdot (\bm x_2 -\bm x_4)}
  - \frac{1}{N_c} \delta^{a' b'}
  [T^e T^e]^{c' d'} e^{-i \bm k \cdot (\bm x_3 -\bm x_4)} 
  \right]
\notag \\
  = & - Q^2_{s,2} \int \frac{d^2 k}{2 \pi} \frac{1}{|\bm k^2|^2}
  \left[ 2 - e^{-i \bm k \cdot (\bm x_1 -\bm x_2)}
  - e^{-i \bm k \cdot (\bm x_3 -\bm x_4)} \right]
\notag \\
  = & - \frac{Q_{s,2}^2}{4} \left[ | \bm x_1 - \bm x_2 |^2 \, 
  \ln \left( \frac{1}{| \bm x_1 - \bm x_2 | \Lambda } \right) + | \bm x_3 - \bm x_4 |^2 \,
  \ln \left( \frac{1}{| \bm x_3 - \bm x_4 | \Lambda } \right) \right] 
\end{align}
and see that it ends up just being equal to $-D_1$, one of the $D_i$'s defined in Eqs.~(\ref{Ds}).

All one has to do now is to find the remaining elements of
matrix $M$ from \eqref{Mdef},
exponentiate it, picking up the ``$11$'' matrix element of the
exponential. Such a calculation, while straightforward, is rather
involved: here we will utilize the large-$N_c$ limit to construct an
approximate result.

Calculating some of the elements of the matrix $M$ and evaluating the
$N_c$-order of the remaining matrix elements yields
\begin{align}
\label{eq:Matrix}
M = \left( \begin{array}{ccccccc}
  D_1 & 0 & 0 & 0 & \frac{D_3-D_2}{\sqrt{N_c^2-1}} & 0 & 0
\\
  0 & M_{22} & {\cal O}(\frac{1}{N_c})
  & {\cal O}(\frac{1}{N_c})  & M_{25} & {\cal O}(\frac{1}{N_c})
  & {\cal O}(\frac{1}{N_c}) 
\\
  0 & {\cal O}(\frac{1}{N_c})  & \ldots & \ldots & {\cal O}(\frac{1}{N_c}) & \ldots & \ldots
\\
  0 & {\cal O}(\frac{1}{N_c})  & \ldots & \ldots & {\cal O}(\frac{1}{N_c})  & \ldots & \ldots
\\
  \frac{D_3-D_2}{\sqrt{N_c^2-1}} & M_{25} & {\cal O}(\frac{1}{N_c}) 
  & {\cal O}(\frac{1}{N_c}) & M_{22}
  & {\cal O}(\frac{1}{N_c}) & {\cal O}(\frac{1}{N_c}) 
\\
  0 & {\cal O}(\frac{1}{N_c})  & \ldots & \ldots & {\cal O}(\frac{1}{N_c})
  & \ldots & \ldots
\\
  0 & {\cal O}(\frac{1}{N_c})  & \ldots & \ldots & {\cal O}(\frac{1}{N_c})
  & \ldots & \ldots
\end{array} \right)
\end{align}
\begin{equation}
\notag
M_{22} = \frac{1}{2}D_1 + \frac{1}{4}(D_2+D_3), \quad M_{25} = \frac{1}{4}(D_3-D_2) 
\end{equation}
with $D_i$ defined in Eqs.~(\ref{Ds}) and ellipsis denoting the matrix
elements we do not need to calculate, as they are at most order-1 in
$N_c$ counting. From \eqref{eq:Matrix} we see that if we start in the
color-singlet state $|P_1 \rangle$ for the top (bottom) gluon pair, a
single interaction can either leave the two gluons in the
color-singlet state, or can flip them into a color-octet state $|P_5
\rangle$. The latter transition comes in with an order-$1/N_c$
suppression factor. In order to evaluate \eqref{corr1} we have to
start and finish with a color-singlet state: to order-$1/N_c^2$ we may
have at most two such transitions: $|P_1 \rangle \rightarrow |P_5
\rangle$ and the inverse, $|P_5 \rangle \rightarrow |P_1
\rangle$. Once the system is in the color-octet $|P_5 \rangle$ state,
it can continue its random walk through color space: however, if we
want to keep the calculation at the order-$1/N_c^2$, the interactions
between the two transitions should be leading-order in $N_c$. This is
why only the leading-$N_c$-order matrix elements $M_{22}$, $M_{25} =
M_{52}$, and $M_{55}$ contribute in the $6 \times 6$ matrix $M_{ij}$
with $i,j = 2, \ldots, 7$: we do not need to calculate the
$1/N_c$-suppressed elements in \eqref{eq:Matrix} or the elements denoted
by ellipsis which can not contribute.

Exponentiating the matrix $M$ from \eqref{eq:Matrix}, picking up the
``$11$'' element of the obtained matrix and expanding the result to
order-$1/N_c^2$ yields
\begin{align}
\label{eq:ans}
& \frac{1}{(N_c^2 -1)^2} \, \langle Tr[ U_{{\bm x}_1} U_{{\bm x}_2}^\dagger ]
  \, Tr[ U_{{\bm x}_3} U_{{\bm x}_4}^\dagger ] \rangle = 
  e^{D_1} + \frac{(D_3-D_2)^2}{N_c^2} \, \left[
  \frac{e^{D_1}}{D_1-D_2} - \frac{2 \, e^{D_1}}{(D_1-D_2)^2} \right.
\notag \\
  & \qquad + \frac{e^{D_1}}{D_1-D_3} - \frac{2 \, e^{D_1}}{(D_1-D_3)^2}
  \left. + \frac{2 \, e^{\frac{1}{2}(D_1+D_2)}}{(D_1-D_2)^2} +
  \frac{2 \, e^{\frac{1}{2}(D_1+D_3)}}{(D_1-D_3)^2} \right] +
  {\cal O} \left( \frac{1}{N_c^4} \right).
\end{align}
Finally, noting that (see \eqref{eq:SG_GM})
\begin{align}
  \label{eq:factDel}
  \frac{1}{(N_c^2 -1)^2} \, \langle Tr[ U_{{\bm x}_1} U_{{\bm
      x}_2}^\dagger ] \rangle \, \langle Tr[ U_{{\bm x}_3} U_{{\bm
      x}_4}^\dagger ] \rangle =  e^{D_1}
\end{align}
and using \eqref{Ddef} we obtain \eqref{Delta_exp}.

%% file: WertepnyPhDThesis.bbl
\begin{thebibliography}{100}

\bibitem{Agashe:2014kda}
K.~A. Olive et~al.
\newblock {Review of Particle Physics}.
\newblock {\em Chin. Phys.}, C38:090001, 2014.

\bibitem{Bassevoimage}
Steffen~A. Bass.
\newblock Cartoon of a ultra-relativistic heavy-ion collision.

\bibitem{Abramowicz:2015mha}
H.~Abramowicz et~al.
\newblock {Combination of measurements of inclusive deep inelastic ${e^{\pm
  }p}$ scattering cross sections and QCD analysis of HERA data}.
\newblock {\em Eur. Phys. J.}, C75(12):580, 2015.

\bibitem{Abelev:2012aa}
Betty Abelev et~al.
\newblock {Long-range angular correlations on the near and away side in $p-Pb$
  collisions at $\sqrt{s_{NN}} = 5.02$~TeV}.
\newblock 2012.

\bibitem{Yagi:2005yb}
K.~Yagi, T.~Hatsuda, and Y.~Miake.
\newblock {Quark-gluon plasma: From big bang to little bang}.
\newblock {\em Camb. Monogr. Part. Phys. Nucl. Phys. Cosmol.}, 23:1--446, 2005.

\bibitem{Kovchegov:2012nd}
Yuri~V. Kovchegov and Douglas~E. Wertepny.
\newblock {Long-Range Rapidity Correlations in Heavy-Light Ion Collisions}.
\newblock {\em Nucl.Phys.}, A906:50--83, 2013.

\bibitem{Kovchegov:2013ewa}
Yuri~V. Kovchegov and Douglas~E. Wertepny.
\newblock {Two-Gluon Correlations in Heavy-Light Ion Collisions: Energy and
  Geometry Dependence, IR Divergences, and $k_T$-Factorization}.
\newblock {\em Nucl.Phys.}, A925:254--295, 2014.

\bibitem{Yang:1954vj}
Chen-Ning Yang and R.~L. Mills.
\newblock {Isotopic spin conservation and a generalized gauge invariance}.
\newblock 1954.

\bibitem{GellMann:1964nj}
Murray Gell-Mann.
\newblock {A Schematic Model of Baryons and Mesons}.
\newblock {\em Phys. Lett.}, 8:214--215, 1964.

\bibitem{Fritzsch:1973pi}
H.~Fritzsch, Murray Gell-Mann, and H.~Leutwyler.
\newblock {Advantages of the Color Octet Gluon Picture}.
\newblock {\em Phys. Lett.}, B47:365--368, 1973.

\bibitem{Gross:1973id}
D.~J. Gross and Frank Wilczek.
\newblock {Ultraviolet Behavior of Non-Abelian Gauge Theories}.
\newblock {\em Phys. Rev. Lett.}, 30:1343--1346, 1973.

\bibitem{Politzer:1973fx}
H.~David Politzer.
\newblock {Reliable Perturbative Results for Strong Interactions?}
\newblock {\em Phys. Rev. Lett.}, 30:1346--1349, 1973.

\bibitem{Weinberg:1973un}
Steven Weinberg.
\newblock {Nonabelian Gauge Theories of the Strong Interactions}.
\newblock {\em Phys. Rev. Lett.}, 31:494--497, 1973.

\bibitem{jackson_classical_1999}
John~David Jackson.
\newblock {\em Classical electrodynamics}.
\newblock Wiley, New York, {NY}, 3rd ed. edition, 1999.

\bibitem{Weinberg:1996kr}
Steven Weinberg.
\newblock {\em {The quantum theory of fields. Vol. 2: Modern applications}}.
\newblock Cambridge University Press, 2013.

\bibitem{Brambilla:2014jmp}
N.~Brambilla et~al.
\newblock {QCD and Strongly Coupled Gauge Theories: Challenges and
  Perspectives}.
\newblock {\em Eur. Phys. J.}, C74(10):2981, 2014.

\bibitem{Gale:2013da}
Charles Gale, Sangyong Jeon, and Bjoern Schenke.
\newblock {Hydrodynamic Modeling of Heavy-Ion Collisions}.
\newblock {\em Int. J. Mod. Phys.}, A28:1340011, 2013.

\bibitem{Accardi:2012qut}
A.~Accardi, J.L. Albacete, M.~Anselmino, N.~Armesto, E.C. Aschenauer, et~al.
\newblock {Electron Ion Collider: The Next QCD Frontier - Understanding the
  glue that binds us all}.
\newblock 2012.

\bibitem{Jalilian-Marian:2005jf}
Jamal Jalilian-Marian and Yuri~V. Kovchegov.
\newblock Saturation physics and deuteron gold collisions at {RHIC}.
\newblock {\em Prog. Part. Nucl. Phys.}, 56:104--231, 2006.

\bibitem{Weigert:2005us}
Heribert Weigert.
\newblock Evolution at small {$x_{bj}$: The Color Glass Condensate}.
\newblock {\em Prog. Part. Nucl. Phys.}, 55:461--565, 2005.

\bibitem{Iancu:2003xm}
Edmond Iancu and Raju Venugopalan.
\newblock The color glass condensate and high energy scattering in {QCD}.
\newblock 2003.

\bibitem{Gelis:2010nm}
Francois Gelis, Edmond Iancu, Jamal Jalilian-Marian, and Raju Venugopalan.
\newblock {The Color Glass Condensate}.
\newblock {\em Ann.Rev.Nucl.Part.Sci.}, 60:463--489, 2010.

\bibitem{KovchegovLevin}
Yuri~V. Kovchegov and E.~Levin.
\newblock {\em Quantum Chromodynamics at High Energy}.
\newblock Cambridge University Press, 2012.

\bibitem{Kuraev:1977fs}
E.~A. Kuraev, L.~N. Lipatov, and Victor~S. Fadin.
\newblock {The Pomeranchuk singlularity in non-Abelian gauge theories}.
\newblock {\em Sov. Phys. JETP}, 45:199--204, 1977.

\bibitem{Balitsky:1978ic}
I.I. Balitsky and L.N. Lipatov.
\newblock {The Pomeranchuk Singularity in Quantum Chromodynamics}.
\newblock {\em Sov.J.Nucl.Phys.}, 28:822--829, 1978.

\bibitem{Balitsky:1996ub}
I.~Balitsky.
\newblock Operator expansion for high-energy scattering.
\newblock {\em Nucl. Phys.}, B463:99--160, 1996.

\bibitem{Balitsky:1998ya}
Ian Balitsky.
\newblock Factorization and high-energy effective action.
\newblock {\em Phys. Rev.}, D60:014020, 1999.

\bibitem{Kovchegov:1999yj}
Yuri~V. Kovchegov.
\newblock Small-x {$F_2$} structure function of a nucleus including multiple
  pomeron exchanges.
\newblock {\em Phys. Rev.}, D60:034008, 1999.

\bibitem{Kovchegov:1999ua}
Yuri~V. Kovchegov.
\newblock Unitarization of the {BFKL} pomeron on a nucleus.
\newblock {\em Phys. Rev.}, D61:074018, 2000.

\bibitem{Jalilian-Marian:1997dw}
Jamal Jalilian-Marian, Alex Kovner, and Heribert Weigert.
\newblock The {Wilson} renormalization group for low x physics: Gluon evolution
  at finite parton density.
\newblock {\em Phys. Rev.}, D59:014015, 1998.

\bibitem{Jalilian-Marian:1997gr}
Jamal Jalilian-Marian, Alex Kovner, Andrei Leonidov, and Heribert Weigert.
\newblock The {Wilson} renormalization group for low x physics: Towards the
  high density regime.
\newblock {\em Phys. Rev.}, D59:014014, 1998.

\bibitem{Iancu:2001ad}
Edmond Iancu, Andrei Leonidov, and Larry~D. McLerran.
\newblock {The renormalization group equation for the color glass condensate}.
\newblock {\em Phys. Lett.}, B510:133--144, 2001.

\bibitem{Iancu:2000hn}
Edmond Iancu, Andrei Leonidov, and Larry~D. McLerran.
\newblock Nonlinear gluon evolution in the color glass condensate. {I}.
\newblock {\em Nucl. Phys.}, A692:583--645, 2001.

\bibitem{McLerran:1994vd}
Larry~D. McLerran and Raju Venugopalan.
\newblock Green's functions in the color field of a large nucleus.
\newblock {\em Phys. Rev.}, D50:2225--2233, 1994.

\bibitem{McLerran:1993ka}
Larry~D. McLerran and Raju Venugopalan.
\newblock Gluon distribution functions for very large nuclei at small
  transverse momentum.
\newblock {\em Phys. Rev.}, D49:3352--3355, 1994.

\bibitem{McLerran:1993ni}
Larry~D. McLerran and Raju Venugopalan.
\newblock Computing quark and gluon distribution functions for very large
  nuclei.
\newblock {\em Phys. Rev.}, D49:2233--2241, 1994.

\bibitem{Adams:2005ph}
J.~Adams et~al.
\newblock {Distributions of charged hadrons associated with high transverse
  momentum particles in $p+p$ and $Au + Au$ collisions at $s_{NN}^{1/2}$ = 200
  GeV}.
\newblock {\em Phys. Rev. Lett.}, 95:152301, 2005.

\bibitem{Adare:2008cqb}
A.~Adare et~al.
\newblock {Dihadron azimuthal correlations in Au+Au collisions at
  $\sqrt(s_{NN})$=200 GeV}.
\newblock {\em Phys. Rev.}, C78:014901, 2008.

\bibitem{Alver:2009id}
B.~Alver et~al.
\newblock {High transverse momentum triggered correlations over a large
  pseudorapidity acceptance in Au+Au collisions at $\sqrt{s_{NN}}$=200 GeV}.
\newblock {\em Phys. Rev. Lett.}, 104:062301, 2010.

\bibitem{Abelev:2009af}
B.I. Abelev et~al.
\newblock {Long range rapidity correlations and jet production in high energy
  nuclear collisions}.
\newblock {\em Phys.Rev.}, C80:064912, 2009.

\bibitem{Khachatryan:2010gv}
Vardan Khachatryan et~al.
\newblock {Observation of Long-Range Near-Side Angular Correlations in
  Proton-Proton Collisions at the LHC}.
\newblock {\em JHEP}, 09:091, 2010.

\bibitem{CMS:2012qk}
Serguei Chatrchyan et~al.
\newblock {Observation of long-range near-side angular correlations in
  proton-lead collisions at the LHC}.
\newblock 2012.

\bibitem{Dumitru:2008wn}
Adrian Dumitru, Francois Gelis, Larry McLerran, and Raju Venugopalan.
\newblock {Glasma flux tubes and the near side ridge phenomenon at RHIC}.
\newblock {\em Nucl. Phys.}, A810:91--108, 2008.

\bibitem{Gavin:2008ev}
Sean Gavin, Larry McLerran, and George Moschelli.
\newblock {Long Range Correlations and the Soft Ridge in Relativistic Nuclear
  Collisions}.
\newblock {\em Phys. Rev.}, C79:051902, 2009.

\bibitem{Armesto:2006bv}
Nestor Armesto, Larry McLerran, and Carlos Pajares.
\newblock {Long Range Forward-Backward Correlations and the Color Glass
  Condensate}.
\newblock {\em Nucl.Phys.}, A781:201--208, 2007.

\bibitem{Armesto:2007ia}
N.~Armesto, M.A. Braun, and C.~Pajares.
\newblock {On the long-range correlations in hadron-nucleus collisions}.
\newblock {\em Phys.Rev.}, C75:054902, 2007.

\bibitem{Dusling:2009ni}
Kevin Dusling, Francois Gelis, Tuomas Lappi, and Raju Venugopalan.
\newblock {Long range two-particle rapidity correlations in A+A collisions from
  high energy QCD evolution}.
\newblock {\em Nucl. Phys.}, A836:159--182, 2010.

\bibitem{Dumitru:2010iy}
Adrian Dumitru, Kevin Dusling, Francois Gelis, Jamal Jalilian-Marian, Tuomas
  Lappi, et~al.
\newblock {The Ridge in proton-proton collisions at the LHC}.
\newblock {\em Phys.Lett.}, B697:21--25, 2011.

\bibitem{Dumitru:2010mv}
Adrian Dumitru and Jamal Jalilian-Marian.
\newblock {Two-particle correlations in high energy collisions and the gluon
  four-point function}.
\newblock {\em Phys. Rev.}, D81:094015, 2010.

\bibitem{Kovner:2010xk}
Alex Kovner and Michael Lublinsky.
\newblock {Angular Correlations in Gluon Production at High Energy}.
\newblock {\em Phys.Rev.}, D83:034017, 2011.

\bibitem{Gelis:2008sz}
Francois Gelis, Tuomas Lappi, and Raju Venugopalan.
\newblock {High energy factorization in nucleus-nucleus collisions. 3. Long
  range rapidity correlations}.
\newblock {\em Phys.Rev.}, D79:094017, 2009.

\bibitem{Kovner:1995ja}
Alex Kovner, Larry~D. McLerran, and Heribert Weigert.
\newblock Gluon production from non{A}belian {W}eizsacker-{W}illiams fields in
  nucleus-nucleus collisions.
\newblock {\em Phys. Rev.}, D52:6231--6237, 1995.

\bibitem{Kovchegov:1997ke}
Yuri~V. Kovchegov and Dirk~H. Rischke.
\newblock Classical gluon radiation in ultrarelativistic nucleus nucleus
  collisions.
\newblock {\em Phys. Rev.}, C56:1084--1094, 1997.

\bibitem{Krasnitz:2003jw}
Alex Krasnitz, Yasushi Nara, and Raju Venugopalan.
\newblock Classical gluodynamics of high energy nuclear collisions: An erratum
  and an update.
\newblock {\em Nucl. Phys.}, A727:427--436, 2003.

\bibitem{Blaizot:2010kh}
J.~P. Blaizot, T.~Lappi, and Y.~Mehtar-Tani.
\newblock {On the gluon spectrum in the glasma}.
\newblock {\em Nucl. Phys.}, A846:63--82, 2010.

\bibitem{Kovchegov:1999ep}
Yuri~V. Kovchegov, Eugene Levin, and Larry~D. McLerran.
\newblock {Large scale rapidity correlations in heavy ion collisions}.
\newblock {\em Phys.Rev.}, C63:024903, 2001.

\bibitem{ALbacete:2010ad}
Javier~L. Albacete and Adrian Dumitru.
\newblock {A model for gluon production in heavy-ion collisions at the LHC with
  rcBK unintegrated gluon densities}.
\newblock 2010.

\bibitem{Kovner:2011pe}
Alex Kovner and Michael Lublinsky.
\newblock {On Angular Correlations and High Energy Evolution}.
\newblock {\em Phys.Rev.}, D84:094011, 2011.

\bibitem{Levin:2011fb}
Eugene Levin and Amir~H. Rezaeian.
\newblock {The Ridge from the BFKL evolution and beyond}.
\newblock {\em Phys.Rev.}, D84:034031, 2011.

\bibitem{Kovner:2012jm}
Alex Kovner and Michael Lublinsky.
\newblock {Angular and long range rapidity correlations in particle production
  at high energy}.
\newblock {\em Int.J.Mod.Phys.}, E22:1330001, 2013.

\bibitem{Dusling:2012iga}
Kevin Dusling and Raju Venugopalan.
\newblock {Azimuthal collimation of long range rapidity correlations by strong
  color fields in high multiplicity hadron-hadron collisions}.
\newblock {\em Phys.Rev.Lett.}, 108:262001, 2012.

\bibitem{Dusling:2012wy}
Kevin Dusling and Raju Venugopalan.
\newblock {Explanation of systematics of CMS p+Pb high multiplicity di-hadron
  data at $\sqrt{s}_{\rm NN} = 5.02$ TeV}.
\newblock 2012.

\bibitem{Dusling:2012cg}
Kevin Dusling and Raju Venugopalan.
\newblock {Evidence for BFKL and saturation dynamics from di-hadron spectra at
  the LHC}.
\newblock 2012.

\bibitem{Gardi:2006rp}
E.~Gardi, J.~Kuokkanen, K.~Rummukainen, and H.~Weigert.
\newblock Running coupling and power corrections in nonlinear evolution at the
  high-energy limit.
\newblock {\em Nucl. Phys.}, A784:282--340, 2007.

\bibitem{Kovchegov:2006vj}
Y.~Kovchegov and H.~Weigert.
\newblock {Triumvirate of Running Couplings in Small-$x$ Evolution}.
\newblock {\em Nucl. Phys. {\bf A}}, 784:188--226, 2007.

\bibitem{Balitsky:2006wa}
I.~I. Balitsky.
\newblock {Quark Contribution to the Small-$x$ Evolution of Color Dipole}.
\newblock {\em Phys. Rev. D}, 75:014001, 2007.

\bibitem{Krasnitz:1999wc}
Alex Krasnitz and Raju Venugopalan.
\newblock The initial energy density of gluons produced in very high energy
  nuclear collisions.
\newblock {\em Phys. Rev. Lett.}, 84:4309--4312, 2000.

\bibitem{Lappi:2003bi}
T.~Lappi.
\newblock Production of gluons in the classical field model for heavy ion
  collisions.
\newblock {\em Phys. Rev.}, C67:054903, 2003.

\bibitem{Faddeev:1967fc}
L.~D. Faddeev and V.~N. Popov.
\newblock {Feynman Diagrams for the Yang-Mills Field}.
\newblock {\em Phys. Lett.}, B25:29--30, 1967.

\bibitem{Chirilli:2015fza}
Giovanni~A. Chirilli, Yuri~V. Kovchegov, and Douglas~E. Wertepny.
\newblock {Regularization of the Light-Cone Gauge Gluon Propagator
  Singularities Using Sub-Gauge Conditions}.
\newblock {\em JHEP}, 12:138, 2015.

\bibitem{Sterman:1994ce}
G.~Sterman.
\newblock {\em An Introduction to quantum field theory}.
\newblock Cambridge University Press, Cambridge, UK, 1993.

\bibitem{Peskin:1995ev}
Michael~E. Peskin and D.~V. Schroeder.
\newblock {\em An Introduction to quantum field theory}.
\newblock Addison-Wesley, Reading, USA, 1995.

\bibitem{Kovchegov:1996ty}
Yuri~V. Kovchegov.
\newblock Non-abelian {Weizs\"{a}cker-Williams} field and a two- dimensional
  effective color charge density for a very large nucleus.
\newblock {\em Phys. Rev.}, D54:5463--5469, 1996.

\bibitem{Mueller:1989st}
Alfred~H. Mueller.
\newblock {Small x Behavior and Parton Saturation: A QCD Model}.
\newblock {\em Nucl. Phys.}, B335:115, 1990.

\bibitem{Kovchegov:1997pc}
Yuri~V. Kovchegov.
\newblock {Quantum structure of the non-Abelian Weizs\"{a}cker-Williams field
  for a very large nucleus}.
\newblock {\em Phys. Rev.}, D55:5445--5455, 1997.

\bibitem{Eggert:1974ek}
K.~Eggert, H.~Frenzel, W.~Thome, B.~Betev, P.~Darriulat, et~al.
\newblock {Angular Correlations Between the Charged Particles Produced in $p p$
  Collisions at ISR Energies}.
\newblock {\em Nucl.Phys.}, B86:201, 1975.

\bibitem{Teaney:2002kn}
Derek Teaney and Raju Venugopalan.
\newblock {Classical computation of elliptic flow at large transverse
  momentum}.
\newblock {\em Phys.Lett.}, B539:53--58, 2002.

\bibitem{Lepage:1980fj}
G.~Peter Lepage and Stanley~J. Brodsky.
\newblock Exclusive processes in perturbative quantum chromodynamics.
\newblock {\em Phys. Rev.}, D22:2157, 1980.

\bibitem{Kovchegov:1998bi}
Yuri~V. Kovchegov and Alfred~H. Mueller.
\newblock Gluon production in current nucleus and nucleon nucleus collisions in
  a quasi-classical approximation.
\newblock {\em Nucl. Phys.}, B529:451--479, 1998.

\bibitem{Kovchegov:2001sc}
Yuri~V. Kovchegov and Kirill Tuchin.
\newblock Inclusive gluon production in dis at high parton density.
\newblock {\em Phys. Rev.}, D65:074026, 2002.

\bibitem{Blaizot:2004wu}
Jean~Paul Blaizot, Francois Gelis, and Raju Venugopalan.
\newblock {High energy p A collisions in the color glass condensate approach.
  I: Gluon production and the Cronin effect}.
\newblock {\em Nucl. Phys.}, A743:13--56, 2004.

\bibitem{Braun:2000bh}
M.~A. Braun.
\newblock {Inclusive jet production on the nucleus in the perturbative QCD with
  $N_c \rightarrow \infty$}.
\newblock {\em Phys. Lett.}, B483:105--114, 2000.

\bibitem{Kharzeev:2003wz}
Dmitri Kharzeev, Yuri~V. Kovchegov, and Kirill Tuchin.
\newblock Cronin effect and high-p(t) suppression in p a collisions.
\newblock {\em Phys. Rev.}, D68:094013, 2003.

\bibitem{JalilianMarian:2004da}
Jamal Jalilian-Marian and Yuri~V. Kovchegov.
\newblock {Inclusive two-gluon and valence quark-gluon production in DIS and p
  A}.
\newblock {\em Phys. Rev.}, D70:114017, 2004.

\bibitem{Chen:1995pa}
Zhang Chen and Alfred~H. Mueller.
\newblock {The dipole picture of high-energy scattering, the BFKL equation and
  many gluon compound states}.
\newblock {\em Nucl. Phys.}, B451:579--604, 1995.

\bibitem{Dominguez:2011gc}
Fabio Dominguez, A.H. Mueller, Stephane Munier, and Bo-Wen Xiao.
\newblock {On the small-x evolution of the color quadrupole and the
  Weizs\'acker-Williams gluon distribution}.
\newblock {\em Phys.Lett.}, B705:106--111, 2011.

\bibitem{Dumitru:2011vk}
Adrian Dumitru, Jamal Jalilian-Marian, Tuomas Lappi, Bjoern Schenke, and Raju
  Venugopalan.
\newblock {Renormalization group evolution of multi-gluon correlators in high
  energy QCD}.
\newblock {\em Phys.Lett.}, B706:219--224, 2011.

\bibitem{Iancu:2011ns}
E.~Iancu and D.N. Triantafyllopoulos.
\newblock {Higher-point correlations from the JIMWLK evolution}.
\newblock {\em JHEP}, 1111:105, 2011.

\bibitem{Grumiller:2008va}
Daniel Grumiller and Paul Romatschke.
\newblock {On the collision of two shock waves in AdS(5)}.
\newblock {\em JHEP}, 0808:027, 2008.

\bibitem{Albacete:2008vs}
Javier~L. Albacete, Yuri~V. Kovchegov, and Anastasios Taliotis.
\newblock {Modeling Heavy Ion Collisions in AdS/CFT}.
\newblock {\em JHEP}, 0807:100, 2008.

\bibitem{Albacete:2009ji}
Javier~L. Albacete, Yuri~V. Kovchegov, and Anastasios Taliotis.
\newblock {Asymmetric Collision of Two Shock Waves in AdS$_5$}.
\newblock {\em JHEP}, 0905:060, 2009.

\bibitem{Chesler:2010bi}
Paul~M. Chesler and Laurence~G. Yaffe.
\newblock {Holography and colliding gravitational shock waves in asymptotically
  AdS$_5$ spacetime}.
\newblock {\em Phys.Rev.Lett.}, 106:021601, 2011.

\bibitem{Fadin:1975cb}
Victor~S. Fadin, E.A. Kuraev, and L.N. Lipatov.
\newblock {On the Pomeranchuk Singularity in Asymptotically Free Theories}.
\newblock {\em Phys.Lett.}, B60:50--52, 1975.

\bibitem{Lappi:2009xa}
T.~Lappi, S.~Srednyak, and R.~Venugopalan.
\newblock {Non-perturbative computation of double inclusive gluon production in
  the Glasma}.
\newblock {\em JHEP}, 1001:066, 2010.

\bibitem{Borghini:2000sa}
Nicolas Borghini, Phuong~Mai Dinh, and Jean-Yves Ollitrault.
\newblock {A New method for measuring azimuthal distributions in
  nucleus-nucleus collisions}.
\newblock {\em Phys.Rev.}, C63:054906, 2001.

\bibitem{Borghini:2001vi}
Nicolas Borghini, Phuong~Mai Dinh, and Jean-Yves Ollitrault.
\newblock {Flow analysis from multiparticle azimuthal correlations}.
\newblock {\em Phys.Rev.}, C64:054901, 2001.

\bibitem{HanburyBrown:1956pf}
R.~Hanbury~Brown and R.Q. Twiss.
\newblock {A Test of a new type of stellar interferometer on Sirius}.
\newblock {\em Nature}, 178:1046--1048, 1956.

\bibitem{Heinz:1999rw}
Ulrich~W. Heinz and Barbara~V. Jacak.
\newblock {Two particle correlations in relativistic heavy ion collisions}.
\newblock {\em Ann.Rev.Nucl.Part.Sci.}, 49:529--579, 1999.

\bibitem{Adams:2003ra}
J.~Adams et~al.
\newblock {Azimuthally sensitive HBT in Au + Au collisions at s(NN)**(1/2) =
  200-GeV}.
\newblock {\em Phys.Rev.Lett.}, 93:012301, 2004.

\bibitem{Kopylov:1972qw}
G.I. Kopylov and M.I. Podgoretsky.
\newblock {Correlations of identical particles emitted by highly excited
  nuclei}.
\newblock {\em Sov.J.Nucl.Phys.}, 15:219--223, 1972.

\bibitem{Capella:1991mp}
A.~Capella, A.~Krzywicki, and E.M. Levin.
\newblock {Pion interferometry and intermittency in heavy ion collisions}.
\newblock {\em Phys.Rev.}, D44:704--716, 1991.

\bibitem{Asakawa:1998cx}
M.~Asakawa, T.~Csorgo, and M.~Gyulassy.
\newblock {Squeezed correlations and spectra for mass shifted bosons}.
\newblock {\em Phys.Rev.Lett.}, 83:4013--4016, 1999.

\bibitem{Csorgo:2007iv}
Tamas Csorgo and Sandra~S. Padula.
\newblock {Disappearance of Squeezed Back-to-Back Correlations: A New signal of
  hadron freeze-out from a supercooled Quark Gluon Plasma}.
\newblock {\em Braz.J.Phys.}, 37:949--962, 2007.

\bibitem{Frankfurt:2003td}
L.~Frankfurt, M.~Strikman, and C.~Weiss.
\newblock {Dijet production as a centrality trigger for $p p$ collisions at
  CERN LHC}.
\newblock {\em Phys.Rev.}, D69:114010, 2004.

\bibitem{Frankfurt:2010ea}
L.~Frankfurt, M.~Strikman, and C.~Weiss.
\newblock {Transverse nucleon structure and diagnostics of hard parton-parton
  processes at LHC}.
\newblock {\em Phys.Rev.}, D83:054012, 2011.

\bibitem{Alver:2010gr}
B.~Alver and G.~Roland.
\newblock {Collision geometry fluctuations and triangular flow in heavy-ion
  collisions}.
\newblock {\em Phys.Rev.}, C81:054905, 2010.

\bibitem{Heinz:2004ir}
Ulrich~W. Heinz and Anthony Kuhlman.
\newblock {Anisotropic flow and jet quenching in ultrarelativistic U + U
  collisions}.
\newblock {\em Phys.Rev.Lett.}, 94:132301, 2005.

\bibitem{Kuhlman:2005ts}
Anthony~J. Kuhlman and Ulrich~W. Heinz.
\newblock {Multiplicity distribution and source deformation in full-overlap U+U
  collisions}.
\newblock {\em Phys.Rev.}, C72:037901, 2005.

\bibitem{Kuhlman:2006qp}
Anthony Kuhlman, Ulrich~W. Heinz, and Yuri~V. Kovchegov.
\newblock {Gluon saturation effects in relativistic U + U collisions}.
\newblock {\em Phys. Lett.}, B638:171--177, 2006.

\bibitem{Wigner:1932eb}
Eugene~P. Wigner.
\newblock {On the quantum correction for thermodynamic equilibrium}.
\newblock {\em Phys.Rev.}, 40:749--760, 1932.

\bibitem{Belitsky:2002sm}
Andrei~V. Belitsky, X.~Ji, and F.~Yuan.
\newblock {Final state interactions and gauge invariant parton distributions}.
\newblock {\em Nucl.Phys.}, B656:165--198, 2003.

\bibitem{Jalilian-Marian:1997xn}
Jamal Jalilian-Marian, Alex Kovner, Larry~D. McLerran, and Heribert Weigert.
\newblock The intrinsic glue distribution at very small x.
\newblock {\em Phys. Rev.}, D55:5414--5428, 1997.

\bibitem{Dominguez:2011wm}
Fabio Dominguez, Cyrille Marquet, Bo-Wen Xiao, and Feng Yuan.
\newblock {Universality of Unintegrated Gluon Distributions at small x}.
\newblock {\em Phys.Rev.}, D83:105005, 2011.

\bibitem{Cvitanovic:2008zz}
Predrag Cvitanovic.
\newblock {\em {Group theory: Birdtracks, Lie's and exceptional groups}}.
\newblock Princeton University Press, 2008. Available online at {\sl
  http://birdtracks.eu/}.

\bibitem{Kovner:2001bh}
Alexander Kovner and Urs~Achim Wiedemann.
\newblock Nonlinear qcd evolution: Saturation without unitarization.
\newblock {\em Phys. Rev.}, D66:051502, 2002.

\bibitem{Levin:2014bwa}
Eugene Levin, Lev Lipatov, and Marat Siddikov.
\newblock {BFKL Pomeron with massive gluons}.
\newblock 2014.

\bibitem{Kharzeev:2001gp}
Dmitri Kharzeev and Eugene Levin.
\newblock {Manifestations of high density QCD in the first RHIC data}.
\newblock {\em Phys. Lett.}, B523:79--87, 2001.

\bibitem{Kharzeev:2001yq}
Dmitri Kharzeev, Eugene Levin, and Marzia Nardi.
\newblock {The onset of classical QCD dynamics in relativistic heavy ion
  collisions}.
\newblock {\em Phys. Rev.}, C71:054903, 2005.

\bibitem{Albacete:2007sm}
Javier~L. Albacete.
\newblock {Particle multiplicities in Lead-Lead collisions at the LHC from
  non-linear evolution with running coupling}.
\newblock {\em Phys. Rev. Lett.}, 99:262301, 2007.

\bibitem{Chirilli:2015tea}
Giovanni~A. Chirilli, Yuri~V. Kovchegov, and Douglas~E. Wertepny.
\newblock {Classical Gluon Production Amplitude for Nucleus-Nucleus Collisions:
  First Saturation Correction in the Projectile}.
\newblock {\em JHEP}, 03:015, 2015.

\bibitem{Kovchegov:2001ni}
Yuri~V. Kovchegov.
\newblock {Diffractive gluon production in proton nucleus collisions and in
  DIS}.
\newblock {\em Phys. Rev.}, D64:114016, 2001.

\bibitem{Marquet:2010cf}
Cyrille Marquet and Heribert Weigert.
\newblock {New observables to test the Color Glass Condensate beyond the
  large-$N_c$ limit}.
\newblock {\em Nucl. Phys.}, A843:68--97, 2010.

\bibitem{Nikolaev:2005zj}
N.N. Nikolaev, W.~Schafer, and B.G. Zakharov.
\newblock {Nonlinear k(perpendicular)-factorization for gluon-gluon dijets
  produced off nuclear targets}.
\newblock {\em Phys.Rev.}, D72:114018, 2005.

\end{thebibliography}
